\documentclass[11pt, a4paper]{article}
\usepackage{jheppub, slashed, color}
\usepackage{multirow}
\usepackage{tikz}
\usepackage{amsmath}
\allowdisplaybreaks[2]
\usepackage{color}
\usepackage[multiple]{footmisc}%for multiple footnotes
\usepackage{textcomp}
\usepackage{amssymb}%for semidirect product symbol
\usetikzlibrary{positioning}
\usepackage{amscd}
\usepackage{amsthm}
\usepackage{array} 

\DeclareFontFamily{U}{MnSymbolC}{}
\DeclareSymbolFont{MnSyC}{U}{MnSymbolC}{m}{n}
\DeclareFontShape{U}{MnSymbolC}{m}{n}{
    <-6>  MnSymbolC5
   <6-7>  MnSymbolC6
   <7-8>  MnSymbolC7
   <8-9>  MnSymbolC8
   <9-10> MnSymbolC9
  <10-12> MnSymbolC10
  <12->   MnSymbolC12}{}
\DeclareMathSymbol{\intprod}{\mathbin}{MnSyC}{'270}
\newcommand{\ov}{\overline}
\newcommand{\C}{\mathbb{C}}
\newcommand{\CP}{\C P}
\newcommand{\CPN}{\CP^{N-1}}
\newcommand{\Z}{\mathbb{Z}}
\newcommand{\R}{\mathbb{R}}

\newcommand{\K}{\mathcal{K}}

\newcommand{\OO}{\mathcal O}
\newcommand{\he}{\til{e}}
\newcommand{\hae}{\hat{e}}

\newcommand{\til}{\widetilde}
\newcommand{\e}{e}
\newcommand{\bQ}{\ov{Q}}

\newcommand{\del}{\partial}
\newcommand{\db}{\ov{\partial}}

\newcommand{\dd}{{ d}}

\newcommand{\g}{\mathfrak g}

\newcommand{\ver}{\ker \dd \pi_E}
\newcommand{\gp}{{\mathfrak g}_P}
\newcommand{\mx}{\mathcal{X}}
\newcommand{\bmx}{\ov{\mathcal{X}}}

\newcommand{\psip}{\psi_{+}}
\newcommand{\psim}{\psi_{-}}

\newcommand{\lam}{\lambda}

\newcommand{\lamp}{\lambda_{+}}
\newcommand{\lamm}{\lambda_{-}}
\newcommand{\lambp}{\ov{\lambda}_+}
\newcommand{\lambm}{\ov{\lambda}_-}
\newcommand{\la}{\xi}

\newcommand{\lampua}{\lambda_{+a}}
\newcommand{\lammua}{\lambda_{-a}}
\newcommand{\lambpua}{\ov{\lambda}_{+a}}
\newcommand{\lambmua}{\ov{\lambda}_{-a}}

\newcommand{\zb}{\bar{z}}
\newcommand{\z}{z}

\newcommand{\epsilonbp}{\bar{\epsilon}_+}
\newcommand{\epsilonbm}{\bar{\epsilon}_-}
\newcommand{\epsilonp}{\epsilon_+}
\newcommand{\epsilonm}{\epsilon_-}

\newcommand{\si}{\sigma}
\newcommand{\sib}{\ov{\sigma}}
\newcommand{\Si}{\Sigma}

\newcommand{\met}{g_{i\overline{\jmath}}}
\newcommand{\jbar}{\overline{\jmath}}

\newcommand{\btheta}{\overline{\theta}}
\newcommand{\cQ}{{\cal Q}}
\newcommand{\bepsilon}{\overline{\epsilon}}
\newcommand{\bPhi}{\overline{\Phi}}
\newcommand{\bphi}{\overline{\phi}}

\newcommand{\bD}{\overline{D}}
\newcommand{\bcQ}{\overline{\cal Q}}

\newcommand{\ts}{{\theta}}
\newcommand{\tib}{\til{b}}
\newcommand{\tic}{\til{c}}
\newcommand{\eff}{{\textit{(eff)}}}
\let\nc\newcommand
\let\renc\renewcommand
\nc{\wbar}{\overline}
\let\td\tilde
\let\wtd\widetilde
\let\wht\widehat
\let\mcl\mathcal

\nc{\ab}{{\bar{a}}} \nc{\at}{\tilde{a}} \nc{\ah}{\hat{a}}
\nc{\bb}{{\bar{b}}} \nc{\bt}{\tilde{b}} \nc{\bh}{\hat{b}}
\nc{\cb}{{\bar{c}}} \nc{\ct}{\tilde{c}} %\nc{\ch}{\hat{c}}
\nc{\fb}{{\bar{f}}} \nc{\ft}{\tilde{f}} \nc{\fh}{\hat{f}}
\nc{\gb}{{\bar{g}}} \nc{\gt}{\tilde{g}} \nc{\gh}{\hat{g}}
\nc{\hb}{{\bar{h}}} \nc{\hh}{\hat{h}} %\nc{\ht}{\tilde{h}}
\nc{\ib}{{\bar{\imath}}} \nc{\ih}{\hat{\imath}} %\nc{\it}{\tilde{\imath}}
\nc{\jb}{{\bar{\jmath}}} \nc{\jt}{\tilde{\jmath}} \nc{\jh}{\hat{\jmath}}
\nc{\kb}{{\bar{k}}} \nc{\kt}{\tilde{k}} \nc{\kh}{\hat{k}}
\nc{\lb}{{\bar{l}}} \nc{\lt}{\tilde{l}} \nc{\lh}{\hat{l}}
\nc{\mb}{{\bar{m}}} \nc{\mt}{\tilde{m}} \nc{\mh}{\hat{m}}
\nc{\nb}{{\bar{n}}} \nc{\nt}{\tilde{n}} \nc{\nh}{\hat{n}}
\nc{\ob}{{\bar{o}}} \nc{\ot}{\tilde{o}} \nc{\oh}{\hat{o}}
\nc{\pb}{{\bar{p}}} \nc{\pt}{\tilde{p}} \nc{\ph}{\hat{p}}
\nc{\qb}{{\bar{q}}} \nc{\qt}{\tilde{q}} \nc{\qh}{\hat{q}}
\nc{\rb}{{\bar{r}}} \nc{\rt}{\tilde{r}} \nc{\rh}{\hat{r}}
\renc{\sb}{{\bar{s}}} \nc{\st}{\tilde{s}} \nc{\sh}{\hat{s}}
\nc{\tb}{{\bar{t}}} \renc{\th}{\hat{t}} %\nc{\tt}{\tilde{t}}
\nc{\ub}{{\bar{u}}} \nc{\ut}{\tilde{u}} \nc{\uh}{\hat{u}}
\nc{\vb}{{\bar{v}}} \nc{\vt}{\tilde{v}} \nc{\vh}{\hat{v}}
\nc{\wb}{{\bar{w}}} \nc{\wt}{\tilde{w}} \nc{\wh}{\hat{w}}
\nc{\xb}{{\bar{x}}} \nc{\xt}{\tilde{x}} \nc{\xh}{\hat{x}}
\nc{\yb}{{\bar{y}}} \nc{\yt}{\tilde{y}} \nc{\yh}{\hat{y}}
\nc{\Ab}{\wbar{A}} \nc{\At}{\wtd{A}} \nc{\Ah}{\wht{A}}
\nc{\Bb}{\wbar{B}} \nc{\Bt}{\wtd{B}} \nc{\Bh}{\wht{B}}
\nc{\Cb}{\wbar{C}} \nc{\Ct}{\wtd{C}} \nc{\Ch}{\wht{C}}
\nc{\Db}{\wbar{D}} \nc{\Dt}{\wtd{D}} \nc{\Dh}{\wht{D}}
\nc{\Eb}{\wbar{E}} \nc{\Et}{\wtd{E}} \nc{\Eh}{\wht{E}}
\nc{\Fb}{\wbar{F}} \nc{\Ft}{\wtd{F}} \nc{\Fh}{\wht{F}}
\nc{\Gb}{\wbar{G}} \nc{\Gt}{\wtd{G}} \nc{\Gh}{\wht{G}}
\nc{\Hb}{\wbar{H}} \nc{\Ht}{\wtd{H}} \nc{\Hh}{\wht{H}}
\nc{\Ib}{\wbar{I}} \nc{\It}{\wtd{I}} \nc{\Ih}{\wht{I}}
\nc{\Jb}{\wbar{J}} \nc{\Jt}{\wtd{J}} \nc{\Jh}{\wht{J}}
\nc{\Kb}{\wbar{K}} \nc{\Kt}{\wtd{K}} \nc{\Kh}{\wht{K}}
\nc{\Lb}{\wbar{L}} \nc{\Lt}{\wtd{L}} \nc{\Lh}{\wht{L}}
\nc{\Mb}{\wbar{M}} \nc{\Mt}{\wtd{M}} \nc{\Mh}{\wht{M}}
\nc{\Nb}{\wbar{N}} \nc{\Nt}{\wtd{N}} \nc{\Nh}{\wht{N}}
\nc{\Ob}{\wbar{O}} \nc{\Ot}{\wtd{O}} \nc{\Oh}{\wht{O}}
\nc{\Pb}{\wbar{P}} \nc{\Pt}{\wtd{P}} \nc{\Ph}{\wht{P}}
\nc{\Qb}{\wbar{Q}} \nc{\Qt}{\wtd{Q}} \nc{\Qh}{\wht{Q}}
\nc{\Rb}{\wbar{R}} \nc{\Rt}{\wtd{R}} \nc{\Rh}{\wht{R}}
\nc{\Sb}{\wbar{S}} \nc{\St}{\wtd{S}} \nc{\Sh}{\wht{S}}
\nc{\Tb}{\wbar{T}} \nc{\Tt}{\wtd{T}} \nc{\Th}{\wht{T}}
\nc{\Ub}{\wbar{U}} \nc{\Ut}{\wtd{U}} \nc{\Uh}{\wht{U}}
\nc{\Vb}{\wbar{V}} \nc{\Vt}{\wtd{V}} \nc{\Vh}{\wht{V}}
\nc{\Wb}{\wbar{W}} \nc{\Wt}{\wtd{W}} \nc{\Wh}{\wht{W}}
\nc{\Xb}{\wbar{X}} \nc{\Xt}{\wtd{X}} \nc{\Xh}{\wht{X}}
\nc{\Yb}{\wbar{Y}} \nc{\Yt}{\wtd{Y}} \nc{\Yh}{\wht{Y}}
\nc{\Zb}{\wbar{Z}} \nc{\Zt}{\wtd{Z}} \nc{\Zh}{\wht{Z}}

\nc{\CA}{\mcl{A}} \nc{\CAb}{\wbar{\CA}} \nc{\CAt}{\wtd{\CA}} \nc{\CAh}{\wht{\CA}}
\nc{\CB}{\mcl{B}} \nc{\CBb}{\wbar{\CB}} \nc{\CBt}{\wtd{\CB}} \nc{\CBh}{\wht{\CB}}
\nc{\cD}{\mcl{D}} \nc{\cDb}{\wbar{\cD}} \nc{\cDt}{\wtd{\cC}} \nc{\cDh}{\wht{\cD}}
\nc{\CE}{\mcl{E}} \nc{\CEb}{\wbar{\CE}} \nc{\CEt}{\wtd{\CE}} \nc{\CEh}{\wht{\CE}}
\nc{\CF}{\mcl{F}} \nc{\CFb}{\wbar{\CF}} \nc{\CFt}{\wtd{\CF}} \nc{\CFh}{\wht{\CF}}
\nc{\CG}{\mcl{G}} \nc{\CGb}{\wbar{\CG}} \nc{\CGt}{\wtd{\CG}} \nc{\CGh}{\wht{\CG}}
\nc{\CH}{\mcl{H}} \nc{\CHb}{\wbar{\CH}} \nc{\CHt}{\wtd{\CH}} \nc{\CHh}{\wht{\CH}}
\nc{\CI}{\mcl{I}} \nc{\CIb}{\wbar{\CI}} \nc{\CIt}{\wtd{\CI}} \nc{\CIh}{\wht{\CI}}
\nc{\CJ}{\mcl{J}} \nc{\CJb}{\wbar{\CJ}} \nc{\CJt}{\wtd{\CJ}} \nc{\CJh}{\wht{\CJ}}
\nc{\CK}{\mcl{K}} \nc{\CKb}{\wbar{\CK}} \nc{\CKt}{\wtd{\CK}} \nc{\CKh}{\wht{\CK}}
\nc{\CL}{\mcl{L}} \nc{\CLb}{\wbar{\CL}} \nc{\CLt}{\wtd{\CL}} \nc{\CLh}{\wht{\CL}}
\nc{\CM}{\mcl{M}} \nc{\CMb}{\wbar{\CM}} \nc{\CMt}{\wtd{\CM}} \nc{\CMh}{\wht{\CM}}
\nc{\CN}{\mcl{N}} \nc{\CNb}{\wbar{\CN}} \nc{\CNt}{\wtd{\CN}} \nc{\CNh}{\wht{\CN}}
\nc{\CO}{\mcl{O}} \nc{\COb}{\wbar{\CO}} \nc{\COt}{\wtd{\CO}} \nc{\COh}{\wht{\CO}}
\nc{\CQ}{\mcl{Q}} \nc{\CQb}{\wbar{\CQ}} \nc{\CQt}{\wtd{\CQ}} \nc{\CQh}{\wht{\CQ}}
\nc{\CR}{\mcl{R}} \nc{\CRb}{\wbar{\CR}} \nc{\CRt}{\wtd{\CR}} \nc{\CRh}{\wht{\CR}}
\nc{\CS}{\mcl{S}} \nc{\CSb}{\wbar{\CS}} \nc{\CSt}{\wtd{\CS}} \nc{\CSh}{\wht{\CS}}
\nc{\CT}{\mcl{T}} \nc{\CTb}{\wbar{\CT}} \nc{\CTt}{\wtd{\CT}} \nc{\CTh}{\wht{\CT}}
\nc{\CU}{\mcl{U}} \nc{\CUb}{\wbar{\CU}} \nc{\CUt}{\wtd{\CU}} \nc{\CUh}{\wht{\CU}}
\nc{\CV}{\mcl{V}} \nc{\CVb}{\wbar{\CV}} \nc{\CVt}{\wtd{\CV}} \nc{\CVh}{\wht{\CV}}
\nc{\CW}{\mcl{W}} \nc{\CWb}{\wbar{\CW}} \nc{\CWt}{\wtd{\CW}} \nc{\CWh}{\wht{\CW}}
\nc{\CX}{\mcl{X}} \nc{\CXb}{\wbar{\CX}} \nc{\CXt}{\wtd{\CX}} \nc{\CXh}{\wht{\CX}}
\nc{\CY}{\mcl{Y}} \nc{\CYb}{\wbar{\CY}} \nc{\CYt}{\wtd{\CY}} \nc{\CYh}{\wht{\CY}}
\nc{\CZ}{\mcl{Z}} \nc{\CZb}{\wbar{\CZ}} \nc{\CZt}{\wtd{\CZ}} \nc{\CZh}{\wht{\CZ}}

\let\eps\epsilon
\let\ups\upsilon
\let\Ups\Upsilon
\let\veps\varepsilon
\let\vtht\vartheta
\let\vsgm\varsigma
\let\vphi\varphi
\let\vrho\varrho

\nc{\alphab}{\bar{\alpha}} \nc{\alphat}{\td{\alpha}} \nc{\alphah}{\hat{\alpha}}
\nc{\betab}{\bar{\beta}}   \nc{\betat}{\td{\beta}}   \nc{\betah}{\hat{\beta}} 
\nc{\gammab}{\bar{\gamma}} \nc{\gammat}{\td{\gamma}} \nc{\gammah}{\hat{\gamma}} 
\nc{\deltab}{\bar{\delta}} \nc{\deltat}{\td{\delta}} \nc{\deltah}{\hat{\delta}} 
\nc{\epsilonb}{\bar{\eps}} \nc{\epsilont}{\td{\eps}} \nc{\epsilonh}{\hat{\eps}} 
\nc{\vepsb}{\bar{\veps}}   \nc{\vepst}{\td{\veps}}   \nc{\vepsh}{\hat{\veps}} 
\nc{\zetab}{\bar{\zeta}}   \nc{\zetat}{\td{\zeta}}   \nc{\zetah}{\hat{\zeta}} 
\nc{\etab}{\bar{\eta}}     \nc{\etat}{\td{\eta}}     \nc{\etah}{\hat{\eta}} 
\nc{\thetab}{\bar{\theta}} \nc{\thetat}{\td{\theta}} \nc{\thetah}{\hat{\theta}} 
\nc{\vthetab}{\bar{\vtht}} \nc{\vthetat}{\td{\vtht}} \nc{\vthetah}{\hat{\vtht}} 
\nc{\lambdab}{\bar{\lambda}} \nc{\lambdat}{\td{\lambda}} \nc{\lambdah}{\hat{\lambda}} 
\nc{\iotab}{\bar{\iota}}   \nc{\iotat}{\td{\iota}}   \nc{\iotah}{\hat{\iota}} 
\nc{\kpp}{\kappa}

\nc{\kappab}{\bar{\kappa}} \nc{\kappat}{\td{\kappa}} \nc{\kappah}{\hat{\kappa}} 
\nc{\lmdb}{\bar{\lmd}}     \nc{\lmdt}{\td{\lmd}}     \nc{\lmdh}{\hat{\lmd}} 
\nc{\mub}{\bar{\mu}}       \nc{\mut}{\td{\mu}}       \nc{\muh}{\hat{\mu}} 
\nc{\nub}{\bar{\nu}}       \nc{\nut}{\td{\nu}}       \nc{\nuh}{\hat{\nu}} 
\nc{\xib}{\bar{\xi}}       \nc{\xit}{\td{\xi}}       \nc{\xih}{\hat{\xi}} 
\nc{\pib}{\bar{\pi}}       \nc{\pit}{\td{\pi}}       \nc{\pih}{\hat{\pi}} 
\nc{\vpib}{\bar{\vpi}}     \nc{\vpit}{\td{\vpi}}     \nc{\vpih}{\hat{\vpi}} 
\nc{\rhob}{\bar{\rho}}     \nc{\rhot}{\td{\rho}}     \nc{\rhoh}{\hat{\rho}} 
\nc{\vrhob}{\bar{\vrho}}   \nc{\vrhot}{\td{\vrho}}   \nc{\vrhoh}{\hat{\vrho}} 
\nc{\sigmab}{\bar{\sigma}} \nc{\sigmat}{\td{\sigma}} \nc{\sigmah}{\hat{\sigma}} 
\nc{\vsigmab}{\bar{\vsgm}} \nc{\vsigmat}{\td{\vsgm}} \nc{\vsigmah}{\hat{\vsgm}} 
\nc{\taub}{\bar{\tau}}     \nc{\taut}{\td{\tau}}     \nc{\tauh}{\hat{\tau}} 
\nc{\upsb}{\bar{\ups}} \nc{\upst}{\td{\ups}} \nc{\upsh}{\hat{\ups}} 
\nc{\phib}{\bar{\phi}}     \nc{\phit}{\td{\phi}}     \nc{\phih}{\hat{\phi}} 
\nc{\varphib}{\bar{\vphi}}   \nc{\varphit}{\td{\vphi}}   \nc{\varphih}{\hat{\vphi}} 
\nc{\chib}{\bar{\chi}}     \nc{\chit}{\td{\chi}}     \nc{\chih}{\hat{\chi}} 
\nc{\omegab}{\bar{\omega}} \nc{\omegat}{\td{\omega}} \nc{\omegah}{\hat{\omega}} 

\nc{\Gammab}{\wbar{\Gamma}}     \nc{\Gammat}{\wtd{\Gamma}}     \nc{\Gammah}{\wht{\Gamma}}
\nc{\Deltab}{\wbar{\Delta}}     \nc{\Deltat}{\wtd{\Delta}}     \nc{\Deltah}{\wht{\Delta}}
\nc{\Thetab}{\wbar{\Theta}}     \nc{\Thetat}{\wtd{\Theta}}     \nc{\Thetah}{\wht{\Theta}}
\nc{\Lambdab}{\wbar{\Lambda}}   \nc{\Lambdat}{\wtd{\Lambda}}   \nc{\Lambdah}{\wht{\Lambda}}
\nc{\Xib}{\wbar{\Xi}}           \nc{\Xit}{\wtd{\Xi}}           \nc{\Xih}{\wht{\Xi}}
\nc{\Pib}{\wbar{\Pi}}           \nc{\Pit}{\wtd{\Pi}}           \nc{\Pih}{\wht{\Pi}}
\nc{\Sigmab}{\wbar{\Sigma}}     \nc{\Sigmat}{\wtd{\Sigma}}     \nc{\Sigmah}{\wht{\Sigma}}
\nc{\Upsilonb}{\wbar{\Upsilon}} \nc{\Upsilont}{\wtd{\Upsilon}} \nc{\Upsilonh}{\wht{\Upsilon}}
\nc{\Phib}{\wbar{\Phi}}         \nc{\Phit}{\wtd{\Phi}}         \nc{\Phih}{\wht{\Phi}}
\nc{\Psib}{\wbar{\Psi}}         \nc{\Psit}{\wtd{\Psi}}         \nc{\Psih}{\wht{\Psi}}
\nc{\Omegab}{\wbar{\Omega}}     \nc{\Omegat}{\wtd{\Omega}}     \nc{\Omegah}{\wht{\Omega}}

\nc{\txd}{d}

\newcommand{\ds}{\del \Sigma}

\title{Open Gauged Sigma Models, Equivariant Branes, and Equivariant Homological Mirror Symmetry}
\author[]{Meer Ashwinkumar}
\author[]{and Meng-Chwan Tan}
\emailAdd{meerashwinkumar@u.nus.edu, mctan@nus.edu.sg}
\affiliation[]{Department of Physics, National University of
  Singapore \\  2 Science Drive 3, Singapore 117551}
\abstract{We describe supersymmetric A-branes and B-branes in open $\mathcal{N}=(2,2)$ dynamically gauged nonlinear sigma models (GNLSM)% on a worldsheet with boundaries
, placing emphasis on toric manifold target spaces. %where the gauge group is abelian. 
For 
a subset of %these 
toric manifolds, %we find their 
these equivariant branes have a mirror description 
as branes in gauged Landau-Ginzburg models with neutral matter. We then study correlation functions in the topological A-twisted version of the %nonabelian 
GNLSM, and identify their values with open Hamiltonian Gromov-Witten invariants. Supersymmetry breaking can occur in the A-twisted GNLSM due to nonperturbative open symplectic vortices, and we canonically BRST quantize the mirror theory to analyze this phenomenon. 
} 
\keywords{}

\arxivnumber{}

\renewcommand{\ell}{l}

\begin{document}
\maketitle
\flushbottom

\section{Introduction}
D-branes are crucial to the nonperturbative dynamics of string theory, and their importance has been well-understood since Polchinski \cite{Polchinski} recognized them as the source of BPS states carrying RR charge, leading to their identification	 with black p-brane solutions of supergravity. From the point of view of mathematics, D-branes are 
essential objects of homological mirror symmetry, first conjectured by Kontsevich \cite{Kontsevich}.
 
%IMPORTANT NOTE: SEE LUBOS' ANSWER AT http://physics.stackexchange.com/questions/4988/what-is-a-d-brane
%IMPORTANT NOTE : D-branes are nonpertubative because their tension is inversely proportional to the string coupling, so in the pertubative limit the brane becomes infinitely massive, and its dynamics can be ignored (as long as we consider finite energy processes). see http://physics.stackexchange.com/questions/241875/non-perturbative-brane-effects
In this paper, we investigate D-branes of dynamically gauged nonlinear sigma models (GNLSMs) with $\mathcal{N}=(2,2)$ supersymmetry, which we shall refer to as equivariant branes. One motivation for this is that GNLSMs with target space $X$ and gauge group $G$ flow in the IR limit to nonlinear sigma models (NLSMs) with target space $X//G$, and hence, we will obtain new descriptions of D-branes in $\mathcal{N}=(2,2)$ NLSMs, including those with Calabi-Yau targets 
%used in
useful for %realistic 
physical compactifications of string theory.
%IMPORTANT NOTE: FOR GLSMS WE USUALLY OBTAIN NONCOMPACT CALABI YAU IN THE IR, TO GET COMPACT CALABI-YAU WE MUST INCLUDE A SUPERPOTENTIAL TO GET A COMPLETE INTERSECTION OF A TORIC MANIFOLD, SEE PAGE 29 OF https://arxiv.org/pdf/hep-th/0108234.pdf AND THE ABSTRACT OF https://arxiv.org/abs/hep-th/0410018. However for a GNLSM the only requirement to get a Calabi-Yau nlsm in the IR is that the UV GNLSM has an equivariant Calabi-Yau target. As Baptista explains on page 14 and 15 of twisting gauged nonlinear sigma models, ANY hyperkaehler manifold with a tri-Hamiltonian G-action is an equivariant Calabi-Yau. So as long as such a hyperkaehler space is compact, then the quotient by G should also be compact, and Calabi-Yau. Such compact hyperkaehler spaces exist, see beginning of Section 2 of https://arxiv.org/pdf/1607.04078.pdf.
%IMPORTANT NOTE: (18TH NOVEMBER 2017) - ALSO SEE https://arxiv.org/pdf/hep-th/9702155.pdf PAGE 12 AND https://arxiv.org/pdf/0806.2612.pdf (ASPINWALL-D-BRANES ON TORIC CALABI-YAU VARIETIES) PAGE 1, SECOND PARAGRAPH- WHEN WE TALK ABOUT COMPACTIFICATION, THE 6D CALABI-YAU MANIFOLD MUST BE COMPACT. BUT "A Calabi–Yau variety cannot be both compact and toric"
%IMPORTANT NOTE: (continuation which i forgot to write earlier ) THUS, WE CAN OBTAIN EQUIVARIANT BRANES WHICH CORRESPOND TO BRANES IN CALABI-YAU MANIFOLDS WHICH ARE COMPACT. BUT THIS IS USING THE DIRECT METHOD. USING GLSMs WE CAN ONLY STUDY NONCOMPACT CALABI-YAU MANIFOLDS. 
 A more mathematically oriented motivation is %proving 
furnishing an \textit{equivariant} generalization of homological mirror symmetry. As we shall see, describing equivariant branes will also allow us to %make contact
define %with the mathematical theory of 
an \textit{open} version of the mathematical theory of Hamiltonian Gromov-Witten invariants \cite{Cieliebak, Mundet, Cieliebak2}.

The $\mathcal{N}=(2,2)$ dynamically gauged nonlinear sigma model (GNLSM) governing maps from a closed Riemann surface into a K\"ahler manifold $X$ with Hamiltonian isometry group $G$ was studied in depth by Baptista \cite{Baptista1,Baptista2}. In particular, it was shown that the A-twisted GNLSM localizes to the moduli space of symplectic vortices, and its correlation functions compute the Hamiltonian Gromov-Witten invariants of $X$. Moreover, for abelian $G$, Baptista used mirror symmetry (as proven by Hori and Vafa \cite{HoriVafa}) to describe the quantum equivariant cohomology ring for toric $X$. 

%On the other hand, %IMPORTANT NOTE: On the other hand means an opposing idea, so should not be used here
Also, D-branes in $\mathcal{N}=(2,2)$ NLSMs on open Riemann surfaces have been studied by Hori, Iqbal and Vafa \cite{HIV,Hori1}, with the mirrors of these D-branes being identified. We are thus led to attempt an understanding of equivariant branes by combining the insights described above, that is, by analyzing  $\mathcal{N}=(2,2)$ GNLSMs on open Riemann surfaces and their mirrors. Since only a subset of the $\mathcal{N}=(2,2)$ supersymmetry can be preserved at the boundaries of these open Riemann surfaces, we are led to two types of equivariant branes, namely equivariant A-branes and equivariant B-branes.

Equivariant B-branes have been previously studied by
Kapustin et al. \cite{Kapustin} within the context of topologically B-twisted GNLSMs, although the mirrors of these branes were not elucidated. %have not been studied before. %However, equivariant A-branes have not been explored prior to the present work. 
On the other hand, equivariant A-branes have only been studied for $G=U(1)$ by Setter \cite{Setter}, using a specialized topologically A-twisted non-dynamical $U(1)$-GNLSM (where the gauge field is identified with the worldsheet spin connection); in this case, their mirrors were not elucidated either. We shall study both types of equivariant branes, and provide the description of their mirrors in a subset of toric target spaces, hence defining equivariant homological mirror symmetry in these contexts. Other proposals for equivariant homological mirror symmetry of equivariant B-branes have appeared in the mathematical literature \cite{Zaslow,Futaki}. 

In addition, understanding equivariant A-branes allows us to define open Hamiltonian Gromov-Witten invariants, which can be understood as integrals over the moduli spaces of open symplectic vortices that describe a map from %the strip or disk 
an open Riemann surface $\Sigma$ to a K\"ahler and Hamiltonian $G$-manifold $X$, whereby the boundaries of %the strip
$\Sigma$ correspond to equivariant A-branes in $X$. We note that closed Hamiltonian Gromov-Witten invariants have been studied extensively in the mathematical literature \cite{Cieliebak,Mundet,Cieliebak2}. However, the open invariants have been largely unexplored, with the exception of the work of Xu \cite{Xu}, which concerns the compactification of the moduli space of open symplectic vortices on the disk for $G=U(1)$, as well as the work of Wang and Xu \cite{Wang} on the relationship between open symplectic vortices for $X$ and open worldsheet instantons for $X//G$ (the open quantum Kirwan map).
 \vspace{0.2cm}
\mbox{}\par\nobreak
\noindent
\textit{An outline of the paper} 
 %An outline of the paper is as follows. 
 
 In Section 2, we introduce the $\mathcal{N}=(2,2)$ supersymmetric dynamically gauged nonlinear sigma model on the infinite strip, focusing on its gauge symmetry and supersymmetry. In Section 3, we first review the mirror symmetry between abelian GNLSMs and gauged Landau-Ginzburg (LG) models with neutral matter. We then derive the explicit reduction of open gauged linear sigma models (GLSMs) to open GNLSMs, making use of the example of $\CPN$. In Section 4, we study equivariant B-branes, paying particular attention to abelian equivariant B-branes in toric manifolds $X$, as well as the LG mirrors of these branes %for 
 when $X$ is Fano. %toric manifolds. %$X=\C^N//U(1)$.
Nonabelian equivariant B-branes in general $G$-manifolds are also analyzed. In Section 5, abelian equivariant A-branes in toric manifolds $X$ are introduced, and their LG mirror description %as D0-branes is shown%for %toric manifolds $X$ with 
%$c_1(X)\geq 0$
is shown for toric manifolds $X$ with $c_1(X)\geq 0$. We also explore nonabelian equivariant A-branes for general $G$-manifolds. In Section 6, we use the %knowledge 
data of equivariant A-branes to study open Hamiltonian Gromov-Witten invariants. %which are given by correlation functions in the open topological gauged A-model. 
The open gauged A-model is first introduced, together with its bulk and boundary observables. The path integrals of these observables are given by classical integrals over the moduli spaces of open symplectic vortices on %the strip
a Riemann surface with boundaries, and these integrals are identified with the open Hamiltonian Gromov-Witten invariants.
  Then, we compute the dimension of these moduli spaces, as well as the related %via the 
 boundary axial R-anomaly. For abelian invariants, we use mirror symmetry to compute the $\hat{Q}_A^2\neq 0$ anomaly, 
 %which obstructs %the compactness of 
 %integration over the moduli space, and also implies supersymmetry breaking.
 which implies supersymmetry breaking, and indicates an obstruction to integration over the moduli spaces. 
  We shall find the condition whereby the anomaly vanishes and supersymmetry is manifest.  Finally, we show how mirror symmetry can be used to compute the abelian invariants themselves. 
 
The reader who is interested in equivariant B-branes should read Sections 2, 3 and 4, whereas the reader who is  interested in equivariant A-branes and open Hamiltonian Gromov-Witten invariants should read Sections 2, 3, 5 and 6.
%they are related to those of NLSMs, and also give us equivariant homological mirror symmetry. 
%Introduction is to be written last, but the main point shall be that since GNLSMs on $X$ are related to NLSMs on $X//G$, we have actually found a new description of the usual A-branes and B-branes of string theory.
\section{The Gauged Nonlinear Sigma Model with Boundaries}
%In this section we shall review the gauged nonlinear sigma model (GNLSM) action, and find the necessary condition for it to be supersymmetric on a worldsheet with boundary. 
The $\mathcal{N}=(2,2)$ supersymmetric nonlinear sigma model (NLSM) has the isometry group, $G$ of its K\"{a}hler target space, $X$, as a global symmetry. This global symmetry can be gauged, by allowing the isometry transformations of the target space to depend on the local coordinates of the worldsheet. As in the case of Yang-Mills theory, invariance under this local symmetry requires the introduction of a gauge field, $A_{\mu}$. Supersymmetry then requires the introduction of gaugino and %Higgs 
scalar fields, denoted as $\lambda$ and $\sigma$. In addition, we can introduce kinetic terms for the gauge field and its superpartners, and as a result we have a dynamically gauged supersymmetric nonlinear sigma model, or GNLSM.
%IMPORTANT NOTE \footnote{The superfield formulation of this action was studied in \cite{hitchin1987}. NOTE THAT TO GET STANDARD DphiDphi ACTION FROM SUPERFIELDS MUST DO INTEGRATION BY PARTS. THIS WILL GIVE BOUNDARY TERMS. MUST TAKE THIS INTO ACCOUNT WHEN USING BAPTISTAS ARGUMENT REDUCING GLSM TO GNLSM IN SUPERFIELD LANGUAGE. IN ANY CASE WE ARE SHOWING IT EXPLICITLY LATER} 

%In two dimensions, globally supersymmetric theories are defined only on flat spacetimes, so in this section we take $\Sigma$
%to be either the complex plane, the cylinder or the torus.
We shall take the worldsheet $\Sigma$ to be the infinite strip $I\times \R$ (where the interval is $I=[0,\pi]$) equipped with a flat Minkowski metric $\eta=\text{diag}(-1,1)$. The spatial coordinate along the interval will be denoted $x^1$ and the time coordinate parametrizing $\R$ will be denoted $x^0$. The main fields of the GNLSM are a connection, $A$, on a principal $G$-bundle 
\begin{equation}
P\rightarrow \Sigma,
\end{equation}
and a section
\begin{equation}
\phi: \Sigma \rightarrow E
\end{equation}
of the associated bundle $E := P\times_G X$,
%The two main fields of the gauged sigma-model are a connection $A$ on a principal $G$-bundle $P \rightarrow \Sigma$
%and a section $\phi: \Sigma \rightarrow E$ of the associated bundle $E := P\times_G X$
 where $X$ is a K\"{a}hler manifold. 
%A local patch of $E$ looks
%On a local patch of
Locally on $\Sigma$,
%IMPORTANT NOTE: MADE THE ABOVE CHANGE ON 4TH APRIL 2017, SEE SETTER PAGE 98 AND 144.
 $E$ looks like the product $\Sigma \times X$, which implies that locally the section $\phi$ looks like a map $\phi:\Sigma \rightarrow X$,\footnote{When $P$ is the trivial $G$-bundle, this becomes true \it{globally}.} as one finds in non-gauged NLSMs.

The other fields of the GNLSM are sections of the following bundles:
\begin{align}
 \psi_{\pm}  &\in  \Gamma (\Sigma ;  S_{\pm} \otimes \phi^\ast \ver )  &     F &\in \Gamma (\Sigma ; \phi^\ast \ver )    \label{2.1}   \\
 \si &\in \Gamma (\Sigma ; \gp^\C ) &      D  &\in \Gamma (\Sigma ; \gp ) \nonumber  \\
\lam_{\pm}  &\in  \Gamma (\Sigma ;  S_{\pm} \otimes \gp^\C )   \nonumber
\end{align}
%Here, as in the rest of the paper, the notation $\Omega^p_{\pm} (\Sigma ; V)$ represents the space of $p$-forms
%on $\Sigma$ with values on the bundle $V \rightarrow \Sigma$; 
%where we have used (+) and (-) to distinguish bosonic and fermionic fields. 
Here,  $S_{\pm} = K^{\pm 1/2}$ are the spinor bundles of $\Sigma$, 
whereby $K = \Lambda^{1,0}\Sigma$ is the canonical bundle of $\Sigma$
%; the bundle 
%$\ver \rightarrow E$, which locally looks like $\Sigma \times TX \rightarrow \Sigma \times X$, and is just the
%sub-bundle of $TE \rightarrow E$ defined as  the kernel of the derivative of the projection $\pi_E : E \rightarrow 
%\Sigma$; and finally $\phi^\ast (\ver) \rightarrow \Sigma$, the pull-back of $\ver$ by the 
%section $\phi$. 
%with $K = \Lambda^{1,0}\Sigma$ being the canonical bundle of $\Sigma$ 
(the bundle of one-forms of type (1,0));  
$\ver \rightarrow E$ is a bundle which locally looks like 
%$\Sigma \times TX \rightarrow \Sigma \times X$,
%IMPORTANT NOTE: MADE THE FOLLOWING CHANGE ON 4TH APRIL 2017, SEE SETTER PAGE 98 AND 144.
$TX$ and is the sub-bundle of $TE \rightarrow E$ %defined as  
corresponding to the kernel of the derivative $\text{d}\pi_E:TE\rightarrow T\Sigma$ of the projection $\pi_E : E \rightarrow 
\Sigma$; $\phi^\ast (\ver) \rightarrow \Sigma$ is the pullback of $\ver$ with respect to the  
section $\phi$; and $\gp := P\times_{\text{Ad}} \mathfrak{g}$ is the associated adjoint bundle (where 
$\mathfrak{g}$ is the Lie algebra of $G$) with $\gp^\C$ denoting its complexification. 

%In other words, the GNLSM has one adjoint scalar Higgs field $\si$, four fermionic fields $\psi_\pm$ and
%$\lambda_\pm$, and two scalar auxiliary fields $F$ and $D$.
The GNLSM action is
\begin{equation}\label{action}
S=\frac{1}{2\pi}(S_{\text{matter}}+S_{\text{gauge}}+S_{B}+S_{\theta}),
\end{equation}
where
\begin{equation}\label{matteraction}
\begin{aligned}
S_{\text{matter}} =\int_{\Sigma}\txd^2x\Big(& 
-g_{j\ov{k}}\del^A_{\mu}\phi^j\del^{A\mu}\ov{\phi}^{\ov{k}} 
\:+\: \,\frac{i}{2}\, g_{j\ov{k}} \,\ov{\psi}_-^{\ov{k}}  \, (\phi^*\overleftrightarrow{\nabla}^A)_{+} \psim^j \:+\: 
\,\frac{i}{2}\, g_{j \ov{k}}\, \ov{\psi}_+^{\ov{k}}  \, (\phi^*\overleftrightarrow{\nabla}^A)_- \psip^j   \\
&\:-\: \frac{1}{2}g_{j\ov{k}}\si^a \he^j_a\ov{\si}^b\ov{\he}^{\ov{k}}_b  \:-\: \frac{1}{2}g_{j\ov{k}}\ov{\si}^a \he^j_a\si^b\ov{\he}^{\ov{k}}_b \:+\: \,  \, i g_{j\ov{k}} \, (\nabla_l \, \he_a^j)\, (\si^a \ov{\psi}_-^{\ov{k}} \psip^l  \:+\: 
\sib^a \ov{\psi}_+^{\ov{k}} \psim^l )   \\
& \:+\:   \, g_{j\ov{k}} \,(\ov{\lam}_+^a\, \he_a^j \, \ov{\psi}_-^{\ov{k}} \:-\: \ov{\lam}_-^a\, \he_a^j\,  \ov{\psi}^{\ov{k}}_+ 
\:-\: \lamp^a\, \ov{\he}_a^{\ov{k}}\, \psim^j \:+\: \lamm^a\, \ov{\he}_a^{\ov{k}}\,  \psip^j)    \\
&  \:+\:  R_{i\ov{\jmath}k\ov{l}} \, \psip^i \psim^k \ov{\psi}^{\ov{\jmath}}_-\, \ov{\psi}^{\ov{l}}_+ \:+\:  g_{j\ov{k}} (F^j \:-\: \Gamma^j_{il}\, \psip^i \psim^l) (\ov{F}^{\ov{k}} \:-\: \Gamma^{\ov{k}}_{\ov{m}\ov{n}}\, \,
\ov{\psi}^{\ov{m}}_-\, \,\ov{\psi}^{\ov{n}}_+)   \ \Big),
\end{aligned}
\end{equation}

\begin{equation}\label{gaugeaction}
\begin{split}
S_{\text{gauge}}  =\frac{1}{2e^2}\int_{\Sigma}\txd^2x&
  \Big(\, F^a_{01} F_{01a}\: - \:   \nabla_{\mu}^A \si^a \nabla^{A\mu} \sib_a \: + \: \frac{1}{4} \, 
 [\si , \sib]^a[\si , \sib]_a\: +\: \, D^aD_a \:- \: 2\, e^2 \, \phi^\ast \mu_a \, D^a \\
\:&+  \,\frac{i}{2}\, (\lambm)_a \overleftrightarrow{\nabla}^A_{+} \lamm^a\,\:   + \:\frac{i}{2} \,(\lambp)_a \overleftrightarrow{\nabla}_{-}^A \lamp^a \: \; 
+\;\,i\, \lambm^a [\si , \lamp]_a  \: +\: \, i\, \lambp^a [\sib , \lamm]_a  \Big)   \ ,
\end{split}
\end{equation}
\begin{equation}\label{bfield}
S_{B} =     - \int_{\Sigma}\, \phi^\ast B \; + \int_{\del \Sigma}   \phi^*C_a A^a ,
\end{equation}
and
\begin{equation}\label{theta}
S_{\theta}=\int_{\Sigma}\txd^2x\;  \, (\theta ,  F_{01} )
\end{equation}
%IMPORTANT NOTE: CORRECTED SIGN IN FRONT OF FOUR SIGMA TERM ON 3/3/17
with $\txd^2x= \txd x^1 \wedge \txd x^0$.\footnote{We only consider the case where $G$ is compact, in order to ensure positive-definiteness of the gauge multiplet kinetic terms.
%IMPORTANT NOTE: ALSO SEE INTRODUCTION TO WITTEN'S QUANTIZATION OF CHERN-SIMONS GAUGE THEORY WITH COMPLEX GAUGE GROUP
} Here, $F^j$ and $D^a$ are auxiliary fields. The covariant derivatives induced by the connection $A$ on the 
 bundles $E$, $\phi^\ast \ver$ and $\gp$ over $\Sigma$ appear above in their local forms, which are given explicitly as
\begin{equation}
\begin{aligned}
\del_{\mu}^A \phi^k  &= \del_{\mu} \phi^k  +  A_{\mu}^a \, \he^k_a   \label{2.13}   \\
(\phi^\ast \nabla^A) _{\mu}\psi^k &=  \del_{\mu} \psi^k  +  A_{\mu}^a \psi^j   \del_j \he^k_a  + \Gamma^k_{jl} (\del^A_{\mu}\phi^j) \psi^l  \\
\nabla_{\mu}^A \sigma^a  &=  \del_{\mu}\sigma^a  +  [A _{\mu}, \sigma]^a  \\
\nabla_{\mu}^A \lambda^a  &=  \del_{\mu}\lambda^a  +  [A _{\mu}, \lambda]^a  \\
\end{aligned}
\end{equation}
where $\phi$ and $\sigma$ are locally regarded as maps  $\Sigma \rightarrow X$ and $\Sigma \rightarrow \g$; $\psi$ and $\lambda$ are locally regarded as 
(fermionic) maps $\Sigma \rightarrow \phi^\ast TX$ and $\Sigma \rightarrow \g$; and $A$ is regarded as a local 1-form on $\Sigma$. 
We have also used the notations
\begin{equation}
\begin{aligned}
\nabla_+&=\nabla_0+\nabla_1\\
\nabla_-&=\nabla_0-\nabla_1,
\end{aligned}
\end{equation}
as well as 
\begin{equation}\label{overarrow}
A\overleftrightarrow{\nabla}B=A\nabla B-\nabla B A.
\end{equation} 
The notation \eqref{overarrow} appears in all the fermionic kinetic terms, since they must have this symmetrized form in order to preserve the %hermiticity 
reality of the action in the presence of worldsheet boundaries.

Let us now explain the quantities that appear in the action. Firstly, note that the components of the Killing vector fields 
\begin{equation}
\he_a=\he^i_a\frac{\del}{\del\phi^i}+\ov{\he}^{\ov{\jmath}}_a\frac{\del}{\del\ov{\phi}^{\ov{\jmath}}}
\end{equation}
which generate the $G$-isometry of $X$ appear in the action. These components are holomorphic/antiholomorphic,
\begin{equation}
\frac{\del\ov{\he}^{\ov{\jmath}}_a}{\del\phi^i}=\frac{\del\he^i_a}{\del\ov{\phi}^{\ov{\jmath}}}=0,
\end{equation}
and this constraint can be shown to be a consequence of the complex structure of $X$ being $G$-invariant, i.e., $\mathcal{L}_{\he}J=0$. Furthermore, they realize an antihomomorphism of the Lie algebra, $\mathfrak{g}$, via their Lie bracket
%The statement that the Killing vector fields generate an action on $X$ implies that each vector field $V^a$ corresponds to an element $T^a$ of the Lie algebra of $G$, and that they realise an antihomomorphism of that Lie algebra, i.e.,
\begin{equation}\label{antihomomorphism}
\begin{aligned}
\lbrack\he_a,\he_b\rbrack&=-{f_{ab}}^{c} \he_c,\\
\lbrack\ov{\he}_a,\ov{\he}_b\rbrack &=-{f_{ab}}^{c} \ov{\he}_c
\end{aligned}
\end{equation} 
%IMPORTANT NOTE: THE MINUS SIGN IS EXPLAINED ON PAGE 85 OF MCDUFF-SALAMON, INTRODUCTION TO SYMPLECTIC TOPOLOGY. ALSO SEE SETTER'S THESIS, PAGE 147
(the generators of $\mathfrak{g}$ satisfy $[T_a,T_b]={f_{ab}}^c T_c$, where ${f_{ab}}^c$ are the real structure constants of $\mathfrak{g}$). Also note that the covariant derivative of the holomorphic component, $\nabla_l \, \he_a^j$ appears in the action.
%IMPORTANT NOTE: ON 24 MARCH 2017 I TOOK OUT THE FACTOR of i in front of f_{ab}^c above, this is because the Lie algebra used by Baptista is actually [T_a,T_b]=f_{ab}^cT_c, with no factor of i. This can be seen by checking the reality of the action, as well as by checking how the supersymmetry transformations of the gauginos change under complex conjugation. The structure constants here are real. This is because we are restricting ourselves to compact Lie groups/Lie algebras. Compact Lie algebras are real, see https://en.wikipedia.org/wiki/Compact_Lie_algebra (here they say that this definition exclude tori, but structure constants are anyway relevant for nonabelian groups). A real Lie algebra must have real structure constants, see https://www.physicsforums.com/threads/structure-constants-in-lie-algebra.722008/. This is because an element of of a real Lie algebra is a linear combination of the basis elements of the Lie algebra with real coefficients. The COMMUTATOR is such an element. ALSO, the structure constants must be real because the complex conjugate equation of [\he_a,\he_b]=-f_{ab}^c \he_c also holds (see prof tan's gauged sigma model paper), but with the structure constants unchanged.
 
The moment map $\mu_a$, which is a map $X\rightarrow \g^*$ (where $\g^*$ is the dual of $\g$) also appears in \eqref{gaugeaction}, and obeys
\begin{equation}
\begin{aligned}
\del_{\ov{k}}\mu_a&=ig_{j\ov{k}}\he_a^j\\
\del_{k}\mu_a&=-ig_{\ov{j}k}\ov{\he}_a^{\ov{\jmath}}
\end{aligned}
\end{equation}
or
\begin{equation}\label{momentmapequation}
d\mu_a=\iota_{\he_a}\omega,
\end{equation}
where $\omega=\frac{1}{2}\omega_{IJ}d\phi^I\wedge d\phi^J=i\met d\phi^i\wedge d\ov{\phi}^{\ov{\jmath}}$ is the K\"ahler form.\footnote{In \cite{Baptista1}, the the moment map equation is given as $2\del_{\ov{k}}\mu_a=ig_{j\ov{k}}\he_a^j$, since the K\"ahler form is defined as $\omega=\frac{i}{2}\met d\phi^i\wedge d\ov{\phi^{\jmath}}$.} This indicates that the $G$-isometry of $X$ is Hamiltonian. 

The fact that only the \textit{derivative} of $\mu_a$ enters \eqref{momentmapequation} ostensibly implies that the moment map $\mu=\mu_aT^a$ is only defined up to a constant in $\mathfrak{g}^*$. In fact, the moment map is defined up to a constant in $[\g,\g]^0$, the subspace of $\g^\ast$ that annihilates commutators;% where  $[\g,\g]^0$ indicates the annihilator of $[\g,\g]$. 
\footnote {$[\g,\g]^0$ can be further identified with the centre of $\mathfrak{g}$, via the identification $\g^*\cong \g$ provided by the inner product $\g^*\times\g\rightarrow\R $.} %However, 
as in \cite{Baptista1}, we shall follow the convention of (\cite{mcduff1998introduction}, page 164) where the definition of a Hamiltonian $G$-action includes the additional condition 
\begin{equation}\label{Momcon2}
\rho_g^\ast \, \mu = {\rm Ad}_g^\ast \circ \mu 
\end{equation}
 for all elements $g
  \in G$. Here, $\rho$ is the $G$-action on $X$ and ${\rm Ad}^\ast $ is the coadjoint representation 
  on $\g^\ast$. This then implies (\cite{libermann1987symplectic}, page 190) that the moment map is only defined up to a constant, $r$, in $[\g,\g]^0$.
  %, the subspace of $\g^\ast$ that annihilates commutators.% where  $[\g,\g]^0$ indicates the annihilator of $[\g,\g]$. 
%\footnote {$[\g,\g]^0$ can be further identified with the centre of $\mathfrak{g}$, via the identification $\g^*\cong \g$ provided by the inner product $\g^*\times\g\rightarrow\R $.}
 In fact, this freedom to redefine the moment map as
\begin{equation}
\mu\rightarrow\mu+r
\end{equation} 
  is manifest in the action; we may add the term 
 \begin{equation}\label{FIterm}
S_r=\frac{1}{2\pi}\frac{1}{2e^2}\int_{\Sigma}\txd^2x( -2e^2r_aD^a)=-\frac{1}{2\pi}\int_{\Sigma}\txd^2x\textrm{ } r_aD^a
 \end{equation}
to \eqref{action}, which results in 
\begin{equation}
\mu_a\rightarrow\mu_a+r_a
\end{equation} 
in \eqref{gaugeaction}.
From \eqref{FIterm}, we see that the constant $r$ now plays the role of 
%$\textrm{dim } G$ 
a $[\g,\g]^0$-valued Fayet-Iliopoulos (FI) parameter. 

In \eqref{bfield}, the $B$-field action is given.
Here, $B$ is an arbitrary
%but fixed
 $G$-invariant 
(i.e., $\mathcal{L}_{\he}B=0$) and closed 2-form on $X$, and $ \phi^\ast B$ denotes the pullback of $B$.
% with derivatives covariantized. 
Explicitly, the B-field term is denoted as 
\begin{equation}
-\phi^* B =-\frac{1}{2}B_{IJ}(\del_1\phi^I\del_0\phi^J-\del_0\phi^I\del_1\phi^J)dx^1\wedge dx^0,
%B_{IJ}\del_0\phi^I\del_1\phi^Jdx^1\wedge dx^0
\end{equation}
%IMPORTANT NOTE: THE form given in the HIDDEN LINE ABOVE IS USEFUL FOR CALCULATIONS
where $B=\frac{1}{2}B_{IJ}d\phi^I\wedge d\phi^J$, and $\phi^I$ are real coordinates on $X$.
%\footnote{In fact the supersymmetric ${\mathcal N} =(2,2)$ 
%theory admits a more general $H$-flux term, instead of the B-field term presented here.
%This is related to the fact that it also admits more general targets $X$, namely (twisted) generalized K\"ahler manifolds, 
%instead of just the K\"ahler targets to which we have restricted ourselves here. For these matters see \cite{K-T} and the 
%references therein.}
A boundary term %\cite{figueroa} 
is also included in the $B$-field action, with
\begin{equation}
 \phi^*C_a A^a= \phi^*C_a A_0^a dx^0,
\end{equation}
where $C$ is a map $X\rightarrow {\g}^*$  which obeys 
\begin{equation}\label{Cmapequation}
dC_a=\iota_{\he_a}B,
\end{equation}
as well as 
\begin{equation}\label{Momcon2Cfield}
\rho_g^\ast \, C = {\rm Ad}_g^\ast \circ C 
\end{equation}
 for all elements $g
  \in G$.
The necessity of this boundary term will be explained below when we investigate the gauge invariance of the $B$-field action. 

Finally, the $\theta$-term is given in \eqref{theta}, where $\theta$ is a constant
%\footnote{Recall that the moment map $\mu$ is also 
%defined only up to a constant in  $[\g, \g]^0$, so that both these constants can be combined into an element of the complexified 
%space $[\g, \g]^0_\CC$. This complex constant, as usual, is the important parameter of the quantum theory. Note, moreover, that the inner product 
%$\kappa$ allows the identification of $[\g, \g]^0$ with the centre of $\g$.}  
in $[\g ,\g]^0\subset\g^*$, while $(\cdot , \cdot )$ is the inner product
%natural pairing 
$\g^\ast \times \g \rightarrow {\mathbb R}$. We can then combine $r$ with $\theta$ as
\begin{equation}
r-i\theta=t\in [\g,\g]_{\mathbb{C}}^0,
\end{equation} 
to obtain the complex FI-theta parameter, $t$, valued in $[\g,\g]_{\mathbb{C}}^0$ (the complexification of $[\g,\g]^0$).
%\item[(ii)]$\rho_g^\ast \, \mu = {\rm Ad}_g^\ast \circ \mu $ for all $g
 % \in G$, where $\rho$ denotes the $G$-action on $X$ and ${\rm Ad}^\ast $ is the coadjoint representation 
  %on $\g^\ast$.
%\end{itemize}
%If a moment map $\mu$ exists, it is not in general unique, but all the
%other moment maps have the form $\mu + r$, where $r \in [\g ,\g ]^0
%\subset \g^\ast $ is a constant in the annihilator of $[\g, \g]$. Under the identification $\g^\ast \simeq \g$
%provided by $\kappa$, the inner product, the annihilator $[\g, \g ]^0 $ is identified with the centre of $\g$. 
%The constant $r$ is then the Fayet-Iliopoulos parameter of the supersymmetric theory.

%Moment map must introduce.
%The nablas with plus and minus dont contain a 1/2 in their defition.

%This is known as the moment map equation. Must mention holomorphicity for Killing vector fields. Must mention that moment map is only defined up to a constant is the commuting subspace. 
%\vspace{3cm}
%\newpage
\mbox{}\par\nobreak
\noindent
\textit{Gauge and Supersymmetry Invariance}

The action is invariant under the following gauge symmetry transformations (where the parameter $\alpha^a$ is a local function on the worldsheet)\footnote{Note that one needs to use various identities in order to show gauge invariance, including the Jacobi identity and the Killing equation $\mathcal{L}_{\he}g_{i\ov{\jmath}}=0$, which implies $\mathcal{L}_{\he}\Gamma^i_{jk}=0$ and $\mathcal{L}_{\he}R_{i\ov{\jmath}k\ov{l}}=0$ (\cite{yano1957theory}, page 52).}
\begin{equation}
\begin{aligned}\label{gaugetrans1}
\delta\phi^k&=\alpha^a\he^{k}_a\\
\delta\ov{\phi}^{\ov{k}}&=\alpha^a\ov{\he}_a^{\ov{k}}\\
\delta\psi_{\pm}^k&=\alpha^a\psi_{\pm}^i\del_i\he^{k}_a\\
\delta\ov{\psi}^{\ov{k}}_{\pm}&=\alpha^a\ov{\psi}_{\pm}^{\ov{\imath}}\del_{\ov{\imath}}\ov{\he}_a^{\ov{k}} \\
\delta F^k&=\alpha^a F^i\del_i\he^{k}_a\\
\delta\ov{F}^{\ov{k}}&=\alpha^a\ov{F}^{\ov{\imath}}\del_{\ov{\imath}}\ov{\he}_a^{\ov{k}}
\end{aligned}
\end{equation}

\begin{equation}
\begin{aligned}\label{gaugetrans2}
\delta A^a_{\mu}&=[\alpha,A_\mu]^a-\del_\mu\alpha^a=-\nabla^A_{\mu}\alpha^a\\
\delta \si^a&=[\alpha,\si]^a\\
\delta \ov{\si}^a&=[\alpha,\ov{\si}]^a\\
\delta \lambda_{\pm}^a&=[\alpha,\lambda_{\pm}]^a\\
\delta \ov{\lambda}_{\pm}^a&=[\alpha,\ov{\lambda}_{\pm}]^a\\
\delta D^a&=[\alpha,D]^a
\end{aligned}
\end{equation}
%The above is based on qin and yuans note titled 'lagrangian for chiral multiplet'. There they give $\delta A_\mu^a=\nabla_\mu^A\varepsilon^a$ and $\delta\Phi=i[\epsilon,\Phi]$. Note from the definitions of the covariant derivative of $\phi$ in the GNLSM and GLSM, that the second term which gives rise to covariance is always the variation of $\phi$ but with the parameter replaced by the gauge field. So this implies that the gauge transformation of the vector multiplet fields should just be the second term of their covariant derivative, but with the gauge field replaced by the parameter. The above choice also gives
%$\delta(\nabla^A_\mu\sigma^a)=[\varepsilon,\nabla^A_\mu\sigma]^a$ (with use of the Jacobi identity), i.e., the covariant derivarive transforms covariantly, as it should. If instead we chose $\delta\sigma^a=i[\varepsilon,\sigma]^a$, then only  $\delta A^a_{\mu}=i[\varepsilon,A_\mu]^a-i\del_\mu\varepsilon^a$ will give rise to $\delta(\nabla^A_\mu\sigma^a)=i[\varepsilon,\nabla^A_\mu\sigma]^a$, but in the abelian case this reduces to  $\delta A^a_{\mu}=-i\del_\mu\varepsilon^a$, which does not agree with the mirror symmetry book, which we are following.
  
%Gauge invariance does not require integration by parts, so there are not boundary terms that should vanish. It however requires the lie derivative of the metric to vanish.
%Note that 
%unlike supersymmetry, 
Let us first explain how gauge invariance requires the boundary term in the $B$-field action \eqref{bfield}.
%, following (\cite{figueroa}, page 28).
 For a closed worldsheet, the term containing the closed two-form $\phi^*B$ is gauge invariant. However, on an open worldsheet, a gauge transformation of this term generates a boundary term, and in order to restore gauge invariance, the boundary term containing $\phi^*C$ must be added to the action.\footnote{Note that the proof of this requires the use of the identities \eqref{Cmapequation} and \eqref{Momcon2Cfield}, with the latter implying that $\alpha^b\mathcal{L}_{\til{e}_b}C_a=[\alpha,C]_a$. }

With the exception of this $B$-field action \eqref{bfield}, the gauge symmetry of the action 
%holds regardless of 
is insensitive to the presence of boundaries. To understand this, note that for a \textit{global} $G$-isometry of $X$, the corresponding symmetry variation of the (non-gauged) NLSM scalar kinetic term is
\begin{equation}\label{isometry}
\begin{aligned}
&\int_{\Sigma}\txd^2x\textrm{ }\delta(g_{i\ov{\jmath}}\del_{\mu}\phi^i\del^{\mu}\ov{\phi}^{\ov{\jmath}})\\
=&\int_{\Sigma}\txd^2x\textrm{ }(\del_{k}g_{i\ov{\jmath}}\he^k_a+\del_{\ov{k}}g_{i\ov{\jmath}}\ov{\he}^{\ov{k}}_a+g_{k\ov{\jmath}}\del_i\he_a^k+g_{i\ov{k}}\del_{j}\ov{\he}^{\ov{k}}_a)\alpha^a\del_{\mu}\phi^i\del^{\mu}\ov{\phi}^{\ov{\jmath}}\\
=&\int_{\Sigma}\txd^2x\textrm{ }\mathcal{L}_{\he}g_{i\ov{\jmath}}\del_{\mu}\phi^i\del^{\mu}\ov{\phi}^{\ov{\jmath}}\\
=&0,
\end{aligned}
\end{equation}
where we have used the Killing equation $\mathcal{L}_{\he}g_{i\ov{\jmath}}=0$. In the computation above, we have not used integration-by-parts (which would introduce a nonzero boundary term), so the presence of boundaries is inconsequential for global $G$-symmetry of NLSMs. 

A crucial step in \eqref{isometry} is that the worldsheet derivative of $\phi^i$ transforms under the global $G$-symmetry as a target space vector field, i.e., 
\begin{equation}
\delta(\del_{\mu}\phi^i)=\alpha^a\del_{\mu}\phi^j\del_j\he^i_a.
\end{equation}
When the $G$-symmetry is \textit{local}, it is the \textit{covariant} derivative that transforms in the above manner, i.e.,
\begin{equation}
\begin{aligned}
\delta(\del^A_\mu\phi^i)&=\del_{\mu}(\alpha^a\he_a^i)-\nabla^A_{\mu}\alpha^a\he_a^i+A_{\mu}^a\alpha^b\he_a^j\del_je^i_b+A^a_{\mu}[\alpha,\he^i]_a\\
&=\alpha^a\del^A_{\mu}\phi^j\del_j\he^i_a,
\end{aligned}
\end{equation}
where we have used \eqref{antihomomorphism}.
Then, the same steps in the computation \eqref{isometry} hold, with $g_{i\ov{\jmath}}\del_{\mu}\phi^i\del^{\mu}\ov{\phi^\jmath}$ replaced by $g_{i\ov{\jmath}}\del^A_{\mu}\phi^i\del^{A\mu}\ov{\phi^\jmath}$, and the scalar kinetic term of the GNLSM  is gauge invariant without using integration-by-parts.
In a similar manner, all the other terms in \eqref{matteraction} %as well as the $B$-field action in \eqref{bfield} 
are gauge invariant without the generation of nonzero boundary terms via integration-by-parts. Furthermore, the gauge action \eqref{gaugeaction}, the $\theta$-term in  \eqref{theta}, and the FI term $\eqref{FIterm}$ are also gauge invariant, without using integration-by-parts.\footnote{Proving the gauge invariance of the term containing the moment map in \eqref{gaugeaction} requires the use of the identity  \eqref{Momcon2}, which implies that $\alpha^b\mathcal{L}_{\til{e}_b}\mu_a=[\alpha,\mu]_a$.
%IMPORTANT NOTE: THE LATTER CAN BE DEDUCED FROM salamon and mcduff-intro to symplectic topology. See for e.g., equation 5.6 and Lemma 5.16. Also appears in other places in the book. The equation we use to prove gauge invariance seems to be the infinitesimal version of \eqref{Momcon2}. This can be shown by noting that the pullback of the moment map by a G transformation is the moment map of as a function of the G-transformation of the coordinates. An infinitesimal G-transformation is given by the original coordinate plus an infinitesimal parameter sum with the Killing vector fields, just like the gauge transformation in equation 2.26. This is just \iota d \mu. Since \iota \mu=0, then we arrive at the equation mentioned in this footnote. THE PROPERTY IS IN FACT NOTHING BUT THE PROPERTY OF G-EQUIVARIANCE OF THE MOMENT MAP. 
}
% and the presence of a boundary is inconsequential.
%In this way, in the GNLSM, gauge symmetry holds regardless of the presence of boundaries, and no nonzero boundary terms are generated via gauge transformations.
In this way, the GNLSM action (\eqref{action}+\eqref{FIterm}) is gauge invariant, and no nonzero boundary terms are generated via gauge transformations.
  
  However, this is not the case for supersymmetry. For a closed worldsheet, the action (\eqref{action}+\eqref{FIterm}) would be invariant under the following ${\mathcal N} =(2,2)$ supersymmetry transformations\footnote{The action \eqref{action} and supersymmetry transformations agree with those of Baptista \cite{Baptista1}, upon analytical continuation to a Minkowski worldsheet following the Appendix in {\it loc. cit.}, symmetrization of the fermion kinetic terms using \eqref{overarrow}, an overall sign flip, rescaling $\sigma$, $\lambda$ and the supersymmetry transformation parameters by $\frac{1}{\sqrt{2}}$, %IMPORTANT NOTE:rescaling the coupling constant $e^2$ by 2 DOES NOT HAPPEN,
 rescaling the theta parameter as $(-\frac{1}{2\pi})\theta_{(there)}=\theta_{(here)}$, as well as a rescaling of the moment map, i.e., $2\mu^a_{(there)}=\mu^a_{(here)}$. Our action \eqref{action} is also given an overall factor of $\frac{1}{2\pi}$.}\footnote{
 %We may include the term \eqref{FIterm} without affecting supersymmetry invariance on a closed worldsheet.
  Note that if $r$ was a constant in $\g^*$ instead of $[\g,\g]^0$, the supersymmetry invariance of \eqref{FIterm} on a closed worldsheet would not hold. This is %the
 another reason we %needed 
 need the condition \eqref{Momcon2}.  }
%The supersymmetric lagrangian  (\ref{2.2}) is a ${\mathcal N} =(2,2)$ lagrangian, and so is invariant (up to total derivatives)
%under four independent fermionic symmetries, whose parameters are denoted $\epsilon_\pm$ and $\ov{\epsilon}_\pm$.
%The general supersymmetry transformations are, for the matter fields,  
\begin{equation}\label{susytrans}
\begin{aligned}
\delta \phi^k &=  (\epsilon_{+} \psim^k - \epsilon_{-} \psip^k )     \\
\delta \ov{\phi}^{\ov{k}} &= -  (\epsilonbp \ov{\psi}_-^{\ov{k}} - \epsilonbm \ov{\psi}_+^{\ov{k}} )     \\
\delta \psip^k &=  i \epsilonbm (\del^A_{0} +\del^A_{1})\phi^k  +  \epsilon_+ F^k +   i \epsilonbp \sib^a \he_a^k     \\ 
\delta \ov{\psi}_+^{\ov{k}} &= -  i \epsilon_- (\del^A_{0} +\del^A_{1}) \ov{\phi}^{\ov{k}}  +  \epsilonbp \ov{F}^{\ov{k}} 
-  i \epsilon_+ \si^a \ov{\he}^{\ov{k}}_a      \\
\delta \psim^k &= - i \epsilonbp (\del^A_{0} -\del^A_{1}) \phi^k  +  \epsilon_- F^k -   i \epsilonbm \si^a \he_a^k     \\ 
\delta \ov{\psi}_-^{\ov{k}} &=  i \epsilon_+ (\del^A_{0} -\del^A_{1}) \ov{\phi}^{\ov{k}}  +  \epsilonbm \ov{F}^{\ov{k}} 
+  i \epsilon_- \sib^a \ov{\he}^{\ov{k}}_a     \\
\delta F^k  &=  - i \epsilonbp (\partial_- \psip^k + A^a_- (\partial_j \he_a^k) \psip^j )   
-  i \epsilonbm (\partial_{+} \psim^k + A^a_{+} (\partial_j \he_a^k) \psim^j )     \\
&\ \ +  \epsilonbm \ov{\lam}^a_+ \, \he_a^k -    \epsilonbp \ov{\lam}^a_-\, \he_a^k - i \epsilonbp \sib^a (\partial_j \he_a^k) \psim^j 
-  i \epsilonbm \si^a (\partial_j \he_a^k) \psip^j     \\
\delta \ov{F}^{\ov{k}}  &= -  i \epsilonp (\partial_- \ov{\psi}_+^{\ov{k}} + A^a_- (\partial_{\ov{\jmath}} \ov{\he}^{\ov{k}}_a ) \ov{\psi}_+^{\ov{\jmath}} )   
-  i \epsilonm (\partial_{+} \ov{\psi}_-^{\ov{k}} + A^a_{+} (\partial_{\ov{\jmath}} \ov{\he}^{\ov{k}}_a ) \ov{\psi}_-^{\ov{\jmath}} )     \\
&\ \ -  \epsilonm \lam^a_+ \, \ov{\he}^{\ov{k}}_a  +   \epsilonp \lam^a_- \, \ov{\he}^{\ov{k}}_a  - i \epsilonp \si^a (\partial_{\ov{\jmath}} \ov{\he}^{\ov{k}}_a ) \ov{\psi}^{\ov{\jmath}}_- 
-  i \epsilonm \sib^a (\partial_{\ov{\jmath}} \ov{\he}^{\ov{k}}_a ) \ov{\psi}^{\ov{\jmath}}_+ \ , 
\end{aligned}
\end{equation}
%The gauge fields, at the same time, transform as
\begin{equation}\label{susytrans2}
\begin{aligned}
\delta A_{+}^a &=  i \epsilonp \lambp^a + i \epsilonbp \lamp^a      \\
\delta A_-^a  &=   i \epsilonm \lambm^a  + i \epsilonbm \lamm^a         \\
\delta \si^a &=  -   i \epsilonbp \lamm^a -  i \epsilonm \lambp^a      \\
\delta \sib^a &=  -   i \epsilonp \lambm^a -  i \epsilonbm \lamp^a      \\
\delta \lamp^a  &=   \epsilonm (\nabla_{+}^A \sib^a ) + \epsilonp (-F_{01}^a  + \frac{1}{2}[\si , \sib]^a + i D^a  )     \\
\delta \lambp^a  &=   \epsilonbm (\nabla_{+}^A \si^a ) + \epsilonbp (-F_{01}^a -\frac{1}{2} [\si , \sib]^a - i D^a  )     \\
\delta \lamm^a  &=   \epsilonp (\nabla_{-}^A \si^a ) + \epsilonm (F_{01}^a  - \frac{1}{2} [\si , \sib]^a + i D^a  )     \\
\delta \lambm^a  &=   \epsilonbp (\nabla_{-}^A \sib^a ) + \epsilonbm (F_{01}^a  + \frac{1}{2}[\si , \sib]^a - i D^a  )     \\
\delta D^a  &= \frac{1}{2}( -\epsilonbp (\nabla^A_- \lamp^a ) -  \epsilonbm ( \nabla^A_{+} \lamm^a ) +\epsilonp (\nabla^A_- \lambp^a )  
+ \epsilonm (\nabla^A_{+} \lambm^a )      \\
& \ \ +  \epsilonp [\si , \lambm]^a  +   \epsilonm [\sib , \lambp]^a 
-   \epsilonbp [\sib , \lamm]^a  -  \epsilonbm [\si , \lamp]^a )   \ ,   
\end{aligned}
\end{equation}
where $A_{\pm}=A_0\pm A_1$, $\del_{\pm}=\del_0\pm \del_1$ and
\begin{equation}\label{susydelta}
\delta = \epsilonp Q_- - \epsilonm Q_+ - \epsilonbp \ov{Q}_- + \epsilonbm \ov{Q}_+  \ 
\end{equation}
in terms of the supercharges of the $\mathcal{N}=(2,2)$ supersymmetry algebra.
On the other hand, since our theory is defined on $I\times \R$, supersymmetry is not preserved at the boundaries, and we find
\begin{equation}\label{dSmatw}
\begin{aligned}
\delta S_{matter}=\frac{1}{2\pi}\int_{\partial\Sigma}dx^0\Big\{-&\epsilon_+\Big(\frac{1}{2}\met(\partial_0^A+\partial_1^A)\overline{\phi}^{\overline{\jmath}}\psi_-^i+\frac{1}{2}\met\sigma^a \ov{\he}^{\jbar}_a\psi^i_+ -\frac{i}{2}\met F^i\ov{\psi}_+^{\ov{\jmath}}+\frac{i}{2}\met \Gamma^i_{jl}\psi_+^j\psi_-^l\ov{\psi}_+^{\ov{\jmath}}\Big)\\
+&\ov{\epsilon}_+\Big(\frac{1}{2}\met(\partial_0^A
+\partial_1^A)\phi^i \ov{\psi}_-^{\jbar}+\frac{1}{2}\met\ov{\sigma}^a \he^{i}_a\ov{\psi}^{\jbar}_++\frac{i}{2}\met \ov{F}^{\ov{\jmath}}\psi_+^{i}-\frac{i}{2}\met {\Gamma}^{\ov{\jmath}}_{\ov{\imath}\ov{l}}\ov{\psi}^{\ov{\imath}}_-\ov{\psi}_+^{\ov{l}}\psi_+^{i}\Big)\\ 
-&\epsilon_-\Big(\frac{1}{2}\met(\partial_0^A
-\partial_1^A)\overline{\phi}^{\overline{\jmath}}\psi_+^i+\frac{1}{2}\met\ov{\sigma}^a  \ov{\he}^{\jbar}_a\psi^i_-+\frac{i}{2}\met F^i\ov{\psi}_-^{\ov{\jmath}}-\frac{i}{2}\met \Gamma^i_{jl}\psi_+^j\psi_-^l\ov{\psi}_-^{\ov{\jmath}}\Big)\\
+&\ov{\epsilon}_-\Big(\frac{1}{2}\met(\partial_0^A
-\partial_1^A)\phi^i \ov{\psi}_+^{\jbar}+  \frac{1}{2}\met\sigma^a \he^{i}_a\ov{\psi}^{\jbar}_--\frac{i}{2}\met \ov{F}^{\ov{\jmath}}\psi_-^{i}+\frac{i}{2}\met {\Gamma}^{\ov{\jmath}}_{\ov{\imath}\ov{l}}\ov{\psi}^{\ov{\imath}}_-\ov{\psi}_+^{\ov{l}}\psi_-^{i}\Big)\Big\}
\end{aligned}
\end{equation}
\begin{equation}\label{dSgauthe}
\begin{aligned}
\delta (S_{gauge}+S_r+&S_\theta)=\\\frac{1}{2\pi}\frac{1}{2e^2}\int_{\del\Sigma}dx^0\Big\{&\eps_+\Big(\frac{i}{2}\lambmua(\nabla_1^A+\nabla_0^A)\si^a-\lambpua\Big(\frac{i}{2}F^a_{01}+\frac{i}{4}[\si,\ov{\si} ]^a+\frac{D^a}{2}\Big)+e^2(\phi^*\mu_a+r_a-i\theta_a)\overline{\lambda}_+^a\Big)\\
+&\ov{\eps}_+\Big(\frac{i}{2}\lammua(\nabla_1^A+\nabla_0^A)\ov{\si}^a-\lampua\Big(\frac{i}{2}F^a_{01}-\frac{i}{4}[\si,\ov{\si} ]^a-\frac{D^a}{2}\Big)-e^2(\phi^*\mu_a+r_a+i\theta_a)\lambda_+^a\Big)\\
+&\eps_-\Big(\frac{i}{2}\lambpua(\nabla_1^A-\nabla_0^A)\ov{\si}^a-\lambmua\Big(\frac{i}{2}F^a_{01}+\frac{i}{4}[\si,\ov{\si} ]^a-\frac{D^a}{2}\Big)-e^2(\phi^*\mu_a+r_a+i\theta_a)\ov{\lambda}_-^a\Big)\\
+&\ov{\eps}_-\Big(\frac{i}{2}\lampua(\nabla_1^A-\nabla_0^A)\si^a-\lammua\Big(\frac{i}{2}F^a_{01}-\frac{i}{4}[\si,\ov{\si} ]^a+\frac{D^a}{2}\Big)+e^2(\phi^*\mu_a+r_a-i\theta_a){\lambda}^a_-\Big)\Big\}
\end{aligned}
\end{equation}
\begin{equation}\label{dSB}
\begin{aligned}
\delta S_B=\frac{1}{2\pi}\int_{\del\Sigma}dx^0\Big\{&(B_{ij}\del^A_0\phi^i+B_{\ov{\imath}j}\del^A_0\ov{\phi}^{\ov{\imath}})(\eps_+\psi^j_--\eps_-\psi^j_+)+(B_{i\ov{\jmath}}\del^A_0\phi^i+B_{\ov{\imath}\ov{\jmath}}\del^A_0\ov{\phi}^{\ov{\imath}})(-\ov{\eps}_+\ov{\psi}^{\ov{\jmath}}_-+\ov{\eps}_-\ov{\psi}^{\ov{\jmath}}_+)\\ +&\frac{i}{2}(\eps_+\ov{\lam}^a_++\ov{\eps}_+{\lam}^a_++\eps_-\ov{\lam}^a_-+\ov{\eps}_-{\lam}^a_-)\phi^*C_a \Big\}.
\end{aligned}
\end{equation}
%IMPORTANT NOTE: TOOK OUT COVARIANT DERIVARIVES FOR THE B-FIELD TERMS, THEY SHOULD NOT BE THERE
%IMPORTANT NOTE(Must make notation for matter fields uniform with earlier part). 
In deriving the above, we have used various identities from K\"ahler geometry, %IMPORTANT NOTE, WE ACTUALLY USE THE NUMBERLESS EQUATION ABOVE 8.74 (WHICH I THINK CAN BE CONSIDERED TO BE THE FIRST BIANCHI IDENTITY), AS WELL AS THE SECOND BIANCHI IDENTITY (SEE CHAPTER 7). SO WE'RE MOSTLY USING IDENTITIES FROM COMPLEX GEOMETRY, KAHLER GEOMETRY COMES IN ONLY FROM THE SYMMETRY OF THE LOWER INDICES OF THE CHRISTOFFEL SYMBOLS
 as well as the Killing equation in the form
\begin{equation}
g_{j\ov{k}}\nabla_{\ov{l}}\ov{\he^k_a}+g_{k\ov{l}}\nabla_j\he^k_a=0.
\end{equation}
We have also used the identity \eqref{Cmapequation} to arrive at the form of $\delta S_B$ given in \eqref{dSB}.

In order to restore supersymmetry at the boundaries, we need to choose an appropriate set of boundary conditions on the fields, and these conditions must themselves be supersymmetric. %with this set itself being invariant under supersymmetry.
 In fact, only an $\mathcal{N}=2$ subset of the four supersymmetries can be preserved at the boundaries, because of the following reasons.  Translation symmetry on the worldsheet is broken at the boundaries, where the worldsheet momentum is no longer conserved. 
Since the $\mathcal{N}=(2,2)$ supersymmetry algebra includes
\begin{equation}
\{Q_{\pm},\overline{Q}_{\pm}\}=H\pm P,
\end{equation}
the previous statement implies that on an open worldsheet, some of the supersymmetries are broken at the boundaries.
In particular, only certain linear combinations of the original supersymmetries are preserved at the boundaries, i.e., those whose algebra do not include the worldsheet momentum. There are two such combinations (\cite{hori2003mirror}, Chapter 39)
\begin{equation}\label{atype}
\begin{aligned}[c]
Q_A=\overline{Q}_+ +e^{i\beta}Q_-,
\end{aligned}
\\
\begin{aligned}[c]
\textrm{   }\\
\textrm{   }\\
\end{aligned}
\\
\begin{aligned}[c]
\textrm{   }\\
\textrm{   }\\
\end{aligned}
\\
\begin{aligned}[c]
Q^\dagger_A=Q_++e^{-i\beta}\overline{Q}_-\end{aligned},
\end{equation}
or
\begin{equation}\label{btype}
\begin{aligned}[c]
Q_B=\overline{Q}_+ +e^{i\beta}\overline{Q}_-,
\end{aligned}
\\
\begin{aligned}[c]
\textrm{   }\\
\textrm{   }\\
\end{aligned}
\\
\begin{aligned}[c]
\textrm{   }\\
\textrm{   }\\
\end{aligned}
\\
\begin{aligned}[c]
Q^\dagger_B=Q_++e^{-i\beta}Q_-\end{aligned},
\end{equation}
where $\beta$ is a real parameter, and these satisfy
\begin{equation}\label{qmalgebra}
\begin{aligned}
%Q^2=(Q^\dagger)^2&=0,\\
\{Q,Q^\dagger\}&=2H,
\end{aligned}
\end{equation}
%IMPORTANT NOTE: IN THE GAUGED CASE, Q^2 can be not equal to zero, or i think IS not equal to zero that's why i took Q^2=0 out
which is in fact the supersymmetry algebra of supersymmetric quantum mechanics. %on $X$. 
These combinations are known as A-type and B-type supersymmetry.\footnote{The supercharges $Q_A$ and $Q_B$ also happen to correspond to the scalar supercharges preserved in the A-twisted and B-twisted topological sigma models, when $\beta=0$.} 

%Note that our In the following sections, we shall attempt to find A-type and B-type boundary conditions for GNLSMs with \it{multiple} $U(1)$ gauge groups, with toric target spaces. These shall be referred to as quiver GNLSMs, 
 
Now, note that one can further generalize the action (\eqref{action}+\eqref{FIterm}) by considering \textit{quiver} GNLSMs, i.e., GNLSMs gauged by a direct product $G_1\times G_2\times G_3\ldots$ . %This corresponds to $G=G_1\times G_2...$ in the above formulae. 
This corresponds to having several copies of (\eqref{gaugeaction}+\eqref{FIterm}+\eqref{theta}), each corresponding to one gauge group $G_i$ together with its own coupling constant $e_i$, as well as coupling the matter in (\eqref{matteraction}+\eqref{bfield}) to all of these gauge groups.
%In particular, there will be each $G_i$ to have its own coupling $e_i$. 
In Sections 4 and 5 of this paper, we shall focus on finding the A-type and B-type supersymmetric boundary conditions for quiver {\it abelian} GNLSMs on toric manifolds, that is, GNLSMs with gauge group $G=U(1)^N$. These boundary conditions will ensure that \eqref{dSmatw}, \eqref{dSgauthe} and \eqref{dSB} vanish, restoring A-type or B-type supersymmetry at the boundaries. Furthermore, we will identify boundary interactions compatible with these conditions. These boundary conditions and boundary interactions shall correspond to equivariant generalizations of D-branes, which we shall refer to as equivariant A-branes and equivariant B-branes.

Before ending this section, we note that the action $S+S_r$ (\eqref{action} + \eqref{FIterm}) and supersymmetry variations \eqref{dSmatw}, \eqref{dSgauthe} and \eqref{dSB} %above 
reduce to the familiar formulae given in \cite{HIV} for GLSMs and NLSMs with boundary, in the appropriate limits. We can recover the $\mathcal{N}=(2,2)$ NLSM by setting the group $G$ to be trivial. Alternatively, by choosing $X=\mathbb{C}^N$, we find the usual formulae for a $\mathcal{N}=(2,2)$ GLSM, with the gauge group $G$ being an isometry group of $\mathbb{C}^N$. For example, the $U(1)^N$ model with target space $\C^N$, i.e., an abelian quiver GLSM, corresponds to the action $S+S_r$ with the $B$-field and all Lie algebra commutators vanishing, the moment map
%\footnote{As explained in page 22 of \cite{mcduff1998introduction}, for a single $U(1)$ group, this moment map in fact corresponds to the Hamiltonian for an $N$-dimensional simple harmonic oscillator (SHO), whereby $\C^N$ is the phase space. The expression \eqref{momentmapequation} is nothing but Hamilton's equations of motion. For example, a particular constant energy Hamiltonian flow in the phase space of a 1D SHO is just a circle, which is the group manifold of $U(1)$.}
%IMPORTANT NOTE: MUST PUT THE ABOVE FOOTNOTE IN THESIS
%\begin{equation}
%\mu^a=-\frac{1}{2}(\sum^{\text{dim }X}_iQ^a_i|\phi^i|%^2-r^a)
%\end{equation}
%\begin{equation}
%\mu^a=-(\sum^{N}_iQ^a_i|\phi^i|^2-r^a)
%\end{equation}
\begin{equation}
\mu_a=-(\sum^{N}_i \mathcal{Q}_{ia}|\phi_i|^2),
\end{equation}
%I think avoid writing moment map for now. Just write the following 
and the Killing vector fields
\begin{equation}
\begin{aligned}
\he_a^i&=i\mathcal{Q}_{ia}\phi_i\\
\ov{\he}_a^{\ov{\imath}}&=-i\mathcal{Q}_{ia}\ov{\phi}_i
\end{aligned}
\end{equation}
where we have chosen the flat metric $g_{i\ov{\jmath}}d\phi^{i}\otimes d\ov{\phi}^{\ov{\jmath}}=\delta_{ij}d\phi_{i}\otimes d\ov{\phi}_{j}$, and where $\mathcal{Q}_{ia}$ are the $U(1)$ charges of $\phi_i$ with $a=1,\ldots,N$.

\section{GNLSMs from GLSMs and Mirror Symmetry}
It is well-known that $\mathcal{N}=(2,2)$ NLSMs with target spaces being toric manifolds can be obtained in the IR limit of $\mathcal{N}=(2,2)$ quiver abelian GLSMs \cite{WittenPhases}. Hori and Vafa \cite{HoriVafa} made use of this to prove the mirror symmetry of manifolds with non-negative first Chern class in terms of Landau-Ginzburg theories. 
  This proof of mirror symmetry was later applied to worldsheets with boundaries, whereby the Landau-Ginzburg mirrors of B-branes \cite{HIV} and A-branes \cite{Hori1} were found.

As shown by Baptista \cite{Baptista2}, it is also possible to obtain quiver abelian GNLSMs on closed worldsheets with toric target spaces by taking a different limit of quiver abelian GLSMs. Moreover, Baptista found %using the mirror symmetry of Hori and Vafa, 
 the mirror Landau-Ginzburg theories of these GNLSMs. This then suggests a natural generalization of Baptista's proof to worldsheets with boundaries, in order to find equivariant A-branes and B-branes in abelian GNLSMs, as well as the Landau-Ginzburg mirrors of these branes. We shall pursue this line of thought in Sections 4 and 5.

%Before doing so, let us review Hori and Vafa's proof of mirror symmetry 
%between NLSMs and Landau-Ginzburg models
% as well as Baptista's generalization of mirror symmetry for GNLSMs. 
%We shall assume that we have a closed worldsheet, for the time being. 
%We shall not specify whether the worldsheet is open or closed, but we shall point out the step where the boundary affects the analysis.
%We shall assume a closed worldsheet for now, but we shall point out the steps where presence of boundaries would affect the analysis.
Before doing so, let us review %Hori and Vafa's proof of mirror symmetry for closed manifolds,
%between NLSMs and Landau-Ginzburg models
% as well as
  Baptista's generalization of mirror symmetry for GNLSMs on closed manifolds. In superfield language, the action of a $ U(1)^N$-GLSM with target space $\C^N$ is 
\begin{equation}\label{GLSMsuper}
\begin{aligned} 
S_{\rm GLSM} \ = \ &\frac{1}{2\pi}\int \txd^2x\int \dd^4 \theta \ \bigg\{  \sum_{j=1}^{N} \ov{\Phi}_j  \Big( e^{{\, \hat{Q}_j^{b} \,\hat{V}_b} + {\, \til{Q}^c_j\, \til{V}_c}}\Big) 
\Phi_j \ - \  \sum_{b=1}^{N-k} \Big( \frac{1}{2\hae_b^2}\ov{\hat{\Sigma}}_b \, \hat{\Sigma}_b \Big)   - \sum_{c=1}^{k} \Big(\frac{1}{2\til{e}_c^2}\ov{\til{\Sigma}}_c \, \til{\Sigma}_c \Big)   \bigg\}\  \\ 
&
 +\ \frac{1}{2\pi} \frac{1}{2} \int \txd^2x\Big(\int \dd^2 \tilde{\theta}\  (-\hat{t}^b\, \hat{\Sigma}_b  - \til{t}^c\, \til{\Sigma}_c)  + {\rm c.c.} \Big)  \ ,  
\end{aligned}  
\end{equation}
where $U(1)^N=U(1)^{(N-k)}\times U(1)^k$, with the indices $b=1,\ldots,N-k$ and $c=1,\ldots,k$.
%; and where $\hat{t}^b=\hat{r}^b-i
%\hat{\theta}^b$ and  $\widetilde{t}^c=\til{r}^c-i\til{\theta}^c$ are the usual combinations of FI parameters and theta angles. 
%According to Hori and Vafa \cite{HoriVafa,hori2003mirror} 
The mirror of this theory is the following Landau-Ginzburg sigma model with twisted chiral superfields $Y^i$ (whose imaginary parts are periodic, with period $2\pi$) and action
%In the present section, we wish to review Baptista's proof of obtaining GNLSMs as a limit of GLSMs, as well as their Landau-Ginzburg mirrors.
\begin{equation}\label{mirrorsuper}
\begin{aligned}
S_{\rm dual} \ = &\frac{1}{2\pi}\int \txd^2x\int \dd^4 \theta\  \bigg\{ -\frac{1}{2} \sum_{j=1}^{N} (Y^j + \ov{Y}^j){\rm log}(Y^j + \ov{Y}^j) \   - \  \sum_{b=1}^{N-k} \Big( \frac{1}{2\hae_b^2}\ov{\hat{\Sigma}}_b \, \hat{\Sigma}_b \Big)   - \sum_{c=1}^{k} \Big(\frac{1}{2\til{e}_c^2}\ov{\til{\Sigma}}_c \, \til{\Sigma}_c \Big)   
\bigg\}    \\
&+\frac{1}{2\pi}\frac{1}{2} \int \txd^2x\bigg\{ \int \dd^2 \tilde{\theta}\  \: \Bigg(\Big(\hat{Q}_j^b Y^j - \hat{t}^b \Big)\hat{\Sigma}_b\ +\  \Big(\til{Q}^c_j Y^j - \til{t}^c \Big) \til{\Sigma}_c
\ +\ \sum_{j=1}^{N} e^{-Y^j} \Bigg)\ +\ {\rm c.c.} \bigg\}  . 
\end{aligned}
\end{equation} 
% Taking quantum effects into account, the target space for this Landau-Ginzburg sigma model is the algebraic torus ${(\C^\times)}^N$ \cite{HoriVafa}.  
%by considering the target space coordinates to be $e^{Y^i}$.E
%\footnote{In fact, it is possible to show that the real part of $Y^j$ is positive-definite, implying a boundary in the target space. However, quantum effects remove this boundary via field renormalization (\cite{hori2003mirror}, page 467). }
%In deriving the duality between the two mirror theories, it can be shown that
It may seem from the logarithm in this action that the real part of $Y^j$ is positive-definite, implying a boundary in the target space where the metric becomes singular. However, quantum effects remove this boundary via the field renormalization $Y_0^j=Y^j+\textrm{log}(\Lambda_{UV}/\mu)$  (\cite{hori2003mirror}, page 467).\footnote{The field renormalization of $Y^i$ arises due to the presence of the twisted superpotential in the action \eqref{mirrorsuper}, which can be written (modulo the exponential term) as 
\begin{equation}\nonumber
\til{W}=\sum_a^N\Big(\mathcal{Q}_{ia} Y^i_{ 0}- {t}_{ a0} \Big){\Sigma}_a
\end{equation}
where $Y_{ i0}$ and ${t}_{ a0}=r_{a0 }-i\theta_{a0}$ are bare quantities. Taking into account the renormalization of the FI parameter ($r_{a0}=r_a+\sum_i^N\mathcal{Q}_{ia}\textrm{log}(\Lambda_{UV}/\mu)$)
%$=\varrho_{ i0}-i\vartheta_{ i0}+\theta^+\ov{\chi}_{ i+0}+\ov{\theta}^-\chi_{ i-0}+E_{i0}$ and ${t}_{ a0}=r_{a0 }-i\theta_{a0}$ are bare quantities.
% Taking into account \eqref{qeffect1}
\cite{hori2003mirror} (where $\Lambda_{UV}$ is the UV cutoff, and $\mu$ is a finite energy scale), in order for the superpotential to be finite, we must renormalize the bare fields $Y_0^j$ as $Y_0^j=Y^j+\textrm{log}(\Lambda_{UV}/\mu)$. }
The corresponding K\"ahler metric for this renormalized field is
%IMPORTANT NOTE= MINUS SIGN IN DEFINITION OF METRIC IS BECAUSE WE THE MIRROR ACTION IS IN TERMS OF TWISTED CHIRAL SUPERFIELDS, SEE 12.63 OF MIRROR SYMMETRY BOOK
%\footnote{Here, $y^i$ is the lowest component of $Y^i$, $\Lambda_{UV}$ is the UV cutoff, and $\mu$ is a finite energy scale.}
\begin{equation}
\dd s^2=\sum_{i=1}^N\frac{|\dd y^i|^2}{
2(2\textrm{log}(\Lambda_{UV}/\mu)+y^i+\overline{y}^i)}\approx
\frac{1}{ 4\textrm{log}(\Lambda_{UV}/\mu)}\sum_{i=1}^N\,|\dd y^i|^2
\label{Kmet0}
\end{equation}
(where $y^i$ is the lowest component of $Y^i$), which becomes flat in the continuum limit $\Lambda_{UV}\rightarrow \infty$. Taking into account the periodicity of the imaginary part of $Y^i$, this implies that the target space for this Landau-Ginzburg sigma model is the algebraic torus ${(\C^\times)}^N$ \cite{HoriVafa}.  

The duality between these actions is shown as follows. The GLSM action \eqref{GLSMsuper} can be obtained by integrating out $Y^j$ from the following action 
\begin{equation}\label{masterAction}
\begin{aligned} 
S_{0} \ = \ &\frac{1}{2\pi}\int \txd^2x\int \dd^4 \theta \ \bigg\{  \sum_{j=1}^{N}   \Big( e^{{\, \hat{Q}_j^b \,\hat{V}_b} + {\, \til{Q}^c_j\, \til{V}_c+B^j}} 
 \ - \frac{1}{2}(Y^j+\ov{Y}^j)B^j\Big)\\ 
&-  \  \sum_{b=1}^{N-k} \Big( \frac{1}{2\hae_b^2}\ov{\hat{\Sigma}}_b \, \hat{\Sigma}_b \Big)   - \sum_{c=1}^{k} \Big(\frac{1}{2\til{e}_c^2}\ov{\til{\Sigma}}_c \, \til{\Sigma}_c \Big)  \bigg\}\  
 +\ \frac{1}{2\pi}\frac{1}{2} \int \txd^2x\Big(\int \dd^2 \tilde{\theta}\  (-\hat{t}^b\, \hat{\Sigma}_b  - \til{t}^c\, \til{\Sigma}_c ) + {\rm c.c.} \Big)  \ .
\end{aligned}  
\end{equation}
Alternatively, one can integrate out $B^j$, to obtain \eqref{mirrorsuper}, modulo the term
\begin{equation}
\sum_{j=1}^{N} e^{-Y^j}+c.c..
\end{equation}
This term appears when we take into account quantum effects, and is generated by 
%quantum effects due to 
vortices (\cite{hori2003mirror}, page 469).
The $\mathcal{N}=(2,2)$ supersymmetric actions \eqref{GLSMsuper} and \eqref{mirrorsuper} are mirror theories, since the former is in terms of chiral superfields and the latter is in terms of twisted chiral superfields. 
Furthermore, by comparing the different expressions one obtains for $B^j$ when deriving the two mirror theories, it can be shown that
\begin{equation}\label{fieldduality}
Y^j+\overline{Y}^j=2\overline{\Phi}_j e^{ \hat{Q}^b_{j}\hat{V}_b +\til{Q}^c_{j}\til{V}_c}\Phi_j,
\end{equation}
which is an explicit relationship between the fields of the mirror theories.
%\footnote{It may seem from this expression that real part of $Y^j$ is positive-definite, implying a boundary in the target space. However, quantum effects remove this boundary via field renormalization (\cite{hori2003mirror}, page 467). } 
%This expression shall be useful for us in the following sections. 
%It shall be important to keep in mind that mirror symmetry between these theories stems from T-duality on the phase of the charged chiral superfields $\Phi_i$ \cite{HoriVafa}.

%, i.e., we have assumed a closed worldsheet. 
%These boundary terms 
%The boundary terms which arise in the proof of mirror symmetry for open worldsheets can be removed by choosing appropriate boundary conditions, as we shall see in in Sections 4 and 5, when we use the above technique to find the mirrors of equivariant branes.

%The boundary terms which arise in the proof of mirror symmetry for open worldsheets can be removed by choosing appropriate boundary conditions \cite{hori2003mirror}, and the relationship \eqref{fieldduality} still holds in this case. Moreover, as we shall see in in Sections 4 and 5 when we use the above technique to find the mirrors of equivariant branes, one may also generalize the above duality to include a boundary action in the GLSM.   

%The presence of boundaries can be taken into account in Sections 4 and 5, when we use the above technique to find the mirrors of equivariant branes. 

Note that we have neglected the boundary terms which appear when integrating out $Y^j$, as well as the boundary terms that come about %in the derivation of  
when deriving \eqref{mirrorsuper} from \eqref{masterAction}. %IMPORTANT NOTE, MORE BOUNDARY TERMS ARE GENERATED WHEN WE EXPAND THE SUPERFIELDS TRY TO GET THE STANDARD KINETIC TERMS
The above technique of proving mirror symmetry can be generalized to the situation whereby the worldsheet has boundaries, where the fields obey boundary conditions and where additional boundary interactions could occur. We shall see this in Sections 4 and 5, when we use the above technique to find mirrors of abelian equivariant branes.

To obtain a GNLSM from \eqref{GLSMsuper}, we take the limit where $\hat{e}_b\rightarrow \infty$. The $\hat{\Sigma}_b$ kinetic terms vanishes, and the remaining fields belonging to $\hat{V}_b$ become auxiliary fields, and are integrated out.
The resulting sigma model has
%toric manifold 
$\C^N//U(1)^{N-k}$ as target space, but is still gauged, since the vector superfields  $\til{V}_c$ are still present in the action.   However, note that in order to obtain a K\"ahler target space with $k$ complex dimensions, the FI parameters $\hat{r}_b$ must be within a K\"ahler cone. In this way, we obtain the $\mathcal{N}=(2,2)$ $U(1)^k$-gauged nonlinear sigma model (GNLSM) with K\"ahler $\C^N//U(1)^{N-k}$ target space.
%where $\C^N//U(1)^{N-k}$ corresponds to a particular K\"ahler toric manifold.
%In other words, we have obtained the $U(1)^k$-gauged nonlinear sigma model (GNLSM) with $\C^N//U(1)^{N-k}$ target space.
% and remove the remaining terms containing the vector superfield $\til{V}$ and superfield strength $\til{\Sigma}$. 

Taking the same limit $\hae_b\rightarrow \infty$ in the dual Landau-Ginzburg sigma model \eqref{mirrorsuper} makes the $\hat{\Sigma}_b$ kinetic terms vanish.  $\hat{\Sigma}_b$ is then an auxiliary superfield, which we can integrate out to impose the constraints 
\begin{equation}
\hat{Q}_j^b Y^j - \hat{t}^b=0.
\end{equation}
These constraints have the solution
\begin{equation}\label{mirrorparamet}
Y^j=\hat{s}^j+\sum^k_{c=1}v_c^j\Theta^c,
\end{equation}
where $\Theta^1,\ldots,\Theta^k$ are new twisted chiral fields, the complex constants $\hat{s}^1,\ldots,\hat{s}^N\in \C$ are any solution of the algebraic relation $\hat{Q}^b_j\hat{s}^j=\hat{t}^b$, and $v_c^j$ are $N$ primitive vectors $v^1,\ldots,v^{N}\in \Z^{k}$ (which generate the regular fan associated with $\C^N//U(1)^{N-k}$) that span $\Z^k$ and satisfy $\sum^N_j \hat{Q}_j^b \: v^j \ =\  0$.
%follow page 4 of Baptista as well as page 52 of Hori Vafa.
Thus, the  $\hae_b\rightarrow \infty$ limit gives the following $U(1)^k$-gauged Landau-Ginzburg sigma model
%Finally, inserting (\ref{3.4}) into the limit of $L_{\rm dual}$ and denoting by $\langle \cdot , \cdot \rangle$ the 
%canonical inner product on $\R^k$, we get that as $e' \rightarrow \infty$ the lagrangian $L_{\rm dual}$ reduces to
\begin{equation}\label{GNLSMmirror}
\begin{aligned}
\hat{S}_{\rm dual} \ = &\ \frac{1}{2\pi} \int \txd^2x\int \dd^4 \theta\  \Big[  -\frac{1}{2} \sum_{j=1}^{N} (\hat{s}^j + \ov{\hat{s}}^j+\langle v^j,\Theta+\ov{\Theta}\rangle){\rm log}(\hat{s}^j + \ov{\hat{s}}^j+\langle v^j,\Theta+\ov{\Theta}\rangle)
- \sum_{c=1}^{k} \Big(\frac{1}{2\til{e}_c^2}\ov{\til{\Sigma}}_c \, \til{\Sigma}_c \Big)\Big]     \\
&+\, \frac{1}{2\pi}\frac{1}{2}\int \txd^2x \bigg\{ \int \dd^2 \tilde{\theta}\ \Bigg( \langle \til{\Sigma} , \til{Q}_j  \rangle \Big(\langle v^j , \Theta \rangle  
+  \hat{s}^j \Big)\; -\; \langle \til{\Sigma}  ,  \til{t} \rangle\;  +\;  \sum_{j=1}^{N} e^{- \langle v^j , \Theta \rangle - \hat{s}^j}  \ 
\Bigg)+\  {\rm c.c.} \bigg\} ,
\end{aligned}
\end{equation}
where $\langle \cdot , \cdot \rangle$ is the 
canonical inner product on $\R^k$.
%This lagrangian is expected to be dual to the $e' \rightarrow \infty$ limit of $L$, i.e. dual to the lagrangian
%of the gauged sigma-model with target $X$ and group $T^k$. 
This gauged Landau-Ginzburg theory has the holomorphic twisted superpotential 
\begin{equation}\label{holtwistsup} 
\til{W}(\Theta,\til{\Sigma})\ = \ \langle \til{\Sigma} , \til{Q}_j  \rangle \Big(\langle v^j , \Theta \rangle  +  \hat{s}^j \Big) 
\ -\; \langle \til{\Sigma}  ,  \til{t} \rangle\;   \ +\  \sum_{j=1}^{N} e^{- \langle v^j , \Theta \rangle - \hat{s}^j} \ .
\end{equation}
The parametrization \eqref{mirrorparamet} implies that the target space of this gauged Landau-Ginzburg sigma model, which is mirror to the GNLSM on the $k$-complex dimensional toric manifold $\C^N//U(1)^{N-k}$, is the $k$-complex dimensional algebraic torus, ${(\C^\times)}^k$.% with coordinates $e^{\Theta}$.

It is important to recognize that even though 
the mirror action 
%\eqref{mirrorsuper} and its $e_a\rightarrow \infty$ limit 
\eqref{GNLSMmirror} has $U(1)^k$ gauge symmetry, 
its matter kinetic terms do not contain the vector superfields $\til{V}^c$, and therefore, the components of the twisted chiral superfields $\Theta^c$ are not coupled to those in the vector superfields, except in the superpotential. %interaction terms
 In other words, the matter fields in the gauged Landau-Ginzburg theory \eqref{GNLSMmirror}, including the scalar fields which parametrize the ${(\C^\times)}^k$ target space, are \textit{neutral} under the $U(1)^k$ gauge symmetry.

%However, note that the kinetic terms for $Y^j$ are not coupled to $V$

The above calculation, due to Baptista, is essentially the same as Hori and Vafa's. The only difference in Hori and Vafa's case is that they start with a $U(1)^{N-k}$ GLSM with $\C^N$ target space instead of one with $U(1)^{N-k}\times U(1)^k$ gauge symmetry, and in the $\hae_b\rightarrow \infty$ limit, the %$U(1)^{N-k}$ 
gauge symmetry completely disappears in both the GLSM and its mirror, leaving us with an NLSM with target space $\C^N//U(1)^{N-k}$ mirror to a (non-gauged) Landau-Ginzburg sigma model with target space ${(\C^\times)}^k$ and holomorphic twisted superpotential 
\begin{equation}
\til{W} \ = \  \sum_{j=1}^{N} e^{- \langle v^j , \Theta \rangle - \hat{s}^j}.
\end{equation}
Put differently, the K\"ahler quotient $\C^N//U(1)^{N-k}$ has a $U(1)^k$ isometry which descends from the %unquotiented
residual $U(1)^k$ isometry of $\C^N$ (\cite{hori2003mirror}, page 362), and this can manifest either as a local or global symmetry in a $\C^N//U(1)^{N-k}$ sigma model. Baptista started with a GLSM fully gauged by the $U(1)^N$ abelian isometry group of $\C^N$, and ended up with a local $U(1)^k$ symmetry when taking $\hae_b\rightarrow \infty$, whereas Hori and Vafa started with a GLSM gauged by only $U(1)^{N-k}$, with the remaining $U(1)^k$ isometry forming a global symmetry of the GLSM, which descends to a global symmetry of the NLSM in the $\hae_b\rightarrow \infty$ limit.

Baptista's technique of obtaining GNLSMs from GLSMs is an extremely powerful one, as it allows us to obtain multiple GNLSMs from a single GLSM, by choosing which coupling constants we wish to send to infinity. This implies the equivalence of several GNLSMs with different K\"ahler target manifolds, as well as the equivalence of equivariant branes contained in these manifolds. %VERY AME BREAKING
 These branes will be the main objective of our study in the following sections. Furthermore, once a particular GNLSM is obtained, even its gauge group can be modified, by demoting some of its $U(1)$ gauge symmetries to global symmetries. These points shall be useful to keep in mind %for us in 
when reading the following sections, where we attempt to study abelian equivariant branes in as much generality as possible. 
This shall be achieved by generalizing Baptista's technique to obtain %for
 GNLSMs on open worldsheets with toric target spaces, while %and
  studying both boundary conditions and boundary interactions.
  
  \vspace{1.25cm}
\mbox{}\par\nobreak
\noindent
\textit{Explicit reduction of GLSMs to GNLSMs in the case with boundaries} 

%However, 
We have only discussed the method of obtaining GNLSMs  from GLSMs in superfield language for closed worldsheets thus far. 
We will now derive this reduction in component form for an \textit{open} worldsheet, so that we may eventually find explicit boundary actions and boundary conditions in GNLSMs. %\footnote{Hereon, we shall assume that the worldsheet is open.} 
% Integrating \eqref{GLSMsuper} over the fermionic directions in superspace, the GLSM action is
The $U(1)^N=U(1)^{N-k}\times U(1)^k$ GLSM action with boundaries is given explicitly as\footnote{It is important to keep in mind that the superfield action \eqref{GLSMsuper} is only equal to this action upon integration by parts, which give rise to boundary terms. However, we shall be concerned with this action, (which has the standard kinetic terms), as well as the GNLSMs we can obtain from it, and their mirrors. %When we use the proof of mirror symmetry in the following sections to find  the mirrors of equivariant branes, all boundary terms will be accounted for. %only refer to the explicit action \eqref{big1} in the remainder of the paper.
}%MAKE SURE TO CHANGE DERIVATIVES TO EXPLICIT FORM, TO AVOID CONFUSION WITH SUPERDERIVATIVES WHICH COME LATER  
\begin{align}
&S_{GLSM}\nonumber\\=&\frac{1}{2\pi}\sum_i^N\int \txd^2x \Big\{- D_{\mu}\overline{\phi}_i D^{\mu} \phi_i
+\frac{i}{2} \overline{\psi}_{-i}(\overleftrightarrow{D_0}+\overleftrightarrow{D_1})\psi_{-i} 
+\frac{i}{2} \overline{\psi}_{+i}
( \overleftrightarrow{D_0}-\overleftrightarrow{D_1})
\psi_{+i} \nonumber \\
&  - (\sum^N_a\mathcal{Q}_{ia}\si_a)(\sum^N_a\mathcal{Q}_{ia}\ov{\si}_a)\overline{\phi}_i
\phi_i- \sum^N_a\mathcal{Q}_{ia}(\overline{\sigma}_a\overline{\psi}_{+i}\psi_{-i}+\sigma_a
\overline{\psi}_{-i}\psi_{+i})  \nonumber\\ 
& -\sum^N_a i \mathcal{\mathcal{Q}}_{ia}\phi_i
(\overline{\lambda}_{-a}
\overline{\psi}_{+i}-\overline{\lambda}_{+a}\overline{\psi}_{-i} )-\sum^N_a i\mathcal{Q}_{ia}
\overline{\phi}_i(\psi_{-i}
\lambda_{+a}-\psi_{+i} \lambda_{-a}) +|F_i|^2  \Big\}\nonumber\\ 
&+{1 \over 2\pi}\sum^N_a{1 \over 2e_a^2}\int \txd^2x\Big\{(F_{01a})^2 
 - \partial_{\mu}\sigma_a \partial^{\mu} 
\overline{\sigma}_a   +(D_a)^2  +2e_a^2D_a(\sum_i^N \mathcal{Q}_{ia}\overline{\phi}_i\phi_i-r_a)\nonumber\\
&+ {i \over 2} \overline{\lambda}_{+a}
(\overleftrightarrow{\del_0}-\overleftrightarrow{\del_1})
\lambda_{+a} +{i \over 2} \overline{\lambda}_{-a}
(\overleftrightarrow{\del_0}+\overleftrightarrow{\del_1})
\lambda_{-a} +2e_a^2\theta_a F_{01a} \Big\}\pagebreak[3]  \nonumber \\=&\frac{1}{2\pi}\sum_i^N\int \txd^2x \Big\{- D_{\mu}\overline{\phi}_i D^{\mu} \phi_i
+\frac{i}{2} \overline{\psi}_{-i}( \overleftrightarrow{D_0}+\overleftrightarrow{D_1})\psi_{-i} 
+\frac{i}{2} \overline{\psi}_{+i}
( \overleftrightarrow{D_0}-\overleftrightarrow{D_1})
\psi_{+i} \nonumber \\ 
&  - (\sum^{N-k}_b \hat{Q}_{ib}\hat{\si}_b+\sum^{k}_c \til{Q}_{ic}\til{\si}_c)(\sum^{N-k}_b \hat{Q}_{ib}\ov{\hat{\si}}_b+\sum^{k}_c \til{Q}_{ic}\ov{\til{\si}}_c)\ov{\phi}_i
\phi_i\nonumber\\&- \sum^{N-k}_b \hat{Q}_{ib}(\overline{\hat{\sigma}}_b\overline{\psi}_{+i}\psi_{-i}+\hat{\sigma}_b
\overline{\psi}_{-i}\psi_{+i}) - \sum^{k}_c \til{Q}_{ic}(\overline{\til{\sigma}}_c\overline{\psi}_{+i}\psi_{-i}+\til{\sigma}_c
\overline{\psi}_{-i}\psi_{+i}) \nonumber\\ 
& -\sum^{N-k}_b i \hat{Q}_{ib}\phi_i
(\overline{\hat{\lambda}}_{-b}
\overline{\psi}_{+i}-\overline{\hat{\lambda}}_{+b}\overline{\psi}_{-i} )-\sum^{k}_c i \til{Q}_{ic}\phi_i
(\overline{\til{\lambda}}_{-c}
\overline{\psi}_{+i}-\overline{\til{\lambda}}_{+c}\overline{\psi}_{-i} )\nonumber\\&-\sum^{N-k}_b i\hat{Q}_{ib}
\overline{\phi}_i(\psi_{-i}\hat{\lambda}_{+b}-\psi_{+i} \lambda_{-b})-\sum^{k}_c i\til{Q}_{ic}
\overline{\phi}_i(\psi_{-i}\til{\lambda}_{+c}-\psi_{+i} \lambda_{-c})
+|F_i|^2  \Big\}\nonumber\\&+{1 \over 2\pi}\int \txd^2x \Big\{\sum^{N-k}_b (\hat{D}_b( \sum_i^N\hat{Q}_{ib}\ov{\phi}_i\phi_i-\hat{r}_b)+\hat{\theta}_b \hat{F}_{01b}) +\sum_c^{k}(\til{D}_c( \sum_i^N\til{Q}_{ic}\overline{\phi}_i\phi_i-\til{r}_c)+\sum^{k}_c\til{\theta}_c \til{F}_{01c})\Big\}\nonumber\\
&+{1 \over 2\pi}\sum^{N-k}_b{1 \over 2\hat{e}_b^2}\int \txd^2x\Big\{(\hat{F}_{01b})^2 
 - \partial_{\mu}\hat{\sigma}_b \partial^{\mu} 
\overline{\hat{\sigma}}_b   +(\hat{D}_b)^2  + {i \over 2} \overline{\hat{\lambda}}_{+b}
(\overleftrightarrow{\del_0}-\overleftrightarrow{\del_1})
\hat{\lambda}_{+b} +{i \over 2} \overline{\hat{\lambda}}_{-b}
(\overleftrightarrow{\del_0}+\overleftrightarrow{\del_1})
\hat{\lambda}_{-b}  \Big\}\nonumber\\
&+{1 \over 2\pi}\sum^{k}_c{1 \over 2\til{e}_c^2}\int \txd^2x\Big\{(\til{F}_{01c})^2 
 - \partial_{\mu}\til{\sigma}_c \partial^{\mu} 
\overline{\til{\sigma}}_c   +(\til{D}_c)^2  + {i \over 2} \overline{\til{\lambda}}_{+c}
(\overleftrightarrow{\del_0}-\overleftrightarrow{\del_1})
\til{\lambda}_{+c} +{i \over 2} \overline{\til{\lambda}}_{-c}
(\overleftrightarrow{\del_0}+\overleftrightarrow{\del_1})
\til{\lambda}_{-c}  \Big\},\label{big1}
 \end{align}
 where the covariant derivatives are 
\begin{equation}\label{covder}
\begin{aligned}
D_{\mu}\phi_i&=(\del_{\mu}+i\sum_b^{N-k} \hat{Q}_{ib}\hat{A}_{\mu b}+i\sum_c^{k} \til{Q}_{ic}\til{A}_{\mu c})\phi_i,\\
D_{\mu}\ov{\phi}_i&=(\del_{\mu}-i\sum_b^{N-k} \hat{Q}_{ib}\hat{A}_{\mu b}-i\sum_c^{k} \til{Q}_{ic}\til{A}_{\mu c})\ov{\phi}_i,\\
D_{\mu}\psi_{\pm i}&=(\del_{\mu}+i\sum_b^{N-k} \hat{Q}_{ib}\hat{A}_{\mu b}+i\sum_c^{k} \til{Q}_{ic}\til{A}_{\mu c})\psi_{\pm i},\\
D_{\mu}\ov{\psi}_{\pm i}&=(\del_{\mu}-i\sum_b^{N-k} \hat{Q}_{ib}\hat{A}_{\mu b}-i\sum_c^{k} \til{Q}_{ic}\til{A}_{\mu c})\ov{\psi}_{\pm i}.\\
\end{aligned}
\end{equation} 
In \eqref{big1}, we have first written the GLSM it its familiar form, and then split the terms containing vector multiplet components as in \eqref{GLSMsuper}. We shall now take the $\hat{e}_b\rightarrow\infty$ limit in \eqref{big1}, whereby the vector multiplet kinetic terms as well as the term $(\hat{D}_b)^2$ vanish. This means that all the components of the vector superfields $\hat{V}_b=\{\hat{A}_{\mu b},\hat{\si}_b,\hat{\lambda}_b,\hat{D}_b\}$ become auxiliary.
% and can be integrated out of the action via their equations of motion.

Consequently, the equations of motion of $\hat{A}_{\mu b}$ and $\hat{\si}_b$ give the following constraints on themselves\footnote{\label{15}To be precise, the equation of motion for $\hat{A}_{0 b}$ (which is the first equation in \eqref{generalv}) will be modified by a boundary term proportional to $\hat{\theta}_b$, unless appropriate boundary conditions and/or boundary terms are used. We shall assume that this is the case for now, and in the following sections we will study boundary actions whereby the $\hat{A}_{0 b}$ equation of motion in \eqref{generalv} is precise both in the bulk and at the boundaries.} 
%IMPORTANT NOTE: THE FOLLOWING CONSTRAINTS ARE N-k equations in N-k unknowns, so they can be solved (in most cases, see https://en.wikipedia.org/wiki/System_of_linear_equations).
%Equations of motion for $A_0$ and $A_1$.
\begin{equation}\label{generalv}	
\begin{aligned}
\sum^N_i\hat{Q}_{ib}[i(\ov{\phi}_iD_0\phi_i-\phi_iD_0\ov{\phi}_i)+\ov{\psi}_{-i}\psi_{-i}+\ov{\psi}_{+i}\psi_{+i}]=0\\
{\sum^N_i\hat{Q}_{ib}[i(\ov{\phi}_iD_1\phi_i-\phi_iD_1\ov{\phi}_i)-\ov{\psi}_{-i}\psi_{-i}+\ov{\psi}_{+i}\psi_{+i}]}=0,
\end{aligned}
\end{equation}
%Equations of motion for $\sigma$ and $\ov{\sigma}$.
\begin{equation}\label{generalsi}
\begin{aligned}
{\sum^N_{i}[-\hat{Q}_{ib}(\sum_b^{N-k}\hat{Q}_{id}\hat{\si}_d+\sum_c^k\til{Q}_{ic}\widetilde{\sigma}_c)\phi_i\ov{\phi}_i-\hat{Q}_{ib}\ov{\psi}_{+i}{\psi}_{-i}]}&=0\\
{\sum^N_{i}[-\hat{Q}_{ib}(\sum_d^{N-k}\hat{Q}_{id}\ov{\hat{\si}}_d+\sum_c^k\til{Q}_{ic}\ov{\til{\si}}_c)\phi_i\ov{\phi}_i-\hat{Q}_{ib}\ov{\psi}_{-i}{\psi}_{+i}]}&=0,
\end{aligned}
\end{equation}
%\begin{equation}
%\ov{\hat{\si}}_b=\frac{\sum^N_{i}[-\hat{Q}_{ib}\sum_c^k(\til{Q}_{ic}\ov{\til{\si}}_c)\phi_i\ov{\phi}_i-\hat{Q}_{ib}\ov{\psi}_{-i}{\psi}_{+i}]}{\sum^N_{j}\hat{Q}_{jb}^2\phi_j\ov{\phi}_j}
%\end{equation}
while integrating out the gauginos ($\hat{\lambda}_b$) gives the constraints
\begin{equation}\label{lamconstraint}
\sum_i^N \hat{Q}_{ib} \ov{\phi}_i\psi_{\pm i}=0, 
\end{equation}
and finally, integrating out $\hat{D}_b$ gives
\begin{equation}\label{Dconstraint}
\sum^N_i\hat{Q}_{ib}\ov{\phi}_i\phi_i-\hat{r}_b=0,
\end{equation}
for $b=1,\ldots, N-k$.\footnote{These constraints are consistent with the $\mathcal{N}=(2,2)$ supersymmetry of the GLSM action.}
 
In order to derive the explicit action for the $U(1)^k$-GNLSM with a {\it K\"ahler} target space $\C^N//U(1)^{N-k}$, we need to set the FI parameters $\hat{r}_b$ to be in a particular K\"ahler cone, before taking the $\hat{e}_b\rightarrow\infty$ limit.\footnote{In fact choosing the FI parameters $\hat{r}_b$ to be in a K\"ahler cone actually triggers spontaneous symmetry breaking of $U(1)^{N-k}$ via a Higgs mechanism. This can be seen by integrating out the auxiliary fields $\hat{D}_b$. Then, the $U(1)^{N-k}$ vector multiplets and the scalar fields transverse to the K\"ahler manifold defined by this K\"ahler cone become massive, with masses given in terms of $\hat{e}_b$. Then, taking the $\hat{e}_b\rightarrow\infty$ limit decouples these massive modes and gives us a theory of massless fields, without $U(1)^{N-k}$ symmetry. %IMPORTANT NOTE:ACTUALLY THE MASS OF PHI WILL BE IN TERMS OF ALL THE \hat{e}_b, I THINK IT WILL BE A SUM OVER b WITH EACH TERM PROPORTIONAL TO \hat{e}_b,
} Next, we need to find parametrizations for the scalar fields $\phi_i$ which satisfy \eqref{Dconstraint}, as well as parametrizations for $\psi_{\pm}$ which satisfy \eqref{lamconstraint}. Then, $\hat{A}_{\mu b}$ and $\hat{\si}_b$ must be integrated out of the action using \eqref{generalv} and \eqref{generalsi}. Finally, we need to replace the matter auxiliary field term
\begin{equation}
\sum_i^N|F_i|^2
\end{equation}
with the matter auxiliary field term of the GNLSM in \eqref{matteraction}. 

This procedure is simplest for the case of $N-k=1$, where \eqref{generalv} and \eqref{generalsi} reduce to 
%Equations of motion for $v_0$ and $v_1$.
\begin{equation}\label{generalv2}	
\begin{aligned}
\hat{A}_0=\frac{\sum^N_i\hat{Q}_i[i(\ov{\phi}_i\til{D}_0\phi_i-\phi_i\til{D}_0\ov{\phi}_i)+\ov{\psi}_{-i}\psi_{-i}+\ov{\psi}_{+i}\psi_{+i}]}{2\sum^N_j\hat{Q}_j^2|\phi_j^2|}\\
\hat{{A}}_1=\frac{\sum^N_i\hat{Q}_i[i(\ov{\phi}_i\til{D}_1\phi_i-\phi_i\til{D}_1\ov{\phi}_i)-\ov{\psi}_{-i}\psi_{-i}+\ov{\psi}_{+i}\psi_{+i}]}{2\sum_j\hat{Q}_j^2|\phi_j^2|}
\end{aligned}
\end{equation}
%Equations of motion for $\hat{\sigma}$ and $\ov{\hat{\sigma}}$.
and
\begin{equation}\label{generalsi2}
\begin{aligned}
\hat{\sigma}&=\frac{\sum^N_{i=1}[-\hat{Q}_i(\sum_c^k\widetilde{Q}_{ic}\widetilde{\sigma}_c)\phi_i\ov{\phi}_i-\hat{Q}_i\ov{\psi}_{+i}{\psi}_{-i}]}{\sum^N_{j=1}\hat{Q}_j^2\phi_j\ov{\phi}_j}\\
\ov{\hat{\sigma}}&=\frac{\sum^N_{i=1}[-\hat{Q}_i(\sum_c^k\widetilde{Q}_{ic}\ov{\til{\si}}_c)\phi_i\ov{\phi}_i-\hat{Q}_i\ov{\psi}_{-i}{\psi}_{+i}]}{\sum^N_{j=1}\hat{Q}_j^2\phi_j\ov{\phi}_j},
\end{aligned}
\end{equation}
where 
\begin{equation}
\begin{aligned}
\til{D}_{\mu}\phi_i&=(\del_{\mu}+i\sum_c^{k} \til{Q}_{ic}\til{A}_{\mu c})\phi_i,\\
\til{D}_{\mu}\ov{\phi}_i&=(\del_{\mu}-i\sum_c^{k} \til{Q}_{ic}\til{A}_{\mu c})\ov{\phi}_i.
\end{aligned}
\end{equation}
%Susy variation of the above. Must generalize the above to several unbroken $U(1)$ gauge groups.
%Equations of motion implied by integrating out gauginos.
%Let us understand how it works for the example of $\C P^{N-1}$. 
A good example is that of $X=\C P^{N-1}$, which corresponds to the quotient
\begin{equation}
\C^N//U(1)
\end{equation}
with charges $\hat{Q}_i=1$, and the FI parameter $\hat{r}>0$, which is the K\"ahler cone of $\C P^{N-1}$. Thus, we should begin with the GLSM \eqref{big1} with $N-k=1$, $\hat{Q}_i=1$ and $\hat{r}>0$. In this case, the constraint \eqref{Dconstraint} is
\begin{equation}\label{Dconstraint2}
\sum^N_i\ov{\phi}_i\phi_i=\hat{r},
\end{equation}
and is solved by 
%Parametrization for $CP^N$
\begin{equation}\label{para}
\begin{aligned}
\phi_i&=\frac{Z^i\sqrt{\hat{r}}e^{it}}{\sqrt{1+\sum_k^{N-1}|Z^k|^2}},\textrm{   }i=1,\ldots,N-1\\
\phi_N&=\frac{\sqrt{\hat{r}}e^{it}}{\sqrt{1+\sum_k^{N-1}|Z^k|^2}},
\end{aligned}
\end{equation}
where 
\begin{equation}\label{inhomogeneous}
Z^i=\frac{\phi_i}{\phi_N}
\end{equation}
correspond to the inhomogeneous coordinates which parametrize a local patch of $\C P^{N-1}$.
%IMPORTANT NOTE:\footnote{Note that in general, the ($U(1)^{N-k}$-invariant) inhomogeneous coordinates which locally parametrize a K\"ahler toric manifold $\C^N//U(1)^{N-k}$ take the form $Z^c=\prod_j\phi_j^{v^j_c}$, where $v^j_c$ are the corresponding  primitive vectors.%IMPORTANT NOTE, THIS FORM ALSO TELLS US THE GNLSM CHARGES, WHICH AS WE SAW APPEARED IN THE MIRROR THEORY } % IMPORTANT NOTE: in the thesis we can talk about the complex Hopf fibration
Furthermore, the fermionic constraint \eqref{lamconstraint} can be solved by 
\begin{equation}\label{lamconstraint2}
\begin{aligned}
\psi_{i\pm}&=\frac{\psi^{Z_i}_{\pm}\sqrt{\hat{r}}e^{it}}{(1+\sum_k^{N-1}|Z^k|^2)^{\frac{1}{2}}}-\frac{Z^i(\sum_l^{N-1}\psi^{Z^l}_{\pm}\ov{Z}^l)\sqrt{\hat{r}}e^{it}}{(1+\sum_k^{N-1}|Z^k|^2)^{\frac{3}{2}}},\textrm{   }i=1,\ldots,N-1\\
\psi_{N\pm}&=\frac{-\sum_j^{N-1}(\psi_{\pm}^{Z^j}\overline{Z}^j)\sqrt{\hat{r}}e^{it}}{(1+\sum_k^{N-1}|Z^k|^2)^{\frac{3}{2}}}
\end{aligned}
\end{equation}
where $\psi^{Z^i}_{\pm}$ correspond to Grassmann-valued vector fields defined on the aforementioned patch of $\C P^{N-1}$. 

Using \eqref{generalv2} and \eqref{para}, we can show that the terms containing only bosonic fields in the scalar kinetic term of the GLSM ($-\sum_i^N D_{\mu}\phi_i D^{\mu}\phi_i$) are
%IMPORTANT NOTE:MADE THE CHANGE ON 3RD APRIL 2017, SEE NOTE '3-A34', EVEN FOR THE REDUCTION FROM GLSM TO NLSM, THE GLSM SCALAR KINETIC TERM STILL CONTAINS TERMS WHICH HAVE FERMIONIC FIELDS
%In the cp1 case it is
%where 
%cpn case we get this metric (must put in covariant  derivatives, and point out killing vector fields)

\begin{equation}\label{cpn}
\boxed{%-\sum_i^N D_{\mu}\phi_i D^{\mu}\phi_i=
-\frac{\hat{r}(\sum_j^{N-1}\del^A_{\mu}Z^j\del^{A\mu}\ov{Z}^j)}{(1+\sum_i^{N-1}|Z^i|^2)}+\frac{\hat{r}(\sum_j^{N-1}Z^j\del^A_{\mu}\ov{Z}^j)(\sum_k^{N-1}\ov{Z}^k\del^{A\mu}Z^k)}{(1+\sum_i^{N-1}|Z^i|^2)^{2}},}
\end{equation}
where 
\begin{equation}\label{cpncovariant}
\del^A_{\mu}Z^j=\del_{\mu}Z^j+i\sum_c^{N-1}(\til{Q}_{jc}-\til{Q}_{Nc})\til{A}_{\mu c}Z^j,\textrm{     (No sum over $j$)}.
\end{equation}
%On the right hand side of
In \eqref{cpn}, we find the scalar kinetic term given in \eqref{matteraction} for $X=\C P^{N-1}$ and $G=U(1)^{N-1}$, with the metric on $\C P^{N-1}$ being the standard Fubini-Study metric.\footnote{\label{18}Here, the FI parameter plays the role of the modulus which parametrizes the size of $\C P^{N-1}$.}
Comparing the covariant derivative \eqref{cpncovariant} with the general form given in \eqref{2.13}, we find that the holomorphic Killing vector fields corresponding to the $U(1)^{N-1}$ isometry on $\C P^{N-1}$ are given by\footnote{The fact that the $U(1)^{N-1}$ charges of the inhomogeneous coordinates $Z^j$ are given by ($\til{Q}_{jc}-\til{Q}_{Nc}$) can also be deduced from \eqref{inhomogeneous}. 
%IMPORTANT NOTE: Likewise, for a general K\"ahler  toric manifold $\C^N//U(1)^{N-k}$ with inhomogeneous coordinates $Z^j=\prod_i\phi_i^{v^i_j}$, the $U(1)^k$ charges can be deduced to be $\sum_i^N\til{Q}_{ic}v^i_j$.
}
\begin{equation}
\he^j_c=i(\til{Q}_{jc}-\til{Q}_{Nc})Z^j.
\end{equation}
%IMPORTANT NOTE THE FOLLOWING ADDED LATER FROM HERE
The term proportional to $\til{D}_c$ in the GLSM is found to contain the moment map for the $U(1)^{N-1}$ isometry of $\CPN$,
\begin{equation}\label{cpN-1momentmap}
\til{\mu}_c=\frac{-\hat{r}(\sum_i^{N-1}\til{Q}_{ic}|Z^i|^2+\til{Q}_{Nc})}{(1+\sum_k^{N-1}|Z|^2)},
\end{equation}
via \eqref{para}, and thereby we retrieve the moment map term of \eqref{gaugeaction}. Similarly, using \eqref{generalv2} and \eqref{para}, we find that the $\hat{\theta}$ term gives rise to the the $B$-field and $C$-field terms of the GNLSM, with
\begin{equation}
B=-\frac{\hat{\theta}}{\hat{r}}\omega,
\end{equation}
(where $\omega$ is the Fubini-Study K\"ahler form) and
%while the $C$-field is proportional to the moment map
\begin{equation}\label{cpropmu}
C_c=-\frac{\hat{\theta}}{\hat{r}}\til{\mu}_c,
\end{equation}
as well as the boundary term 
\begin{equation}\label{extra}
-\frac{1}{2\pi}\int_{\del \Sigma}\frac{\hat{\theta}}{2\hat{r}}\sum_i^{N}(\ov{\psi}_{-i}\psi_{-i}+\ov{\psi}_{+i}\psi_{+i}),
\end{equation}
where we have maintained its GLSM form for convenience. %fermionic fields. 
We shall comment more on this term below. 
%UNTIL HERE

%Analogously
In a similar manner, we may continue the procedure explained below \eqref{Dconstraint} with the help of \eqref{generalv2}, \eqref{generalsi2},  \eqref{para} and  \eqref{lamconstraint2} in order to obtain the complete GNLSM action given in \eqref{action}+\eqref{FIterm} for $X=\C P^{N-1}$ and $G=U(1)^{N-1}$. For conciseness, we shall only write out the resulting action for $N=2$, i.e., the $U(1)$-GNLSM with $\C P^1$ target:
\begin{equation}
\begin{aligned}
S=&\frac{1}{2\pi}\int_{\Sigma} \txd^2x \Big\{-\hat{r}\frac{\partial^A_\mu Z\partial^{A\mu} \ov{Z}}{(1+|Z|^2)^2}+\frac{i}{2}\hat{r}\frac{\ov{\psi}^{\ov{Z}}_+\overleftrightarrow{\nabla}^A_-\psi^Z_+}{(1+|Z|^2)^2}+\frac{i}{2}\hat{r}\frac{\ov{\psi}^{\ov{Z}}_-\overleftrightarrow{\nabla}^A_+\psi^Z_-}{(1+|Z|^2)^2}\\
&-\frac{\hat{r}|\til{\si}|^2|\til{e}^Z|^2}{(1+|Z|^2)^2}+\frac{i\hat{r}}{(1+|Z|^2)^2}\nabla_Z\til{e}^Z(\til{\si}\ov{\psi}_-^{\ov{Z}}\psi^Z_++\ov{\til{\si}}\ov{\psi}_+^{\ov{Z}}\psi^Z_-)\\
&+\frac{\hat{r}}{(1+|Z|^2)^2}\Big(\ov{\til{\lam}}_+\til{e}^Z\ov{\psi}^{\ov{Z}}_--\ov{\til{\lam}}_-\til{e}^Z\ov{\psi}^{\ov{Z}}_+ -\til{\lam}_+\til{e}^{\ov{Z}}{\psi}^{{Z}}_-+\til{\lam}_-\til{e}^{\ov{Z}}{\psi}^{{Z}}_+\Big)\\
&-\frac{2\hat{r}}{(1+|Z|^2)^4}\psi_+^Z\psi_-^Z\ov{\psi}_-^{\ov{Z}}\ov{\psi}_+^{\ov{Z}}+\:   \frac{1}{(1+|Z|^2)^2}(F^Z +\frac{2\ov{Z}}{(1+|Z|^2)}\, \psip^Z \psim^Z) (\ov{F}^{\ov{Z}} +\frac{2Z}{(1+|Z|^2)}\, \ov{\psi}_-^{\ov{Z}} \ov{\psi}_+^{\ov{Z}})\Big\}
\\&+{1 \over 2\pi}{1 \over 2\til{e}^2}\int_{\Sigma} \txd^2x\Big\{(\til{F}_{01})^2  - \partial_{\mu}\til{\sigma} \partial^{\mu} 
\overline{\til{\sigma}}   +(\til{D})^2  + {i \over 2} \overline{\til{\lambda}}_{+}
(\overleftrightarrow{\del_-})
\til{\lambda}_{+} +{i \over 2} \overline{\til{\lambda}}_{-}
(\overleftrightarrow{\del_+})
\til{\lambda}_{-}  -2\til{e}^2(Z^*\til{\mu}+\til{r})\til{D}\Big\}\\
&- \frac{1}{2\pi}\int_{\Sigma}\, \phi^\ast B \; + \frac{1}{2\pi}\int_{\del \Sigma}  \Big\{ \phi^*C_a A^a -\frac{\hat{\theta}}{2\hat{r}}\sum_i^{N=2}(\ov{\psi}_{-i}\psi_{-i}+\ov{\psi}_{+i}\psi_{+i})\Big\}\\ 
&+{1 \over 2\pi}\int_{\Sigma} \txd^2x (\til{\theta} \til{F}_{01} ),
\end{aligned}
\end{equation}  
where the holomorphic Killing vector field is
\begin{equation}
\he^Z=i(\til{Q}_1-\til{Q}_2)Z
\end{equation}
with covariant derivative
\begin{equation}
\nabla_Z\til{e}^Z=i(\til{Q}_1-\til{Q}_2)\Big(\frac{1-|Z|^2}{1+|Z|^2}\Big),
\end{equation}
and the moment map
\begin{equation}\label{cp1momentmap}
\til{\mu}=\frac{-\hat{r}(\til{Q}_1|Z|^2+\til{Q}_2)}{(1+|Z|^2)}.
\end{equation}
%The $B$-field we obtain is in fact proportional to the K\"ahler form $\omega$, i.e.,
%\begin{equation}
%B=-\frac{\hat{\theta}}{\hat{r}}\omega,
%\end{equation}
%while the $C$-field is proportional to the moment map
%\begin{equation}
%C=-\frac{\hat{\theta}}{\hat{r}}\til{\mu}.
%\end{equation}

The  action we have obtained for the $U(1)^{N-1}$-GNLSM with $\CP^{N-1}$ target space corresponds exactly to the general action \eqref{action}+\eqref{FIterm}, with the exception of the (spurious) boundary term \eqref{extra}.
%It is important to note that the boundary term appears in the same form shown in \eqref{extra} when doing the reduction procedure to obtain the $U(1)^{N-1}$-GNLSM with $\C P^{N-1}$ target space, while for other toric manifolds which can be obtained from a GLSM by breaking only a single $U(1)$ group, it is
This term takes the form 
\begin{equation}\label{extrator1}
-\frac{1}{2\pi}\int_{\del \Sigma}\frac{\hat{\theta}}{2}\frac{\sum_i^{N}\hat{Q}_i(\ov{\psi}_{-i}\psi_{-i}+\ov{\psi}_{+i}\psi_{+i})}{\sum_j^N \hat{Q}^2_j|\phi_j|^2},
\end{equation}
when we start with GLSMs with arbitrary $\hat{Q}_i$, for which the reduction procedure gives GNLSMs with toric target space $X=\C^{N}//U(1)$.
This boundary term also occurs in the reduction of GLSMs to NLSMs \cite{Govinda}, and can be removed in several ways, including the addition of a boundary term to the GLSM we start with \cite{Govinda}, or by a judicious choice of boundary conditions on the fermionic fields. In the following sections, we shall explain how this term is removed 
%in the reduction of GLSMs to GNLSMs 
when investigating the cases of A-type and B-type supersymmetry at the boundaries.

Before ending this section, we would like to point out that the classical procedure of obtaining GNLSMs from GLSMs which we have explained above is valid at the quantum level, since the $\hat{e}_b\rightarrow \infty$ limit can be taken for the path integral of the GLSM, and functional integration over the auxiliary components of $\hat{V}_b$ is equivalent to imposing their algebraic equations of motion as constraints. However, taking renormalization of the FI parameters into account, it can be shown that we may only obtain quantum GNLSMs for K\"ahler targets with $c_1(X)\geq 0$. This is because the RG flow at the one-loop level of the bare FI parameters $\hat{r}_{0b}$ is
\begin{equation}
\hat{r}_{0b}=\hat{r}_b(\mu)+\sum_{i=1}^N\hat{Q}_{ib}\log\Big(\frac{\Lambda_{\rm UV}}{\mu}\Big),
\end{equation}
 (where $\Lambda_{\rm UV}$ is an ultraviolet cut-off and $\mu$ is a finite energy scale).
 % and the values of $\sum_{i=1}^N\hat{Q}_{ib}$ for a space which does not satisfy $c_1(X)\geq 0$ are such that the parameters $\hat{r}_{0b}$ will not be in the K\"ahler cone of $X$ in the continuum limit where $\Lambda_{\rm UV}\rightarrow\infty$. The argument for this statement is as follows. 
 As shown in \cite{hori2003mirror}, for a basis $e_b$ of $H_2(X,\Z)$, we have $\sum_i^N\hat{Q}_{ib}=c_1(X)\cdot e_b$. Then, for a holomorphic curve $m=\sum_b^{N-k} m_b e_b$ in $X$ (i.e., an element of the Mori cone of $X$) 
 \begin{equation}\label{kahlerconecheck}
\sum_b^{N-k} m_b\hat{r}_{0b}=\sum_b^{N-k} m_b\hat{r}_b(\mu)+\sum_b^{N-k} m_b(c_1(X)\cdot e_b)\log\Big(\frac{\Lambda_{\rm UV}}{\mu}\Big).
\end{equation}
For the bare FI parameters to be in the K\"ahler cone of $X$, the LHS of \eqref{kahlerconecheck} ought to be greater than zero. In the continuum limit ($\Lambda_{\rm UV}\rightarrow\infty$), this is %only 
impossible if $c_1(X)\geq 0$ is not satisfied.

\section{Equivariant B-branes and their Mirrors}

We shall apply the techniques discussed above to find boundary actions and boundary conditions in abelian GNLSMs with toric target spaces, $X$, as well as their mirror descriptions. These boundary actions and boundary conditions will correspond to branes %wrapping submanifolds of
in $X$, which we refer to as equivariant branes. We first study the case where B-type supersymmetry is preserved at the boundaries of the $I\times \R$ worldsheet, since this leads us to make contact with a result found previously by Kapustin et al. \cite{Kapustin,Vyas}, while in the next section, we shall use similar techniques to study equivariant A-branes. After gaining insights from the study of abelian equivariant B-branes, we shall then proceed to analyze equivariant B-branes for nonabelian GNLSMs.

The combination of supercharges which define B-type supersymmetry are given by \eqref{btype}. In the following, we shall set $\beta=0$ for simplicity, though it is straightforward to study the $\beta\neq 0$ generalization by the same techniques. In other words, we assume that the supercharges conserved at the boundaries are 
\begin{equation}\label{btype2}
\begin{aligned}[c]
Q_B=\overline{Q}_+ +\overline{Q}_-,
\end{aligned}
\\
\begin{aligned}[c]
\textrm{   }\\
\textrm{   }\\
\end{aligned}
\\
\begin{aligned}[c]
\textrm{   }\\
\textrm{   }\\
\end{aligned}
\\
\begin{aligned}[c]
Q^\dagger_B=Q_++ Q_-\end{aligned}.
\end{equation}
From \eqref{susydelta}, we find that the corresponding relations among the supersymmetry transformation parameters are  
\begin{equation}
\begin{aligned}
\eps&=\eps_+=-\eps_-\\
\ov{\eps}&=\ov{\eps}_+=-\ov{\eps}_-.
\end{aligned}
\end{equation}
We shall also make use of superfields when discussing boundary conditions, and to this end, the concept of `boundaries' in superspace \cite{Hori1} is useful. For the case at hand, the relevant boundary in superspace is known as `B-boundary', and corresponds to
\begin{equation}
\begin{aligned}
\theta&=\theta^+=\theta^-\\
\ov{\theta}&=\ov{\theta}^+=\ov{\theta}^-.
\end{aligned}
\end{equation}

Let us first review what is known of ordinary B-branes. For $\mathcal{N}=(2,2)$ NLSMs%with B-type supersymmetry preserved at the boundaries
, the boundary condition needed to preserve B-type supersymmetry at the boundaries maps each boundary to a holomorphically embedded complex submanifold of the target space \cite{HIV}. In addition, we may include the following boundary action 
\begin{equation}\label{Bbrane}
S_{\partial\Sigma}=\int\limits_{\partial\Sigma}
\dd x^0\,\left\{\,
A^X_M\partial_0X^M-\frac{i}{4}F^X_{MN}(\psi_+^M+\psi_-^M)(\psi_+^N+\psi_-^N)
\,\right\},
\end{equation}
which is B-type supersymmetric if 
%and only if 
$F^X_{mn}=F^X_{\ov{m}\ov{n}}=0$ (here, we use ($M,N,\ldots$) as coordinate indices on the holomorphically embedded branes), %Here, we have used real coordinates for the K\"ahler target space $X$, 
where $A^X_M$ 
%is the
%takes the form of a
corresponds to a connection of a line ($U(1)$) %IMPORTANT NOTE: IN GSW, Witten seems to say they are the same on page 460 
bundle on %the
each B-brane, and  $F^X_{MN}$ the corresponding curvature. The conditions $F^X_{mn}=F^X_{\ov{m}\ov{n}}=0$ indicate that each line bundle is holomorphic \cite{GSW2}. This boundary action is in fact a supersymmetric Wilson line, and since we have two boundary components, we 
%can 
actually have two different Wilson lines along each boundary, corresponding to two different B-branes supporting holomorphic line bundles, each with different connections and curvatures (\cite{HIV}, page 21). %Moreover, one must also impose boundary conditions compatible with this boundary action, which preserve B-type supersymmetry of the bulk action at the boundary. The relevant boundary conditions are Neumann boundary conditions on $X^M$, which correspond to space-filling B-branes.
%IMPORTANT NOTE: IS THIS BOUNDARY ACTION COMPATIBLE WITH BOUNDARY CONDITIONS OTHER THAN THE SPACE-FILLING ONE? YES, ACCORDING TO HIV, IT JUST NEEDS TO BE A HOLOMORPHICALLY EMBEDDED COMPLEX SUBMANIFOLD, AND IT WILL BE COMPATIBLE WITH BOUNDARY ACTION ABOVE. THE ONLY THING IS THAT SOME OF THE \del_0X's WILL VANISH

An alternative formulation of B-branes exists \cite{HIV}, where mixed Dirichlet-Neumann boundary conditions are imposed on some of the target space coordinates, and the boundary action is 
\begin{equation}\label{BbraneAlt}
S_{\partial\Sigma}=\int\limits_{\partial\Sigma}
\dd x^0 
A^X_M\partial_0X^M,
\end{equation}
which is an ordinary Wilson line along each boundary component.
This formulation %is equivalent to that of \eqref{Bbrane}, %since 
%in the sense that both 
leads us to the same spacetime theory as \eqref{Bbrane} \cite{nappi}. %IMPORTANT NOTE I actually know this point from page 46 of HORI Linear models.
Moreover, B-type supersymmetry is preserved at the boundaries if 
%and only if 
$F^X_{mn}=F^X_{\ov{m}\ov{n}}=0$, indicating a holomorphic line bundle on each B-brane.

We are interested in the generalizations of \eqref{Bbrane} and \eqref{BbraneAlt} (and their corresponding boundary conditions) for GNLSMs. One method of obtaining such a generalization would be to replace ordinary worldsheet derivatives by covariant ones, and attempt to maintain supersymmetry and gauge symmetry by adding additional terms. However, it is known that \eqref{Bbrane} and \eqref{BbraneAlt} and their corresponding boundary conditions can be obtained from GLSM boundary actions and boundary conditions \cite{HIV}. This suggests the more elegant method of obtaining the GNLSM generalizations from GLSM boundary conditions and a GLSM boundary action, using the methods of Section 3.
In the following, we shall attempt to generalize the boundary action \eqref{Bbrane} to the case of $U(1)^{k}$-GNLSMs with K\"ahler toric target space, before proceeding to do the same for the boundary action \eqref{BbraneAlt}. %$\C^N//U(1)$. 

\vspace{0.2cm}
\mbox{}\par\nobreak
\noindent
\textit{B-branes on $\C^N//U(1)$ from GLSM} 

Let us first recall how B-type supersymmetric boundary conditions and the boundary action \eqref{Bbrane} for an NLSM with $\C^N//U(1)$ target space can be obtained from %B-type supersymmetric 
boundary conditions and the boundary action of a $U(1)$-GLSM with $\C^N$ target \cite{HIV}. We shall focus on obtaining NLSM boundary conditions corresponding to space-filling branes. To this end, we must impose B-type supersymmetric boundary conditions at the GLSM level which include Neumann boundary conditions on the chiral superfields.\footnote{\label{19}In general, the B-type NLSM also admits boundary conditions which correspond to lower dimensional branes, the only restriction is that such a brane corresponds to a complex submanifold holomorphically embedded in the target space %$\C P^{N-1}$
 \cite{HIV,aspinwall}. These boundary conditions %can also 
 %should be 
 are obtainable in the $\hat{e}\rightarrow\infty$ limit from %B-type 
 GLSM boundary %conditions 
 terms which %include
 effectively impose Dirichlet boundary conditions on some of the chiral multiplets \cite{Hori1}. %These, however, 
%However, the latter conditions modify the boundary action \eqref{nlsmuvb}.
%IMPORTANT NOTE: I HAVE TAKEN OUT ' the latter conditions modify the boundary action \eqref{nlsmuvb}', SINCE THIS IS ONLY TRUE IN A COORDINATE SYSTEM WHICH SEPARATES THE DIRECTIONS TANGENT AND NORMAL TO THE BRANE. actually no, this is true for NLSMs, i'm talking about GLSMs above
%IMPORTANT NOTE, HOLOMORPHICALLY EMBEDDED MEANS THAT THE COMPLEX STRUCTURES ARE ALIGNED, SEE PAGE 408 OF BECKER-BECKER-SCHWARZ 
}

%In addition, we must impose B-type supersymmetric boundary conditions, which leave the GLSM action supersymmetric at the boundaries, yet do not trivialize any part of the boundary action \eqref{nlsmuvb}. 
Using the language of superfields, these conditions are \cite{HIV,Hori1}
\begin{equation}\label{Bsuperbc}
\begin{aligned}
\hat{\mathcal{D}}_+\Phi_i&=\hat{\mathcal{D}}_-\Phi_i\\
\hat{\Sigma}&=\ov{\hat{\Sigma}}
\end{aligned}
\end{equation}
at B-boundary,\footnote{A boundary condition imposed on a superfield automatically implies that its components obey a set of boundary conditions  which are closed under supersymmetry.} where $\hat{\mathcal{D}}_{\pm}=e^{-\hat{Q}_i\hat{V}}D_{\pm}e^{\hat{Q}_i\hat{V}}$; while in components, they are given as  \begin{equation}\label{BtypeGLSMtoNLSMbc}
\begin{aligned}
\psi_{+i}-\psi_{-i}&=0\\
F_i&=0\\
\hat{D}_1\phi_i&=0\\
\hat{D}_1(\psi_{+i}+\psi_{-i})&=0
\end{aligned}
\end{equation}
and
\begin{equation}\label{NLSMBgauge}
\begin{aligned}
%%A_1&=\textrm{constant}\\
%%\del_1 A_0&=0\\
\textrm{Im}(\hat{\si})&=0\\
{\hat{\lambda}}_{+}+{\hat{\lambda}}_{-}&=0\\
\del_1\textrm{Re}(\hat{\si})+\hat{F}_{01}&=0\\
%\ov{\hat{\lambda}}_{+}+\ov{\hat{\lambda}}_{-}&=0\\
%\del_1({\hat{\lambda}}_{+}-{\hat{\lambda}}_{-})&=0\\
%\del_1(\ov{\hat{\lambda}}_{+}-\ov{\hat{\lambda}}%_{-})&=0\\
%%\del_1\textrm{Re}(\hat{\si})&=0\\
%\del_1(\hat{D}+\del_1\textrm{Im}(\hat{\si}))&=0,
\end{aligned}
\end{equation}%Moreover, we impose the condition%\footnote{This condition ensures that we have a sufficient number of boundary conditions for all that the equation of motion of the gauge field is local} 
%In \eqref{nlsmuvb}, 
where the covariant derivative of the scalar fields is 
\begin{equation}
\hat{D}_{\mu}\phi_i=\del_{\mu}\phi_i+i\hat{Q}_i\hat{A}_{\mu}\phi_i.
\end{equation}
However, the boundary conditions of some of the vector multiplet fields remain to be specified, and %to this end 
therefore we impose \cite{HIV,Hori1}
\begin{equation}\label{Flocal}
\hat{F}_{01}=-\hat{e}^2\hat{\theta}\\,
\end{equation}
which further implies 
\begin{equation}\label{furtherimplied}
\begin{aligned}
%F_{01}&=-\hat{e}^2\theta\\
\del_1({\hat{\lambda}}_{+}-{\hat{\lambda}}_{-})&=0\\
%\del_1(\ov{\hat{\lambda}}_{+}-\ov{\hat{\lambda}}_{-})&=0\\
%%\del_1\textrm{Re}(\hat{\si})&=0\\
\del_1(\hat{D}+\del_1\textrm{Im}(\hat{\si}))&=0
\end{aligned}
\end{equation}
via B-type supersymmetry.
These conditions are also invariant under $U(1)$ gauge transformations. %IMPORTANT NOTE : Even if we did not impose the extra condition on F_01, we would still have B-type supersymmetry at the boundary, the only reason we impose it is to get boundary conditions for D and \lambda_+-\lambda_- 
 %while in components, they are given as
%\begin{equation}\label{BtypeGLSMtoNLSMbc}
%\begin{aligned}
%\psi_{+i}-\psi_{-i}&=0\\
%F_i&=0\\
%\hat{D}_1\phi_i&=0\\
%\hat{D}_1(\psi_{+i}+\psi_{-i})&=0
%\end{aligned}
%\end{equation}
%and
%\begin{equation}\label{NLSMBgauge}
%\begin{aligned}
%%A_1&=\textrm{constant}\\
%%\del_1 A_0&=0\\
%\del_1\textrm{Re}(\hat{\si})+\hat{F}_{01}&=0\\
%{\hat{\lambda}}_{+}+{\hat{\lambda}}_{-}&=0\\
%\ov{\hat{\lambda}}_{+}+\ov{\hat{\lambda}}_{-}&=0\\
%\del_1({\hat{\lambda}}_{+}-{\hat{\lambda}}_{-})&=0\\
%\del_1(\ov{\hat{\lambda}}_{+}-\ov{\hat{\lambda}}_{-})&=0\\
%%\del_1\textrm{Re}(\hat{\si})&=0\\
%\textrm{Im}(\hat{\si})&=0\\
%\del_1(\hat{D}+\del_1\textrm{Im}(\hat{\si}))&=0.
%\end{aligned}
%\end{equation}
We also need to add a boundary term to cancel the B-type supersymmetry variation of the bulk theta term, i.e., the expression 
\begin{equation}\label{thetasusy}
\frac{\hat{\theta}}{2\pi}\int_{\Sigma}\txd^2x \textrm{ }  \hat{F}_{01}+\frac{\hat{\theta}}{2\pi}\int_{\del\Sigma}\txd x^0\frac{(\hat{\si}+\ov{\hat{\si}})}{2}
\end{equation}
is B-type supersymmetry invariant.
Having imposed the above boundary conditions and the boundary term in \eqref{thetasusy},
%These boundary conditions %which 
%leave 
the GLSM action is B-type supersymmetric at the boundaries.

%i.e., 
In addition, we include the boundary action
\begin{equation}\label{nlsmuvb}
S_{\partial\Sigma}={\hat{\theta}\over 4\pi \hat{r}}\int\limits_{\partial\Sigma}
\txd x^0 \sum_i^N\Big(\, i\hat{D}_0\ov{\phi}_i\,\phi_i-i\ov{\phi}_i \hat{D}_0\phi_i
 \:+\: (\psi_{+i}+\psi_{-i})(\ov{\psi}_{+i}+\ov{\psi}_{-i})
-\hat{Q}_{i}(\hat{\si}+\ov{\hat{\si}})|\phi_i|^2
\Big),
\end{equation}
which is B-type supersymmetric on its own.\footnote{\label{21}If we were to discard the expression \eqref{nlsmuvb}, %these
the boundary conditions would ensure the locality of the equations of motion of the worldsheet fields. %However,
 Its inclusion  renders the equations of motion for $\phi_i$, $\psi_i$, $\hat{A}_{\mu}$ and $\hat{\si}$ nonlocal \cite{HIV}, i.e., they contain boundary terms.%IMPORTANT NOTE: WHEN WE DONT HAVE \eqref{nlsmuvb}, FOR SIGMA FIELD, THE BOUNDARY CONDITION =e^2\theta ensures that the equation of motion is local even with the \si+\ov{\si} boundary term. The F_01 boundary condition is referred to as the Gauss law constraint by Hori in phases, and makes its equation of motion local, as explained in HIV, page 55. The others remain local as well, see eqn 3.2 of HIV and my eqbranes notes. Equation of motion for D is always local, since its action has no derivatives.
}
%as well as
%\begin{equation}\label{thetasusy}
%\frac{\hat{\theta}}{2\pi}\int_{\Sigma}\txd^2x \textrm{ }  \hat{F}_{01}+\frac{\hat{\theta}}{2\pi}\int_{\del\Sigma}\txd x^0\frac{(\hat{\si}+\ov{\hat{\si}})}{2},
%\end{equation}
%which is the bulk theta term, together with a boundary term which results in the B-type supersymmetry invariance of the entire expression \eqref{thetasusy}. 
Its inclusion is necessary to obtain the boundary action \eqref{Bbrane}, which plays the role of elucidating the geometry of the branes.

Now, recall from \cite{HIV} that the bulk theta term can be converted into a boundary term %without violating gauge invariance 
in some circumstances. In particular, we have 
\begin{equation}\label{thetaconvert}
\frac{\hat{\theta}}{2\pi}\int_{\Sigma}\txd^2x \textrm{ }  \hat{F}_{01}=-\frac{\hat{\theta}}{2\pi}\int_{\del\Sigma}\txd x^0 \textrm{ }  \hat{A}_{0},
\end{equation}
via Stoke's theorem, but this violates gauge invariance. %\cite{HIV}. %at the Lagrangian level. 
%When the boundary components of our $I\times \R$ worldsheet are compactified on circles \cite{HIV},
%IMPORTANT NOTE: THE BOUNDARIES OF THE STRIP AREN'T ACTUALLY COMPACTIFIED TO BECOME CIRCLES, SEE POLCHINSKI VOLUME 1 EQUATION 8.61 AND 8.62 
 The violation is
\begin{equation}\label{vio}
\frac{\hat{\theta}}{2\pi}\int_{\del\Sigma}\txd x^0 \textrm{ }  \del_0\alpha=\frac{\hat{\theta}}{2\pi}\int\txd \alpha=\frac{\hat{\theta}}{2\pi}2\pi m
\end{equation}
where $m\in \Z$. However, if $\hat{\theta}\in 2\pi \Z$, then \eqref{vio} implies that $\textrm{exp}	(-i\frac{\hat{\theta}}{2\pi}\int_{\del \Sigma} \hat{A})$ is gauge invariant, and hence, the {\it path integral} remains gauge invariant. We shall assume that $\hat{\theta}\in 2\pi \Z$ hereon, by setting $\hat{\theta}=2 \pi n$, where $n\in \Z$.%and in these 

Doing so, we may then write \eqref{thetasusy} and \eqref{nlsmuvb} as
\begin{equation}\label{nlsmuvb2}
\begin{aligned}
S_{\partial\Sigma}'=&{n\over 2 \hat{r}}\int\limits_{\partial\Sigma}
\txd x^0 \Big(\sum_i^N\, i\del_0\ov{\phi}_i\,\phi_i-i\ov{\phi}_i \del_0\phi_i
 \:+\: \sum_i^N(\psi_{+i}+\psi_{-i})(\ov{\psi}_{+i}+\ov{\psi}_{-i})\\&
 +2\hat{A}_0(\sum_i^N\hat{Q}_{i}|\phi_i|^2-\hat{r})
-(\hat{\si}+\ov{\hat{\si}})(\sum_i^N\hat{Q}_{i}|\phi_i|^2-\hat{r})
\Big)
\end{aligned}
\end{equation}
Taking the $\hat{e}\rightarrow \infty$ (NLSM) limit, the components of the vector multiplet become auxiliary. Integrating $\hat{D}$ out of the bulk action enforces the constraint
\begin{equation}\label{nlsmconstraint1}
\sum_i^N \hat{Q}_{i}|\phi_i|^2-\hat{r}=0
\end{equation}
and consequently, the second line of \eqref{nlsmuvb2} vanishes. Integrating out the rest of the vector multiplet gives several more constraints, the one relevant to the boundary action being
\begin{equation}\label{nlsmconstraint2}
\sum_i^N \hat{Q}_{i} \ov{\phi}_i\psi_{\pm i}=0. 
\end{equation}
Thus, the boundary action \eqref{nlsmuvb2} reduces to 
\begin{equation}\label{nlsmuvb3}
\begin{aligned}
S_{\partial\Sigma}'=&{n\over 2 \hat{r}}\int\limits_{\partial\Sigma}
\txd x^0 \sum_i^N\Big(\, (i\del_0\ov{\phi}_i\,\phi_i-i\ov{\phi}_i \del_0\phi_i)
 \:+\: (\psi_{+i}+\psi_{-i})(\ov{\psi}_{+i}+\ov{\psi}_{-i})
\Big)
\end{aligned}
\end{equation}
with \eqref{nlsmconstraint1} and \eqref{nlsmconstraint2} strictly imposed. As explained in \cite{HIV}, the first term in \eqref{nlsmuvb3} is nothing but the hermitian connection 
\begin{equation}\label{hermi}
A^X_I\txd X^I=
-n{i\over 2}{\displaystyle~\sum \!{}_{i=1}^N
\ov{\phi}_i\overleftrightarrow{\txd} \phi_i~\over
\displaystyle \sum \!{}_{i=1}^N\hat{Q}_i|\phi_i|^2}.
\end{equation}
of %the holomorphic line bundle 
$\mathcal{O}_X(-n)$ on the toric manifold $X=\C^N//U(1)$, since it transforms under $U(1)$ gauge transformations ($\phi_i\rightarrow e^{i\hat{Q}_i\alpha}\phi_i$)as 	
\begin{equation}\label{boundgauge}
A^X_I\txd X^I\rightarrow A^X_I\txd X^I -(-n)d\alpha.
\end{equation}
Here, $\mathcal{O}_X(-n)$ is the holomorphic line bundle on $X$ %defined $-n$ is an integer 
with $\int_X c_1(\mathcal{O}_{X}(-n))=-n$.
% IMPORTANT NOTE: recall from Nakahara that the first chern class of a bundle is proportional to the (lie algebra-valued) curvature of the bundle. Also see aspinwall page 34 and HIV page 61. on page 426 of nakahara, it is indicated that the first chern class is a TWO-FORM on the BASE SPACE MANIFOLD on which the bundle is defined, not the total space. I have said 'the' holomorphic line bundle, because for CP^n, the holomorphic line bundles labelled by an integer are the only possible line bundles over CP^N, see green schwarz witten page 460

Using parametrizations for $\phi_i$ and $\psi_{\pm i}$ which satisfy \eqref{nlsmconstraint1} and \eqref{nlsmconstraint2}, the explicit NLSM boundary action \eqref{Bbrane} is obtained, with $F^X_{jk}=F^X_{\ov{\jmath}\ov{k}}=0$, together with the compatible boundary conditions. For example, for $X=\C P^{N-1}$, we may use the parametrizations \eqref{para} and \eqref{lamconstraint2} to satisfy \eqref{nlsmconstraint1} and \eqref{nlsmconstraint2}, %whereupon 
and with the help of the constraints which come from integrating out the vector multiplet,
the boundary conditions \eqref{BtypeGLSMtoNLSMbc} become
\begin{equation}\label{BsuperbcNLSM}
\begin{aligned}
\psi^{Z^i}_+-\psi^{Z^i}_-&=0\\
F^{Z^i}&=0\\
\del_1Z^i&=0\\
\del_1(\psi^{Z^i}_++\psi^{Z^i}_-)&=0,
\end{aligned}
\end{equation}
where the purely Neumann boundary conditions on $Z^i$ indicate that the B-brane is a space-filling brane;
 while the boundary action \eqref{nlsmuvb3} is reexpressed as %the following NLSM boundary action, which agrees with \eqref{Bbrane}
\begin{equation}\label{BbraneCpn}
\begin{aligned}
S_{\partial\Sigma}'=&\frac{n}{2\hat{r}}\int\limits_{\partial\Sigma}
\dd x^0\Bigg\{
\frac{-i\hat{r}\sum_i^{N-1}(\ov{Z}^i\del_0Z^i-Z^i\del_0\ov{Z}^i)}{(1+\sum^{N-1}_k|Z^k|^2)}+2\hat{r}\del_0t+\frac{\hat{r}\sum_i^{N-1}(\psi_+^{Z^i}+\psi_-^{Z^i})(\ov{\psi}_+^{\ov{Z}^i}+\ov{\psi}_-^{\ov{Z}^i})}{(1+\sum^{N-1}_k|Z^k|^2)}
\\&-\frac{\hat{r}\sum_i^{N-1}\ov{Z}^i(\psi_+^{Z^i}+\psi_-^{Z^i})\sum_j^{N-1}{Z}^j(\ov{\psi}_+^{\ov{Z}^j}+\ov{\psi}_-^{\ov{Z}^j})}{(1+\sum^{N-1}_k|Z^k|^2)^2}
\Bigg\},
\\=&\int\limits_{\partial\Sigma}
\dd x^0\,\left\{\,
A^X_j\partial_0 X^j+A^X_{\ov{\jmath}}\partial_0 \ov{X}^{\ov{\jmath}}+n\del_0t-\frac{i}{2}F^X_{j\ov{k}}(\psi_+^j+\psi_-^j)(\ov{\psi}_+^{\ov{k}}+\ov{\psi}_-^{\ov{k}})
\,\right\},
\end{aligned}
\end{equation}
where $X^j=Z^j$, $\psi^j=\psi^{Z^j}$,
\begin{equation}\label{connectioncom}
\begin{aligned}[c]
A^X_j=-n\frac{i}{2}\frac{\ov{Z}^j}{(1+\sum^{N-1}_k|Z^k|^2)},
\end{aligned}
\\
\begin{aligned}[c]
\textrm{   }\\
\textrm{   }\\
\end{aligned}
\\
\begin{aligned}[c]
A^X_{\ov{\jmath}}=n\frac{i}{2}\frac{Z^j}{(1+\sum^{N-1}_k|Z^k|^2)}
\end{aligned}
\end{equation}
are the components of the connection of $\mathcal{O}_{\C P^{N-1}}(-n)$, 
%IMPORTANT NOTE: the 'n' here sort of plays the role of the charge of the target space field, see my stackexchange question
while its curvature is
\begin{equation}\label{curvaturecom}
F_{j\ov{k}}=n\omega_{j\ov{k}},
\end{equation}
where $\omega_{j\ov{k}}=ig_{j\ov{k}}$ are the components of the normalized\footnote{Unlike \eqref{FSkahlerform}, the Fubini-Study metric that appears in the bulk NLSM action contains the FI parameter, $\hat{r}$, which is the size modulus of the Fubini-Study metric; see \eqref{cpn}.
% and footnote \ref{18}.
 } Fubini-Study K\"ahler form of $\C P^{N-1}$, 
\begin{equation}\label{FSkahlerform}
\omega=i\frac{(\sum_j^{N-1}dZ^j\wedge d\ov{Z}^j)}{(1+\sum_k^{N-1}|Z^k|^2)}-i\frac{(\sum_l^{N-1}\ov{Z}^l dZ^l)\wedge(\sum_j^{N-1}Z^j d\ov{Z}^j)}{(1+\sum_k^{N-1}|Z^k|^2)^{2}}.
\end{equation}
The expression \eqref{BbraneCpn} agrees with \eqref{Bbrane} up to the term $n\del_0t$. This term merely reflects the fact that the $U(1)$ gauge symmetry is broken at the boundaries, and it can be removed via the gauge transformation \eqref{boundgauge}, with $\alpha=-t$. At the path integral level, we may instead use $\int_{\del \Sigma} dx^0 n\del_0t =2\pi n m$ (for some $m\in \mathbb{Z}$) to remove the term. Thus, the boundary conditions \eqref{BsuperbcNLSM} and the boundary action \eqref{BbraneCpn} indicate that the worldsheet boundaries are mapped to a space-filling B-brane on $\CP^{N-1}$, which supports the holomorphic line bundle $\mathcal{O}_{\C P^{N-1}}(-n)$.

%The B-type GLSM boundary conditions \eqref{Bsuperbc} also reduce to B-type NLSM boundary conditions when taking the $\hat{e}\rightarrow\infty$. For the example of $\C P^{N-1}$, these are 

%\begin{equation}\label{BsuperbcNLSM}
%\begin{aligned}
%\psi^{Z^i}_+-\psi^{Z^i}_-&=0\\
%F^{Z^i}&=0\\
%\del_1Z^i&=0\\
%\del_1(\psi^{Z^i}_++\psi^{Z^i}_-)&=0.
%\end{aligned}
%\end{equation}
%The purely Neumann boundary conditions on $Z^i$ indicate that the brane which supports $\mathcal{O}_{\C P^{N-1}}(-n)$ is a space-filling brane.\footnote{In general, the B-type NLSM also admits boundary conditions which correspond to lower dimensional branes, the only restriction is that such a brane corresponds to a complex submanifold holomorphically embedded in $\C P^{N-1}$ \cite{aspinwall}. These boundary conditions can also be obtained in the $\hat{e}\rightarrow\infty$ limit from B-type GLSM boundary conditions which include Dirichlet boundary conditions on some of the chiral multiplets. These, however, modify the boundary action \eqref{nlsmuvb}.
%IMPORTANT NOTE, HOLOMORPHICALLY EMBEDDED MEANS THAT THE COMPLEX STRUCTURES ARE ALIGNED, SEE PAGE 408 OF BECKER-BECKER-SCHWARZ }

%\mbox{}\par\nobreak
%\noindent
%\textit{Equivariant B-branes on $\C^N//U(1)$ from GLSM} 
\subsection{Equivariant B-branes on $\C^N//U(1)$ from GLSM}

Having recalled how the B-type supersymmetric boundary action and boundary conditions for NLSMs with toric target spaces of the form $\C^N//U(1)$ are obtained from a GLSM, we shall now proceed to obtain the B-type supersymmetric boundary action and boundary conditions for abelian GNLSMs with the same target spaces. 

%In addition
Firstly, we impose the following B-type supersymmetric boundary conditions, which are invariant under $U(1)^{N}$ gauge symmetry, and which include Neumann boundary conditions on the chiral superfields, %and preserve B-type supersymmetry of the GLSM at the boundaries of the strip, 
i.e.,
\begin{equation}
\label{gnlsmBsuperbc}
\begin{aligned}
{\mathcal{D}}_+\Phi_i&={\mathcal{D}}_-\Phi_i\\
{\Sigma}_a&=\ov{{\Sigma}}_a
\end{aligned}
\end{equation}
at B-boundary, where 
$\mathcal{D}_{\pm}=e^{-\sum_a^{N}{\mathcal{Q}}_{ia} V_a}D_{\pm} e^{\sum_a^{N}{\mathcal{Q}}_{ia} V_a}$,
as well as 
\begin{equation}\label{Flocal2}
F_{01a}=-{e_a}^2{\theta_a}\\.
\end{equation} 
In components, these boundary conditions are
%IMPORTANT NOTE: THE FOLLOWING SHOULD BE CORRECT, BUT TO BE SAFE I TOOK IT OUT, BECAUSE IN THE ORIGNAL REFERENCE (DOESN'T SEEM TO BE IN THE ONE CITED BELOW, SEEMS TO ONLY BE IN HORI ROMO) THEY ARE DEALING WITH NONABELIAN GAUGE THEORY, AND THE GAUGE FIELDS HAVE DIFFERENT BOUNDARY CONDITION. 
%\footnote{If we analytically continue our worldsheet to be Euclidean, then the requirement of B-type supersymmetry of the boundary conditions \eqref{BtypeGLSMtoGNLSMbc} and \eqref{Bgauge} implies an {\it infinite} number of boundary conditions \cite{HoriPhases}.}
\begin{equation}\label{BtypeGLSMtoGNLSMbc}
\begin{aligned}
\psi_{+i}-\psi_{-i}&=0\\
F_i&=0\\
D_1\phi_i&=0\\
D_1(\psi_{+i}+\psi_{-i})&=0.
\end{aligned}
\end{equation}
and
\begin{equation}\label{Bgauge}
\begin{aligned}
%%A_1&=\textrm{constant}\\
%%\del_1 A_0&=0\\
\textrm{Im}(\si_a)&=0\\
{\lambda}_{+a}+{\lambda}_{-a}&=0\\
\del_1\textrm{Re}(\si_a)&={e_a}^2{\theta_a}\\
F_{01a}&=-{e_a}^2{\theta_a}\\
%\ov{\lambda}_{+a}+\ov{\lambda}_{-a}&=0\\
\del_1({\lambda}_{+a}-{\lambda}_{-a})&=0\\
%\del_1(\ov{\lambda}_{+a}-\ov{\lambda}_{-a})&=0\\
%%\del_1\textrm{Re}(\si^a)&=0\\
\del_1(D_a+\del_1\textrm{Im}(\sigma_a))&=0.
\end{aligned}
\end{equation}
It is crucial to note that these boundary conditions are compatible with the constraints \eqref{generalv2} and \eqref{generalsi2} which are imposed when taking the $\hat{e}\rightarrow\infty$ limit to reduce the GLSM to a GNLSM. We also ought to supersymmetrize the bulk theta terms as in \eqref{thetasusy}, which gives
\begin{equation}\label{thetasusygnlsm}
\frac{\hat{\theta}}{2\pi}\int_{\Sigma}\hat{F}_{01}d^2x+\frac{\hat{\theta}}{2\pi}\int_{\del\Sigma}\frac{(\hat{\si}+\ov{\hat{\si}})}{2}dx^0+\sum_c^{N-1}\Big(\frac{\widetilde{\theta}_c}{2\pi}\int_{\Sigma}\widetilde{F}_{01c}d^2x+\frac{\widetilde{\theta}_c}{2\pi}\int_{\del\Sigma}\frac{(\widetilde{\si}_c+\ov{\widetilde{\si}}_c)}{2}dx^0\Big).
\end{equation}
With the above boundary conditions and boundary terms, the GLSM action is B-type supersymmetric at the boundaries.

In addition%in order to obtain the boundary action
, we must generalize the $U(1)$-GLSM boundary action \eqref{nlsmuvb} to a boundary action for the $U(1)^N$-GLSM given in \eqref{big1}, with $N-k=1$. This is given by 
\begin{equation}\label{gnlsmuvb}
S_{\partial\Sigma}={\hat{\theta}\over 4\pi \hat{r}}\int\limits_{\partial\Sigma}
\txd x^0 \sum_i^N\Big(\, i{D}_0\ov{\phi}_i\,\phi_i-i\ov{\phi}_i {D}_0\phi_i
 \:+\: (\psi_{+i}+\psi_{-i})(\ov{\psi}_{+i}+\ov{\psi}_{-i})
-\sum_a^N\mathcal{Q}_{ia}(\si_a+\ov{\si}_a)|\phi_i|^2
\Big),
\end{equation}
(where the covariant derivatives of the scalar fields are given by \eqref{covder}) and is B-type supersymmetric on its own. 

Next, we set $\hat{\theta}=2\pi n$, which allows us to write \eqref{thetasusygnlsm} and \eqref{gnlsmuvb} in a form which generalizes \eqref{nlsmuvb2}, i.e.,
%However, we shall make a slightly different choice for $\til{\theta}_c$, that is, $\widetilde{\theta}_c=2\pi n\til{Q}_{Nc}$. %Taking $e^2\rightarrow \inf$, we have  
\begin{equation}\label{gnlsmuvb2}
\begin{aligned}
S_{\partial\Sigma}'=&{n\over 2 \hat{r}}\int\limits_{\partial\Sigma}
\txd x^0 \Big(\, \sum_i^N(i\del_0\ov{\phi}_i\,\phi_i-i\ov{\phi}_i \del_0\phi_i)
 \:+\: \sum_i^N(\psi_{+i}+\psi_{-i})(\ov{\psi}_{+i}+\ov{\psi}_{-i})\\&
 +2\hat{A}_0(\sum_i^N\hat{Q}_{i}|\phi_i|^2-\hat{r})
-(\hat{\si}+\ov{\hat{\si}})(\sum_i^N\hat{Q}_{i}|\phi_i|^2-\hat{r})\\&
 +2\sum^{N-1}_c\til{A}_{0c}\sum_i^N\til{Q}_{ic}|\phi_i|^2
-\sum^{N-1}_c(\til{\si}_c+\ov{\til{\si}}_c)\sum_i^N\til{Q}_{ic}|\phi_i|^2\Big)\\&
+\sum_c^{N-1}\Big(\frac{\widetilde{\theta}_c}{2\pi}\int_{\Sigma}\widetilde{F}_{01c}d^2x+\frac{\widetilde{\theta_c}}{2\pi}\int_{\del\Sigma}\frac{(\widetilde{\si}_c+\ov{\widetilde{\si}}_c)}{2}dx^0\Big).
\end{aligned}
\end{equation}
As in the NLSM case, taking the $\hat{e}\rightarrow\infty$ limit results in the vector multiplet components becoming auxiliary (see Section 3). 
%imposes the constraints \eqref{nlsmconstraint1} and 
Integrating $\hat{D}$ out of the bulk action imposes the condition \eqref{nlsmconstraint1}, and this results in the second line of \eqref{gnlsmuvb2} vanishing. Integrating out the rest of the vector multiplet components imposes \eqref{nlsmconstraint2}, as well as \eqref{generalv2} and \eqref{generalsi2} (the latter are no longer relevant to the boundary action once the second line of \eqref{gnlsmuvb2} vanishes%to the boundary action
).\footnote{%Recall that the constraint \eqref{nlsmconstraint1} is imposed by integrating out $\hat{D}$.
 %One must be careful
 It is important to integrate out  $\hat{D}$ \textit{before} integrating out $\hat{A}_0$, otherwise the algebraic equation of motion of $\hat{A}_0$ given in \eqref{generalv2} will be modified by a boundary term, see footnote \ref{15}.}
 %IMPORTANT NOTE: This is because the bulk theta term, which usually gives rise to the offending term, has now been pulled into the boundary action.
%the components of the vector multiplet become auxiliary. Integrating $\hat{D}$ out of the bulk action enforces the constraint
%\begin{equation}\label{nlsmconstraint1}
%\sum_i^N \hat{Q}_{i}|\phi_i|^2-\hat{r}=0
%\end{equation}
%and consequently, the second line of \eqref{nlsmuvb2} vanishes. Integrating out the rest of the vector multiplet gives several more constraints, the one relevant to the boundary action being

Next, to find the explicit B-type GNLSM boundary conditions and boundary action, we must find parametrizations which satisfy \eqref{nlsmconstraint1} and \eqref{nlsmconstraint2}. Let us study our usual example of $\C P^{N-1}$. 
%For $\CP^{N-1}$, we use \eqref{para} and \eqref{lamconstraint2}, which gives
 Using the parametrizations \eqref{para} and \eqref{lamconstraint2}, as well as the constraints \eqref{generalv2} and \eqref{generalsi2}, the boundary conditions become
\begin{equation}
\boxed{
\begin{aligned}
\psi^{Z^i}_+-\psi^{Z^i}_-&=0\\
F^{Z^i}&=0\\
\del^A_1Z^i&=0\\
\phi^*\nabla^A_1(\psi^{Z^i}_++\psi^{Z^i}_-)&=0
\end{aligned}}
\end{equation}
and 
\begin{equation}\label{Bgauge2}
\boxed{
\begin{aligned}
\textrm{Im}(\til{\si}_c)&=0\\
{\til{\lambda}}_{+c}+{\til{\lambda}}_{-c}&=0\\
\del_1\textrm{Re}(\til{\si}_c)&={\til{e}_c}^2{\til{\theta}_c}\\
\til{F}_{01c}&=-{\til{e}_c}^2{\til{\theta}_c}\\
%\ov{\til{\lambda}}_{+c}+\ov{\til{\lambda}}_{-c}&=0\\
\del_1({\til{\lambda}}_{+c}-{\til{\lambda}}_{-c})&=0\\
%\del_1(\ov{\til{\lambda}}_{+c}-\ov{\til{\lambda}}_{-c})&=0\\
\del_1(\til{D}_c+\del_1\textrm{Im}(\til{\si}_c))&=0,
\end{aligned}}
\end{equation}
which are invariant under the $U(1)^{N-1}$ gauge symmetry, and satisfy the B-type 
supersymmetry transformations obtained from \eqref{susytrans} and \eqref{susytrans2}. Moreover, these boundary conditions result in the vanishing of the expressions \eqref{dSmatw} and \eqref{dSgauthe}, thus ensuring the preservation of B-type supersymmetry at the boundaries. Note that the expression \eqref{dSB} does not occur when performing a supersymmetry variation, since the $B$-field and $C$-field terms do not appear in the action of the GNLSM, as we have used the bulk $\hat{\theta}$ term of the corresponding GLSM in the construction of our boundary action via \eqref{thetaconvert}. The spurious boundary term \eqref{extra} also does not occur, for the same reason.
In analogy with the NLSM case, the Neumann boundary conditions on $Z^i$ imply that the equivariant B-brane %which supports a equivariant holomorphic line bundle 
wraps the entire target space, $\CP^{N-1}$, i.e., it is space-filling.

Next, let us find the explicit form of the boundary action. The parametrizations \eqref{para} and \eqref{lamconstraint2} give 
\begin{equation}\label{equivariantBbraneCpn}
\begin{aligned}
S_{\partial\Sigma}'=&\int\limits_{\partial\Sigma}
\dd x^0
\Bigg\{A^X_j\partial_0 X^j+A^X_{\ov{\jmath}}\partial_0 \ov{X}^{\ov{\jmath}}-\sum_c^{N-1}iR_c\til{\mathcal{A}}_{c}+n\del_0t-\frac{i}{2}F^X_{j\ov{k}}(\psi_+^j+\psi_-^j)(\ov{\psi}_+^{\ov{k}}+\ov{\psi}_-^{\ov{k}})\Bigg\}
\\&
+\sum_c^{N-1}\Big(\frac{\widetilde{\theta}_c}{2\pi}\int_{\Sigma}\widetilde{F}_{01c}d^2x+\frac{\widetilde{\theta}_c}{2\pi}\int_{\del\Sigma}\frac{(\widetilde{\si}_c+\ov{\widetilde{\si}}_c)}{2}dx^0\Big),
\end{aligned}
\end{equation}
where, as in \eqref{BbraneCpn}, $X^j=Z^j$, $\psi^j_{\pm}=\psi^{Z^j}_{\pm}$, and the components of the connection $A$ and curvature $F$ of $\mathcal{O}_{\C P^{N-1}}(-n)$ are given by \eqref{connectioncom} and \eqref{curvaturecom}, respectively. Besides the supersymmetrized $\til{\theta}$ terms, the only other new term (with respect to \eqref{BbraneCpn}) is 
\begin{equation}
-\sum_c^{N-1}iR_c\til{\mathcal{A}}_{c},
\end{equation}
where
\begin{equation}\label{complexgaugefield}
\til{\mathcal{A}}_{c}=-i\big(\til{A}_{0c}-\frac{(\til{\si}_c+\ov{\til{\si}}_c)}{2}\big),
\end{equation}
 and 
 \begin{equation}\label{Rc}
 R_c=\frac{-n(\sum_i^{N-1}\til{Q}_{ic}|Z^i|^2+\til{Q}_{Nc})}{(1+\sum_k^{N-1}|Z|^2)}.
 \end{equation}
The expression \eqref{equivariantBbraneCpn} is gauge invariant under the unbroken $U(1)^{N-1}$ symmetry.\footnote{Note that, from the local parametrizations \eqref{para} and \eqref{lamconstraint2}, we can see that the $U(1)^{N-1}$ gauge transformation of $t$ is $\delta t=\til{Q}_{Nc}\alpha$, since we know that the $U(1)^{N-1}$ charge of $Z^i$ is $\til{Q}_{1c}-\til{Q}_{Nc}$.} Now, we must remove the $n\del_0t$ term as we did in the NLSM case, since $t$ is not a coordinate of the $\C P^{N-1}$ target space, but rather locally parametrizes the Hopf fiber over $\C P^{N-1}$ which gives rise to the sphere $S^{N+1}$ defined by equation \eqref{Dconstraint2}. Furthermore, it is not a field which appears in the bulk theory, and has no supersymmetry transformation, leaving us unable to test the supersymmetry of the boundary action. Thus, we shall %gauge away
remove $n\del_0t$. However, doing so will break the classical $U(1)^{N-1}$ symmetry of our GNLSM at the boundaries. Now, this $U(1)^{N-1}$ gauge symmetry is not broken if we only require that it holds at the path integral level.
%and not necessarily the action. %Even in that case, the following steps
Nevertheless, attempting to restore the classical symmetry will help make the geometric properties of the brane obvious. %which will eventually aid us in deriving the description of nonabelian equivariant B-branes.

%Thus, we shall gauge away $n\del_0t$. However, doing so will break the $U(1)^{N-1}$ symmetry of our GNLSM at the boundaries. SEE SHORT VERSION FOR UPDATED VERSION OF ARGUMENT. A POINT WHICH IS NOT WRITTEN DOWN THERE IS THAT WE DO NOT EVEN NEED TO GAUGE AWAY $n\del_0t$, ItS INTEGRAL IS JUST EQUAL TO $2\pi n m$, FOR SOME INTEGER $m$, SO IT IS INCONSEQUENTIAL TO THE PATH INTEGRAL.

The cure to this broken symmetry is via the supersymmetrized $\til{\theta}$ terms, as follows. Setting $\widetilde{\theta}_c=2\pi n\til{Q}_{Nc}$, we have 
\begin{equation}
\begin{aligned}
&\sum_c^{N-1}\Big(\frac{\widetilde{\theta}_c}{2\pi}\int_{\Sigma}\widetilde{F}_{01c}d^2x+\frac{\widetilde{\theta}_c}{2\pi}\int_{\del\Sigma}\frac{(\widetilde{\si}_c+\ov{\widetilde{\si}}_c)}{2}dx^0\Big)\\
=&\sum_c^{N-1}\Big(-n\til{Q}_{Nc}\int_{\del\Sigma}\widetilde{A}_{0c}dx^0+n\til{Q}_{Nc}\int_{\del\Sigma}\frac{(\widetilde{\si_c}+\ov{\widetilde{\si_c}})}{2}dx^0\Big)\\
=&\sum_c^{N-1}\Big(-in\til{Q}_{Nc}\int_{\del\Sigma}\widetilde{\mathcal{A}}_{c}dx^0\Big)
\end{aligned}
\end{equation}
since both $n$ and the charge $\til{Q}_{Nc}$ are integers, as explained below \eqref{thetaconvert}. Then, the final boundary action takes the form
\begin{equation}\label{equivariantBbraneCpnFINAL}
\boxed{
\begin{aligned}
S_{\partial\Sigma}'=&\int\limits_{\partial\Sigma}
\dd x^0
\Bigg\{A^X_j\partial_0 X^j+A^X_{\ov{\jmath}}\partial_0 \ov{X}^{\ov{\jmath}}-\sum_c^{N-1}i\til{R}_c\til{\mathcal{A}}_{c}-\frac{i}{2}F^X_{j\ov{k}}(\psi_+^j+\psi_-^j)(\ov{\psi}_+^{\ov{k}}+\ov{\psi}_-^{\ov{k}})\Bigg\},
\end{aligned}}
\end{equation}
where\footnote{\label{26}
%As in the analogous NLSM case \cite{HIV}, %these boundary conditions together with 
The boundary conditions \eqref{Bgauge2} and the boundary action \eqref{equivariantBbraneCpnFINAL} result in equations of motion which are modified by boundary terms, for some of the fields.
%c.f. footnote \ref{21}.
%IMPORTANT NOTE: added on 2nd April 2017 IT IS CORRECT TO REFER TO ONLY THE VECTOR MULTIPLET BOUNDARY CONDITIONS IN THIS FOOTNOTE. EVEN WHEN WE DELETE THE BOUNDARY ACTION, THESE BOUNDARY CONDITIONS DO NOT LEAVE THE EQUATIONS OF MOTIONS LOCAL BECAUSE OF THETA PARAMETERS IN TWO OF THEM.THE BOUNDARY ACTION JUST MAKES THE EQUATIONS OF MOTION MORE NONLOCAL.
%IMPORTANT NOTE, SEE EQBRANES NOTES FILE TO SEE WHICH FIELDS HAVE NONLOCAL EOM, SHOULD BE ONLY PHI PSI AND F_01 IF NON MISTAKEN 
 %The same statement holds for the GLSM from which we derived these boundary conditions and boundary action.
 } 
\begin{equation}\label{Rctil}
\begin{aligned}
 \til{R}_c&=\frac{-n(\sum_i^{N-1}(\til{Q}_{ic}-\til{Q}_{Nc})|Z^i|^2)}{(1+\sum_k^{N-1}|Z^k|^2)}\\
&=-A^X_i\til{e}^i_c-A^X_{\ov{\jmath}}\ov{\til{e}}^{\ov{\jmath}}_c\\
&=-\iota_{\til{e}_c}A^X.
\end{aligned}
\end{equation}
%is known as the {\it moment} of the line bundle $\mathcal{O}_{\C P^{N-1}}(-n)$ with $U(1)^{N-1}$-\textit{equivariant} structure \cite{BGV,szabo}. 
%Indeed, 
Invariance  of the boundary action \eqref{equivariantBbraneCpnFINAL} under the B-type supersymmetry transformations %given in
%which follow from 
(given by \eqref{susytrans} and \eqref{susytrans2} for $\eps_+=-\eps_-$) %at B-boundary 
holds %if
since 
\begin{equation}\label{eqBianchi}
d\til{R}=\iota_{\til{e}}F^X.
\end{equation}
This is known as the 
%follows from 
the equivariant Bianchi identity, and implies that the line bundle $\mathcal{O}_{\C P^{N-1}}(-n)$ has $U(1)^{N-1}$-\textit{equivariant} structure,\footnote{\label{27}The $G$-equivariant Bianchi identity is equivalent to the $G$-invariance of the connection, $A$, of the bundle (equation \eqref{eqBianchi2}), which implies that the covariant derivative $d+A$ is $G$-invariant, and this defines a $G$-equivariant bundle, see \cite{szabo}, Section %IMPORTANT NOTE also see section 2.4. previously i referred to this section
3.2.}
 for which $\til{R}_c$ is the \textit{moment} \cite{BGV,szabo}.\footnote{For equivariant bundles with \textit{abelian} connection, the equivariant Bianchi identity takes the same form as the moment map equation \eqref{momentmapequation}. For the present case of $\mathcal{O}_{\C P^{N-1}}(-n)$, it is in fact proportional; the curvature is $F=n\omega$, which means that $\til{R}=n\til{\mu}$, where $\til{\mu}$ is the normalized moment map of the $U(1)^{N-1}$-isometry of $\C P^{N-1}$ (the discrepancy with \eqref{cpN-1momentmap} is because the moment map for an abelian $G$-action is only defined up to the addition of a constant, as explained below equation \eqref{Momcon2}).% The moment of a $G$-equivariant bundle and the moment map of the $G$-isometry of a manifold are closely related. 
} 
%\begin{equation}\label{eqBianchi}
%d\til{R}=\iota_{\til{e}}F^X,
%\end{equation}
%which implies that our equivariant B-brane supports a $U(1)^{N-1}$-equivariant line bundle $\mathcal{O}_{\C P^{N-1}}(-n)$.

%We can now show that the boundary action \eqref{equivariantBbraneCpnFINAL} is invariant under the B-type supersymmetry transformations given in \eqref{susytrans} and \eqref{susytrans2}.

The equivariant Bianchi identity is in fact a restatement of the $U(1)^{N-1}$-invariance of the connection,
\begin{equation}\label{eqBianchi2}
\mathcal{L}_{\til{e}}A^X=0,
\end{equation}
 %Furthermore
 Now, rewriting the boundary action \eqref{equivariantBbraneCpnFINAL} as
\begin{equation}\label{equivariantBbraneCpnFINALgaugein}
\begin{aligned}
S_{\partial\Sigma}'=&\int\limits_{\partial\Sigma}
\dd x^0
\Bigg\{A^X_j\partial^A_0 X^j+A^X_{\ov{\jmath}}\partial^A_0 \ov{X}^{\ov{\jmath}}+\sum_c^{N-1}\til{R}_c\frac{(\til{\si}_c+\ov{\til{\si}}_c)}{2}-\frac{i}{2}F^X_{j\ov{k}}(\psi_+^j+\psi_-^j)(\ov{\psi}_+^{\ov{k}}+\ov{\psi}_-^{\ov{k}})\Bigg\}
\end{aligned}
\end{equation}
facilitates the proof that it is invariant under the gauge transformations given in \eqref{gaugetrans1} and \eqref{gaugetrans2} for $G=U(1)^{N-1}$. The variation is
\begin{equation}\label{GAUGEVARIATIONequivariantBbraneCpnFINALgaugein}
\begin{aligned}
&\delta S_{\partial\Sigma}'\\=&\sum_a^{N-1}\alpha_a\int\limits_{\partial\Sigma}
\dd x^0
\Bigg\{\mathcal{L}_{\til{e}_a}A^X_j\partial^A_0 X^j+\mathcal{L}_{\til{e}_a}A^X_{\ov{\jmath}}\partial^A_0 \ov{X}^{\ov{\jmath}}+\sum_c^{N-1}\iota_{\til{e}_a}d\til{R}_c\frac{(\til{\si}_c+\ov{\til{\si}}_c)}{2}-\frac{i}{2}\mathcal{L}_{\til{e}_a}F^X_{j\ov{k}}(\psi_+^j+\psi_-^j)(\ov{\psi}_+^{\ov{k}}+\ov{\psi}_-^{\ov{k}})\Bigg\},
\end{aligned}
\end{equation}
which vanishes using \eqref{eqBianchi2}, as well as the identities $\mathcal{L}_{\til{e}}F^X=0$ and $\mathcal{L}_{\til{e}}\til{R}=\iota_{\til{e}}d\til{R}=0$. 

It may seem that we have picked a random value for $\til{\theta}_c$ in the derivation above. If we only required $U(1)^{N-1}$ gauge invariance of the path integral, then we would have been free to choose $\til{\theta}_c=2\pi m_c$ for any integer $m_c$, and we would still derive a boundary action which is B-type supersymmetry invariant, as well as gauge invariant mod $2\pi \Z$. This freedom is merely a reflection of the fact that the moment in the equivariant Bianchi identity \eqref{eqBianchi} is only defined up to a constant.

We have thus found B-type supersymmetric and $U(1)^{N-1}$ gauge invariant boundary conditions and boundary interactions corresponding to an equivariant B-brane in $\CPN$, which is a space-filling brane supporting the holomorphic line bundle $\mathcal{O}_{\C P^{N-1}}(-n)$ with $U(1)^{N-1}$-equivariant structure. We may follow a procedure analogous to that presented above for $\CPN$ in order to describe an equivariant B-brane in a toric manifold $X=\C^N//U(1)$ (by choosing different values for $\hat{Q}_i$), which would be a space-filling brane supporting the holomorphic line bundle $\mathcal{O}_{X}(-n)$ with $U(1)^{N-1}$-equivariant structure.

The GNLSM boundary action \eqref{equivariantBbraneCpnFINAL} that we have derived from the GLSM expressions \eqref{thetasusygnlsm} and \eqref{gnlsmuvb} is a special case of the more general boundary Wilson line found by Kapustin et al. \cite{Kapustin,Vyas}, using a B-twisted \textit{topological} nonabelian GNLSM, with gauge group $G$ and target space $X$, i.e., a gauged B-model. When the worldsheet is the Euclidean strip $I\times \R$, this Wilson line takes the form of the path integral insertion
\begin{equation}\label{superwilsonline}
W=\textrm{STr}(P\textrm{ }(e^{i\mathcal{N}}))
\end{equation}
with
\begin{equation}\label{equivariantBbraneKapustin}
\begin{aligned}
\mathcal{N}=&\int\limits_{\partial\Sigma}
\dd x^2
\Bigg\{A^X_j\partial_2 X^j+A^X_{\ov{\jmath}}\partial_2 \ov{X}^{\ov{\jmath}}-\sum_c^{\textrm{dim }\mathfrak{g}}\til{R}_c\til{\mathcal{A}}_{2c}-\frac{1}{2}F^X_{j\ov{k}}(\psi_+^j+\psi_-^j)_2(\ov{\psi}_+^{\ov{k}}+\ov{\psi}_-^{\ov{k}})_s+\frac{1}{2}(\psi_+^j+\psi_-^j)_2\nabla^E_jT\Bigg\},
\end{aligned}
\end{equation}
where $x^2$ is the direction along the boundaries, 
\begin{equation}\label{complexgaugefield2}
\til{\mathcal{A}}_{2c}=\til{A}_{2c}+i\frac{(\til{\si}_c+\ov{\til{\si}}_c)_2}{2},
\end{equation}
is a complexified gauge field valued in $G_{\C}$,
and where quantities with subscript `$2$' are components of one-forms along the boundaries, while the quantity with subscript `$s$' is a scalar. Here, $A$ is the {\it superconnection} of a %$\Z$-
{\it graded} $G$-equivariant holomorphic {\it vector} bundle%{\it superbundle}
, $E$, with the covariant derivative $\nabla^E=d+A$; while $T$ is a holomorphic degree-1 endomorphism $T:E\rightarrow E$, which satisfies $T^2=0$. The holomorphic transition functions of $E$ are valued in a %$U(m)$ structure group, where $m$ is the rank of $E$.
structure supergroup.
Moreover, $A$, $F$, $\til{R}$, and $T$ are valued in the Lie superalgebra of this structure supergroup, and obey the equivariant Bianchi identity 
\begin{equation}
\nabla^E\til{R}=\iota_{\til{e}}F^X,
\end{equation}
as well as the identity
\begin{equation}
\til{e}^i\nabla^E_iT+[\til{R},T]=0.
\end{equation}
The expression \eqref{equivariantBbraneKapustin} agrees with our result \eqref{equivariantBbraneCpnFINAL} when we take $E$ to be an ungraded holomorphic line bundle %associated to a principal $U(1)$-bundle. 
with $U(1)$ structure group. To see this, we need to analytically continue the Minkowski strip to Euclidean signature ($x^0=-ix^2$) and B-twist the fields in \eqref{equivariantBbraneCpnFINAL}. Then, the expression \eqref{complexgaugefield} becomes the complexified gauge field \eqref{complexgaugefield2}, and the fermionic fields become scalars or one-forms along the boundaries. Finally, we note that the $\nabla^E_jT$ term in \eqref{equivariantBbraneKapustin} does not occur in \eqref{equivariantBbraneCpnFINAL} since an ungraded %holomorphic vector 
bundle does not admit a degree-1 endomorphism; hence, $T=0$. 
%IMPORTANT NOTE: SEE HORI,HERBST,PAGE PHASES OF 1+1 THEORIES AND ALSO HORI LINEAR MODELS OF SUPERSYMMETRIC D-BRANES FOR DISCUSSION OF THIS T.
% , SHOW EUCLIDEAR COMPLEX GAUGE FIELD THEN ANALYTICALLY CONTINUE TO SHOW IT AGREES WITH THE PREVIOUS ONE
%Upon analytical continuation to Euclidean worldsheet ($x
%^0=-ix^2$) and B-twisting, the expression \eqref{complexgaugefield} becomes a complexified gauge field valued in the Lie algebra of $G_{\C}$, which is the complexification of $G$:
%where it is indicated that the combination ($\til{\si}_c+\ov{\til{\si}}_c$) is now a component of a worldsheet one-form after twisting. Likewise, the combination 
%($\psi_+^j+\psi_-^j$) is now a component of a one-form, while ($\ov{\psi}_+^{\ov{k}}+\psi_-^{\ov{k}}$) becomes a worldsheet scalar.
%The path integral Wilson line \eqref{superwilsonline} reduces to 
 %In other words, there are more general B-branes admissible, and these correspond to graded equivariant holomorphic vector bundles. 
%his implies that 
%I think straight away start comparing with Kapustin
As explained by Kapustin et al. \cite{Kapustin,Vyas}, in some cases, the category of branes defined by \eqref{equivariantBbraneKapustin} is equivalent to $D^b_{G_{\C}}(Coh(X))$, the bounded, derived category of $G_{\C}$-equivariant coherent sheaves on the target space, $X$. This occurs if $X$ has a $G$-resolution property, i.e., any $G$-equivariant coherent sheaf on $X$ has a $G$-equivariant resolution by $G$-equivariant holomorphic vector bundles. This property, however, does not hold for general complex manifolds. Nevertheless, even for such spaces where it does not hold, it is believed that the full category of equivariant B-branes is still $D^b_{G_{\C}}(Coh(X))$, where the GNLSMs for these spaces require more general boundary actions corresponding to  differential graded (DG) modules over the Dolbeault DG-algebra of $X$, instead of holomorphic bundles \cite{Kapustin,Vyas}.   
%IMPORTANT NOTE: SEE GLUING AFFINE VORTICES https://arxiv.org/pdf/1610.09764.pdf BY XU, PAGE 2, HERE HE MENTIONS THAT ON PROJECTIVE MANIFOLDS THE G-ACTION EXTENDS TO AN ALGEBRAIC ACTION BY G_{\C}/
In our construction, we have found abelian equivariant B-branes which wrap toric manifolds given by the quotient $X=\C^N//U(1)$, and which support the $U(1)^{N-1}$-equivariant holomorphic line bundle $\mathcal{O}_{X}(-n)$. In the language of algebraic geometry, $\mathcal{O}_{X}(-n)$ is a locally-free sheaf of rank 1, and is in fact one of the simplest objects of $D^b(Coh(X))$ (\cite{aspinwall}, page 56).
%IMPORTANT NOTE: there is no added information when going from locally free sheaves to coherent sheaves (pg 56 Aspinwall)we only do so because we need an abelian category.Using the derived category adds more informations, since it includes not just locally free sheaves.We need to do it because we do not have enough B-branes to match the number of A-branes.
%IMPORTANT NOTE: 17TH SEPTEMBER IT SEEMS LIKE WE HAVE ASSUMED HERE THAT LINE BUNDLES WITH EQUIVARIANT STRUCTURE, WHERE THE LINE BUNDLE IS A COHERENT SHEAF, CORRESPONDS TO AN EQUIVARIANT SHEAF. IT SHOULD BE CORRECT. IF IT IS CORRECT, THEN WE HAVE SHOWN THAT THE G-RESOLUTION PROPERTY MENTIONED BY KAPUSTIN HOLDS FOR TORIC MANIFOLDS. (TAKING INTO ACCOUNT THE RESULTS FOR \C^N//U(1)^N-k). added later- the ASSUMPTION SEEMS CORRECT SEE https://en.wikipedia.org/wiki/Equivariant_sheaf - WHICH CONTAINS DEFINITIONS FOR BOTH EQUIVARIANT SHEAFS AND EQUIVARIANT VECTOR BUNDLES. IT SEEMS TO BE CORRECT BECAUSE OF THE LINE 'Thus, if V is a vector bundle corresponding to F, then ϕ {\displaystyle \phi } \phi induces isomorphisms between fibers V x → ≃ V g x {\displaystyle V_{x}{\overset {\simeq }{\to }}V_{gx}} V_x \overset{\simeq}\to V_{gx}, which are linear maps.' THE ISOMORPHISM BETWEEN FIBERS SEEMS TO IMPLY AN EQUIVARIANT COHERENT SHEAF, BASED ON THE EARLIER PART OF THE ARTICLE (STALKS ARE LIKE FIBERS). BASICALLY BOTH DEFINITIONS SHOWN THERE REQUIRE AN ISOMORPHISM OF FIBRES/STALKS. IN BOTH CASES THERE IS A NOTION OF LIFTING THE G ACTION FROM X TO THE SHEAF/VECTOR BUNDLE
 The additional $U(1)^{N-1}$-equivariant structure then implies that the equivariant B-branes we have found are objects in $D^b_{(\C^{\times})^{N-1}}(Coh(X))$, the bounded, derived category of $(\C^{\times})^{N-1}$-equivariant coherent sheaves on $X$.\footnote{The algebraic torus $(\C^{\times})^{N-1}$ is the complexification of $U(1)^{N-1}$.} Of course, we have not constructed all the objects in the category.

In particular, we have not constructed 
%IMPORTANT NOTE: THE FOLLOWING IS TRUE ON ITS OWN, BUT THE PREVIOUS SENTENCE SAYS THAT WE HAVE NOT CONSTRUCTED ALL THE OBJECTS IN THE CATEGORY, BUT THE CATEGORY DOES NOT ALWAYS CORRESPOND TO HOLOMORPHIC VECTOR BUNDLES, AS KAPUSTIN HAS SAID. O_x(-n) is a locally free sheaf, so it's okay to say that its is part of the category even for lower dimensional branes.
%equivariant holomorphic vector bundles of rank greater than 1, nor have we constructed
 non-space-filling equivariant B-branes. %of lower dimension. 
 The latter, i.e., equivariant B-branes of lower dimension, should exist, in analogy with the NLSM case (see footnote \ref{19}), although we shall not attempt to derive them from GLSMs here.
 %construct them here. %IMPORTANT NOTE: Dirichlet boundary conditions on some of the chiral multiplets will result in some of the boundary del_0 terms becoming zero
The path to doing so is via Hori's construction of non-space-filling ordinary B-branes from GLSMs \cite{Hori1}. Using the same GLSM used there, but with gauge group generalized to $U(1)^N$, we should be able to derive the relevant GNLSM boundary action and boundary conditions, as we have done for space-filling equivariant B-branes in this section.

%\mbox{}\par\nobreak
%\noindent
%\textit{Equivariant B-branes on $\C^N//U(1)^{N-k}$ from GLSM} 
\subsection{Equivariant B-branes on $\C^N//U(1)^{N-k}$ from GLSM}

The prior discussion can be generalized to the case of general K\"ahler toric manifolds, i.e., $X=\C^N//U(1)^{N-k}$. %Once again, 
%IMPORTANT NOTE: THERE ARE ALSO TORIC ORBIFOLDS, WHICH CORRESPOND TO FINITE GAUGE GROUP THEORIES
We impose the B-type supersymmetric boundary conditions \eqref{gnlsmBsuperbc} and \eqref{Flocal2} on the GLSM (for $N-k>1$), which include the purely Neumann boundary conditions on $\phi_i$, while also supersymmetrizing the GLSM theta terms
\begin{equation}\label{thetasusygnlsmmulti}
\sum_b^{N-k}\Big(\frac{\hat{\theta}_b}{2\pi}\int_{\Sigma}\hat{F}_{01b}d^2x+\frac{\hat{\theta}_b}{2\pi}\int_{\del\Sigma}\frac{(\hat{\si}_b+\ov{\hat{\si}}_b)}{2}dx^0\Big)+\sum_c^{k}\Big(\frac{\widetilde{\theta}_c}{2\pi}\int_{\Sigma}\widetilde{F}_{01c}d^2x+\frac{\widetilde{\theta}_c}{2\pi}\int_{\del\Sigma}\frac{(\widetilde{\si}_c+\ov{\widetilde{\si}}_c)}{2}dx^0\Big).
\end{equation} This preserves B-type supersymmetry at the boundaries. In addition, the B-type supersymmetric GLSM boundary action needed is 
\begin{equation}\label{gnlsmuvbmulti}
S_{\partial\Sigma}={\theta'\over 4\pi r'}\int\limits_{\partial\Sigma}
\txd x^0 \sum_i^N\Big(\, i{D}_0\ov{\phi}_i\,\phi_i-i\ov{\phi}_i {D}_0\phi_i
 \:+\: (\psi_{+i}+\psi_{-i})(\ov{\psi}_{+i}+\ov{\psi}_{-i})
-\sum_a^N\mathcal{Q}_{ia}(\si_a+\ov{\si}_a)|\phi_i|^2
\Big),
\end{equation}
where $\theta'=2\pi n'$ ($n'\in \Z$) and $r'\in \R$. %$n'\in \Z$ and $r'\in \R$, 
In addition, we ought to set $\hat{\theta}_b=2\pi \hat{n}_b$, where $\hat{n}_b\in \Z$, and we need to impose the condition
%IMPORTANT NOTE:PREVIOUSLY WE DID NOT DEFINE \theta' in this subsection
%\begin{equation}\label{multiconstraint}
%\frac{n'}{r'}=\frac{\hat{n}_b}{\hat{r}_b}
%\end{equation} 
\begin{equation}\label{multiconstraint}
\frac{\theta'}{r'}=\frac{\hat{\theta}_b}{\hat{r}_b}
\end{equation}
for all values of $b$. 

This allows us to write \eqref{gnlsmuvbmulti} and \eqref{thetasusygnlsmmulti} as
\begin{equation}\label{gnlsmuvb2multi}
\begin{aligned}
S_{\partial\Sigma}'=&{n'\over 2 r'}\int\limits_{\partial\Sigma}
\txd x^0 \Big(\, \sum_i^N(i\del_0\ov{\phi}_i\,\phi_i-i\ov{\phi}_i \del_0\phi_i)
 \:+\: \sum_i^N(\psi_{+i}+\psi_{-i})(\ov{\psi}_{+i}+\ov{\psi}_{-i})\\&
 +2\sum^{N-k}_b \hat{A}_{0b}(\sum_i^N\hat{Q}_{ib}|\phi_i|^2-\hat{r}_b)
-\sum^{N-k}_b(\hat{\si}_b+\ov{\hat{\si}}_b)(\sum_i^N\hat{Q}_{ib}|\phi_i|^2-\hat{r}_b)\\&
 +2\sum^{k}_c\til{A}_{0c}\sum_i^N\til{Q}_{ic}|\phi_i|^2
-\sum^{k}_c(\til{\si}_c+\ov{\til{\si}}_c)\sum_i^N\til{Q}_{ic}|\phi_i|^2\Big)\\&
+\sum_c^{k}\Big(\frac{\widetilde{\theta}_c}{2\pi}\int_{\Sigma}\widetilde{F}_{01c}d^2x+\frac{\widetilde{\theta_c}}{2\pi}\int_{\del\Sigma}\frac{(\widetilde{\si}_c+\ov{\widetilde{\si}}_c)}{2}dx^0\Big)
\end{aligned}
\end{equation}
Taking the $\hat{e}_b\rightarrow \infty$ limit allows us to integrate $\hat{D}_b$ out of the action, which imposes the constraints \eqref{Dconstraint}, and the second line in \eqref{gnlsmuvb2multi} vanishes. Integrating out the other components of the vector multiplets, $\hat{V}_b$, then imposes \eqref{lamconstraint} on the entire action, as well \eqref{generalv} and \eqref{generalsi} on the bulk action. Then, to find the explicit boundary action, one needs to use parametrizations which satisfy \eqref{lamconstraint} and \eqref{Dconstraint}. The explicit boundary conditions are also found using these parametrizations, together with \eqref{generalv} and \eqref{generalsi}.

We would then be able to identify the first term
%We can identify the first term 
in \eqref{gnlsmuvb2multi} as the Hermitian connection of the holomorphic line bundle $\mathcal{O}(k_1,\ldots,k_{N-k})$
%\begin{equation}\label{hermimulti}
%A^X_I\txd X ^I=
%-{n' \over r'}{i\over 2}{\displaystyle~\sum \!{}_{i=1}^N
%\ov{\phi}_i\overleftrightarrow{\txd} \phi_i~},
%\end{equation}
%of the holomorphic line bundle $\bigotimes_{b=1}^{N-k}\mathcal{O}_X(-\hat{n}_b)=\mathcal{O}_X(\sum_{b=1}^{N-k}(-\hat{n}_b))$ 
over $\C^N//U(1)^{N-k}$, where $k_1,\ldots,k_{N-k}\in \Z$ would be integers related to $\hat{n}_1,\ldots,\hat{n}_{N-k}$.
%, since it transforms under $U(1)^{N-k}$ gauge transformations ($\phi_i\rightarrow e^{i\sum_{b=1}^{N-k}\hat{Q}_{ib}\alpha_b}\phi_i$) as 	
%\begin{equation}\label{gboundgauge}
%A^X_I\txd X^I\rightarrow A^X_I\txd X^I -\sum_{b=1}^{N-k}(-\hat{n}_b)d\alpha_b,
%\end{equation}
%and setting $\alpha_b=\alpha$, we retrieve the $U(1)$ gauge transformation of the connection of $\bigotimes_{b=1}^{N-k}\mathcal{O}_X(-\hat{n}_b)$.
Both supersymmetry invariance and gauge invariance under the residual $U(1)^k$ gauge symmetry of the GNLSM would then require that this line bundle has $U(1)^k$-equivariant structure. Moreover, 
%upon topological B-twisting,
 we would be able to identify the equivariant B-branes we have found as objects in $D^b_{(\C^{\times})^{k}}(Coh(X))$, the bounded, derived category of $(\C^{\times})^{k}$-equivariant coherent sheaves on $X$.

The simplest example would be that of the $U(1)^2$-equivariant holomorphic line bundle $\mathcal{O}_X(-\hat{n}_1,-\hat{n}_2)$ over $X=\CP^{1}\times \CP^{1}$, which just corresponds to two copies of the boundary action given in \eqref{equivariantBbraneCpnFINAL}, with $N=2$.\footnote{In fact, for toric manifolds which are Cartesian products like $X=\CP^{1}\times \CP^{1}$, the complete decoupling of the two boundary actions means that we no longer need the constraint \eqref{multiconstraint}.} %IMPORTANT NOTE: WHEN X IS A TRIVIAL PRODUCT OF TWO TORIC MANIFOLDS WE NO LONGER REQUIRE THE CONSTRAINT WHICH RELATES THE DIFFERENT n_b's! 
 One can even consider equivariant B-branes on %more complicated toric manifolds which involve 
 fibrations of $\CP^1$ over $\CP^1$ known as Hirzebruch surfaces, using GLSMs with appropriately charged scalar fields \cite{hori2003mirror}. It is worth noting that the derived categories of $\C^{\times}$-equivariant coherent sheaves over $\CP^1$, Hirzebruch surfaces, $\CP^1$ fibered over Hirzebruch surfaces etc. provide a construction of Khovanov homology \cite{cautiskamnitzer}. %IMPORTANT NOTE, WITTEN MENTIONS THE 4D PHYSICAL INTERPRETATION OF THIS CONSTRUCTION 

%\mbox{}\par\nobreak
%\noindent
%\textit{Alternative Formulation} 
\subsection{Alternative Formulation}
%Analogous to the case of ordinary B-branes \cite{HIV}, there should exist an alternative formulation of equivariant B-branes, where the classical boundary conditions on $\phi_i$ fields of the GLSM are mixed Dirichlet-Neumann instead of pure Neumann. We shall now show that such an alternative formulation exists. 
We shall now derive the alternative formulation of abelian equivariant B-branes, in terms of a boundary action which generalizes \eqref{BbraneAlt}, as well as the relevant boundary conditions. Let us first recall the derivation of the NLSM boundary action \eqref{BbraneAlt} for a space-filling B-brane on $X=\C^N//U(1)$ \cite{HIV}. The $U(1)$-GLSM boundary action from which \eqref{BbraneAlt} can be derived is 
\begin{equation}\label{nlsmuvbAlt}
S_{\partial\Sigma}={\hat{\theta}\over 4\pi \hat{r}}\int\limits_{\partial\Sigma}
\txd x^0 \sum_i^N\Big(\, i\hat{D}_0\ov{\phi}_i\,\phi_i-i\ov{\phi}_i \hat{D}_0\phi_i
\Big).
\end{equation}
%The relevant boundary conditions necessary for the preservation of B-type supersymmetry at the boundary are 
To preserve B-type supersymmetry at the boundaries, we must impose \cite{HIV}
\begin{equation}\label{BsuperbcAlt}
\begin{aligned}
e^{-i\hat{\gamma}}\hat{\mathcal{D}}_+\Phi_i&=e^{i\hat{\gamma}}\hat{\mathcal{D}}_-\Phi_i\\
e^{i\hat{\gamma}}\hat{\Sigma}&=e^{-i\hat{\gamma}}\ov{\hat{\Sigma}}
\end{aligned}
\end{equation}
at B-boundary, where $\hat{\mathcal{D}}_{\pm}=e^{-\hat{Q}_i\hat{V}}D_{\pm}e^{\hat{Q}_i\hat{V}}$, and where $\hat{\gamma}$ is the phase of $\hat{t}=\hat{r}-i\hat{\theta}=|\hat{t}|e^{i\hat{\gamma}}$. In components, these are 
 \begin{equation}\label{BtypeGLSMtoNLSMbcAlt}
\begin{aligned}
e^{-i\hat{\gamma}}\psi_{+i}-e^{i\hat{\gamma}}\psi_{-i}&=0\\
F_i&=0\\
\textrm{cos}(\hat{\gamma})\hat{D}_1\phi_i-i\textrm{sin}(\hat{\gamma}) \hat{D}_0\phi_i&=0\\
\textrm{cos}(\hat{\gamma})\hat{D}_1(\psi_{+i}+\psi_{-i})-i\textrm{sin}(\hat{\gamma}) \hat{D}_0(\psi_{+i}+\psi_{-i})-\textrm{cos}(\hat{\gamma})({\hat{\lambda}}_{+}+{\hat{\lambda}}_{-})\phi_i&=0
\end{aligned}
\end{equation}
and
\begin{equation}\label{NLSMBgaugeAlt}
\begin{aligned}
%%A_1&=\textrm{constant}\\
%%\del_1 A_0&=0\\
\textrm{Im}(e^{i\hat{\gamma}}\hat{\si})&=0\\
{e^{-i\hat{\gamma}}\hat{\lambda}}_{+}+{e^{i\hat{\gamma}}\hat{\lambda}}_{-}&=0\\
\del_1\textrm{Re}(e^{i\hat{\gamma}}\hat{\si})+\textrm{cos}(\hat{\gamma})\hat{F}_{01}-\textrm{sin}(\hat{\gamma})\hat{D}&=0\\
%\ov{\hat{\lambda}}_{+}+\ov{\hat{\lambda}}_{-}&=0\\
%\del_1({\hat{\lambda}}_{+}-{\hat{\lambda}}_{-})&=0\\
%\del_1(\ov{\hat{\lambda}}_{+}-\ov{\hat{\lambda}}%_{-})&=0\\
%%\del_1\textrm{Re}(\hat{\si})&=0\\
%\del_1(\hat{D}+\del_1\textrm{Im}(\hat{\si}))&=0.
\end{aligned}
\end{equation}
which includes the mixed Dirichlet-Neumann boundary condition on the scalar fields $\phi_i$. These conditions are not sufficient for B-type supersymmetry of the GLSM action at the boundaries, and in addition, we must impose the boundary condition 
%IMPORTANT NOTE, WE MUST HAVE THE FOLLOWING CONDITION TO GET 6.23 OF HIV. IN THE PREVIOUS FORMULATION, THIS WAS NOT NECESSARY, AS THE BOUNDARY \si + \ov{\si} term helped make T real in eqn 6.7 of HIV.
\begin{equation}\label{HIVf01condition}
\frac{\hat{F}_{01}}{\hat{e}^2}=-\hat{\theta}+\hat{\theta}\frac{\sum_i^N\hat{Q}_i|\phi_i|^2}{\hat{r}},
\end{equation}
as well as integrate $\hat{D}$ out of the action to obtain its algebraic equation of motion
\begin{equation}\label{HIVDcondition}
\frac{\hat{D}}{\hat{e}^2}=\hat{r}-\sum_i^N\hat{Q}_i|\phi_i|^2,
\end{equation}
which holds on the entire worldsheet.\footnote{The constraints \eqref{HIVf01condition} and \eqref{HIVDcondition} result in the third equation of \eqref{NLSMBgaugeAlt} becoming $\del_1\textrm{Re}(e^{i\hat{\gamma}}\hat{\si})=0$.} Note that the B-type supersymmetry transformations of \eqref{HIVf01condition} further implies the boundary conditions
%IMPORTANT NOTE, (kiv) THE BOTTOM TWO CONDITIONS ARE NOT ACTUALLY NECESSARY FOR B-TYPE SUSY INVARIANCE AT THE BOUNDARY 
\begin{equation}
\begin{aligned}
\frac{i}{2\hat{e}^2}\big((\del_0-\del_1)\ov{\lam}_++(\del_0+\del_1)\ov{\lam}_-\big)&=\frac{\hat{\theta}}{\hat{r}}\sum^N_i\hat{Q}_i(\psi_{-i}+\psi_{+i})\ov{\phi}_i,\\
\frac{1}{\hat{e}^2}\big(\del_1(\hat{D}+\del_1\textrm{Im}(\hat{\si}))+\del_0(i\hat{F}_{01}-\del_0\textrm{Im}(\hat{\si}))\big)=&\frac{\hat{\theta}}{\hat{r}}\sum^N_i\hat{Q}_i\big(2iD_0\phi_i\ov{\phi}_i+2\textrm{Re}(\hat{\si})\hat{Q}_i|\phi_i|^2\\&-(\psi_{+i}+\psi_{-i})(\ov{\psi}_{+i}+\ov{\psi}_{-i}) \big).
\end{aligned}
\end{equation}
% via B-type supersymmetry.
 Unlike the formulation presented earlier, the boundary conditions given above ensure the locality of the equations of motion derived from the (bulk+boundary) action. %It will be useful for later purposes to 
%Note that in $\hat{r}\rightarrow \infty$ limit, which implies $\hat{{\gamma}}\rightarrow 0$, these boundary conditions reduce to those given by \eqref{BtypeGLSMtoNLSMbc}, \eqref{NLSMBgauge}, \eqref{Flocal} and \eqref{furtherimplied}.
%IMPORTANT NOTE: THE ABOVE LINE IS BLANKED OUT BECAUSE BY USING THE D EQUATION OF MOTION, we get \del_1Re(e^{i\hat{\gamma}}\hat{\si})=0, not \del_1Re(e^{i\hat{\gamma}}\hat{\si})=e^2_a\theta_a

%We shall briefly recall the derivation \cite{HIV} for $X=\C//U(1)$, before generalizing it for GNLSMs. 

Converting the theta term to a boundary term as in \eqref{thetaconvert}, with $\hat{\theta}=2 \pi n$, the complete boundary action is
\begin{equation}\label{nlsmuvb2Alt}
\begin{aligned}
S_{\partial\Sigma}'=&{n\over 2 \hat{r}}\int\limits_{\partial\Sigma}
\txd x^0 \Big(\, \sum_i^N(i\del_0\ov{\phi}_i\,\phi_i-i\ov{\phi}_i \del_0\phi_i)
 +2\hat{A}_0(\sum_i^N\hat{Q}_{i}|\phi_i|^2-\hat{r}) 
\Big).
\end{aligned}
\end{equation} 
The term proportional to $\hat{A}_0$ vanishes in the $\hat{e}\rightarrow \infty$ limit, whereby \eqref{nlsmconstraint1} is strictly imposed via \eqref{HIVDcondition}%after integrating $\hat{D}$ out
, and the remaining term is just the hermitian connection \eqref{hermi} of the holomorphic line bundle $\mathcal{O}_X(-n)$ on the toric manifold $X=\C^N//U(1)$, and therefore we obtain \eqref{BbraneAlt} for a space-filling B-brane. Integrating out the rest of the vector multiplet components imposes additional constraints which only affect the bulk action but not the boundary action. These constraints, together with the appropriate parametrizations for $\phi_i$ and $\psi_i$, are useful for finding the corresponding NLSM boundary conditions.

Now, to derive the boundary action for a GNLSM with $X=\C^N//U(1)$, we start with the $U(1)^N$-GLSM boundary action 
\begin{equation}\label{gnlsmuvbAlt}
S_{\partial\Sigma}={\hat{\theta}\over 4\pi \hat{r}}\int\limits_{\partial\Sigma}
\txd x^0 \sum_i^N\Big(\, i{D}_0\ov{\phi}_i\,\phi_i-i\ov{\phi}_i {D}_0\phi_i
\Big),
\end{equation}
where the covariant derivatives of the scalar fields are given by \eqref{covder}. %The compatible boundary conditions necessary for 
B-type supersymmetry invariance of the $U(1)^N$-GLSM at the boundaries of the worldsheet firstly requires that we impose
\begin{equation}
\label{gnlsmBsuperbcAlt}
e^{-i\hat{\gamma}}{\mathcal{D}}_+\Phi_i=e^{i\hat{\gamma}}{\mathcal{D}}_-\Phi_i\\
\end{equation}
\begin{equation}
\label{gnlsmBsuperbcAlt2}
e^{i\gamma_a}{\Sigma}_a=e^{-i\gamma_a}\ov{{\Sigma}}_a
\end{equation}
at B-boundary, where 
$\mathcal{D}_{\pm}=e^{-\sum_a^{N}{\mathcal{Q}}_{ia} V_a}D_{\pm} e^{\sum_a^{N}{\mathcal{Q}}_{ia} V_a}$,
 while $\hat{\gamma}$ and ${\gamma_a}$ are the phases of $\hat{t}=|\hat{t}|e^{i\hat{\gamma}}$ and ${t_a}=|{t_a}|e^{i{\gamma_a}}$ respectively. %To ensure B-type supersymmetry at the boundaries
 Secondly , we also ought to impose % Furthermore, we ought to impose 
\begin{equation}\label{assumption1}
\frac{\hat{\theta}}{\hat{r}}=\frac{{\theta_a}}{{r_a}},
\end{equation}
%in order to ensure B-type supersymmetry at the boundaries,
%IMPORTANT NOTE: THE ABOVE IMPLIES THE BELOW, BUT THE BELOW DOES NOT IMPLY THE ABOVE BECAUSE OF THE PROPERTIES OF THE TAN^{-1} FUNCTION
% which in turn implies
and
\begin{equation}\label{hatgammagammaa}
\hat{\gamma}=\gamma_a.
\end{equation}
Then, in components, \eqref{gnlsmBsuperbcAlt} and \eqref{gnlsmBsuperbcAlt2} become
 \begin{equation}\label{BtypeGLSMtoGNLSMbcAlt}
\begin{aligned}
e^{-i\hat{\gamma}}\psi_{+i}-e^{i\hat{\gamma}}\psi_{-i}&=0\\
F_i&=0\\
\textrm{cos}(\hat{\gamma}){D}_1\phi_i-i\textrm{sin}(\hat{\gamma}) {D}_0\phi_i&=0\\
\textrm{cos}(\hat{\gamma}){D}_1(\psi_{+i}+\psi_{-i})-i\textrm{sin}(\hat{\gamma}) {D}_0(\psi_{+i}+\psi_{-i})-\textrm{cos}(\hat{\gamma})\sum_a^NQ_{ia}({{\lambda}}_{+a}+{{\lambda}}_{-a})\phi_i&=0
\end{aligned}
\end{equation}
and
%VERY IMPORTANT NOTE, WE HAVE REPLACED \gamma_a by \hat{\gamma} in the following. The first condition which becomes Im(e^{i\hat{\gamma}}\si) then results in the last two equations above taking their present form, otherwise there will be terms proportional to Im(e^{i\hat{\gamma}}\si) appearing in both
%\begin{equation}\label{GNLSMBgaugeAlt}
%\begin{aligned}
%\textrm{Im}(e^{i{\gamma_a}}{\si_a})&=0\\
%{e^{i{\gamma_a}}{\lambda}}_{+a}+{e^{-i{\gamma_a}}{\lambda}}_{-a}&=0\\
%\del_1\textrm{Re}(e^{i{\gamma_a}}{\si_a})+\textrm{cos}({\gamma_a}){F}_{01a}-\textrm{sin}({\gamma_a}){D_a}&=0\\
%\end{aligned}
%end{equation}
\begin{equation}\label{GNLSMBgaugeAlt}
\begin{aligned}
%%A_1&=\textrm{constant}\\
%%\del_1 A_0&=0\\
\textrm{Im}(e^{i{\hat{\gamma}}}{\si_a})&=0\\
{e^{-i{\hat{\gamma}}}{\lambda}}_{+a}+{e^{i{\hat{\gamma}}}{\lambda}}_{-a}&=0\\
\del_1\textrm{Re}(e^{i{\hat{\gamma}}}{\si_a})+\textrm{cos}({\hat{\gamma}}){F}_{01a}-\textrm{sin}({\hat{\gamma}}){D_a}&=0\\
%\ov{\hat{\lambda}}_{+}+\ov{\hat{\lambda}}_{-}&=0\\
%\del_1({\hat{\lambda}}_{+}-{\hat{\lambda}}_{-})&=0\\
%\del_1(\ov{\hat{\lambda}}_{+}-\ov{\hat{\lambda}}%_{-})&=0\\
%%\del_1\textrm{Re}(\hat{\si})&=0\\
%\del_1(\hat{D}+\del_1\textrm{Im}(\hat{\si}))&=0.
\end{aligned}
\end{equation}
which includes the mixed Dirichlet-Neumann boundary condition on the scalar fields $\phi_i$. Finally, for complete boundary B-type supersymmetry invariance, we must impose the boundary condition 
\begin{equation}\label{Flocal3}
\frac{{F}_{01a}}{e_a^2}=-{\theta_a}+\hat{\theta}\frac{\sum_i^N{\mathcal{Q}}_{ia}|\phi_i|^2}{\hat{r}},
\end{equation}
as well as integrate ${D}_a$ out of the action to obtain its algebraic equation of motion
\begin{equation}\label{GHIVDcondition}
\frac{{D}_a}{{e}_a^2}={r}_a-\sum_i^N{\mathcal{Q}}_{ia}|\phi_i|^2,
\end{equation}
which holds on the entire worldsheet.\footnote{The constraints \eqref{hatgammagammaa}, \eqref{Flocal3} and \eqref{GHIVDcondition} result in the third equation of \eqref{GNLSMBgaugeAlt} becoming $\del_1\textrm{Re}(e^{i\hat{\gamma}}{\si_a})=0$.} %Note that the B-type supersymmetry transformations of \eqref{HIVf01condition} further implies the boundary conditions
%for complete boundary B-type supersymmetry invariance, and 
The condition \eqref{Flocal3} further implies two more boundary conditions via B-type supersymmetry.
%\begin{equation}
%\begin{aligned}
%\frac{i}{2\hat{e}^2}\big((\del_0-\del_1)\ov{\lam}_++(\del_0+\del_1)\ov{\lam}_-\big)&=\frac{\hat{\theta}}{\hat{r}}\sum^N_i\hat{Q}_i(\psi_{-i}+\psi_{+i})\ov{\phi}_i\\
%\frac{1}{\hat{e}^2}\big(\del_1(\hat{D}+\del_1\textrm{Im}(\hat{\si}))+\del_0(i\hat{F}_{01}-\del_0\textrm{Im}(\hat{\si}))\big)=&\frac{\hat{\theta}}{\hat{r}}\sum^N_i\hat{Q}_i\big(2iD_0\phi_i\ov{\phi}_i+2\textrm{Re}(\hat{\si})\hat{Q}_i|\phi_i|^2\\&-(\psi_{i+}+\psi_{i-})(\ov{\psi}_{i+}+\ov{\psi}_{i-}) \big)
%\end{aligned}
%\end{equation}
 %via B-type supersymmetry. 
As expected, all the boundary conditions above ensure the locality of the equations of motion derived from the action. %It will be useful for later purposes to note that in $\hat{r}\rightarrow \infty$ limit, which implies $\hat{{\gamma}}\rightarrow 0$, these boundary conditions reduce to those given by \eqref{BtypeGLSMtoNLSMbc}, \eqref{NLSMBgauge} and \eqref{Flocal}.
%as well as 
%\begin{equation}\label{Flocal2Alt}
%F_{01a}=-{e_a^2}{\theta_a}\\.
%\end{equation}

Now, setting $\hat{\theta}=2\pi n$, %and $\widetilde{\theta}_c=2\pi n\til{Q}_{Nc}$, which allows us to convert all the theta terms into boundary terms, and 
the relevant action which consists of \eqref{gnlsmuvbAlt} together with the theta terms is 
\begin{equation}\label{gnlsmuvb2Alt}
\begin{aligned}
S_{\partial\Sigma}'=&{n\over 2 \hat{r}}\int\limits_{\partial\Sigma}
\txd x^0 \Big(\, \sum_i^N(i\del_0\ov{\phi}_i\,\phi_i-i\ov{\phi}_i \del_0\phi_i)
 +2\hat{A}_0(\sum_i^N\hat{Q}_{i}|\phi_i|^2-\hat{r})
\\&
 +2\sum^{N-1}_c\til{A}_{0c}\sum_i^N\til{Q}_{ic}|\phi_i|^2\Big)
+\sum_c^{N-1}\Big(\frac{\widetilde{\theta}_c}{2\pi}\int_{\Sigma}\widetilde{F}_{01c}d^2x\Big).
\end{aligned}
\end{equation}
The term proportional to $\hat{A}_0$ vanishes in the $\hat{e}\rightarrow \infty$ limit %by integrating out 
using the equation of motion for $\hat{D}$ given in \eqref{GHIVDcondition}, while the constraints that arise from subsequently integrating out the rest of the vector multiplet $\hat{V}$ do not affect the boundary action. For $X=\CP^{N-1}$, we can use the parametrizations \eqref{para}, and \eqref{gnlsmuvb2Alt} becomes 
\begin{equation}\label{equivariantBbraneCpnAlt}
\begin{aligned}
S_{\partial\Sigma}'=&\int\limits_{\partial\Sigma}
\dd x^0
\Bigg\{A^X_j\partial_0 X^j+A^X_{\ov{\jmath}}\partial_0 \ov{X}^{\ov{\jmath}}-\sum_c^{N-1}iR_c\til{{A}}_{c}+n\del_0t \Bigg\}
+\sum_c^{N-1}\Big(\frac{\widetilde{\theta}_c}{2\pi}\int_{\Sigma}\widetilde{F}_{01c}d^2x \Big),
\end{aligned}
\end{equation}
with $A_I$ given in \eqref{connectioncom}, and $R_c$ given in \eqref{Rc}. Then, gauging away the $n\del_0t$ term, and setting $\widetilde{\theta}_c=2\pi n\til{Q}_{Nc}$, we arrive at the boundary action
\begin{equation}\label{equivariantBbraneCpnAlt2}
\boxed{
\begin{aligned}
S_{\partial\Sigma}'=&\int\limits_{\partial\Sigma}
\dd x^0
\Bigg\{A^X_j\partial_0 X^j+A^X_{\ov{\jmath}}\partial_0 \ov{X}^{\ov{\jmath}}-\sum_c^{N-1}i\til{R}_c\til{\mathcal{A}}_{c}\Bigg\},
\end{aligned}}
\end{equation}
where $\til{R}_c$ is the moment %of the $U(1)^{N-1}$-equivariant line bundle $\mathcal{O}_{\CP^{N-1}}$ 
given by \eqref{Rctil}. The boundary action can be rewritten concisely as 
\begin{equation}\label{equivariantBbraneCpnFINALgaugeinAlt}
\begin{aligned}
S_{\partial\Sigma}'=&\int\limits_{\partial\Sigma}
\dd x^0
\Bigg\{A^X_j\partial^A_0 X^j+A^X_{\ov{\jmath}}\partial^A_0 \ov{X}^{\ov{\jmath}}\Bigg\},
\end{aligned}
\end{equation}
and gauge invariance follows since \eqref{eqBianchi2} is obeyed, which implies that the line bundle $\mathcal{O}_{\CP^{N-1}}(-n)$ supported by the equivariant B-brane %supports a 
has $U(1)^{N-1}$-equivariant structure.
%IMPORTANT NOTE: ONLY gauge invariance of the boundary action requires use of the moment map equation, and hence only gauge invariance implies that the line bundle is equivariant. Supersymmetry invariance follows from the boundary conditions below.

The boundary conditions for the GNLSM with $\CP^{N-1}$ target can similarly be found; for the $U(1)^{N-1}$ vector multiplets, the boundary conditions follow from  
\eqref{GNLSMBgaugeAlt} and \eqref{Flocal3}, while for the matter fields, the boundary conditions are 
\begin{equation}
\boxed{
\begin{aligned}
g_{\ov{\jmath}i}(\psi^{Z^i}_--\psi^{Z^i}_+)+2\pi F^X_{\ov{\jmath}i}(\psi^{Z^i}_++\psi^{Z^i}_-)&=0\\
g_{\ov{\jmath}i}\del^A_1Z^i-2\pi F^X_{\ov{\jmath}i}\del^A_0Z^i&=0,\\
\end{aligned}}
\end{equation}
and their B-type supersymmetric completions, where $g$ is the Fubini-Study metric and $F$ is the curvature of $\mathcal{O}_{\CP^{N-1}}(-n)$ given in \eqref{curvaturecom}. 
%TO CHECK
%IMPORTANT NOTE: 4TH APRIL 2017, CHECKED
%IMPORTANT NOTE: ON 3RD APRIL 2017, CORRECTED FIRST EQUATION. THESE EQUATIONS ARE DIFFERENT FROM THOSE GIVEN IN HIV BY A FACTOR OF -2\pi on F. This is because their case actually corresponds to the B-field term having a negative sign in the action, at least for our convention for the measure (i know this from Nonabelian equivariant A-brane analysis, we get the same result as them). The 2\pi is explained below.
%IMPORTANT NOTE: i think the 2\pi is because the boundary action is defined without 1/2\pi, while the bulk action is multiplied by 1/2\pi. In HIV there is no 1/2\pi when discussing NLSMs.

An alternative formulation also exists for $U(1)^k$-GNLSMs with $X=\C^N//U(1)^{N-k}$, i.e., general K\"ahler toric manifolds. The boundary action for the $U(1)^{N-k}\times U(1)^k$ GLSM is
\begin{equation}\label{gnlsmuvbmultiAlt}
S_{\partial\Sigma}={\theta'\over 4\pi r'}\int\limits_{\partial\Sigma}
\txd x^0 \sum_i^N\Big(\, i{D}_0\ov{\phi}_i\,\phi_i-i\ov{\phi}_i {D}_0\phi_i
\Big),
\end{equation}
where $\theta'=2\pi n'$ ($n'\in \Z$) and $r'\in \R$, together with the theta terms
\begin{equation}\label{thetasusygnlsmmultiAlt}
\sum_b^{N-k}\Big(\frac{\hat{\theta}_b}{2\pi}\int_{\Sigma}\hat{F}_{01b}d^2x \Big)+\sum_c^{k}\Big(\frac{\widetilde{\theta}_c}{2\pi}\int_{\Sigma}\widetilde{F}_{01c}d^2x \Big).
\end{equation}
Setting 
\begin{equation}\label{assumption2}
\frac{{\theta}'}{{r}'}=\frac{{\hat{\theta}_b}}{{\hat{r}_b}}=\frac{{\til{\theta}_c}}{{\til{r}_c}},
\end{equation}
%which implies
and 
\begin{equation}
{\gamma'}=\hat{\gamma}_b=\til{\gamma}_c,
\end{equation}
the relevant boundary conditions are \eqref{BtypeGLSMtoGNLSMbcAlt}, \eqref{GNLSMBgaugeAlt} and \eqref{Flocal3}, with $\frac{\hat{\theta}}{\hat{r}}$ replaced by $\frac{\theta'}{r'}$ and $\hat{\gamma}$ replaced by $\gamma'$. In addition, the $D_a$ equation of motion is also necessary for complete B-type supersymmetry at the boundaries.%as well as the restriction of the $D_a$ equation of motion to the boundary.

By taking the $\hat{e}_b\rightarrow \infty$ limit and repeating the familiar procedure, we can obtain the GNLSM boundary action which includes the Hermitian connection of a $U(1)^k$-equivariant holomorphic line bundle over $\C^N//U(1)^{N-k}$	, as well as the relevant GNLSM boundary conditions.

%In addition, we ought to set $\hat{\theta}_b=2\pi \hat{n}_b$, where $\hat{n}_b\in \Z$, and we need to impose the condition
%\begin{equation}\label{multiconstraintAlt}
%\frac{n}{r}=\frac{\hat{n}_b}{\hat{r}_b}
%\end{equation} 
%for all values of $b$. 
  
%CHANGED THE FOLLOWING PARAGRAPH IN THESIS  
  
An important advantage of the alternative formulation of equivariant B-branes over the first one is that because of the constraints \eqref{assumption1} and \eqref{assumption2}, the form of the GLSM boundary action does not depend on which gauge symmetries we are breaking to obtain the GNLSM. This 
%is necessary for 
%IMPORTANT NOTE: I CHANGED IS NECESSARY TO IMPLIES, BECAUSE THE EQUIVALENCE OF THE EQUIVARIANT BRANES IS SOMETHING WE SHOW, NOT SOMETHING WE REQUIRE
implies the equivalence of equivariant B-branes in different toric targets of GNLSMs obtained from a single GLSM. In order to ensure that 
%this holds for 
the first formulation also does not depend on which gauge symmetries we are breaking, we can impose the same constraints for it. %In this way, 
 %This will ensure a unified GLSM description of equivariant B-branes in 
%\mbox{}\par\nobreak
%\noindent
%\textit{Quantum Corrections}   
\subsection{Quantum Corrections} 
We have heretofore analyzed the boundary conditions of the classical $U(1)^{N-k}\times U(1)^k$ GLSM, and  the respective GNLSM limits of these conditions, in two equivalent formulations. 
%So far we have been analyzing the boundary condition of
%the classical theory.
We shall now investigate quantum effects for the alternative formulation of equivariant B-branes given in Section 4.3,\footnote{We shall not study the quantum effects for the first formulation, since the main quantum correction is the running of the FI parameters, and the FI parameters do not enter the boundary conditions in that formalism.} since we shall use this formulation for the proof of mirror symmetry in the following section.\footnote{The following is a generalization of the analysis given in Section 6 of \cite{HIV} to the case of multiple $U(1)$ gauge groups. }

There are two quantum effects of the $U(1)^{N-k}\times U(1)^k$ GLSM with $\sum_{i=1}^N\mathcal{Q}_{ia}\ne 0$ which are important. The first of these is
 the running of the FI parameters
\begin{equation}\label{qeffect1}
r_{0a}=r_a(\mu)+\sum_{i=1}^N\mathcal{Q}_{ia}\log\Big(\frac{\Lambda_{\rm UV}}{\mu}\Big),
\end{equation}
 where $r_{0a}$ denotes bare parameters, $\Lambda_{UV}$ is an ultraviolet cut-off, and $\mu$ is a finite energy scale. By integrating the beta functions of the FI parameters, $\beta_a=\mu \frac{dr_a}{d\mu}$, the $\mu$-dependence is found to be
\begin{equation}\label{running}
r_a(\mu)=\sum_{i=1}^N\mathcal{Q}_{ia}\textrm{log}\Bigg(\frac{\mu}{\Lambda}\Bigg),
\end{equation}
where $\Lambda$ is the renormalization group invariant dynamical scale.  The running of $r_a$ implies that
the phase,
$e^{i\gamma_a}=t_a/|t_a|$,
which appears in the boundary conditions we have used,
changes with the renormalization group flow. The second quantum effect is the anomaly of the $U(1)$ axial R-symmetry, whereby axial R-rotations 
$\psi_{\pm i}\to e^{\pm i\beta/2}\psi_{\pm i}$,
$\sigma_a\to e^{-i\beta}\sigma_a$ and
$\lambda_{\pm a}\to e^{\pm i\beta/2}\lambda_{\pm a}$ no longer leave the action invariant,
but result in a shift of the theta angles, i.e.,%{\it together with the shift of the Theta angle}
\begin{equation}\label{qeffect2}
\theta_a\to\theta_a+\sum_{i=1}^N\mathcal{Q}_{ia}\beta.
\end{equation}

%In particular, if $\sum_{i=1}^N\mathcal{Q}_{ia}>0$,
%%(which corresponds to an asymptotic free sigma model),I REMOVED THIS ASYMPTOTIC FREE POINT FIRST
%the bare phase
%${\gamma_{0a}}\to 0$
%in the continuum limit $\Lambda_{\rm UV} \to\infty$.
%I DONT WANT TO MENTION CONTINUUM LIMIT YET IN LIGHT OF LAST POINT

%Both effects \eqref{qeffect1} and \eqref{qeffect2}
These effects should be apparent in a quantum effective description, whereby the lowest components $\si_a$ of the superfields $\Sigma_a$ are chosen to be slowly varying and to 
%have large expectation values 
be large compared to the energy scale $\mu$ at which we look at the effective theory. This imparts large masses to the charged matter superfields $\Phi_i$, which can then be integrated out as long as we are studying the theory at some finite energy scale $\mu$. From a path integral computation \cite{hori2003mirror}, the superpotential of the effective action, which corresponds to a Landau-Ginzburg model,\footnote{
To be precise, the theory involves a gauge field, whose only effect is a vacuum energy
% $\sum_a^Ne_a^2({\rm Im}\,t_{\it{(eff)}a})^2/2$ 
\cite{hori2003mirror}.
% in the absence of light or tachyonic charged matter field,
%the effect of the gauge field is simply to create the
%vacuum energy ,
%as the standard auxiliary field does.
%There is actually a (minor) subtlety; If the theory is formulated on $\R^2$,
%the physics is periodic in $\theta$ which is identified as the
%constant electric field (divided by $e^2$). This is because of the pair
%creation of the electron and positron \cite{coleman} which
%run away to opposite infinity in the space.
%However, if the theory is formulated on a strip, $\R\times [0,\pi]$,
%the electron positron pair, even if they are pair-created,
%can never run away to infinity. Thus, the physics is not periodic in
%$\theta$.
} %action 
is
%These effects should be visible in a quantum effective description.
%Here we look at the effective action in terms of $\Sigma$-field
%whose scalar component is chosen to have large expectation values. 
%This is obtained by integrating out the charged matter fields
%and is given (for $Q_i=1$ case)
%by
\begin{equation}
\widetilde{W}_{\textit{(eff) }}=-\sum^N_{a=1}\Bigg[\sum_i^N\mathcal{Q}_{ia}\Big(\log\Big(\frac{\sum^N_{a'}\mathcal{Q}_{ia'}\Sigma_{a'}}{\mu}\Big)-1\Big)\Bigg]\Sigma_a-\sum^N_{a=1}t_a(\mu)\Sigma_a,
\label{SlogS}
\end{equation}
wherefrom the effective FI-Theta parameter
\begin{equation}\label{effFItheta}
t_{\textit{(eff)}a }=t_a(\mu)+\sum_i^N\mathcal{Q}_{ia}\Big(\log\Big(\frac{\sum^N_{a'}\mathcal{Q}_{ia'}\Sigma_{a'}}{\mu}\Big)\Big),
\end{equation}
is obtained.
Now, by performing an ordinary axial R-rotation $\Sigma_a\to e^{-i\beta}\Sigma_a$ in \eqref{effFItheta}, we can retrieve the shift \eqref{qeffect2}. 

Now, it is known from \cite{HIV} that a D-brane which preserves the B-type supercharges $Q_B=\ov{Q}_++\ov{Q}_-$ and $Q_B^{\dagger}={Q}_++{Q}_-$ is a Lagrangian submanifold of the space $\C^N$ defined by the fields $\si_a$. In addition, this D-brane ought to be the preimage of a horizontal straight line in the $\til{W}_\eff$-plane, i.e., $\textrm{Im}\big(\til{W}_\eff(\si)\big)=\textrm{constant}$. If we were to solve these constraints in terms of $\si_a$, then we will obtain the quantum corrected boundary condition for $\si_a$. In general, these constraints are difficult to solve. However, when the parameters $\theta_a=0$, then there is the solution $\si_a=|\si_a|$, which satisfies $\textrm{Im}(\si_a)=0$ and $\textrm{Im}\big(\til{W}_\eff(\si)\big)=0$.

In order to obtain a less trivial solution, we can perform an axial R-rotation, which includes the shift of  $\theta_a=0$ to $\theta_a=\sum_{i=1}^N\mathcal{Q}_{ia}\beta$, due to the aforementioned anomaly. Then, we obtain the solution $\si_a=e^{i\beta}|\si_a|$, which satisfies $\textrm{Im}(e^{-i\beta}\si_a)=0$ and the straight line equation $\textrm{Im}\big(e^{-i\beta}\til{W}_\eff(\si)\big)=0$. These conditions are compatible with the constraints of the B-type supercharges  
\begin{equation}
Q_B=\bQ_++\e^{i\beta}\bQ_-
\label{Qcomb}
\end{equation}
and $Q_B^{\dagger}={Q}_++\e^{-i\beta}{Q}_-$ found in \cite{HIV}, i.e., the D-brane ought to be a Lagrangian submanifold of the field space $\C^N$, and it ought to be the preimage of a straight line in the $\til{W}_\eff$-plane with slope $\textrm{tan}(\beta)$, i.e., $\textrm{Im}\big(e^{-i\beta}\til{W}_\eff(\si)\big)=\textrm{constant}$.

%Indeed the image of
%$\sigma=e^{i\beta}|\sigma|$
%in the $\widetilde{W}$-plane is a straight line only when this shift
%is made.
Hence, we find that there is a %one parameter 
family of
explicit solutions which include
\begin{equation}\label{quantumbc}
\begin{array}{l}
\sigma_a=e^{i\beta}|\sigma_a|,\\[0.1cm]
\e^{i\beta/2}\lambda_{+a}+\e^{-i\beta/2}\lambda_{-a}=0,\\[0.1cm]
\e^{-i\beta/2}\ov{\lam}_{+a}+\e^{i\beta/2}\ov{\lam}_{-a}=0,
\end{array}
~~~\mbox{at}~~\partial\Sigma,
%\label{qcond}
\end{equation}
parametrized by $\beta=\theta_a/\sum_{i=1}^N\mathcal{Q}_{ia}$,\footnote{$\sigma_a=e^{i\beta}|\sigma_a|$ implies the boundary condition $\textrm{Im}(e^{-i\beta}\sigma_a)=0$.}
%These 
which preserve the B-type supercharges
$Q_B=\bQ_++\e^{i\beta}\bQ_-$
and $Q_B^{\dagger}={Q}_++\e^{-i\beta}{Q}_-$.
%IMPORtation.
Other solutions, including those with $\beta\ne \theta_a/\sum_{i=1}^N\mathcal{Q}_{ia}$, should exist,
but in these cases the quantum corrections are non-trivial,
and therefore they are difficult to determine, and we shall not consider them.
%It is easy to extend the above solutions to the general $Q_i$'s:
%replace $N$ in these formulae by $\sum_{i=1}^NQ_i$.

%IMPORTAe proro, see equation 4.97
Now, note that we have $\beta=\theta_a/\sum_{i=1}^N\mathcal{Q}_{ia}$ for all $a=1,\ldots,N$. Using \eqref{qeffect1} and \eqref{running}, we have 
$r_0=\sum_{i=1}^N\mathcal{Q}_{ia}\log\Big(\frac{\Lambda_{\rm UV}}{\Lambda}\Big)$, which implies
%Furthermore, in the continuum limit $\Lambda_{\textrm{UV}}\to \infty$, $r_{0a}\approx \sum_{i=1}^N\mathcal{Q}_{ia}\log\Big(\frac{\Lambda_{\rm UV}}{\mu}\Big)$.
\begin{equation}\label{allthesame}
\frac{\theta_a}{r_{0a}}=\frac{\sum_{i=1}^N\mathcal{Q}_{ia}\beta}{\sum_{i=1}^N\mathcal{Q}_{ia}\log\Big(\frac{\Lambda_{\rm UV}}{\Lambda}\Big)}=\frac{\beta}{\log\Big(\frac{\Lambda_{\rm UV}}{\Lambda}\Big)},
\end{equation}
i.e., we find that $\theta_a/r_{0a}$ are equal for all values of $a$.\footnote{Na\"ively, it may seem that the boundary action \eqref{gnlsmuvbAlt} vanishes in the continuum limit ($\Lambda_{UV}\rightarrow \infty$) due to \eqref{allthesame}. However, this is not the case, at least for $\sum_{i=1}^N\mathcal{Q}_{ia}>0$, as we shall see in the next section.} This agrees with the constraints \eqref{assumption1} and \eqref{assumption2}. In other words, we find that these constraints, which we previously imposed by hand at the classical level, emerge naturally as a result of quantum effects. 
 
%\mbox{}\par\nobreak
%\noindent
%\textit{Mirrors of Equivariant B-branes} 
\subsection{Mirrors of Equivariant B-branes}
%Having found the boundary conditions and boundary actions of GNLSMs representing equivariant B-branes in toric manifolds, 
In this section, we shall use the alternative formulation for equivariant B-branes, given in Section 4.3, to derive the Landau-Ginzburg mirrors of equivariant B-branes,
%let us now use mirror symmetry to find the Landau-Ginzburg mirrors of these branes, 
following the exposition in Section 3, as well as the results of \cite{HIV}. We shall assume in the following that 
\begin{equation}\label{fanoassump}
b_{1a}=\sum^N_i \mathcal{Q}_{ia}>0.
\end{equation}
In particular, $\hat{b}_{1b}=\sum^N_i \hat{Q}_{ib}>0$ implies that we are studying the mirrors of GNLSMs with Fano target spaces. 

%We shall first find the mirrors of equivariant B-branes in Fano manifolds of the form $\C^N//U(1)$. 
Let us start with the mirrors of equivariant B-branes on Fano manifolds of the form $X=\C^N//U(1)$. 
We focus on the family of boundary conditions \eqref{quantumbc}. The corresponding boundary conditions of the matter fields include
\begin{equation}
\begin{array}{l}
\cos(\hat{\gamma}_0)\, D_1\phi_i-i\sin(\hat{\gamma}_0)\, D_0\phi_i=0,\\[0.2cm]
\e^{-i\hat{\gamma}_0+i\beta/2}\psi_{+i}=\e^{i\hat{\gamma}_0-i\beta/2}\psi_{-i},
\\[0.2cm]
\e^{i\hat{\gamma}_0-i\beta/2}\ov{\psi}_{+i}=\e^{-i\hat{\gamma}_0+i\beta/2}\ov{\psi}_{-i},
\end{array}
%~~~~\mbox{at}~~\partial\Sigma.
\label{bcth}
\end{equation}
%I of page 58 of HIV.
where the axial R-rotations on the fermionic fields have been taken into account. These boundary conditions preserve the B-type supercharge $Q_B=\bQ_++\e^{i\beta}\bQ_-$ and its conjugate. Now, in the continuum limit $\Lambda_{\textrm{UV}}\to \infty$ whereby $\hat{r}_0= \hat{b}_{1}\textrm{log}(\Lambda_{\textrm{UV}}/\Lambda)\to \infty$, we have $\hat{\gamma}_0\to 0$. As a result, the mixed Dirichlet-Neumann boundary conditions on $\phi_i$ reduce to pure Neumann boundary conditions. 
%I are studying. 

%Moreover, we shall assume that all the complex FI-theta parameters run as  
%\begin{equation}
%t_a(\mu)=b_a\textrm{log}\Bigg(\frac{\mu}{\Lambda_{\C}}\Bigg),
%\end{equation}
%where $\Lambda_{\C}=\Lambda e^{i\lambda}$ is a fixed complex number, which is a complexified RG-invariant dynamical scale. 
%In particular, this implies that 
%\begin{equation}
%\lambda=\frac{\theta_a}{b_a},
%\end{equation}
%for all values of $a$, where $\lambda \in \R$.

%Now, the mirror symmetry we have been discussing is in fact nothing but T-duality on the the phase of $\phi_i$, denoted $\varphi_i$. As such, we may focus on the terms in the bulk and boundary action containing $\varphi_i$. 
With these facts in mind, let us shift our attention to the boundary action
\begin{equation}
\begin{aligned}
S_{\it \del \Sigma}&=
{\hat{\theta}\over 4\pi \hat{r}_0}\int\limits_{\partial\Sigma}
\sum_{i=1}^N\Big(\, iD_0\ov{\phi}_i\,\phi_i-i\ov{\phi}_i D_0\phi_i\,\Big)
\,d x^0
\\
&={\hat{\theta}\over 2\pi \hat{r}_0}
\int\limits_{\partial\Sigma}
\sum_{i=1}^N\,|\phi_i|^2(\partial_0\varphi_i+\sum_a^N\mathcal{Q}_{ia}A_{0a})
\,d x^0.
\end{aligned}
\end{equation}
%In terms of renormalized dual fields, $|\phi_i|^2=\frac{\hat{r}_0}{\hat{b}_1}+\varrho_i$, and in the continuum limit, this implies that
Now, by integrating over the modes of $\phi_i$ in the frequency range $\mu \leq |k| \leq \Lambda_{\textrm{UV}}$
in the path integral, $|\phi_i|^2$ is replaced by $\langle|\phi_i|^2\rangle=\textrm{log}({\Lambda_{\textrm{UV}}}/{\mu})$. %TA symmetry
 Since ${\hat{r}_0}/{\hat{b}_{1}}= \textrm{log}({\Lambda_{\textrm{UV}}}/{\mu})+{\hat{r}}/{\hat{b}_{1}}$, taking the continuum limit $\Lambda_{\textrm{UV}}\to \infty$ gives us $|\phi_i|^2\approx {\hat{r}_0}/{\hat{b}_{1}}$, which implies that
\begin{equation}
S_{\it \del \Sigma}=
{\hat{\theta}\over 2\pi}
\int\limits_{\partial\Sigma}
\left({1\over \hat{b}_1}\sum_{i=1}^N\partial_0\varphi_i\,+\,\hat{A}_0+\sum_c^{N-1}\frac{\til{b}_{1c}}{\hat{b}_1}\til{A}_{0c}\,\right) d x^0.
\end{equation}

The relevant portion of the action with regard to the dualization of mirror symmetry is then
\begin{equation}
S_{\varphi}
={1\over 2\pi}\int\limits_{\Sigma}
\sum_{i=1}^N\frac{\hat{r}_0}{\hat{b}_1}|\dd\varphi_i+\sum_a^N\mathcal{Q}_{ia}A_a|^2
-{i\hat{\theta}\over 2\pi}\int\limits_{\partial\Sigma}
\Bigl({1\over \hat{b}_1}\sum_{i=1}^N\dd\varphi_i\,+\,\hat{A}+\sum_c^{N-1}\frac{\til{b}_{1c}}{\hat{b}_1}\til{A}_c\Bigr),
\label{origi}
\end{equation}
where we have considered Euclidean signature on the worldsheet for simplicity,\footnote{In the following derivation, we use the notation $|A|^2= A \wedge \star A$.}
and where the terms with fermionic fields which are not essential
in the present analysis have been ignored.
Let us consider another action with one-form fields
${\cal B}_i={\cal B}_{i\mu}\dd x^{\mu}$ given by
\begin{equation}
\begin{aligned}
S^{\prime}=&\sum_{i=1}^N\left[
{\hat{b}_1\over 8\pi \hat{r}_0}\int\limits_{\Sigma}{\cal B}_i\wedge *{\cal B}_i
+{i\over 2\pi}\int\limits_{\Sigma}
{\cal B}_i\wedge (\dd\varphi_i+\sum_a^N\mathcal{Q}_{ia}{A}_a)\right]
\\&-{i\hat{\theta}\over 2\pi}\int\limits_{\partial\Sigma}
\Bigl({1\over \hat{b}_1}\sum_{i=1}^N\dd\varphi_i\,+\,\hat{A}+\sum_c^{N-1}\frac{\til{b}_{1c}}{\hat{b}_1}\til{A}_c\Bigr).
\label{dualiz}
\end{aligned}
\end{equation}
The one-form fields ${\cal B}_i$ have the boundary condition
\begin{equation}
{\cal B}_i|_{\partial\Sigma}=0,
\label{bcB}
\end{equation}
i.e., their inner products with tangent vectors of the boundaries vanish. %they vanish against the tangent vectors of the boundary.
If we were to first integrate out
${\cal B}_i$, the constraint
${\cal B}_i=i2(\hat{r}_0/\hat{b}_1)*(\dd\varphi_i+\sum_a^NQ_{ia}A_a)$
is obtained (whereby the boundary condition \eqref{bcB} is consistent with the boundary condition $D_1\phi_i=0$ obtained
in the continuum limit) and the original action
(\ref{origi}) is obtained. 
Alternatively, if we were to first integrate out $\varphi_i$,
the constraint
\begin{equation}
{\cal B}_i=\dd\vartheta_i
\label{Bth}
\end{equation}
is obtained, where the fields $\vartheta_i$ are periodic with period $2\pi$.\footnote{For details on why $\vartheta_i$ ought to be periodic, see (\cite{hori2003mirror}, page 250).}
The boundary conditions (\ref{bcB}) then imply that
$\vartheta_i$ are constants at the boundaries of the worldsheet.
The %vanishing of all 
boundary terms %proportional to $\int_{\partial\Sigma}\dd(\delta\varphi_i)$ 
containing $\del_2(\delta\varphi_i)$ obtained when integrating  out $\varphi_i$
%of the action (\ref{dualiz}) 
%requires   \beta
cancel if these constants are
\begin{equation}\label{varthetabc}
\vartheta_i={\hat{\theta}/\hat{b}_1}~~~\mbox{at}~~\partial\Sigma,
%\label{bcthea}
\end{equation}
%P because of the boundary condition Im(e^{-i\beta}\si)=0
for all $i$, where ${\hat{\theta}/\hat{b}_1}=\beta=\theta_a/b_{1a}$. Now, using the constraint (\ref{Bth}) in (\ref{dualiz}), the mirror action
\begin{equation}
\begin{aligned}
S_{\vartheta}
&=\sum_{i=1}^N\frac{1}{2\pi}\left[
{\hat{b}_1\over 4 \hat{r}_0}\int\limits_{\Sigma}|\dd\vartheta_i|^2
+{i}\int\limits_{\Sigma}\dd\vartheta_i\wedge (\sum^N_a\mathcal{Q}_{ia}A_a)\,\right]
-{i\hat{\theta}\over 2\pi}\int\limits_{\partial\Sigma}(\hat{A} +\sum_c^{N-1}\frac{\til{b}_{1c}}{\hat{b}_1}\til{A}_c)
\\
&=\sum_{i=1}^N \frac{1}{2\pi} \left[
{\hat{b}_1\over 4 \hat{r}_0}\int\limits_{\Sigma}|\dd\vartheta_i|^2
-{i}\int\limits_{\Sigma}\sum_a^N\mathcal{Q}_{ia}\vartheta_i d A_a\,\right]
\\&+{i\over 2\pi}\int\limits_{\partial\Sigma}\big[
\Bigl(\sum_{i=1}^N\hat{Q}_i\vartheta_i-\hat{\theta}\,\Bigr)\hat{A} +\sum_c^{N-1}(\sum_{i=1}^N\til{Q}_{ic}\vartheta_i-\frac{\til{b}_{1c}}{\hat{b}_1}\hat{\theta})\til{A}_c\big],
\end{aligned}
\end{equation}
is obtained. Finally, the boundary term in this action vanishes when we use the boundary condition \eqref{varthetabc},
%\begin{equation}
%\vartheta=\frac{\hat{\theta}}{\hat{b}_1},
%\end{equation}
and the dualization process ends with only a bulk action.
%proceeds exactly as in the bulk theory.

In particular, the relationship (reviewed in Section 3) between the fields of the mirror theories, i.e.,
\begin{equation}\label{fieldduality2}
Y_i+\overline{Y}_i=2\overline{\Phi}_i e^{\sum_a^{N} \mathcal{Q}_{ia}V_a}\Phi_i,
\end{equation}
holds, and we have the following relationships between superfield components:
\begin{equation}
\begin{aligned}\label{bcchi}
y_i&=\varrho_i-i\vartheta_i,
\left\{
\begin{array}{l}
\varrho_i=|\phi_i|^2,\\
\partial_{\pm}\vartheta_i=
\pm 2\Bigl(-|\phi_i|^2
(\partial_{\pm}\varphi_i+\sum_a^N\mathcal{Q}_{ia}A_{\pm a})
+\ov{\psi}_{\pm i}\psi_{\pm i}\Bigr),
\end{array}\right.\\
\chi_{i+}&=2\ov{\psi}_{+i}\phi_i,\textrm{  }\chi_{i-}=-2\ov{\psi}_{-i}\phi_i,
\\
\ov{\chi}_{+i}&=2\ov{\phi}_i\psi_{+i},\textrm{  }\ov{\chi}_{-i}=-2\ov{\phi}_i\psi_{-i},\\
E_i&=-2\ov{\psi}_{-i}\psi_{+i}-2|\phi_i|^2\sum_a\mathcal{Q}_{ia}\ov{\si}_a,
\end{aligned}
\end{equation}
where %$\varphi_i$ is the phase of $\phi_i$,
$\del_ {\pm}=\del_0\pm \del_1$, $Y_i=y_i+\theta^+\ov{\chi}_{+i}+\ov{\theta}^-\chi_{-i}+\theta^+\ov{\theta}^-E_i$.
% and $\varphi_i$ is the $2\pi$-periodic phase of  $\phi_i=|\phi_i|\,e^{i\varphi_i}$.
%It is now evident
The relationship between the periodic fields $\vartheta_i$ and $\varphi_i$ is in fact evidence that mirror symmetry of the two theories stems from 
T-duality on the phase of the charged chiral superfields $\Phi_i$, whereby the neutral twisted chiral superfields $Y_i$ are periodic, i.e., $Y_i\equiv Y_i + 2\pi i$ \cite{hori2003mirror}. 

Furthermore, the K\"ahler metric 
of the %$(\C^{\times})^N$
 target space of the mirror Landau-Ginzburg sigma model 
%we have obtained 
is 
given by
\begin{equation}
\dd s^2={\hat{b}_1\over 4\hat{r}_0}\sum_{i=1}^N\,((d\varrho_i)^2 +(d\vartheta_i)^2),
\label{Kmet}
\end{equation}
which is the flat cylinder metric on $(\C^{\times})^N$.
As %explained 
in Section 3, taking the $\hat{e}\to \infty$ limit allows us to integrate $\hat{\Sigma}$ out  of the action,
%I separately.
 and imposes the constraint 
\begin{equation}\label{glgconstraint}
\sum_j^N\hat{Q}_j Y_j - \hat{t}=0,
\end{equation}
giving us the gauged Landau-Ginzburg theory with holomorphic twisted superpotential 
\begin{equation}\label{twsuperp}
\til{W} =\sum_c^{N-1}\Big(\sum_{j=1}^{N}\til{Q}_{jc} Y_j - \til{t}_c \Big) \til{\Sigma}_c
\ +\ \sum_{j=1}^{N} e^{-Y_j}.
\end{equation} 
We recall that the constraint \eqref{glgconstraint} fixes the target space of the gauged Landau-Ginzburg theory to be the algebraic torus $(\C^{\times})^{N-1}$. 
 
The boundary conditions \eqref{varthetabc} imply that $e^{-y_i}$ have a common phase which is fixed. In other words, the boundaries of the worldsheet are mapped by $e^{-y_i}$ to a cycle $\gamma_{\hat{\theta}}$ in ${(\C^{\times})}^{N-1}$ which has $N-1$ real dimensions. This cycle is given by
\begin{equation}
(\e^{-y_1},\ldots,\e^{-y_N})
=(\e^{-\varrho_1+i\hat{\theta}/b_1},\ldots,\e^{-\varrho_N+i\hat{\theta}/b_1}),
\label{real}
\end{equation}
where $\varrho_i$ are 
constrained by $\sum_{i=1}^N\hat{Q}_i\varrho_i=\hat{r}$.
In the continuum limit, the pure Neumann boundary condition we obtain for $\phi_i$
from (\ref{bcth}),
implies the Neumann boundary condition
\begin{equation}
\partial_1\varrho_i=0
%~~~\mbox{at}~~\partial\Sigma,
\end{equation}
for the coordinates $\varrho_i$ tangent to $\gamma_{\hat{\theta}}$.
Using (\ref{bcchi}) and (\ref{bcth}), we may also obtain  boundary conditions on the fermionic dual fields, which are
\begin{equation}
\begin{array}{l}
\e^{-i\beta/2}\chi_{+i}+\e^{i\beta/2}\chi_{-i}=0,\\[0.1cm]
\e^{i\beta/2}\ov{\chi}_{+i}+\e^{-i\beta/2}\ov{\chi}_{-i}=0.
\end{array}
\label{bcfc}
\end{equation}
These boundary conditions correspond to a D-brane wrapped on the cycle $\gamma_{\hat{\theta}}$.
%MPRY.
%; the phases $\e^{\pm i\beta/2}$ in \eqref{bcfc} signify %that we have performed 
%an
%R-rotation.
%These are the standard boundary condition on the worldsheet
%fields corresponding to the D-brane wrapped on .
 
The cycle $\gamma_{\hat{\theta}}$ is a Lagrangian submanifold of $(\C^{\times})^{N-1}$.
%$\gamma_{\hat{\theta}}$ in the $\til{W}$-plane is  
The A-brane wrapping this Lagrangian submanifold is the mirror of the space-filling B-brane supporting the holomorphic line bundle $\mathcal{O}_{X}(-n)$ with $U(1)^{N-1}$-equivariant structure, where $X$ is a Fano toric manifold of the form $\C^N//U(1)$. 

Let us investigate this A-brane further, by studying the image of the cycle  $\gamma_{\hat{\theta}}$ in the $\til{W}$-plane. In particular, we would like to find the mirror of the $U(1)^{N-1}$-equivariant structure on the B-brane. The twisted superpotential \eqref{twsuperp} can be rewritten as 
\begin{equation}\label{separate}
\til{W}=\til{W}_{equiv}+\til{W}_{{X}},
\end{equation}
where the first and second term of \eqref{twsuperp} correspond respectively to the first and second term of  \eqref{separate}.
The image of  $\gamma_{\hat{\theta}}$ in the $\til{W}_X$-plane is 
\begin{equation}
\widetilde{W}_X|_{\del \Sigma}=\e^{i\beta}\sum_{i=1}^N|\e^{-y_i}|,
\end{equation}
which is the mirror condition found in \cite{HIV} when studying the mirrors of B-branes without equivariant structure. 
%In particular, $\gamma_{\hat{\theta}}$ is a `wavefront trajectory' emanating from a critical point $p_0$ of $\til{W}_X$ (a point in the target space which satisfies $\del_{y_i}\til{W}_X=0$). This wavefront trajectory corresponds to the totality of all possible soliton solutions (of the non-gauged LG model mirror to the NLSM with target $X$) starting from $p_0$, whose image in the $\til{W}_X$-plane 
In particular, it is a straight line %with slope
which makes an angle $\beta=\hat{\theta}/\hat{b}_1$ with respect to the real axis. 
 
%The mirror of the $U(1)^k$-equivariant structure is found in $\til{W}_{equiv}$. In particular, it contains the factor
% Shifting our focus to the image of $\gamma_{\hat{\theta}}$ in $\til{W}_{equiv}$, we find that it contains the factor
%\begin{equation}
%\sum_{j=1}^{N}\til{Q}^c_j y_j=\sum_{j=1}^{N}\til{Q}^c_j \varrho_j-i\sum_{j=1}^{N}\til{Q}^c_j \frac{\hat{\theta}}{\hat{b}_1},
%\end{equation}
%where the first term on the right hand side is a map 
%\begin{equation}
%Eq : \gamma_{\hat{\theta}}\to  \mathfrak{u}(1)^{N-1},
%\end{equation}
%where $\mathfrak{u}(1)^{N-1}$ is the Lie algebra of $U(1)^{N-1}$. Thus, the map $Eq$ from the cycle $\gamma_{\hat{\theta}}$ (on which an A-brane is wrapped) to $\mathfrak{u}(1)^{N-1}$ is the mirror of $U(1)^{N-1}$-equivariant structure on the B-brane.

Shifting our focus to the boundary value of $\til{W}_{equiv}$, we find that it is given by 
\begin{equation}
\begin{aligned}
\til{W}_{equiv}|_{\del \Sigma}&=\sum_c^{N-1}\textrm{Re}(e^{-i\beta}\til{\si}_c)e^{i\beta}\Big(\sum_j^N\til{Q}_{jc}\varrho_j-\til{r}_c\Big)\\
&=\sum_c^{N-1}\textrm{Re}(e^{-i\beta}\til{\si}_c)Eq_c(\varrho),
\end{aligned}
\end{equation}
where we have used the boundary conditions $\vartheta_i=\hat{\theta}/\hat{b}_1=\beta$ and $\textrm{Im}(e^{-i\beta}\til{\si}_c)=0$ as well as the identity $\beta=\til{\theta}_c/\sum_{i=1}^N\til{Q}_{ic}$. Here, $Eq_c$ is a complex-valued map
\begin{equation}
\boxed{
Eq : \gamma_{\hat{\theta}}\to  \mathfrak{u}(1)^{N-1},
}
\end{equation}
where $\mathfrak{u}(1)^{N-1}$ is the Lie algebra of $U(1)^{N-1}$. In particular, for a given  value of $c$, $\gamma_{\theta}$ is mapped to a straight line in the $Eq_c$-plane, which makes an angle $\beta$ with respect to the real axis. Thus, this map $Eq$ from the cycle $\gamma_{\hat{\theta}}$ (on which the A-brane is wrapped) to $\mathfrak{u}(1)^{N-1}$ is the mirror of $U(1)^{N-1}$-equivariant structure on the B-brane. In addition, we note that the boundary value of the total twisted superpotential is
\begin{equation}
\boxed{
\til{W}|_{\del \Sigma}=e^{i\beta}\Bigg(\sum_{i=1}^N\e^{-\varrho_i}+\sum_c^{N-1}\textrm{Re}(e^{-i\beta}\til{\si}_c)\Big(\sum_j^N\til{Q}_{jc}\varrho_j-\til{r}_c\Big)\Bigg),
}
\end{equation}
which is a map from $\textrm{Re}(e^{-i\beta}\til{\si}_c)$ and $\varrho_i$ to a straight line in the $\til{W}$-plane which makes an angle   $\beta$ with respect to the real axis. Since we have set $\hat{\theta}=2\pi n$ earlier, and $\int_X c_1(\mathcal{O}_{X}(-n))=-n$,
%recall from Nakahara that the first chern class of a bundle is proportional to the (lie algebra-valued) curvature of the bundle % OIe 460 
 the slope of this straight line depends on the first Chern class %($ \int_X c_1=-n$) 
of the holomorphic line bundle $\mathcal{O}_{X}(-n)$ supported by the %mirror 
B-brane.

%It is expected that 
The mirrors of equivariant B-branes on Fano toric manifolds of the form $X=\C^N//U(1)^{N-k}$ can similarly be found using the above method. These mirror A-branes correspond to Lagrangian submanifolds ($\gamma_{\theta'}$) of the cylinder $(\C^{\times})^{k}$ which is defined by 

\begin{equation}\label{glgconstraint2}
\sum_j^N\hat{Q}_{jb} Y_j - \hat{t}_b=0,
\end{equation}
with the additional data of the superpotential
\begin{equation}\label{twsuperp2}
\til{W} =\sum_c^{k}\Big(\sum_{j=1}^{N}\til{Q}_{jc} Y_j - \til{t}_c \Big) \til{\Sigma}_c
\ +\ \sum_{j=1}^{N} e^{-Y_j}.
\end{equation} 
The first term on the right hand side of \eqref{twsuperp2}, when restricted to its boundary value, contains the mirror data of equivariant structure on the holomorphic line bundle (which is supported by the space-filling B-brane), which is a map   
\begin{equation}
Eq : \gamma_{\theta'}\to  \mathfrak{u}(1)^k.
\end{equation}
%IMPORTANT NOTE: 9TH APRIL 2017, THE ONLY DIFFERENCE IN THE PROOF OF MIRROR SYMMETRY FOR GENERAL TORIC MANIFOLDS IS THAT we set  $\theta'/r'=\hat{\theta}_1/\hat{r}_1$ (i.e., we choose one of the U(1) groups which were broken), and then the same proof of mirror symmetry follows (i.e. starting with a master action and integrating out in two different ways.). The same results for the superpotential above hold with $N-1$ replaced by $k$. In particular the straight lines in the W and Eq planes make a of slope of angle \beta. \beta=\hat{\theta}_b/(\sum_i\hat{Q}_{ib} ), so the slope depends on all the $\hat{n}_b$. 
\subsection{Nonabelian Equivariant B-branes}

Although Kapustin et al. \cite{Kapustin} introduced the nonabelian equivariant B-brane boundary Wilson line via a gauged B-model, the compatible boundary conditions were not derived explicitly. This  motivates us to derive boundary conditions corresponding to nonabelian equivariant B-branes in our \textit{untwisted} GNLSM.
% and we shall close this gap in knowledge in this subsection. %In this subsection, w
We shall use the insights obtained from studying abelian equivariant B-branes to %infer 
find the complete description of nonabelian equivariant B-branes. 
%In this subsection, we shall use the insights obtained from studying abelian equivariant B-branes to %infer 
%find the description of nonabelian equivariant B-branes. 
%In other words, we shall find the B-type boundary conditions necessary for the nonabelian GNLSM studied in Section 2 to be supersymmetric at the boundaries, and state the admissible boundary interactions. %i.e., for the supersymmetry variations \eqref{dSmatw} and \eqref{dSgauthe} to vanish (
 This will be achieved by generalizing the first formulation studied in this section for abelian gauge groups (c.f. Sections 4.1-4.2) to nonabelian gauge groups. Note that the GNLSM notation of Section 2 is used in this subsection. For simplicity, we shall only consider the case where the $B$-field, $C$-field and $\theta$-parameter of the GNLSM given in \eqref{bfield} and \eqref{theta} are zero. 
We shall first investigate the 
%necessary 
boundary conditions required for B-type supersymmetry, before proceeding to discuss the admissible boundary action. 
%We first consider the case with no boundary action. %Firstly, we 
Now, note that all the terms in $\delta (S_{gauge}+S_r)$ (equation \eqref{dSgauthe}) vanish using the following 
%gauge invariant 
boundary conditions 
%\begin{equation}
%\begin{aligned}
%A_{1a}&=0,\\
%\del_1 A_{0a}&=0,\\
%\lampua-\lambmua&=0,\\
%\lammua-\lambpua&=0,\\
%\nabla^A_1(\lammua+\lambpua)&=0,\\
%\nabla^A_1(\lampua+\lambmua)&=0,\\
%\nabla^A_1\si_a&=0,\\
%\nabla^A_1\ov{\si}_a&=0,\\
%\mu_a&=-r_a.
%\end{aligned}
%\end{equation}
\begin{equation}\label{nonabelianBcond}
\begin{aligned}
\textrm{Im}(\si_a)&=0,\\
\lampua+\lammua&=0,\\
%\lambmua+\lambpua&=0,\\
\del_1\textrm{Re}(\si_a)&=0,\\
A_{1a}&=0,\\
\del_1 A_{0a}&=0,\\
\del_1(\lammua-\lampua)&=0,\\
%\del_1(\lambpua-\lambmua)&=0,\\
\del_1({D}_a+\del_1\textrm{Im}({\si}_a))&=0.
\end{aligned}
\end{equation}
%IMPORTANT NOTE The nonabelian commutator term [\si,\ov{\si}] vanishes since the imaginary part of sigma is zero at the boundary, and the commutator of the real part of sigma with itself is zero, which can be shown using the antisymmetry of the structure constants
These conditions are a %direct
%IMPORTANT NOTE: TOOK OUT DIRECT BECAUSE IN THE ABELIAN CASE THE THETA PARAMETER APPEARS IN THE BOUNDARY CONDITION
 generalization of the conditions given in \eqref{Bgauge2} for the example of $\CP^{N-1}$, except that the boundary condition for $F_{01a}$ is replaced by the stricter conditions $A_{1a}=0$ and $\del_1A_{0a}=0$, and the boundary condition for $\textrm{Re}(\si_a)$ becomes $\del_1 \textrm{Re}(\si_a)=0$. %IMPORTANT NOTE, IN THE CPN EXAMPLE, EVEN AFTER THERE ARE NO THETA TERMS LEFT EXPLICITLY IN THE BULK AND BOUNDARY ACTION (they are however still implicitly present in the boundary action), THE BOUNDARY CONDITION STILL CONTAINS THETA. THE $F_{01c}=-e_c^\theta_c2$ IS USUALLY NECESSARY FOR LOCALITY OF THE EQUATIONS OF MOTION FOR THE GAUGE FIELDS. HOWEVER, IN ANY CASE, THE PRESENCE OF THE BOUNDARY ACTION IMPLIES THAT THE EQUATIONS OF MOTION ARE NOT LOCAL
These stricter conditions are necessary since we now require that the boundary conditions preserve the locality of the relevant equations of motion when no additional boundary action is added, 
%IMPORTANT NOTE: FOR THE ABELIAN CASE THIS DOES NOT HOLD, BUT WE WANT THIS REQUIREMENT BECAUSE WE USE IT LATER TO DERIVE THE BOUNDARY CONDITIONS FOR THE MATTER FIELDS. THE ONLY DIFFERENCE WHEN WE DON'T IMPOSE THIS IS THAT \del_1Re(\si_a)-\del_1(A_0a)=0 is the boundary condition, with each term being a constant in the centre of \mathfrak{g}. This generalizes what we saw earlier for the abelian case. See the long IMPORTANT NOTE ABOVE
%The stricter boundary conditions on the gauge fields are necessary since 
and because the 
%field strength $F_{01a}$ now 
supersymmetry transformations now contain nonabelian terms, which causes B-type supersymmetry invariance of the set of boundary conditions to not hold unless we use the stricter conditions on the gauge fields.\footnote{If we relax the requirement of locality of equations of motion, then the boundary conditions on $A_{0a}$ and $\textrm{Re}(\si_a)$ become $\del_1A_{0a}=\tau_a$ and $\del_1\textrm{Re}(\si_a)=\tau_a$, where $\tau$ is a constant valued in the centre of $\g$.} The boundary conditions 
%on the gauge fields 
%IMPORTANT NOTE: ALL BOUNDARY CONDITIONS REQUIRE THAT \del_1\alpha_a=0
in fact imply that gauge transformations have to be restricted such that the transformation parameter $\alpha^a$ has vanishing derivative with respect to $x^1$ at the boundaries, in order for these boundary conditions to be gauge invariant.

%IMPORTANT NOTE: REMEMBER THAT IN THE NONABELIAN CASE THE GAUGE TRANSFORMATIONS ARE MODIFIED FROM THE ABELIAN CASE, WHERE THERE IS A NONABELIAN TERM, SEE SECTION 2 %Just like in the example of $\CPN$, 
%The boundary conditions given in \eqref{nonabelianBcond} also preserve the locality of the relevant equations of motion, as long as no additional boundary action is added.% Now, because of \eqref{requirement} the only terms left to consider in \eqref{dSB} %the supersymmetry variation of the action 
%are the $B$-field terms. These vanish by requiring that the $B$-field on $X$ vanishes when restricted to the submanifold $\gamma$.
%IMPORTANT NOTE, FOR NLSMS ONE CAN HAVE MORE GENERAL BOUNDARY CONDITIONS THAN LAGRANGIAN SUBMANIFOLD IN THE PRESENCE OF B-FIELDS, SEE PAGE 804 OF MIRROR SYMMETRY BOOK

Next, we turn to the boundary conditions for the matter fields. %We first recall that for $\CP^{N-1}$, the equivariant A-brane corresponded to a Lagrangian torus $T^{N-1}$, which was invariant under the $U(1)^{N-1}$ isometry of $\CP^{N-1}$. 
Let us first consider the $\mathcal{N}=1$ subalgebra of B-type supersymmetry, which corresponds to $\eps_+=i\til{\eps}$, $\ov{\eps}_+=-i\til{\eps}$, $\eps_-=-i\til{\eps}$ and $\ov{\eps}_-=i\til{\eps}$, where $\til{\eps}$ is a real parameter. In this case, after integrating out the auxiliary fields $F^i$ and $\ov{F}^{\ov{\imath}}$, we find that \eqref{dSmatw} is 
\begin{equation}\label{n1matwBtype}
\begin{aligned}
\delta S_{matter} =-\frac{1}{2\pi}\frac{i\til{\eps}}{2}\int_{\del\Sigma}dx^0 \{& g_{IJ} \del_0^A \phi^I(\psi^J_- - \psi^J_+)
+ g_{IJ}\del_1^A\phi^I(\psi^J_-+\psi^J_+)
\\&+g_{IJ}(\psi^I_+ - \psi^I_-)\textrm{Re}(\si^a)\he^J_a+
\omega_{IJ}(\psi^I_+ + \psi^I_-)\textrm{Im}(\si^a)\he^J_a\},
\end{aligned}
\end{equation}
where $g_{IJ}X^I Y^J=g_{i\ov{\jmath}}(X^i Y^{\ov{\jmath}}+ X^{\ov{\jmath}}Y^i)$ and $\omega_{IJ}X^I Y^J=ig_{i\ov{\jmath}}(X^i Y^{\ov{\jmath}}- X^{\ov{\jmath}}Y^i)$, and where $(I,J,K,\ldots)$ are indices corresponding to real coordinates on $X$.
% are respectively the metric and K\"ahler form on $X$.
In addition, if we insist on locality of the matter equations of motion%(like in the $\CPN$ case)
, we require that 
\begin{equation}\label{localeomsssB}
\begin{aligned}
g_{IJ}\delta \phi^I \del_1^A\phi^J=0\\g_{IJ}(\psi^I_-\delta_{\nabla}\psi^J_--\psi^I_+\delta_{\nabla}\psi^J_+)=0
\end{aligned}
\end{equation}
%IMPORTANT NOTE: WE GET TWO SEPARATE EQUATIONS BECAUSE THERE IS A FACTOR of i/2 which multiplies the second term when we do the variation. First term is multiplied by a factor of -1.
at the boundaries, where $\delta_{\nabla}\psi^J=\delta\psi^J+\Gamma^{J}_{KL}\delta\phi^K\psi^L$.
%\footnote{Here, we generalize the approach of \cite{HIV} for non-equivariant B-branes. However, as noted in (\cite{Albertsson}, page 17), this is not the most general approach.}
%IMPORTANT NOTE: REMOVED FOOTNOTE BECAUSE I HAVE NOW INCLUDED THE CHRISTOFFEL SYMBOL TERMS. BUT IT IS IMPORTANT TO REMEMBER THAT WE STILL DO NOT CLAIM TO HAVE THE MOST GENERAL SOLUTION, AND WHAT IS IN THE FOLLOWING NOTE SHOULD STILL HOLD.
%IMPORTANT NOTE: THE REASON THIS FOOTNOTE IS INCLUDED IS BECAUSE in THE EQUATIONs ABOVE WE ARE IGNORING BOUNDARY TERMS WHICH CONTAIN THE CHRISTOFFEL SYMBOL IN  AND FERMETRY.
 %The %first constraint of 
 %vanishing of
 %Requiring the first term of \eqref{localeomsssB} to vanish then implies that $\del_1\phi^J$ is normal to $\gamma$, since $A_{1a}=0$ at the boundaries.
 An equivariant B-brane shall wrap a submanifold (denoted as $\gamma$)  of  %target 
 $X$, to which a boundary of the worldsheet is mapped via $(\phi^i,\ov{\phi}^{\ov{\imath}})$. Now, any allowed variation of $\phi$ (denoted $\delta \phi^I$ for the real coordinate $\phi^I$) along the boundary, and the derivative along the boundary,
%IMPORTANT NOTE: THE PHRASE THE BOUNDARY HERE IS CORRECT, DO NOT CHANGE! 
 $\del_0 \phi^I$, ought to be tangent to $\gamma$. The first constraint of \eqref{localeomsssB} then implies that $\del_1\phi^J$ is normal to $\gamma$, since $A_{1a}=0$ at the boundaries. Then, taking into account the facts that $\textrm{Im}(\si_a)=0$ and $A_{1a}=0$ at the boundaries, we find that \eqref{n1matwBtype} vanishes if
 %if and only if 
 %IMPORTANT NOTE: TOOK OUT IF AND ONLY IF BECAUSE THERE MIGHT BE MORE GENERAL SOLUTIONS, AS THE ALBERTSSON PAPER SAYS, SEE  FOOTNOTE WHERE I CITE IT.
 $ \psi^I_- - \psi^I_+$ and $ \psi^I_- + \psi^I_+$ are respectively normal and tangent to $\gamma$, and $\til{e}^I_a$ is tangent to $\gamma$, which implies that $\gamma$ is $G$-invariant. % and the K\"ahler form vanishes against tangent vectors of $\gamma$.
 %The last condition  
% In addition, we note that 
%$i\eps (\psi^I_- - \psi^I_+)$ being normal to $\gamma$ implies that $\psi^I_--\psi^I_+=0$ along $\gamma$ and along the worldsheet boundary, which satisfies the second constraint of  \eqref{localeomsssB}.
In addition, we note that 
$\psi^I_- - \psi^I_+$ being normal to $\gamma$ and $\psi^I_- + \psi^I_+$ being tangent to $\gamma$ implies that 
%$\psi^I_--\psi^I_+=0$ 
\begin{equation}
\begin{aligned}
\psi^I_--\psi^I_+&=0,\textrm{ }I\textrm{ : tangent to $\gamma$},\\
\psi^I_-+\psi^I_+&=0,\textrm{ }I\textrm{ : normal to $\gamma$},
\end{aligned}
\end{equation}
%along $\gamma$ and along the worldsheet boundary, 
(for a choice of coordinates which separates the normal and tangent directions) which satisfies the second constraint of  \eqref{localeomsssB}.

%Next, under B-type supersymmetry, $\phi^i$ transforms as
%\begin{equation}
%\begin{aligned}
%\delta\phi^i=&\epsilon_+(\psi_-^i+\psi_+^i)
%\\
%=&\epsilon_1(\psi_-^i+\psi_+^i)
%+i\epsilon_2(\psi_-^i+\psi_+^i),
%\end{aligned}
%\end{equation}
%where $\epsilon_+=\epsilon_1+i\epsilon_2$. Hence, $\til{\eps}(\psi_-^i+\psi_+^i)$
%and $i\til{\eps}(\psi_-^i+\psi_+^i)$ are holomorphic components of
%vectors which are tangent to $\gamma$, where $\til{\eps}$ is a real parameter. % %However, from the previous paragraph, we know that $i\epsilon(\psi_-^i-\psi_+^i)$ is the holomorphic component of a vector which is normal to $\gamma$. 
%This implies that the tangent space of $\gamma$ is preserved by multiplication by $i$ on holomorphic components of vectors. %into normal vectors of $\gamma$, and vice versa. 
%Therefore, $\gamma$ is a holomorphically embedded complex submanifold of $X$. This complex submanifold also happens to be $G$-invariant, which we know from the previous paragraph.

Next, the $\mathcal{N}=(2,2)$ supersymmetry transformation of $\phi^I$ is
\begin{equation}
\delta\phi^I=i(\eps_{+2}\psi^I_--\eps_{+1}J^I_{\textrm{ }K}\psi^K_--\eps_{-2}\psi^I_++\eps_{-1}J^I_{\textrm{ }K}\psi^K_+),
\end{equation}
%IMPORTANT NOTE: OBTAINED FROM MELNIKOV, PLESSER AND RINKE, with J rescaled by a minus sign 
where $\eps_+=\eps_{+1}+i\eps_{+2}$ and $\eps_-=\eps_{-1}+i\eps_{-2}$, and where $J$ is the almost complex structure of $X$ locally 
%IMPORTANT NOTE: I HAVE SAID LOCALLY HERE BECAUSE MELNIKOV, PLESSER AND RINKE SAY THAT "Since the target space is a complex manifold, and we are working locally, we can choose a set of coordinates where the complex structure J^A_B is constant, in addition to being covariantly constant." Also see Aspinwall, equation 40, where he introduces almost complex structure instead. Hori just mentions complex structure on page 801 of the book, but does not claim that it is constant. Also see Witten-topological sigma models for a nice explanation of complex structure. Witten does not show a constant form of the complex structure but seems to use the constant form to obtain equation 2.18 from equation 2.16. Nakahara shows that the form of J given in equation 8.18 is independent of the chart, see right above equation 8.18. But i think this does not hold for the form of J given in equation  8.22 since this depends on how we complexify the coordinates. But then at the beginning of page 320 nakahara says that J is defined independently of charts. But then again at this point he is still talking about almost complex structure, not complex structure.
given by $J^i_{\textrm{ }k}=i\delta^i_{\textrm{ }k}$ and $J^{\ov{\imath}}_{\textrm{ }\ov{k}}=-i\delta^{\ov{\imath}}_{\textrm{ }\ov{k}}$. B-type supersymmetry corresponds to $\eps_{+1}=-\eps_{-1}$ and $\eps_{+2}=-\eps_{-2}$, whereby
\begin{equation}
\delta\phi^I=i\big(\eps_{+2}(\psi^I_-+\psi^I_+)-\eps_{+1}J^I_{\textrm{ }K}(\psi^K_-+\psi^K_+)\big).
\end{equation}
Hence, $\psi^I_-+\psi^I_+$ and $J^I_{\textrm{ }K}(\psi^K_-+\psi^K_+)$ are tangent to $\gamma$, %However, from the previous paragraph, we know that $\psi^I_--\psi^I_+$ is normal to $\gamma$. In addition, $J^M_{\textrm{ }I}J^I_{\textrm{ }K}=-\delta^M_{\textrm{ }k}$. 
which implies that the application of the almost complex structure, $J$, preserves the tangent space of $\gamma$.
%IMPORTANT NOTE: BORROWED TERMINOLOGY OF APPLICATION FROM ASPINWALL
Therefore, $\gamma$ is a holomorphically embedded complex submanifold of $X$. This complex submanifold also happens to be $G$-invariant, which we know from the previous paragraph.
 %converts normal vectors of $\gamma$ into tangent vectors of $\gamma$, and vice versa. Thus, $\gamma$ is a middle-dimensional Lagrangian submanifold of $X$. This Lagrangian submanifold also happens to be $G$-invariant, which we know from the previous paragraph.

%IMPORTANT NOTE, (repeated) HOLOMORPHICALLY EMBEDDED MEANS THAT THE COMPLEX STRUCTURES ARE ALIGNED, SEE PAGE 408 OF BECKER-BECKER-SCHWARZ 
%IMPORTANT NOTE: THE ABOVE BOUNDARY CONDITION, WHICH IS FOR THE MATTER FIELDS, IS GAUGE INVARIANT SINCE IT IMPLIES THAT EACH BOUNDARY HAS A 1D GNLSM, WHERE THE TARGET SPACE IS A HOLOMORPHICALLY EMBEDDED G-INVARIANT SUBMANIFOLD  OF X

Indeed, \eqref{dSmatw} vanishes under this boundary condition; integrating out the auxiliary fields $F^i$  and $\ov{F}^{\ov{\imath}}$, \eqref{dSmatw} can be rewritten (for $\eps_+=-\eps_-=\eps$) as 
\begin{equation}
\begin{aligned}
\delta S_{matter}=\frac{1}{2\pi}\frac{1}{4}\int_{\partial\Sigma}dx^0\Big\{&\eps\Big(-g(\del_0^A\phi,\psi_--\psi_+)-i\omega(\del_0^A\phi,\psi_--\psi_+)\\&-g(\del_1^A\phi,\psi_-+\psi_+)-i\omega (\del_1^A\phi,\psi_-+\psi_+)\\&-\textrm{Re}(\si^a)g(\til{e}_a,\psi_+-\psi_-)-i\textrm{Re}(\si^a)\omega(\til{e}_a,\psi_+-\psi_-)\\&-i\textrm{Im}(\si^a)g(\til{e}_a,\psi_++\psi_-)+\textrm{Im}(\si^a)\omega(\til{e}_a,\psi_++\psi_-)\Big) + c.c. \Big\}
\end{aligned}
\end{equation}
(where $g(X,Y)=g_{IJ}X^IY^J$ and $\omega(X,Y)=\omega_{IJ}X^IY^J$), which vanishes using $\textrm{Im}(\si_a)=0$ and $A_{1a}=0$ as well as the conditions that $\del_0\phi^I$, $\psi_-^I+\psi_+^I$ and $\til{e}_a^I$ are tangent to $\gamma$ while $\del_1\phi^I$ and $\psi_-^I-\psi_+^I$ are normal to $\gamma$.\footnote{Recall that for a tangent vector, $T$, and normal vector, $N$, of a holomorphically embedded complex submanifold, $\gamma$, of the K\"ahler manifold $X$, we have $\omega (T,N)=g(JT,N)=0$.
% on $\gamma$.
 } 
%IMPORTANT NOTE: $\omega (T,N)=g(JT,N)=0$ follows from page 212 and page 805 of the mirror symmetry book, it's different from the definition given on page 18 of Aspinwall's D-branes on Calabi-Yau manifolds
%IMPORTANT NOTE: RECALL THAT we know that it is a 'holomorphically embedded' complex submanifold when the complex structure preserves the tangent and normal directions from Aspinwall page 29

%As in the abelian case, 
We may add the B-type supersymmetric boundary action 
\begin{equation}\label{equivariantBbranenonabelian}
\begin{aligned}
S_{\partial\Sigma}'=&\int\limits_{\partial\Sigma}
\dd x^0 \Bigg\{A^X_m\partial^A_0 \phi^m+A^X_{\ov{m}}\partial^A_0 \ov{\phi}^{\ov{m}}+\til{R}_a\frac{({\si}^a+\ov{{\si}}^a)}{2}-\frac{i}{2}F^X_{m\ov{n}}(\psi_+^m+\psi_-^m)(\ov{\psi}_+^{\ov{n}}+\ov{\psi}_-^{\ov{n}})\Bigg\}\\=&\int\limits_{\partial\Sigma}
\dd x^0
\Bigg\{A^X_m\partial_0 \phi^m+A^X_{\ov{m}}\partial_0 \ov{\phi}^{\ov{m}}-i\til{R}_a{\mathcal{A}}^{a}-\frac{i}{2}F^X_{m\ov{n}}(\psi_+^m+\psi_-^m)(\ov{\psi}_+^{\ov{n}}+\ov{\psi}_-^{\ov{n}})\Bigg\},
\end{aligned}
\end{equation}
where we use ($m,\ov{m},n,\ov{n}$) as coordinate indices on the B-branes, where the curvature of $A^X$ satisfies $F^X_{mn}=F^X_{\ov{m}\ov{n}}=0$, and where ${\mathcal{A}}_{a}=-i\big({A}_{0a}-\frac{({\si}_a+\ov{{\si}}_a)}{2}\big)$ and
\begin{equation}\label{Rctilnonabelian}
\begin{aligned}
 \til{R}_a&=-A^X_m\til{e}^m_a-A^X_{\ov{m}}\ov{\til{e}}^{\ov{m}}_a\\
&=-\iota_{\til{e}_a}A^X.
\end{aligned}
\end{equation}
%is the moment of a $G$-equivariant holomorphic line bundle. % defined on $\gamma$. 
B-type supersymmetry invariance and gauge invariance of this action 
%follow from
require the equivariant Bianchi identity 
\begin{equation}
d\til{R}=\iota_{\til{e}}F^X,
\end{equation}
and this implies that 
%$\gamma$
each B-brane supports a $G$-equivariant holomorphic line bundle (c.f. footnote \ref{27}), for which $\til{R}_a$ is the moment.\footnote{Note that gauge invariance of the boundary action requires the use of the identity $\alpha^b\mathcal{L}_{\til{e}_b}\til{R}_a=[\alpha,\til{R}]_a$. 
%IMPORTANT NOTE: THIS IDENTITY FOLLOW SINCE THE MOMENT IS ALSO G-EQUIVARIANT LIKE THE MOMENT MAP, SEE PAGE 29 OF SZABO
}
The inclusion of this boundary action %As in the analogous NLSM case \cite{HIV}, %these boundary conditions together with 
%the boundary action \eqref{equivariantBbraneCpnFINAL} 
results in some of the equations of motion being modified by boundary terms.
One may generalize this even further (at least in the twisted case), as shown by Kapustin et al. \cite{Kapustin} (c.f. Section 4.1), by instead including a Wilson line which represents a $G$-equivariant graded holomorphic vector bundle.
%IMPORTANT NOTE, SOME OF \del_0\phi^i's actually vanish (in an appropriate coordinate system which separates tangent and normal directions to the brane)because we do not necessarily have a space-filling brane. 
%As in the abelian case, 
% (c.f. footnote \ref{26}).
%IMPORTANT NOTE, TOOK OUT REFERENCE TO ABELIAN CASE, BECAUSE THERE EVEN WITHOUT THE BOUNDARY ACTION THE BOUNDARY CONDITIONS DO NOT GIVE LOCAL EQUATIONS OF MOTION

 In conclusion, we find that in general, \vspace{1em}

 \noindent\fbox{ \parbox{\textwidth}{\linespread{1.1}\selectfont{Equivariant B-branes are $G$-invariant holomorphically embedded complex submanifolds of $X$, which support $G$-equivariant holomorphic vector bundles (which may be graded).}}}
 
 %\noindent
 \vspace{1em} 
\noindent 
As discussed in Section 4.1, at least in some cases, this implies that they are objects in the bounded, derived category of $G_{\C}$-equivariant coherent sheaves on $X$.

%IMPORTANT NOTE: WE HAVE NOT SPECIFIED THE BOUNDARY CONDITION FOR F_i, for this take a look at the important note below the vector multiplet boundary conditions in the subsection describing nonabelian equivariant A-branes.
\section{Equivariant A-branes and their Mirrors}

%We shall apply the techniques discussed above to find boundary actions and boundary conditions in abelian GNLSMs with toric target spaces, $X$, as well as their mirror descriptions. These boundary actions and boundary conditions will correspond to branes wrapping submanifolds of $X$, which we refer to as equivariant branes.
 In this section, we study the A-type supersymmetric boundary actions and boundary conditions in abelian GNLSMs on $I\times \R$ with toric target spaces, $X$, as well as their mirror descriptions. These boundary actions and boundary conditions correspond to equivariant A-branes wrapping submanifolds of $X$. Then, with the insights we find from analyzing these abelian equivariant A-branes, we shall proceed to study equivariant A-branes for nonabelian GNLSMs.
 
 % case where B-type supersymmetry is preserved at the boundary of the $I\times \R$ worldsheet, since this leads us to make contact with a result found previously by Kapustin et al. \cite{Kapustin,Vyas}, while in the next section, we shall use similar techniques to study equivariant A-branes.

A-type supersymmetry is defined by the combination of supercharges given by \eqref{atype}. In what follows, we shall set $\beta=0$ for simplicity, though it is straightforward to study the $\beta\neq 0$ generalization using the same techniques. In other words, we assume that the supercharges conserved at the boundaries are 
\begin{equation}\label{atype2}
\begin{aligned}[c]
Q_A=\overline{Q}_+ +Q_-,
\end{aligned}
\\
\begin{aligned}[c]
\textrm{   }\\
\textrm{   }\\
\end{aligned}
\\
\begin{aligned}[c]
\textrm{   }\\
\textrm{   }\\
\end{aligned}
\\
\begin{aligned}[c]
Q^\dagger_A=Q_++\overline{Q}_-\end{aligned}.
\end{equation}

From \eqref{susydelta}, it can be seen that the corresponding relations among the supersymmetry transformation parameters are  
\begin{equation}
\begin{aligned}
\eps&=\eps_+=\ov{\eps}_-\\
\ov{\eps}&=\ov{\eps}_+=\eps_-.
\end{aligned}
\end{equation}
We shall also make use of superfields when discussing boundary conditions and boundary actions, and to this end, we shall make use of the concept of `boundaries' in superspace \cite{Hori1}. For A-type supersymmetry, the relevant boundary in superspace is known as `A-boundary', and corresponds to
\begin{equation}\label{Aboundary}
\begin{aligned}
\theta&=\theta^+=-\ov{\theta}^-\\
\ov{\theta}&=\ov{\theta}^+=-\theta^-.
\end{aligned}
\end{equation}
%IMPORTANT NOTE: 19TH APRIL 2017, It seems to ME THAT THE MINUS SIGN IS JUST CONVENTION, EVEN WITHOUT THEM, WE STILL GET SUPEROPERATORS WHICH CONTAIN \del_0 and not \del_1 from the bulk superoperators. Note that it seems that in obtaining the boundary superoperators given in 2.22 and 2.23 of Hori, there is some rescaling being done on the supercoordinates and supercharges.

Let us first review what is known of ordinary A-branes. For $\mathcal{N}=(2,2)$ NLSMs%with B-type supersymmetry preserved at the boundaries
, the boundary condition needed to preserve A-type supersymmetry at the boundaries maps each boundary to a middle-dimensional Lagrangian submanifold of the target space \cite{HIV}. With target space coordinates ($X^I$) chosen appropriately, %the relevant boundary conditions are 
this is expressed as Dirichlet boundary conditions on half of the fields $X^I$ with Neumann boundary conditions for the rest. Here, we have used real coordinates for the K\"ahler target space $X$. In addition, we may include the following boundary action %Now, for $\mathcal{N}=(2,2)$ NLSMs with A-type supersymmetry preserved at the boundaries, the following boundary action \cite{HIV}
\begin{equation}\label{Abrane}
S_{\partial\Sigma}=\int\limits_{\partial\Sigma}
\dd x^0 
A^X_M\partial_0X^M=\int\limits_{x^1=\pi}\dd x^0\partial_0\phi^{M_b}A^{X(b)}_{M_b}
-\int\limits_{x^1=0}\dd x^0\partial_0\phi^{M_a}A^{X(a)}_{M_a},
\end{equation}
where $A^{X(a)}$ and $A^{X(b)}$ are the connections of 
$U(1)$ 
%gauge fields 
line bundles on the A-branes
$\gamma_a$ and $\gamma_b$
on which the boundaries $x_1=0$ and $x_1=\pi$ end
%left and the right boundaries end.
%Here, we have used real coordinates for the K\"ahler target space $X$,
(we shall use ($M,N,\ldots$) as coordinate indices on the Lagrangian submanifold branes).
 %with $A^X_M$ the connection of a $U(1)$ line 
%IMPORTANT NOTE: IN GSW, Witten seems to say they are the same on page 460 
%IMPORTANT NOTE: ON 23 MARCH 2017, I CHANGED U(1) BUNDLE TO U(1) LINE BUNDLE, SO THAT THE LATER DISCUSSION ABOUT THE FUKAYA-FLOER CATEGORY IS CLEARER
%bundle on the A-brane. % and  $F_{MN}$ its curvature. 
%which is %admissible 
This boundary action is A-type supersymmetric if $F^X_{MN}=\del_MA^X_N-\del_NA^X_M=0$.
%and only if  
%$F^X=dA^X=0$. 
 % The condition $F^X=0$
 This condition on the curvature of each  $U(1)$ bundle indicates that it is flat. This boundary action %is
 takes the form of a Wilson line, and we see that since we have two boundary components, we %can
 actually have two different Wilson lines along each boundary, corresponding to two different A-branes supporting flat $U(1)$ bundles, each with different connections \cite{HIV}. 
  %Moreover, one must also impose boundary conditions compatible with this boundary action, which preserve A-type supersymmetry of the bulk action at the boundary. With target space coordinates ($X^M$) chosen appropriately, the relevant boundary conditions are Dirichlet boundary conditions on half of the fields $X^M$ with Neumann boundary conditions for the rest, corresponding to A-branes which are middle dimensional Lagrangian submanifolds.

%An alternative formulation of B-branes exists, \cite{HIV} where mixed Dirichlet-Neumann boundary conditions are imposed on the target space coordinates, and the boundary action is 
%\begin{equation}\label{BbraneAlt}
%S_{\partial\Sigma}=\int\limits_{\partial\Sigma}
%\dd x^0 
%A_M\partial_0X^M,
%\end{equation}
%which is an ordinary Wilson line along each boundary component.
%This formulation is equivalent to that of \eqref{Bbrane}, since both lead to the same spacetime theory \cite{nappi}. %IMPORTANT NOTE I actually know this point from page 46 of HORI Linear models.
%Moreover, B-type supersymmetry is preserved at the boundary if and only if $F_{jk}=F_{\ov{\jmath}\ov{k}}=0$, indicating a holomorphic line bundle.

We are interested in the generalizations of Lagrangian boundary conditions and the boundary action \eqref{Abrane} for GNLSMs. A possible method of obtaining such a generalization would be to replace ordinary worldsheet derivatives by covariant ones, and to attempt to maintain supersymmetry and gauge symmetry by adding additional terms, if necessary. However, it is known that the boundary conditions and the boundary action \eqref{Abrane} can be obtained from %GLSM boundary conditions and 
a GLSM boundary action \cite{Hori1}. We are then led to attempt the more elegant method of obtaining the GNLSM boundary conditions and boundary action from a GLSM boundary action, using the methods of Section 3.
In the following, we shall attempt to generalize the NLSM Lagrangian boundary conditions and boundary action \eqref{Abrane} to the case of $U(1)^{k}$-GNLSMs with K\"ahler toric target space. %before proceeding to do the same for the boundary action \eqref{BbraneAlt}. %$\C^N//U(1)$. 

Let us elaborate further on how NLSM boundary data is obtained from GLSM boundary data. 
%Hori
The essential idea of \cite{Hori1}
%proposed
is to obtain Lagrangian boundary conditions and the boundary action \eqref{Abrane} from a GLSM  %by starting with 
using boundary conditions given by
\begin{equation}\label{Asuperfieldbc}
\bPhi_i\e^{\hat{V}}\Phi_i=c_i
\end{equation}
which includes the Dirichlet boundary condition
\begin{equation}\label{dirichletA}
|\phi_i|^2=c_i
\end{equation}
%as well as 
and Neumann boundary conditions on $\varphi_i$ (where $\varphi_i$ is %the phase of the GLSM scalar field 
defined by $\phi_i=|\phi_i|e^{i\varphi_i}$), as well as
the boundary Wilson line
\begin{equation}
S_{a}
=\sum_i^N\int\limits_{\partial\Sigma}\,
{a_i\over 2\pi}\partial_0\varphi_i\,\dd x^0,
\label{aterm}
\end{equation}
(where $a_i$ is a constant).\footnote{Although $a_i$ is a constant, Stoke's theorem does not result in the vanishing of \eqref{aterm}, since $\varphi_i$ is a periodic, multi-valued function \cite{Hori1}.}
%and  
These indicate that at the GLSM level, the A-brane is wrapped 
%is realized 
 on a torus parametrized by the $\varphi_i$'s, which is a Lagrangian submanifold of $\C^N$, and where $c_i$ measures the size of the circle parametrized by $\varphi_i$, while $a_i$ parametrizes the holonomy of the $U(1)$ connection on the D-brane. 
%IMPORTANT NOTE: THE FOLLOWING should be CORRECT, BUT TAKING OUT for the sake of selamat. Atas diambil from Hori. ACTUally i think a_i itself is a connection from the GLSM point of VIEW, SEE NAKAHARA EXPRESSION FOR WILSON LINE.
% while $e^{i a_i}$ is the holonomy of the target space $U(1)$ gauge field along %%the worldvolume of the torus brane.
% the circle parametrized by $\varphi_i$.
%However, unlike B-branes, the constraints \eqref{generalv} and \eqref{generalsi} obtained via the reduction procedure of Section 3 are not compatible with %gauge-invariant 
%A-type supersymmetric boundary conditions corresponding to flat . 
However, 
%if one were to 
instead of imposing the boundary condition \eqref{dirichletA} and its A-type supersymmetric completions at the GLSM level %(e.g. for a $U(1)$ GLSM),
before taking the NLSM ($\hat{e}\to \infty$) limit,
% and integrating out the vector multiplet components would impose constraints analogous to 
%\eqref{generalv2} and \eqref{generalsi2}, which can only be shown to %be incompatible 
%only 
%be compatible with A-type supersymmetry %unless 
%if $c_i=0$.
% which are incompatible with the boundary conditions.
%would trivialize the boundary action above via the constraints  \eqref{generalv} and \eqref{generalsi}. 
%IMPORTANT NOTE(this happens because we would get $\si=0$, which then would require $v_0=0$, which via the constraint would give $\sum_i \hat{Q}_{ib}|\phi_i|^2\del_0\varphi_i=0$.).
% ACTUALLY what happens is that the constraint for \hat{A}_0 reduces to just the part with bosonic fields via the fermionic boundary conditions, and \hat{si} reduces to zero using these same boundary conditions. Now is \hat{si} is zero, \hat{A}_0 should also be zero using A-type susy. The numerator of \eqref{generalv2} can be written as -2\sum_i^N\hat{Q}_i|\phi_i|^2\del_0\varphi_i. The only way this can be zero is if c_i=0, since \del_0\varphi=0 is not one of our boundary conditions. So we say that the constraints are not compatible with A-type susy, unless c_i=0 (which is too restrictive for us).
%To avoid this,
 one uses `boundary superfields' \cite{Hori1}, whereby no boundary conditions are imposed by hand at the GLSM level, but rather they are understood as being derived through boundary interactions involving these boundary superfields. The advantage of this formulation is that the geometric parameters of the NLSM D-brane enter a `boundary F-term', and this aids our understanding of quantum corrections \cite{Hori1}. 
% method of using  to impose boundary conditions via boundary interactions. 

We shall follow this method in deriving the GNLSM generalization of the boundary conditions and the boundary action \eqref{Abrane}. To this end, let us first briefly review the concept of boundary superfields \cite{Hori1}, in particular, those living in A-boundary superspace. The coordinates of A-boundary superspace are $x^0$, $\theta$, $\ov{\theta}$, and boundary superfields are simply functions of these coordinates, and transform under A-type supersymmetry.
Boundary superfields can be of both bosonic and fermionic nature. %and the latter are called boundary Fermi superfields. 

The important differential operators on A-boundary superspace are $\partial_0$ and %the following differential operators
\begin{equation}
\begin{aligned}
\bD&=\bD_++D_-=-{\partial\over\partial\btheta}+i\theta\partial_0,\\
D&=D_++\bD_-={\partial\over\partial\theta}-i\btheta\partial_0,\\
\bcQ&=\bcQ_++\cQ_-=-{\partial\over\partial\btheta}-i\theta\partial_0,
\label{bQA}\\
\cQ&=\cQ_++\bcQ_-={\partial\over\partial\theta}+i\btheta\partial_0.
\end{aligned}
\end{equation}
Boundary superfields in A-boundary superspace transform under A-type supersymmetry by $\delta=\eps\ov{\mathcal{Q}}-\ov{\eps}\mathcal{Q}$. 
Bulk superfields restricted to A-boundary are boundary superfields. Furthermore, boundary \textit{chiral} superfields obey
\begin{equation}
\bD{\it \Phi}=0,
\end{equation}
and are expanded as 
\begin{equation}
{\it \Phi}=\phi(x^0)+\theta\psi(x^0)-i\theta\btheta\partial_0\phi(x^0).
\end{equation}
Following the convention of \cite{Hori1}, fermionic boundary chiral superfields shall be referred to as boundary Fermi superfields. 
%Given a function $J({\cal F}_i)$ of boundary superfields ${\cal F}_i$,
%the following integral is invariant under the supersymmetry
%variation $\delta$
The integral 
\begin{equation}
\int \dd x^0\dd\theta\dd\btheta\,J({\cal F}_i),
\label{bD}
\end{equation}
over a a function $J({\cal F}_i)$ of boundary superfields ${\cal F}_i$ is invariant under A-type supersymmetry transformations.
In addition, the integral 
\begin{equation}
\int \dd x^0\dd\theta ~{\mit\Psi}{\cal V}({\mit\Phi}_i)\,\Bigr|_{\btheta=0},
\label{bF}
\end{equation}
where ${\mit\Psi}$ is a boundary Fermi superfield and ${\cal V}({\mit\Phi}_i)$ is a holomorphic function of boundary chiral superfields ${\mit \Phi}_i$, is also invariant under A-type supersymmetry. Expressions of the form \eqref{bD} and \eqref{bF} are known as boundary D-terms and boundary F-terms respectively.

%IMPORTANT NOTE: THE FOLLOWING FACTS GIVEN BY HORI ARE ALSO IMPORTANT, SEE HORI FOR THE ACTUAL POINTS, SOME POINTS MAY BE MISSING HERE
%is invariant under the supersymmetry $\delta$.
%We call ${\cal V}({\mit\Phi}_i)$ a {\it boundary superpotential}.
%The above term is invariant under the boundary R-rotation if
%${\mit\Phi}{\cal V}({\mit\Phi}_i)$ has boundary R-charge $1$.
%We shall sometimes refer to (\ref{bD}) and (\ref{bF}) as boundary D-term
%and boundary F-term respectively.

%A bulk superfield restricted to the boundary (\ref{bosb})-(\ref{Ab})
%is a boundary superfield.
%The boundary R-rotation comes from the axial R-rotation in the bulk.
%It is easy to see that
%a bulk twisted chiral superfield
%restricted on the A-boundary is a boundary chiral superfield.
%The boundary superpotential must be a holomorphic function
%of the boundary chiral superfields. This strongly constrains
%a possible form of quantum corrections, as in \cite{seiberg},
%as we will see explicitly in several examples.
Finally, we recall that only the axial R-symmetry of the bulk is preserved at the boundaries, since vector R-rotations do not leave the A-boundary (defined by \eqref{Aboundary}) invariant.
\vspace{0.2cm}
\mbox{}\par\nobreak
\noindent
\textit{A-branes on $\C^N//U(1)$ from GLSM} 

We shall first recall from \cite{Hori1} how the boundary conditions and the boundary action \eqref{Abrane} for an NLSM with $\C^N//U(1)$ target space can be obtained from an A-type supersymmetric boundary action of a $U(1)$-GLSM with $\C^N$ target. The boundary action consists of two parts
\begin{equation}\label{nlsmuva1}
S_1={1\over 2\pi}\int\limits_{\partial\Sigma}
\dd x^0\left[\,{1\over 2}\sum_i^N\partial_1|\phi_i|^2
+{i\over 2}\sum_i^N(\overline{F}_i\phi_i-\ov{\phi}_i F_i)
+{i\over 4\hat{e}^2}(\hat{\lam}_-\hat{\lam}_+-\ov{\hat{\lam}}_+\ov{\hat{\lam}}_-)
+\hat{\theta} \hat{A}_0\,\right]
\end{equation}
and 
\begin{equation}\label{nlsmuva2}
S_2={1\over 2\pi}\sum_i^N\int\limits_{\partial\Sigma}
\dd x^0
\left[~\int\dd\theta\dd\btheta\,
\bPhi_i\e^{\hat{V}}\Phi_i(U_i-{\rm Im}\log\Phi_i)
+{\rm Re}
\int \dd\theta
\,s_i\Upsilon_i~\right],
\end{equation}
where $U_i$ is a real, bosonic, boundary auxiliary superfield expanded as 
\begin{equation}
U_i=u_i+\theta\ov{\mathcal{X}}_i-\btheta\mathcal{X}_i+\theta\btheta E_i
\end{equation}
(with the lowest component $u_i$ being a periodic (multivalued) scalar field defined on the boundaries),
$\Upsilon_i=\ov{D}U$ is the `field-strength' of $U_i$, 
expanded as 
\begin{equation}
\Upsilon_i:=\bD \,U_i
=\mx_i+\theta(E_i+i\partial_0 u_i)-i\theta\btheta\partial_0\mx_i,
\end{equation}
and is a boundary Fermi superfield satisfying $\ov{D}\Upsilon_i=0$, while the parameter 
\begin{equation}
s_i=c_i-ia_i
\end{equation}
is the boundary analogue of the complex FI-theta parameter $\hat{t}$. It is important to note that although both $u_i$ and $\varphi_i$ are periodic, multi-valued functions, the presence of the term $\int\dd\theta\dd\btheta\,
\bPhi_i\e^{\hat{V}}\Phi_i(u_i-\varphi_i)$ in \eqref{nlsmuva2} requires that $u_i-\varphi_i$ is single-valued.  
%IMPORTANT NOTE: THIS IS SO THAT THE ACTION, AND THEREFORE THE PATH INTEGRAL IS SINGLE VALUED

The first part \eqref{nlsmuva1}, together with the bulk GLSM action, $S$, have the simple supersymmetry transformation 
%has the following simple transformation property;
\begin{equation}
\delta(S + S_1)={\hat{r}\over 4\pi}\int\limits_{\partial\Sigma}
\dd x^0\left\{\epsilon(\ov{\hat{\lam}}_++{\hat{\lam}}_-)
-\ov{\epsilon}(\ov{\hat{\lam}}_-+\hat{\lam}_+)\right\}.
\label{delSp}
\end{equation}
%is invariant under A-type supersymmetry without the use of any boundary conditions, up to 
 The second part, \eqref{nlsmuva2}, includes interactions which effectively impose boundary conditions on the components of the  chiral multiplets.
%and has the supersymmetry transformation 
%We would like to construct a linear model for
%D-branes in this gauge theory.
%As in the case without the gauge field,
%we introduce the superfield $U$ on the A-boundary
%with the boundary interaction 
%as in (\ref{Action}):
%\begin{equation}
%S_{\it boundary}={1\over 2\pi}\int\limits_{\partial\Sigma}
%\dd x^0
%\left[~\int\dd\theta\dd\btheta\,
%\bPhi\e^V\Phi(U-{\rm Im}\log\Phi)
%+{\rm Re}
%\int \dd\theta
%\,s\Upsilon~\right],
%\end{equation}
%where $\Upsilon=\bD U$ is the fieldstrength.
Let us recall how it transforms under supersymmetry. Gauge invariance of the first term in \eqref{nlsmuva2},
requires that $U_i$ transforms under gauge transformations as 
%as ${\rm Im}\log\Phi$. Thus, it transforms as
\begin{equation}
U_i\longrightarrow U_i+{\hat{Q}_i\over 2}({A}+\overline{{A}}),
\label{GaU}
\end{equation}
in order to cancel the gauge variation of $\textrm{Im log } \Phi_i$. This implies the following modification of the supersymmetry transformations of the
components of $U_i$ in order to preserve the Wess-Zumino gauge
%in the Wess-Zumino gauge as
%IMPORTANT NOTE, SEE EQUATION 15.28 OF MIRROR SYMMETRY BOOK,   WHEN WE TAKE THE SUSY TRANSFORMS OF THE VECTOR SUPERFIELD COMPONENTS,  THE WESS ZUMINO GAUGE WHICH WE IMPOSED ON THE VECTOR SUPERFIELDS IS LOST, SO WE MUST APPEND A GAUGE TRANSFORMATION USING EQUATION 15.28 TO RETURN TO WESS-ZUMINO GAUGE. TO MY UNDERSTANDING, WHEN WE INCLUDE MATTER, THE SAME GAUGE TRANSFORMATION WHICH IS USED TO RETURN TO WESS-ZUMINO GAUGE MUST BE PERFORMED TOGETHER WITH THE SUSY TRANSFORMATIONS OF THE MATTER FIELDS, HENCE GIVING RISE TO THE MODIFIED SUSY TRANSFORMATIONS WE USUALLY SEE. THEN, THE MOMENT WE REALIZE U_i also transforms under gauge transformations (not really correct to say that its charged, since it transforms like the phase of a complex scalar field), we should also append its susy transformations with the appropriate gauge transformations. 
\begin{equation}
\begin{aligned}
\delta u_i&=\epsilon\mx_i-\bepsilon\bmx_i,
\\
\delta \mx_i&=-\bepsilon(E_i+i(\partial_0 u_i+\hat{Q}_i\hat{A}_0))
-i\epsilon\hat{Q}_i\hat{\si},
\\
\delta \bmx_i&=-\epsilon(E_i-i(\partial_0 u_i+\hat{Q}_iA_0))
+i\bepsilon\hat{Q}_i\ov{\hat{\si}},\\
\delta E_i&=i\epsilon\partial_0\mx_i+i\bepsilon\partial_0\bmx_i
-{1\over 2}\epsilon\hat{Q}_i(\hat{\lam}_-+\ov{\hat{\lam}}_+)
+{1\over 2}\bepsilon\hat{Q}_i(\ov{\hat{\lam}}_-+\hat{\lam}_+).
\\
\end{aligned}
\end{equation}
Under these supersymmetry transformations,
the boundary superpotential term
\begin{equation}\label{boundarysuperpotential}
{1\over 2\pi}\sum_i^N\int\limits_{\partial\Sigma}
\dd x^0~{\rm Re}\int\dd\theta
\,s_i\Upsilon_i\,=
{1\over 2\pi}\sum_i^N\int\limits_{\partial\Sigma}
\dd x^0\left(c_iE_i+a_i\partial_0 u_i\right)
%\label{boundt}
\end{equation}
is not invariant,\footnote{The reason for this nonzero  variation is that the boundary Fermi superfield $\Upsilon$ is not
invariant under the gauge transformation \eqref{GaU}.} but rather varies as
\begin{equation}
\delta \left[{1\over 2\pi}\sum_i^N\int\limits_{\partial\Sigma}
\dd x^0\,{\rm Re}\int\dd\theta
\,s_i\Ups_i~
\right]
=-{\sum_i^N\hat{Q}_ic_i\over 4\pi}
\int\limits_{\partial\Sigma}
\dd x^0\left\{\epsilon(\ov{\hat{\lam}}_++\hat{\lam}_-)
-\bepsilon(\ov{\hat{\lam}}_-+{\hat{\lam}}_+)\right\}.
\label{nontV}
\end{equation}
%IMPORTANT NOTE : THE REASON THE BOUNDARY SUPERPOTENTIAL TERM NOT INVARIANT EVEN THOUGH ITS MADE UP OF SUPERFIELDS, AND IN A FORM WHICH SHOULD BE INVARIANT, IS BECAUSE IT IS NOT GAUGE INVARIANT, SEE HORI
Supersymmetry invariance of the entire action then requires that \eqref{delSp} and \eqref{nontV} cancel, which is possible if and only if 
\begin{equation}\label{bubocr}
\sum_i^N\hat{Q}_ic_i=\hat{r}.
\end{equation}

Similarly, the first part \eqref{nlsmuva1} of the boundary 
action is not gauge invariant, but varies as 
\begin{equation}
\delta S_1= \frac{1}{2\pi}\int\limits_{\partial\Sigma} dx^0 \hat{\theta}(-\del_0\alpha),
\end{equation}
while \eqref{boundarysuperpotential} varies under gauge transformations as 
\begin{equation}
\delta \left[{1\over 2\pi}\sum_i^N\int\limits_{\partial\Sigma}
\dd x^0\,{\rm Re}\int\dd\theta
\,s_i\Ups_i~
\right]
=\frac{\sum_i^N\hat{Q}_ia_i}{2\pi}\int\limits_{\partial\Sigma} dx^0\del_0\alpha,
\end{equation}
since the residual gauge transformation $A=\alpha(x)$ of the Wess-Zumino gauge shifts $u_i\to u_i +\hat{Q}_i\alpha$, while leaving $\mx_i$ and $E_i$ invariant.
Thus, gauge invariance of the boundary action follows if\footnote{\label{43}To be precise, \eqref{buboat}  only needs to hold up to the additional term $2\pi m$, where $m\in \Z$, since the \textit{path integral} remains gauge invariant in such cases \cite{Hori1}. However, we shall set $m=0$ in the following for simplicity.}
\begin{equation}\label{buboat}
\sum_i^N\hat{Q}_ia_i=\hat{\theta}.
\end{equation}
Combining \eqref{bubocr} and \eqref{buboat}, we find that we need 
\begin{equation}\label{bubost}
\sum_i^N\hat{Q}_is_i=\hat{t}
\end{equation}
for gauge invariance and A-type supersymmetry invariance of the action. 

To analyze the $\hat{e}\to \infty$ limit, it is advantageous to write the boundary action explicitly\footnote{Note that some terms in \eqref{nlsmuva1} cancel terms in \eqref{nlsmuva2}.}
\begin{equation}
\begin{aligned}
S_{\del\Sigma} &=S_1+S_2\\
&=\frac{1}{2\pi}\int\limits_{\partial\Sigma}dx^0\Big(
{i\over 4\hat{e}^2}(\hat{\lam}_-\hat{\lam}_+-\ov{\hat{\lam}}_+\ov{\hat{\lam}}_-)
+\hat{\theta} \hat{A}_0
\Big)\\
&+{1\over 2\pi}\sum_i^N\int\limits_{\partial\Sigma}
\dd x^0\Big((i \overline{\phi}_i\overleftrightarrow{\hat{D}}_1\phi_i+\overline{\psi}_{+i}\psi_{+i} - \overline{\psi}_{-i}\psi_{-i} + \overline{F}_i\phi_i+\overline{\phi}_iF_i)u'_i \\& +(\overline{\phi}_i\psi_{-i}+\overline{\psi}_{+i}\phi_i)\ov{\mx}_i+{\mx}_i(\overline{\psi}_{-i}\phi_i+\overline{\phi}_i\psi_{+i}) -i{3\over 2}{\ov{\phi}_i\over \phi_i}{\psi}_{+i}{\psi}_{-i}+i{3\over 2}{{\phi}_i\over \ov{\phi}_i}\ov{\psi}_{-i}\ov{\psi}_{+i}\\&-(|\phi_i|^2-c_i)E_i+a_i\partial_0 u_i\Big)
\end{aligned}
\end{equation}
where $u'_i=u_i-\varphi_i$, and where the covariant derivative of the scalar field is $\hat{D}_1\phi_i=\del_0\phi_i+i\hat{Q}_i\hat{A}_1\phi_i$.
 Firstly, we note that Stoke's theorem implies 
%\overline{\phi}_i\psi_{-i}+\overline{\psi}_{+i}\phi_i&=0,\\
%\overline{\psi}_{-i}\phi_i+\overline{\phi}_i\psi_{+i}&=0
\begin{equation}
{a_i\over 2\pi}\int\limits_{\partial\Sigma}\partial_0u_i\dd x^0
={a_i\over 2\pi}\int\limits_{\partial\Sigma}\Bigl\{\partial_0\varphi_i
+\partial_0(u_i-\varphi_i)\Bigr\}\dd x^0
={a_i\over 2\pi}\int\limits_{\partial\Sigma}\partial_0\varphi_i\dd x^0,
\label{reda}
\end{equation}
since $u_i-\varphi_i$ is single-valued, and we %retrieve 
find that one of the terms in the boundary action is the expression \eqref{aterm}. Then, taking $\hat{e}\to\infty$, and integrating out the boundary auxiliary superfields, we are left with 
\begin{equation}\label{penultimate}
\frac{1}{2\pi}\int\limits_{\partial\Sigma}dx^0\Big(
\sum_i^Na_i\partial_0\varphi_i+\hat{\theta} \hat{A}_0
\Big),
\end{equation}
with the boundary conditions defined by \eqref{Asuperfieldbc} imposed at the boundaries. These boundary conditions are given explicitly as
\begin{equation}
\begin{aligned}\label{bcA}
|\phi_i|^2&=c_i,\\
\overline{\phi}_i\psi_{-i}+\overline{\psi}_{+i}\phi_i&=0,\\
\overline{\psi}_{-i}\phi_i+\overline{\phi}_i\psi_{+i}&=0,\\
i \overline{\phi}_i\overleftrightarrow{\hat{D}}_1\phi_i+\overline{\psi}_{+i}\psi_{+i} - \overline{\psi}_{-i}\psi_{-i} + \overline{F}_i\phi_i+\overline{\phi}_iF_i&=0.\\
\end{aligned}
\end{equation} 
Integrating the vector multiplet out of the bulk action sets 
 %IMPORTANT NOTE: FERMI TERMS VANISH VIA BOUNDARY CONDITIONS
\begin{equation}
\hat{A}_0={1\over 2}{\sum_{i=1}^N\hat{Q}_i(
i\bphi_i\overleftrightarrow{\del}_{\!\!\!0}\phi_i+\ov{\psi}_{-i}\psi_{-i}+\ov{\psi}_{+i}\psi_{+i})\over \sum_{j=1}^N\hat{Q}_j^2|\phi_j|^2}
=
-{\sum_{i=1}^N\hat{Q}_ic_i\partial_{0}\varphi_i\over
\sum_{j=1}^N\hat{Q}_j^2c_j},
\end{equation}
 at the boundaries, where \eqref{bcA} has been used in the last step.\footnote{The presence of the boundary term proportional to $\hat{\theta}\hat{A}_0$ ensures that the algebraic equation of motion for $\hat{A}_0$ does not contain a boundary term, see footnote \ref{15}.}
Thus, the final boundary action is
\begin{equation}\label{finalboundaryaction}
S_{\del\Sigma}={1\over 2\pi}\int\limits_{\partial\Sigma}
\left[\,\sum_{i=1}^N a_i\dd\varphi_i
-\hat{\theta}{\sum_{i=1}^N\hat{Q}_ic_id\varphi_i\over
\sum_{j=1}^N\hat{Q}_j^2c_j}
%\,-\,\theta\left(\sum_{i=1}^N\hat{Q}_ic_i\dd\varphi_i%\Biggr/
%%\mbox{\huge$\slash$}
%\sum_{i=1}^N\hat{Q}_i^2c_i\right)
\right].
\end{equation}
%It is easy to see that the holonomy in the gauge orbit direction
%indeed vanishes if the condition
%$\theta=\sum_{i=1}^NQ_ia_i$ in (\ref{sgcond1}) is satisfied.

It will be useful for us to analyze Hori's results \eqref{bcA} and \eqref{finalboundaryaction} for $X=\C P^{N-1}$, which corresponds to $\hat{Q}_i=1$. Firstly, 
the inhomogeneous coordinates \eqref{inhomogeneous} which parametrize a local patch of $\CP^{N-1}$ can be written as  
\begin{equation}
|Z^i|e^{i\gamma^i}=\frac{|\phi_i|e^{i\varphi_i}}{|\phi_N|e^{i\varphi_N}}.
\end{equation} 
In other words, the argument of $Z^i$ is 
\begin{equation}\label{argZ}
\gamma^i=\varphi_i-\varphi_N.
\end{equation}
The A-type supersymmetric boundary conditions of the NLSM which can be obtained from \eqref{bcA} using the parametrizations \eqref{para} and \eqref{lamconstraint2} are\footnote{To be precise, the last condition of \eqref{bcA} is actually trivialized using the algebraic equation of motion of $\hat{A}_1$. The last condition of \eqref{bcAcpn} is in fact obtained via A-type supersymmetry transformations of the fermionic boundary conditions. }
\begin{equation}\label{bcAcpn}
\begin{aligned}
|Z^i|^2&=\frac{c_i}{c_N}\\
\overline{Z}^i\psi^{Z^i}_{-}+Z^i\overline{\psi}^{\ov{Z}^i}_{+}&=0\\
\overline{Z}^i\psi^{Z^i}_{+}+Z^i\overline{\psi}^{\ov{Z}^i}_{-}&=0\\
i(\ov{Z}^i\del_1Z^i-Z^i\del_1\ov{Z}^i)+\ov{\psi}_+^{\ov{Z}^i}\psi^{{Z}^i}_+-\ov{\psi}_-^{\ov{Z}^i}\psi^{{Z}^i}_-+F^{{Z}^i}\ov{Z}^i+\ov{F}^{\ov{Z}^i}Z^i&=0,
\end{aligned}
\end{equation}
where the last condition is in fact a Neumann boundary condition on $\gamma^i$, since $\ov{Z}^i\del_1Z^i-Z^i\del_1\ov{Z}^i=2|Z^i|^2i\del_1\gamma^i$. Thus, the Neumann boundary condition on $\gamma^i$ and Dirichlet boundary condition on $|Z^i|$ in \eqref{bcAcpn} implies that the A-brane wraps a torus $T^{N-1}$ parametrized by $\gamma_i$.\footnote{Here, both boundaries are mapped to the same A-brane. If the boundaries are assigned  unique parameters $s_i^\pi$ and $s_i^0$ in \eqref{nlsmuva2}, then each boundary is mapped to a different A-brane. However, for simplicity, in most of what follows in this section, we shall assume that both boundaries are assigned the same parameter $s_i$.} 
Moreover, this torus is a Lagrangian submanifold of $\CP^{N-1}$ with respect to the Fubini-Study K\"ahler form given by \eqref{FSkahlerform}.
%IMPORTANT NOTE The fourth condition is actually trivialized, using the equations of motion for $F$ which have been integrated out of the bulk action, as well as the equation of motion for $v$.
%To get the final boundary condition, we take the A-type SUSY variation of 
%\eqref{second}, which is  

 We can rewrite the boundary action \eqref{finalboundaryaction} with the help of \eqref{argZ} as 
\begin{equation}\label{finalboundaryactionCPN}
\begin{aligned}
S_{\del\Sigma}&={1\over 2\pi}\int\limits_{\partial\Sigma}
\sum_{i=1}^N\left[\, a_i
-\hat{\theta}{ c_i \over
\sum_{j=1}^Nc_j}
%\,-\,\theta\left(\sum_{i=1}^{N-1}\hat{Q}_ic_i
%\sum_{j=1}^N c_j\right)
\right]\dd\varphi_i\\
&={1\over 2\pi}\int\limits_{\partial\Sigma}
\left[\,\sum_{i=1}^{N-1}\Bigg( a_i
-(\sum_{k=1}^Na_k){ c_i\over
\sum_{j=1}^Nc_j}\Bigg)d\varphi_i+\Bigg( a_N
-(\sum_{k=1}^Na_k){ c_N \over
\sum_{j=1}^Nc_j}\Bigg)d\varphi_N
%\,-\,\theta\left(\sum_{i=1}^N\hat{Q}_ic_i\dd\varphi_i%\Biggr/
%%\mbox{\huge$\slash$}
%\sum_{j=1}^N c_j\right)
\right]\\
&={1\over 2\pi}\int\limits_{\partial\Sigma}
\left[\,\sum_{i=1}^{N-1}\Bigg( a_i
-(\sum_{k=1}^Na_k){ c_i\over
\sum_{j=1}^Nc_j}\Bigg)d(\varphi_i-\varphi_N)\right]\\
&=\int\limits_{\partial\Sigma}\sum_{i=1}^{N-1}A^{X}_id\gamma^i,
\end{aligned}
\end{equation}
%NOT SO IMPORTANT NOTE: IF WE HAVE $\sum_i^NQ_ia_i +2\pi m=\theta$, then the effect is the addition of -2\pi m n to the final result above, where $n$ is the winding number of all the $\varphi_i$ or the $u_i$. i think for $u_i-\varphi_i$ to be single valued both u_i and \varphi_i should have the same winding number. see note for more details 
where we have also used \eqref{buboat}, and defined the constant 
\begin{equation}\label{constantconnect}
A^X_i= {1\over 2\pi}\Big(a_i -(\sum_{k=1}^Na_k) {c_i\over \sum_{j=1}^Nc_j}\Big),
\end{equation}
which is understood to be the connection of a flat $U(1)$ bundle on the Lagrangian torus $T^{N-1}$ parametrized by $\gamma_i$. In other words, the A-brane, defined by the boundary conditions \eqref{bcAcpn} and boundary action \eqref{finalboundaryactionCPN}, wraps a Lagrangian submanifold of $X=\CP^{N-1}$ and supports a flat $U(1)$ bundle. This corresponds to the data of an object in the Fukaya category of $\CP^{N-1}$ \cite{aspinwall2009Dmirror}.
%IMPORTANT NOTE: BUT THERE ITS WRITTEN DOWN FOR CALABI-YAU. ITS ON PAGE 600. AND THERE'S TWO MORE DATA, INCLUDING SPIN STRUCTURE ON THE LAGRANGIAN MANIFOLD WHICH WE SHOULD HAVE
Supersymmetry invariance of the boundary action \eqref{finalboundaryactionCPN} follows since the fermionic superpartners of $\gamma^i$ are not periodic nor multivalued (\cite{hori2003mirror}, page 307), hence, the supersymmetry variation of \eqref{finalboundaryactionCPN} vanishes via Stoke's theorem.

We have thus derived the boundary action \eqref{Abrane} and the pertinent boundary conditions for $X=\CP^{N-1}$ from a GLSM.
Similarly, for other toric manifolds $X=\C^N//U(1)$, we may use the same techniques shown above to find that the A-brane wraps a torus $T^{N-1}$ which is a Lagrangian submanifold of $X$, and supports a flat $U(1)$ bundle \cite{Hori1}.
%This torus is a Lagrangian submanifold of $\CP^{N-1}$ with respect to the Fubini-Study K\"ahler form given by \eqref{FSkahlerform}.

%\vspace{0.2cm}
%\mbox{}\par\nobreak
%\noindent
\subsection{Equivariant A-branes on $\C^N//U(1)$ from GLSM} 

%Having recalled how the A-type supersymmetric boundary action and boundary conditions for NLSMs with toric target spaces of the form $\C^N//U(1)$ is obtained from a GLSM, 
We shall now proceed to obtain the A-type supersymmetric boundary conditions and boundary action for abelian GNLSMs with toric target spaces of the form $\C^N//U(1)$. In order to do so, we must generalize the $U(1)$-GLSM boundary action consisting of \eqref{nlsmuva1} and \eqref{nlsmuva2} to a boundary action for the $U(1)^N$-GLSM given in \eqref{big1}, with $N-k=1$. 

The first step would be the obvious generalization of the terms with vector multiplet fields, i.e., from $U(1)$ to $U(1)^N$.
%which already contained in \eqref{nlsmuva1} and \eqref{nlsmuva2}
%by including several $U(1)$ gauge groups. 
Next, we note that in obtaining a $U(1)^{N-1}$-GNLSM from the $U(1)^N$-GLSM, we do not integrate out all vector multiplets, unlike in the procedure of obtaining the NLSM. However, the boundary action \eqref{nlsmuva2} only imposes boundary conditions on the matter fields in the $\hat{e}\to\infty$ limit. This implies that we ought to include additional boundary interactions at the GLSM level, which impose boundary conditions on the remaining vector multiplet fields in the $\hat{e}\to\infty$ limit. We claim that the $U(1)^N$-GLSM boundary action consists of 
\begin{equation}\label{gnlsmuva1}
S_1={1\over 2\pi}\int\limits_{\partial\Sigma}
\dd x^0\left[\,{1\over 2}\sum_i^N\partial_1|\phi_i|^2
+{i\over 2}\sum_i^N(\overline{F}_i\phi_i-\ov{\phi}_i F_i)
+\sum_a^N{i\over 4{e_a}^2}({\lam}_{-a}{\lam}_{+a}-\ov{{\lam}}_{+a}\ov{{\lam}}_{-a})
+\sum_a^N{\theta_a}{A}_{0a}\,\right]
\end{equation}
and 
\begin{equation}\label{gnlsmuva2}
\begin{aligned}
S_2=&{1\over 2\pi}\sum_i^N\int\limits_{\partial\Sigma}
\dd x^0 \Bigg[
\int\dd\theta\dd\btheta\,
\bPhi_i\e^{\sum_a^N\mathcal{Q}_{ia}V_a}\Phi_i(U_i-{\rm Im}\log\Phi_i)
+{\rm Re}
\int \dd\theta
\,s_i\Upsilon_i\\
&+\sum_a^N\frac{1}{2e_a^2}\int\dd\theta\dd\ov{\theta}\textrm{Re}\big[\Xi_a(D_+\Sigma_a-\ov{D}_-\Sigma_a)\big]
\Bigg],
\end{aligned}
\end{equation}
where we have introduced an A-type supersymmetry invariant boundary D-term for the vector superfields, which contains the complex boundary Fermi superfields
	\begin{equation}
	\begin{aligned}
	\Xi_a&=\xi_a +\theta G_a +\ov{\theta}H_a+  \theta\ov{\theta}\K_a\\
\ov{\Xi}_a&=\ov{\xi}_a +\ov{\theta} \ov{G}_a +{\theta}\ov{H}_a+  {\theta}\ov{\theta}\ov{\K}_a,
\end{aligned}
\end{equation}
where $\xi_a$ and $\K_a$ are fermionic auxiliary fields while $G_a$ and $H_a$ are bosonic auxiliary fields, all defined along the boundaries. The A-type supersymmetry transformations of these fields may be found using the differential operator $\delta=\eps\ov{\mathcal{Q}}-\ov{\eps}\mathcal{Q}$ defined in \eqref{bQA} on the superfields $\Xi_a$ and $\ov{\Xi}_a$. In addition, they are defined to be invariant under gauge transformations. The form of \eqref{gnlsmuva2} is chosen such that the boundary conditions 
\begin{equation}
\begin{aligned}
\bPhi_i\e^{\sum_a^N\mathcal{Q}_{ia}V_a}\Phi_i&=c_i\\
D_+\Sigma_a&=\ov{D}_-\Sigma_a\\
\ov{D_+\Sigma_a}&=\ov{\ov{D}_-{\Sigma}_a}
\end{aligned}
\end{equation}
are effectively imposed via boundary interactions.
%IMPORTANT NOTE: TO UNDERSTAND WHY THE FIRST EQUATION IS EFFECTIVELY IMPOSED SEE ABOVE EQUATION 3.27 OF HORI-LINEAR MODELS
 In components, these are 
\begin{equation}
\begin{aligned}\label{GbcAOri}
|\phi_i|^2&=c_i,\\
\overline{\phi}_i\psi_{-i}+\overline{\psi}_{+i}\phi_i&=0,\\
\overline{\psi}_{-i}\phi_i+\overline{\phi}_i\psi_{+i}&=0,\\
i \overline{\phi}_i\overleftrightarrow{{D}}_1\phi_i+\overline{\psi}_{+i}\psi_{+i} - \overline{\psi}_{-i}\psi_{-i} + \overline{F}_i\phi_i+\overline{\phi}_iF_i&=0\\
\end{aligned}
\end{equation} 
and
\begin{equation}\label{cpngaugebcOri}
\begin{aligned}
%\partial_1v_0&=0,\\
%v_1&=0,\\
{\lam}_{+a}-\ov{{\lam}}_{-a}&=0,\\
\del_1{\si}_a&=0,\\
%\partial_1\overline{\sigma}^a&=0,\\
{F}_{01a}&=0,\\
{D}_a&=0,\\
\partial_1({\lam}_{+a} +\ov{{\lam}}_{-a})&=0,\\
\end{aligned}
\end{equation}
and the complex conjugates of the conditions in \eqref{cpngaugebcOri}.
	%\begin{equation}
%\begin{aligned}
%S_{\it boundary}=&{1\over 2\pi}\int\limits_{\partial\Sigma}
%\dd x^0
%\Bigg[\int\dd\theta\dd\ov{\theta}\,
%\ov{\Phi}\e^V\Phi(U-{\rm Im}\log\Phi)
 % +\sum_a^N\frac{1}{2e_a^2}\int\dd\theta\dd\ov{\theta}5\textrm{Re}\big[\Xi(D_+\Sigma^a-\ov{D}_-\Sigma^a)\big]\\&+{\rm Re}
%\int \dd\theta
%\,s\Upsilon \Bigg],
%\end{aligned}
%\end{equation}
%IMPORTANT NOTE -The above is still tentative, must check. Unlike $U$, $\til{U}$ is complex. CHECKED
%-Can include the above extra term in Hori's case, Taking the $e_a\to \infty$ limit just makes it diappear. Thus, there is no inconsistency in terms of the $e_a\to \infty$ limit imposing constraints which violate the supersymmetric boundary conditions.
%-Likewise in our case we take $e_a\to \infty$ only for the gauge groups we want to break, and boundary conditions are ONLY imposed for the vector multiplets corresponding to the residual gauge symmetries.Once again, no inconsistency.
%-Mirror symmetry should not be affected, since the proof is only concerned with dualizing the $\varphi$ fields.

The supersymmetry transformation of the bulk GLSM action together with \eqref{gnlsmuva1} is 
%has the following simple transformation property;
\begin{equation}
\delta(S + S_1)=\sum_a^N{{r}_a\over 4\pi}\int\limits_{\partial\Sigma}
\dd x^0\left\{\epsilon(\ov{{\lam}}_{+a}+{{\lam}}_{-a})
-\ov{\epsilon}(\ov{{\lam}}_{-a}+{\lam}_{+a})\right\}.
\label{GdelSp}
\end{equation}
%is invariant under A-type supersymmetry without the use of any boundary conditions, up to 
% The second part \eqref{nlsmuva2} includes interactions which effectively impose boundary conditions on the components of the  chiral multiplets.
%and has the supersymmetry transformation 
%We would like to construct a linear model for
%D-branes in this gauge theory.
%As in the case without the gauge field,
%we introduce the superfield $U$ on the A-boundary
%with the boundary interaction 
%as in (\ref{Action}):
%\begin{equation}
%S_{\it boundary}={1\over 2\pi}\int\limits_{\partial\Sigma}
%\dd x^0
%\left[~\int\dd\theta\dd\btheta\,
%\bPhi\e^V\Phi(U-{\rm Im}\log\Phi)
%+{\rm Re}
%\int \dd\theta
%\,s\Upsilon~\right],
%\end{equation}
%where $\Upsilon=\bD U$ is the fieldstrength.
Now, $U(1)^N$ gauge invariance of the first term in \eqref{gnlsmuva2},
requires that $U_i$ transforms under $U(1)^N$ gauge transformations as 
%as ${\rm Im}\log\Phi$. Thus, it transforms as
\begin{equation}
U_i\longrightarrow U_i+\sum_a^N{\mathcal{Q}_{ia}\over 2}({A_a}+\overline{A}_a),
\label{GGaU}
\end{equation}
in order to cancel the gauge variation of $\textrm{Im log } \Phi_i$. This implies the following % leads to the 
modification of the supersymmetry transformations of the
components of $U_i$ in order to preserve the Wess-Zumino gauge
%in the Wess-Zumino gauge as
%IMPORTANT NOTE, SEE EQUATION 15.28 OF MIRROR SYMMETRY BOOK,   WHEN WE TAKE THE SUSY TRANSFORMS OF THE VECTOR SUPERFIELD COMPONENTS,  THE WESS ZUMINO GAUGE WHICH WE IMPOSED ON THE VECTOR SUPERFIELDS IS LOST, SO WE MUST APPEND A GAUGE TRANSFORMATION USING EQUATION 15.28 TO RETURN TO WESS-ZUMINO GAUGE. TO MY UNDERSTANDING, WHEN WE INCLUDE MATTER, THE SAME GAUGE TRANSFORMATION WHICH IS USED TO RETURN TO WESS-ZUMINO GAUGE MUST BE PERFORMED TOGETHER WITH THE SUSY TRANSFORMATIONS OF THE MATTER FIELDS, HENCE GIVING RISE TO THE MODIFIED SUSY TRANSFORMATIONS WE USUALLY SEE. THEN, THE MOMENT WE REALIZE U_i also transforms under gauge transformations (not really correct to say that its charged, since it transforms like the phase of a complex scalar field), we should also append its susy transformations with the appropriate gauge transformations. 
\begin{equation}
\begin{aligned}
\delta u_i&=\epsilon\mx_i-\bepsilon\bmx_i,
\\
\delta \mx_i&=-\bepsilon(E_i+i(\partial_0 u_i+\sum_a^N\mathcal{Q}_{ia}{A}_{0a}))
-i\epsilon \sum_a^N\mathcal{Q}_{ia} {\si_a},
\\
\delta \bmx_i&=-\epsilon(E_i-i(\partial_0 u_i+\sum_a^N\mathcal{Q}_{ia}{A}_{0a}))
+i\bepsilon\sum_a^N\mathcal{Q}_{ia}\ov{{\si}}_a,\\
\delta E_i&=i\epsilon\partial_0\mx_i+i\bepsilon\partial_0\bmx_i
-{1\over 2}\epsilon\sum_a^N\mathcal{Q}_{ia}({\lam}_{-a}+\ov{{\lam}}_{+a})
+{1\over 2}\bepsilon\sum_a^N\mathcal{Q}_{ia}(\ov{{\lam}}_{-a}+{\lam}_{+a}).
\\
\end{aligned}
\end{equation}
The boundary superpotential term in \eqref{gnlsmuva2}
is not invariant under supersymmetry,\footnote{This nonzero  variation occurs because the boundary Fermi superfield $\Upsilon$ is not
invariant under the gauge transformation \eqref{GGaU}.} but rather varies as
\begin{equation}
\delta \left[{1\over 2\pi}\sum_i^N\int\limits_{\partial\Sigma}
\dd x^0\,{\rm Re}\int\dd\theta
\,s_i\Ups_i~
\right]
=-{\sum_a^N\sum_i^N\mathcal{Q}_{ia}c_i\over 4\pi}
\int\limits_{\partial\Sigma}
\dd x^0\left\{\epsilon(\ov{{\lam}}_{+a}+{\lam}_{-a})
-\bepsilon(\ov{{\lam}}_{-a}+{{\lam}}_{+a})\right\}.
\label{GnontV}
\end{equation}
%IMPORTANT NOTE : THE REASON THE BOUNDARY SUPERPOTENTIAL TERM NOT INVARIANT EVEN THOUGH ITS MADE UP OF SUPERFIELDS, AND IN A FORM WHICH SHOULD BE INVARIANT, IS BECAUSE IT IS NOT GAUGE INVARIANT, SEE HORI
Hence, supersymmetry invariance of the entire action requires that \eqref{GdelSp} and \eqref{GnontV} cancel, which is possible if and only if 
\begin{equation}\label{Gbubocr}
\sum_i^N\mathcal{Q}_{ia}c_i={r}_a.
\end{equation}

Likewise, the first part \eqref{gnlsmuva1} of the boundary 
action is not $U(1)^N$-gauge invariant, but varies as 
\begin{equation}
\delta S_1= \frac{1}{2\pi}\int\limits_{\partial\Sigma} dx^0 \sum_a^N{\theta}_a(-\del_0\alpha_a),
\end{equation}
while the boundary superpotential term varies under gauge transformations as 
\begin{equation}
\delta \left[{1\over 2\pi}\sum_i^N\int\limits_{\partial\Sigma}
\dd x^0\,{\rm Re}\int\dd\theta
\,s_i\Ups_i~
\right]
=\frac{\sum_a^N\sum_i^N\mathcal{Q}_{ia}a_i}{2\pi}\int\limits_{\partial\Sigma} dx^0\del_0\alpha_a,
\end{equation}
since the residual gauge transformation $A_a=\alpha_a(x)$ of the Wess-Zumino gauge shifts $u_i\to u_i +\sum_a^N\mathcal{Q}_{ia}\alpha_a$, while leaving $\mx_i$ and $E_i$ invariant.
Therefore, gauge invariance of the boundary action follows if\footnote{As noted in footnote \ref{43}, \eqref{Gbuboat}  only needs to hold up to the additional term $2\pi m$, but we shall set $m=0$ in the following for simplicity.
}
\begin{equation}\label{Gbuboat}
\sum_i^N\mathcal{Q}_{ia}a_i={\theta}_a.
\end{equation}
Combining \eqref{Gbubocr} and \eqref{Gbuboat}, we find that we need 
\begin{equation}\label{Gbubost}
\sum_i^N\mathcal{Q}_{ia}s_i={t_a}
\end{equation}
for gauge invariance and A-type supersymmetry invariance of the action. 
%Susy invariance of boundary action
%Gauge invariance, must mention gauge invariance of the superfields $\Xi_a$ and $\ov{\Xi}_a$.

Expanding the boundary action in components, we have
\begin{equation}\label{gnlsmuvafull}
\begin{aligned}
S_{\del\Sigma} &=S_1+S_2\\
&=\frac{1}{2\pi}\int\limits_{\partial\Sigma}dx^0\Bigg(
\sum_a^N{i\over 4{e_a}^2}({\lam}_{-a}{\lam}_{+a}-\ov{{\lam}}_{+a}\ov{{\lam}}_{-a})
+\sum_a^N{\theta_a}{A}_{0a}
\Bigg)\\
&+{1\over 2\pi}\sum_i^N\int\limits_{\partial\Sigma}
\dd x^0\Bigg((i \overline{\phi}_i\overleftrightarrow{{D}}_1\phi_i+\overline{\psi}_{+i}\psi_{+i} - \overline{\psi}_{-i}\psi_{-i} + \overline{F}_i\phi_i+\overline{\phi}_iF_i)u'_i \\& +(\overline{\phi}_i\psi_{-i}+\overline{\psi}_{+i}\phi_i)\ov{\mx}_i+{\mx}_i(\overline{\psi}_{-i}\phi_i+\overline{\phi}_i\psi_{+i}) -i{3\over 2}{\ov{\phi}_i\over \phi_i}{\psi}_{+i}{\psi}_{-i}+i{3\over 2}{{\phi}_i\over \ov{\phi}_i}\ov{\psi}_{-i}\ov{\psi}_{+i}\\&-(|\phi_i|^2-c_i)E_i+a_i\partial_0 u_i\Bigg)\\
&+\sum_a^N\frac{1}{2e_a^2}\int\limits_{\partial\Sigma}
\dd x^0\frac{1}{2}\Bigg(\xi_a\Big(\del_0(\lam_{-a}-\ov{\lam}_{+a})+2\del_1(\lam_{-a}+\ov{\lam}_{+a})\Big)\\& +i2G_a(\del_1\si_a)-2H_a(D_a-iF_{01a})+i{\K}_a(\lam_{-a}-\ov{\lam}_{+a} )+c.c.\Bigg),
\end{aligned}
\end{equation}
where the covariant derivative of the scalar fields is given in \eqref{covder}. Performing the manipulation given in \eqref{reda}, taking the $\hat{e}\to\infty$ limit, and subsequently integrating out the boundary auxiliary fields, we obtain the boundary action 
\begin{equation}\label{Gpenultimate}
\begin{aligned}
S_{\del\Sigma} &=\frac{1}{2\pi}\int\limits_{\partial\Sigma}dx^0\Big(
\sum_i^Na_i\partial_0\varphi_i+{\hat{\theta}} \hat{A}_{0}+\sum_c^{N-1}{\til{\theta}_c} \til{A}_{0c}
\Big),
\end{aligned}
\end{equation}
 together with boundary conditions 
\begin{equation}
\begin{aligned}\label{GbcA}
|\phi_i|^2&=c_i,\\
\overline{\phi}_i\psi_{-i}+\overline{\psi}_{+i}\phi_i&=0,\\
\overline{\psi}_{-i}\phi_i+\overline{\phi}_i\psi_{+i}&=0,\\
i \overline{\phi}_i\overleftrightarrow{{D}}_1\phi_i+\overline{\psi}_{+i}\psi_{+i} - \overline{\psi}_{-i}\psi_{-i} + \overline{F}_i\phi_i+\overline{\phi}_iF_i&=0\\
\end{aligned}
\end{equation} 
on the matter fields, as well as boundary conditions
\begin{equation}\label{cpngaugebc}
\begin{aligned}
%\partial_1v_0&=0,\\
%v_1&=0,\\
\til{\lam}_{+c}-\ov{\til{\lam}}_{-c}&=0,\\
\del_1\til{\si}_c&=0,\\
%\partial_1\overline{\sigma}^a&=0,\\
\til{F}_{01c}&=0,\\
\til{D}_c&=0,\\
\partial_1(\til{\lam}_{+c} +\ov{\til{\lam}}_{-c})&=0,\\
\end{aligned}
\end{equation}
on vector multiplet fields, and their complex conjugates. In superfield notation, the latter are 
\begin{equation}
\begin{aligned}
\bPhi_i\e^{\sum_a^N\mathcal{Q}_{ia}V_a}\Phi_i&=c_i\\
D_+\til{\Si}_c&=\ov{D}_-\til{\Si}_c\\
\ov{D_+\til{\Si}_c}&=\ov{\ov{D}_-\til{\Si}_c}.
\end{aligned}
\end{equation}
Before proceeding, we note that the boundary conditions on the matter fermion fields in \eqref{GbcA} ensure that the spurious boundary term \eqref{extra} vanishes. 

Now, we shall rewrite \eqref{Gpenultimate} as
\begin{equation}
\frac{1}{2\pi}\int\limits_{\partial\Sigma}dx^0\Big(
\sum_i^Na_i\til{D}_0\varphi_i+{\hat{\theta}} \hat{A}_{0}
\Big),
\end{equation}
where we have used $\sum_i^N\til{Q}_{ic}a_i=\til{\theta}_c$ and where the covariant derivative of $\varphi_i$ is 
\begin{equation}
\til{D}_0\varphi_i=\del_0\varphi_i+\sum_c^{N-1}{\til{Q}_{ic}} \til{A}_{0c},
\end{equation}
which agrees with the general definition for scalar fields given in \eqref{2.13}.
By integrating the vector multiplet out of the bulk action (c.f. \eqref{generalv2}), we obtain 
 %IMPORTANT NOTE: FERMI TERMS VANISH VIA BOUNDARY CONDITIONS
\begin{equation}
\hat{A}_0={1\over 2}{\sum_{i=1}^N\hat{Q}_i(
i\bphi_i\overleftrightarrow{\til{D}}_{\!\!\!0}\phi_i+\ov{\psi}_{-i}\psi_{-i}+\ov{\psi}_{+i}\psi_{+i})\over \sum_{j=1}^N\hat{Q}_j^2|\phi_j|^2}
=
-{\sum_{i=1}^N\hat{Q}_ic_i\til{D}_{0}\varphi_i\over
\sum_{j=1}^N\hat{Q}_j^2c_j},
\end{equation}
 at the boundaries,\footnote{As in the NLSM case, the presence of the boundary term proportional to $\hat{\theta}\hat{A}_0$ ensures that the algebraic equation of motion for $\hat{A}_0$ does not contain a boundary term, see footnote \ref{15}.} where \eqref{GbcA} has been used in the last step.
Hence, the final boundary action is\footnote{To be precise, the complete boundary action includes the $C$-field term given in \eqref{bfield}. However, to simplify the following arguments, we shall consider the $C$-field term to be part of the bulk action, by using Stoke's theorem to promote it to a bulk term.}
\begin{equation}\label{Gfinalboundaryaction}
S_{\del\Sigma}={1\over 2\pi}\int\limits_{\partial\Sigma}dx^0
\left[\,\sum_{i=1}^N a_i\til{D}_0\varphi_i
-\hat{\theta}{\sum_{i=1}^N\hat{Q}_ic_i\til{D}_0\varphi_i\over
\sum_{j=1}^N\hat{Q}_j^2c_j}
%\,-\,\theta\left(\sum_{i=1}^N\hat{Q}_ic_i\dd\varphi_i%\Biggr/
%%\mbox{\huge$\slash$}
%\sum_{i=1}^N\hat{Q}_i^2c_i\right)
\right].
\end{equation}
%It is easy to see that the holonomy in the gauge orbit direction
%indeed vanishes if the condition
%$\theta=\sum_{i=1}^NQ_ia_i$ in (\ref{sgcond1}) is satisfied.

Now, let us investigate the example of $X=\CP^{N-1}$. 
We can derive the A-type supersymmetric boundary conditions of the GNLSM matter fields from \eqref{GbcA} using the parametrizations \eqref{para} and \eqref{lamconstraint2}\footnote{Analogous to the NLSM case, the last condition of \eqref{GbcA} is trivialized using the algebraic equation of motion of $\hat{A}_1$ in \eqref{generalv2}. The last condition of \eqref{GbcAcpn} is obtained via A-type supersymmetry transformations of the fermionic boundary conditions. }
\begin{equation}\label{GbcAcpn}
\boxed{
\begin{aligned}
|Z^i|^2&=\frac{c_i}{c_N}\\
\overline{Z}^i\psi^{Z^i}_{-}+Z^i\overline{\psi}^{\ov{Z}^i}_{+}&=0\\
\overline{Z}^i\psi^{Z^i}_{+}+Z^i\overline{\psi}^{\ov{Z}^i}_{-}&=0\\
i(\ov{Z}^i\del^A_1Z^i-Z^i\del^A_1\ov{Z}^i)+\ov{\psi}_+^{\ov{Z}^i}\psi^{{Z}^i}_+-\ov{\psi}_-^{\ov{Z}^i}\psi^{{Z}^i}_-+F^{{Z}^i}\ov{Z}^i+\ov{F}^{\ov{Z}^i}Z^i&=0,
\end{aligned}}
\end{equation}
where the last condition is in fact a Neumann boundary condition on $\gamma^i$, since $\ov{Z}^i\del^A_1Z^i-Z^i\del^A_1\ov{Z}^i=2|Z^i|^2i\del^A_1\gamma^i$,
where 
%the covariant derivative is
\begin{equation}
\del_{\mu}^A\gamma^i=\del_{\mu}\gamma^i+\sum_c^{N-1}{(\til{Q}_{ic}-\til{Q}_{Nc})} \til{A}_{\mu c}=\del_{\mu}\gamma^i+\sum_c^{N-1}\til{e}_c^{\gamma^i} \til{A}_{\mu c},
\end{equation}
with $\til{e}_c^{\gamma^i}$ being the Killing vector field which generates the $U(1)^{N-1}$ isometry of the torus, $T^{N-1}$, parametrized by $\gamma_i$.
The Neumann boundary condition on $\gamma^i$ together with the Dirichlet boundary condition on $|Z^i|$ implies that the equivariant A-brane wraps this torus%, $T^{N-1}$ %parametrized by $\gamma_i$
. Furthermore, this torus is a Lagrangian submanifold of $\CP^{N-1}$ with respect to the Fubini-Study K\"ahler form given by \eqref{FSkahlerform}.
%IMPORTANT NOTE The fourth condition is actually trivialized, using the equations of motion for $F$ which have been integrated out of the bulk action, as well as the equation of motion for $v$.
%To get the final boundary condition, we take the A-type SUSY variation of 
%\eqref{second}, which is  
The remaining boundary conditions, i.e., for the fields in the vector multiplet of the GNLSM, are given by \eqref{cpngaugebc}. The complete set of GNLSM boundary conditions is invariant under  the $U(1)^{N-1}$ gauge symmetry, and satisfy the supersymmetry transformations given in \eqref{susytrans} and \eqref{susytrans2} for $\eps_+=\ov{\eps}_-$. In addition, the boundary conditions also ensure the locality of the classical equations of motion, i.e., that they contain no boundary terms. 

Next, with the aid of \eqref{argZ}, we can rewrite the boundary action \eqref{Gfinalboundaryaction} as 
\begin{equation}
\begin{aligned}
S_{\del\Sigma}&={1\over 2\pi}\int\limits_{\partial\Sigma}dx^0
\sum_{i=1}^N\left[\, a_i
-\hat{\theta}{ c_i \over
\sum_{j=1}^Nc_j}
%\,-\,\theta\left(\sum_{i=1}^{N-1}\hat{Q}_ic_i
%\sum_{j=1}^N c_j\right)
\right]\til{D}_0\varphi_i\\
&={1\over 2\pi}\int \limits_{\partial\Sigma}dx^0
\left[\,\sum_{i=1}^{N-1}\Bigg( a_i
-(\sum_{k=1}^Na_k){ c_i\over
\sum_{j=1}^Nc_j}\Bigg)\til{D}_0\varphi_i+\Bigg( a_N
-(\sum_{k=1}^Na_k){ c_N \over
\sum_{j=1}^Nc_j}\Bigg)\til{D}_0\varphi_N
%\,-\,\theta\left(\sum_{i=1}^N\hat{Q}_ic_i\dd\varphi_i%\Biggr/
%%\mbox{\huge$\slash$}
%\sum_{j=1}^N c_j\right)
\right]\\
&={1\over 2\pi}\int\limits_{\partial\Sigma}
dx^0\left[\,\sum_{i=1}^{N-1}\Bigg( a_i
-(\sum_{k=1}^Na_k){ c_i\over
\sum_{j=1}^Nc_j}\Bigg)\Big(\del_0(\varphi_i-\varphi_N)+\sum_c^{N-1}(\til{Q}_{ic}-\til{Q}_{Nc})\til{A}_{0c}\Big)\right],
\end{aligned}
\end{equation}
or
\begin{equation}\label{GfinalboundaryactionCPN}
\boxed{
S_{\del\Sigma}=\int\limits_{\partial\Sigma}dx^0\Bigg(\sum_{i=1}^{N-1}A^X_i\del_0\gamma^i-\sum_{c=1}^{N-1}\til{R}_c\til{A}_{0c}\Bigg),}
\end{equation}
%NOT SO IMPORTANT NOTE: IF WE HAVE $\sum_i^NQ_ia_i +2\pi m=\theta$, then the effect is the addition of -2\pi m n to the final result above, where $n$ is the winding number of all the $\varphi_i$ or the $u_i$. i think for $u_i-\varphi_i$ to be single valued both u_i and \varphi_i should have the same winding number. see note for more details 
where we have also used \eqref{buboat}, and where $A^X_i$ is the constant given in \eqref{constantconnect},
%\begin{equation}
%A_i= {1\over 2\pi}\Bigg(a_i -(\sum_{k=1}^Na_k) {c_i\over \sum_{j=1}^Nc_j}\Bigg),
%\end{equation}
%which is understood to be 
which is the connection of a flat $U(1)$ bundle on the Lagrangian torus $T^{N-1}$ parametrized by $\gamma_i$,
and where 
\begin{equation}
\begin{aligned}
\til{R}_c&=-\sum_i^{N-1}(\til{Q}_{ic}-\til{Q}_{Nc})A^X_i\\&=-\sum_i^{N-1}\til{e}_c^{\gamma^i}A^X_i\\
&=-\iota_{\til{e}_c} A^X.
\end{aligned}
\end{equation}
%is known as the \textit{moment} of a flat $U(1)$ bundle with $U(1)^{N-1}$-\textit{equivariant} structure \cite{BGV,szabo}. 
%The moment obeys the equivariant Bianchi identity, 
 As we explain below, A-type supersymmetry invariance %requires the use 
 holds since 
 %and only if 
\begin{equation}\label{eqBianchi2A}
d\til{R}=\iota_{\til{e}}F^X,
\end{equation}
which is equal to zero because $F^X=0$. This is known as the \textit{equivariant} Bianchi identity, and implies that the flat $U(1)$ bundle has $U(1)^{N-1}$-\textit{equivariant} structure,\footnote{The $G$-equivariant Bianchi identity is equivalent to the $G$-invariance of the connection, $A$, of the bundle ($\mathcal{L}_{\til{e}}A=0$), which implies that the covariant derivative $d+A$ is $G$-invariant, and this defines a $G$-equivariant bundle, see \cite{szabo}, Section %IMPORTANT NOTE also see section 2.4. previously i referred to this section
3.2.}
 for which $\til{R}_c$ is the \textit{moment} \cite{BGV,szabo}. %As we explain below, A-type supersymmetry invariance requires the use of \eqref{eqBianchi2A}, hence implying that our equivariant A-brane supports a $U(1)^{N-1}$-equivariant flat $U(1)$ bundle.

 %Thus, the equivariant A-brane defined by the boundary conditions \eqref{bcAcpn} and boundary action \eqref{finalboundaryactionCPN} wraps a Lagrangian submanifold of $\CP^{N-1}$ and supports a flat $U(1)$ bundle. 
%The boundary action can be rewritten as

%Must write full form of $\mu$, $B$ and $C$ for $\CP^{N-1}$ in the beginning. And change footnote 28.

Now, the boundary action \eqref{GfinalboundaryactionCPN} is not invariant under the supersymmetry transformations \eqref{susytrans} and \eqref{susytrans2} for $\eps_+=\ov{\eps}_-$. Instead, the \textit{total action} $S+S_{\del\Sigma}$ is invariant under these transformations at the boundaries, using the boundary conditions \eqref{GbcAcpn} and \eqref{cpngaugebc}, and therefore the sum of the expressions \eqref{dSmatw}, \eqref{dSgauthe} and \eqref{dSB} with the supersymmetry variation of the boundary action vanishes. The proof of this involves the supersymmetry invariance of the constant moment $\til{R}_c$, which is essentially the equivariant Bianchi identity \eqref{eqBianchi2A}, as well as the boundary constraint 
\begin{equation}\label{mupropr}
\til{\mu}_c=-\til{r}_c
\end{equation}
%IMPORTANT NOTE: IF YOU'RE CONFUSED WHY THERE IS A MINUS SIGN, COMPARE THE SIGN OF THE MU TERM IN THE GNLSM AT THE BEGINNING, AND THE SIGN GLSM TERM FROM WHICH THE MU TERMS ORIGINATES
on the moment map, which can be derived from \eqref{bubocr} using $c_i=|\phi_i|^2$ and the parametrization \eqref{para}. Furthermore, the nonzero supersymmetry variation of the boundary action \eqref{GfinalboundaryactionCPN} is cancelled by the $C$-term in \eqref{dSB} and the $\til{\theta}_c$-term in \eqref{dSgauthe} via
\begin{equation}\label{requirement0}
2\pi\til{R}_c=-\til{\theta}_c+C_c,
\end{equation}
which can be shown to hold via \eqref{cpropmu}, \eqref{mupropr}, and \eqref{Gbubost}.
%, while 
Finally, the $B$-field terms in \eqref{dSB} 
%vanishes since 
(where the $B$-field is proportional to the K\"ahler form), vanish using the boundary conditions given in \eqref{GbcAcpn}. %which vanishes when %contracted with the pullback of tangent vectors on 
%which vanishes when restricted to the Lagrangian submanifold on which the equivariant A-brane is wrapped. 
%IMPORTANT NOTE: this is a known mathematical fact, see Wikipedia : https://en.wikipedia.org/wiki/Symplectic_manifold under the section Lagrangian_and_other_submanifolds. What was previously written, i.e., that the B-field terms vanish because the Kahler form vanishes when contracted with the pullback of tangent vectors on the Lagrangian submanifold is not correct because the the fermions in the boundary term can also belong to the normal bundle to the lagrangian submanifold, see my notes on 23rd March 2017 for more info. (Lagrangian submanifold of R^2 diagram onwards)

Next, writing the boundary action as 
\begin{equation}\label{gaugedWilsonline}
S_{\del\Sigma}=\int\limits_{\partial\Sigma}dx^0\Bigg(\sum_{i=1}^{N-1}A^X_i\del^A_0\gamma^i\Bigg),
\end{equation}
it becomes obvious that it is invariant under the gauge transformations given in \eqref{gaugetrans1} and \eqref{gaugetrans2}, since $A^X_i$ is a constant and the expression $\del^A_0\gamma^i$ is invariant under gauge transformations. 
%Must change the word covariant earlier.

%Must mention $\mu=-r$ condition.
% Moreover, these boundary conditions result in the vanishing of the expressions \eqref{dSmatw} and \eqref{dSgauthe}, thus ensuring the preservation of B-type supersymmetry at the boundary. Note that the expression \eqref{dSB} does not occur when performing a supersymmetry variation, since the $B$-field and $C$-field terms do not appear in the bulk action, as we have used the bulk $\hat{\theta}$ term in the construction of our boundary action via \eqref{thetaconvert}. The spurious boundary term \eqref{extra} also does not occur, for the same reason.

%Gauge invariance of boundary action.

We have thus found A-type supersymmetric and $U(1)^{N-1}$ gauge invariant boundary conditions and boundary interactions corresponding to an equivariant A-brane in $\CP^{N-1}$, which wraps a Lagrangian submanifold $T^{N-1}$ which supports a $U(1)^{N-1}$-equivariant flat $U(1)$ bundle. We may follow a procedure analogous to that presented above for $\CP^{N-1}$ in order to describe an equivariant A-brane in a toric manifold $X=\C^N//U(1)$ (by choosing different values for $\hat{Q}_i$), which would again be a Lagrangian submanifold $T^{N-1}$ supporting a flat $U(1)$ bundle with $U(1)^{N-1}$-equivariant structure. 

%TO UNDERSTAND WHY WE STILL GET TORI AS BRANES IN THE TORIC MANIFOLD, SEE PAGE 104 OF MIRROR SYMMETRY BOOK. THE COMPLEXIFIED TORUS IN $C^N$ MOD THE GAUGE  GROUP WE ARE BREAKING GIVES A DENSE OPEN TORUS IN THE QUOTIENT, MAKING THE QUOTIENT A TORIC MANIFOLD. ON MATHOVERFLOW, I GOT THE ANSWER FOR WHY TORIC FIBERS OF A TORIC MANIFOLD ARE LAGRANGIAN SUBMANIFOLDS. THE ABOVE APPLIES TO THE NEXT SECTION AS WELL.
\subsection{Equivariant A-branes on $\C^N//U(1)^{N-k}$ from GLSM}

We can generalize further, since the examples above have been solely for equivariant A-branes on $X=\C^N//U(1)^{N-k}$ where $N-k=1$. For general values of $N-k$, we may derive the relevant boundary conditions and boundary action from the GLSM boundary action \eqref{gnlsmuvafull}, but instead of taking the $\hat{e}\rightarrow \infty$ limit for a single gauge group, we take $\hat{e}_b\rightarrow \infty$, where $b=1,\ldots,N-k$. Integrating out auxiliary fields, and using parametrizations analogous to \eqref{para} and \eqref{lamconstraint2}, we will be able to derive the $U(1)^k$-GNLSM boundary conditions and boundary action which represent an equivariant A-brane wrapping a Lagrangian torus $T^k$, which supports a flat $U(1)$ bundle with $U(1)^k$-equivariant structure.   

Kapustin et al. (\cite{Kapustin}, page 58) have conjectured that the category of $G$-equivariant A-branes is some sort of $G$-equivariant version of the Fukaya category (which includes Lagrangian submanifolds which support flat unitary vector bundles as objects). Indeed, if we generalize the definition of the equivariant Fukaya category given for \textit{finite} groups by Cho and Hong (\cite{ChoHong}, page 68) to $G=U(1)^k$, we see that the equivariant A-branes which we have found are 
%the simplest
 objects in the $U(1)^k$-equivariant Fukaya category, and therefore, we have partially verified the conjecture of Kapustin et al. The other objects in the category which we have not constructed correspond to Lagrangian submanifolds which support equivariant flat unitary vector bundles. 
\subsection{Quantum Corrections} 

There are two important quantum effects of the bulk $U(1)^{N-k}\times U(1)^k$ GLSM, which affect the FI parameters $r_a$ and theta angles $\theta_a$ \cite{hori2003mirror}.
%\footnote{The following is a generalization of the results of Section 4.3 of \cite{Hori1} to the case of multiple $U(1)$ gauge groups.%IMPORTANT NOTE: I've written results here instead of analysis like in section 4 because I've used a different analysis to find the quantum correction from Hori
% }
  The first effect is the renormalization of the FI parameters,  
\begin{equation}\label{qeffect1A}
r_{0a}=r_a(\mu)+\sum_{i=1}^N\mathcal{Q}_{ia}\log\Big(\frac{\Lambda_{\rm UV}}{\mu}\Big),
\end{equation}
 where $r_{0a}$ denotes bare parameters, $\Lambda_{UV}$ is an ultraviolet cut-off, and $\mu$ is a finite energy scale. Via integration of the beta functions of the FI parameters, $\beta_a=\mu \frac{dr_a}{d\mu}$, the $\mu$-dependence is found to be 
\begin{equation}\label{runningA}
r_a(\mu)=\sum_{i=1}^N\mathcal{Q}_{ia}\textrm{log}\Bigg(\frac{\mu}{\Lambda}\Bigg),
\end{equation}
where $\Lambda$ is the renormalization group invariant dynamical scale.  The second quantum effect is the anomaly of the bulk $U(1)$ axial R-symmetry, whereby axial R-rotations 
$\psi_{\pm i}\to e^{\mp i\beta}\psi_{\pm i}$,
$\sigma_a\to e^{2i\beta}\sigma_a$ and
$\lambda_{\pm a}\to e^{\mp i\beta}\lambda_{\pm a}$ no longer leave the action invariant,
but result in a shift of the theta angles, i.e.,%{\it together with the shift of the Theta angle}
\begin{equation}\label{qeffect2A}
\theta_a\to\theta_a-2\sum_{i=1}^N\mathcal{Q}_{ia}\beta.
\end{equation}
%IMPORTANT NOTE: WE USE DIFFERENT CONVENTIONS HERE FOR AXIAL R-ROTATION COMPARED TO THE QUANTUM CORRECTIONS SECTION OF SECTION 4, WHICH FOLLOWS HIV. Here we follow Baptista, the Mirror Symmetry book, and Hori's linear models.
%Since 
%These parameters

The FI parameters are closely related to the boundary parameters $c_i$,
%and $a_i$
 via \eqref{Gbubocr},
% and \eqref{Gbuboat},  %we may deduce the quantum effects on the boundaries
and the 
%boundary parameters
latter undergo similar 
%quantum corrections
renormalization to that of \eqref{qeffect1A} \cite{Hori1},
%. Firstly,
i.e., the 
%running of the FI parameters \eqref{runningA} together with \eqref{Gbubocr} imply that
parameters $c_i$ run as  
\begin{equation}\label{boundaryqeffect1A}
c_{i}(\mu)=\log\Big(\frac{\mu}{\Lambda}\Big).
\end{equation}
 %Similarly, the effect of the bulk axial R-anomaly, which is the shift \eqref{qeffect2A},  can be used together with \eqref{Gbuboat} to find the effect of 
%IMPORTANT NOTE: THE RG FLOW OF c_i CAN BE SEEN FROM THE DISCUSSION BELOW EQUATION 3.34 of HORI's Linear Models (note that the gauge groups don't affect this computation). FOR THE BOUNDARY AXIAL ANOMALY WHICH FOLLOWS, THE BOUNDARY ZERO MODES MUST BE STUDIED, AND AFTER DOING THE MANIPULATION \eqref{reda}, we can follow a calculation similar to those leading up to equations (3.20) and (3.21) of Hori. However here it seems that here we have to assume that there are not fermi zero modes for the gauginos. This is the same assumption made in the mirror symmetry book on page 354/355 when computing the axial anomaly. But note that now the derivatives acting on the fermions  are covariant!  
 %Next, the boundary axial R-anomaly
 %, which is
 %causes a shift of the parameters $a_i$, i.e.,
 %\begin{equation}\label{boundaryqeffect2A}
%a_i\to a_i+\beta.
%\end{equation}
 Note that 
 %these 
 this quantum effect is nontrivial even when $\sum_{i=1}^N \mathcal{Q}_{ia}=0$, unlike the %quantum correction to
 running of $r_a(\mu)$. 
 %and $\theta_a$
  In particular, \eqref{boundaryqeffect1A} implies that the size of the equivariant A-brane in the toric manifold $X$ could depend on the energy scale $\mu$.
  %, since the GLSM boundary condition is $|\phi_i|^2=c_i$. 
%  the equivariant A-brane could disappear from the theory at low energies, since the Dirichlet boundary condition is $|Z^i|^2=c_i/c_N$. 
 However, for $\CPN$, this is not the case, because the Dirichlet boundary condition is 
  $|Z^i|^2=c_i/c_N$, and hence the equivariant A-brane stays the same size regardless of the energy scale. On the other hand, when $\sum_{i=1}^N \mathcal{Q}_{ia}>0$, the manifold $X$ becomes large at high energies due to \eqref{runningA}, since $\hat{r}_b$ are the size moduli    
  %IMPORTANT NOTE: WHEN WE ARE CONSIDERING QUANTUM CORRECTIONS, WE ONLY CONSIDER THE CASE WHERE X HAS SEMIPOSITIVE FIRST CHERN CLASS, AS EXPLAINED BEFORE THE START OF SECTION 4.
of $X$ (for $\CP^{N-1}$, this is obvious from \eqref{cpn}). 
%This means that an equivariant A-brane in $\CPN$ eventually disappears from the theory at high energies. 
Finally, it is expected that in addition to the bulk axial R-anomaly, a \textit{boundary} axial R-anomaly also occurs \cite{hori2003mirror}.
\subsection{Mirrors of Equivariant A-branes} 

%We first note that during the dualization procedure
%we take $|\phi_i|^2$ to be non-zero and 
%$\varphi_i={\rm Im}\log\phi_i$ is well-defined.
%This allows us to shift the $U_i$ field
%as $U_i=\varphi_i+U'_i$
%so that $U_i'$ is a single valued superfield.
%In terms of the shifted variables
% the boundary term (\ref{Bount}) is expressed as
%\begin{equation}
%S_{\it boundary}={1\over 2\pi}\sum_{i=1}^N\int\limits_{\partial\Sigma}
%\dd x^0
%\left[\int\dd\theta\dd\btheta\,
%\bPhi_i\e^{Q_iV}\Phi_iU'_i
%+{\rm Re}
%\int \dd\theta
%\,S_i\Upsilon'_i
%+a_i\partial_0\varphi_i\,
%\right].
%\label{Bount2}
%\end{equation}
Having described equivariant A-branes in toric manifolds, we shall now use mirror symmetry to find the Landau-Ginzburg mirrors of these branes, following the exposition in Section 3, as well as the results of \cite{Hori1}. We shall obtain the mirrors of branes in toric manifolds which obey $c_1(X)\geq 0$, since mirror symmetry is a \textit{quantum} duality (which holds after taking all pertubative and nonpertubative quantum effects into account), and we can only obtain quantum GNLSMs for K\"ahler targets with $c_1(X)\geq 0$ from GLSMs (c.f. Section 3.1).
%IMIT.

%With these relations in mind, let us investigate  mirror symmetry when the boundary action consisting of \eqref{gnlsmuva1} and \eqref{gnlsmuva2} is included.  
%In the following analysis, we shall take $|\phi_i|^2$ to be non-zero.
%, and  $\varphi_i=\textrm{Im log} \phi_i$ is understood to be well-defined.
%Must say that we assume $|\phi_i|^2\neq 0$ in the following.
%Since, as mentioned in Section 3, mirror symmetry arises from T-duality on $\varphi_i$, only 
%We first note that during the dualization procedure
%we take $|\phi_i|^2$ to be non-zero and 
%\varphi_i={\rm Im}\log\phi_i$ is well-defined.
%PORlued.
%This allows us to shift the $U_i$ field
%as $U_i=\varphi_i+U'_i$
%so that $U_i'$ is a single valued superfield.
%In terms of the shifted variables
% Performing the shift indicated in \eqref{reda}, the expression \eqref{gnlsmuva2} is expressed as
%\begin{equation}
%\begin{aligned}
%S_{\it boundary}=&{1\over 2\pi}\sum_{i=1}^N\int\limits_{\partial\Sigma}
%\dd x^0
%\Bigg[\int\dd\theta\dd\btheta\,
%\bPhi_i\e^{\sum_a^N\mathcal{Q}_{ia}V_a}\Phi_iU'_i
%+{\rm Re}
%\int \dd\theta
%\,s_i\Upsilon'_i
%+a_i\partial_0\varphi_i\\&+\sum_a^N\frac{1}{2e_a^2}\int\dd\theta\dd\ov{\theta}\textrm{Re}\big[\Xi_a(D_+\Sigma_a-\ov{D}_-\Sigma_a)\big]\Bigg],
%\label{Bount2}
%\end{aligned}
%\end{equation}
%where $U'_i=U_i-\varphi_i$, and $\Upsilon_i'=\ov{D}U'_i$ is its 'field-strength'.
The boundary action of the $U(1)^N$ GLSM which we wish to dualize is given by \eqref{gnlsmuvafull}, with $\sum_i^Na_i\del_0u_i$ replaced by $\sum_i^Na_i\del_0\varphi_i$ via \eqref{reda}.\footnote{In the following analysis, we shall take $|\phi_i|^2$ to be non-zero, and $\varphi_i=\textrm{Im log } \phi_i$ is %taken 
understood to be well-defined, permitting us to set $U_i=\varphi_i+U_i'$ 
whereby $U_i'$ is a boundary superfield which is single-valued.
}
%RTties.
The terms in the full $U(1)^N$ GLSM action 
%(which includes the boundary actions \eqref{gnlsmuva1} and \eqref{gnlsmuva2}) 
%involving $\varphi_i$ will be 
relevant for the dualization are those which involve $\varphi_i$:
\begin{equation}
\begin{aligned}
S_{\varphi}=&-{1\over 2\pi}\sum_{i=1}^N\int\limits_{\Sigma}
|\phi_i|^2(\partial_{\mu}\varphi_i+\sum_a^N\mathcal{Q}_{ia}A_{\mu a})(\partial^{\mu}\varphi_i+\sum_a^N\mathcal{Q}_{ia}A^{\mu}_{a})\dd^2x\\
&+{1\over 2\pi}\int\limits_{\partial\Sigma}
\left[\,\,\sum_{i=1}^N\left(
-2u_i'|\phi_i|^2(\partial_1\varphi_i+\sum_a^N\cQ_{ia}A_{1a})+a_i\partial_0\varphi_i\right)
+\sum_a^N \theta_a A_{0a}\,\,
\right]\dd x^0,~~~
\label{dure}
\end{aligned}
\end{equation}
where $-2|\phi_i|^2(\partial_1\varphi_i+\sum_a^N\cQ_{ia}A_{1a})=i \overline{\phi}_i\overleftrightarrow{{D}}_1\phi_i$.\footnote{The dualization portion of the subsequent analysis follows from that given in \cite{Hori1}.
%, the only difference being that we have generalized the gauge group from $U(1)$ to $U(1)^N$.
%(the D-term we introduced in \eqref{gnlsmuva2} containing $\Xi_a$ is inconsequential to the dualization). 
%where $u_i'$ is the lowest component of $U_i'$.
%Since only the fields $\varphi_i$ are relevant for dualization,
}
Here, the boundary theta term
$$\sum_a^N{\theta_a\over 2\pi}\int_{\partial\Sigma}A_{0a}\dd x^0$$
 has been included in order to maintain the
gauge invariance, i.e., the gauge
transformations $\varphi_i\to\varphi_i+\sum_a^N\mathcal{Q}_{ia}\alpha_a$,
$A_{\mu a}\to A_{\mu a}-\partial_{\mu}\alpha_a$ leave the expression \eqref{dure} invariant (as long as \eqref{Gbuboat} holds). 
%provided
%the condition $\sum_{i=1}^NQ_ia_i=\theta$ from (\ref{sgcond1})
%holds the expression (\ref{dure}) itself
%is invariant under.
All other terms, including those involving fermions, have been suppressed for simplicity.  
% (This
%is merely for simplicity and there is no obstacle to include them
%into the discussion below.)

Now, let us consider a system of $N$ one form fields $(B_i)_{\mu}$, as well as
$N+N$ periodic scalar fields consisting of $\vartheta_i$ and $\widetilde{u}_i$
%ORG
with the action
\begin{equation}
S'={1\over 2\pi}\sum_{i=1}^N\Biggl[~
\int\limits_{\Sigma}\left(
-|\phi_i|^2B_{i\mu}B_i^{\mu}\dd^2x -B_i\wedge \dd\vartheta_i
+\sum_a^N\cQ_{ia}\vartheta_iF_a\right)
+\int\limits_{\partial\Sigma}
(a_i-\vartheta_i)\partial_0\widetilde{u}_i\dd x^0
~\Biggr],
\label{Sdu}
\end{equation}
where $F_a$ is the curvature of $A_a=A_{\mu a}dx^{\mu}$, $F_a=\dd A_a$.
In addition, the boundary condition
\begin{equation}
(B_i)_1=0
\label{btan}
\end{equation}
is imposed.
Integrating out $\vartheta_i$ gives rise to the constraints
\begin{equation}
\begin{aligned}
\dd B_i&=\sum_a^N \cQ_{ia}F_a~~~\mbox{on $\Sigma$}
\\
(B_i)_0&=\partial_0\widetilde{u}_i~~~\mbox{along $\partial\Sigma$}.
\end{aligned}
\end{equation}
The first of these constraints is solved by $B_i=\dd\varphi_i+\sum_a^N \cQ_{ia}A_a$,
where $\varphi_i$ is a periodic scalar field of period $2\pi$.\footnote{For details on why $\varphi_i$ ought to be periodic, see \cite{hori2003mirror}, page 250.}
Then, the second constraint together with
the boundary condition (\ref{btan})
implies the relations 
\begin{equation}
\begin{aligned}
\partial_0\varphi_i+\sum_a^N \cQ_{ia}A_{0a}&=\partial_0\widetilde{u}_i,
\\
\partial_1\varphi_i+\sum_a^N \cQ_{ia}A_{1a}&=0.
\label{lattercond}
\end{aligned}
\end{equation}
on the boundaries.
Inserting the first expression of \eqref{lattercond} into (\ref{Sdu}) we obtain the action
(\ref{dure}) without the $u_i'$-dependent terms (using
$\sum_{i=1}^N \cQ_{ia}a_i=\theta_a$).
The second condition in (\ref{lattercond}) is equivalent to the presence of the $u_i'$-dependent terms, since integrating out $u_i'$ imposes the second equation of  (\ref{lattercond}).

Alternatively, integrating out the fields $B_i$ imposes 
\begin{equation}
(B_{i})_0=\frac{-\del_1\vartheta_i}{2|\phi_i|^2},
\end{equation}
\begin{equation}
(B_{i})_1=\frac{-\del_0\vartheta_i}{2|\phi_i|^2},
\end{equation}
and we obtain
\begin{equation}
S_{\vartheta}={1\over 2\pi}\sum_{i=1}^N\Biggl[~
\int\limits_{\Sigma}\left(
-\frac{1}{4|\phi_i|^2}{\del_{\mu}\vartheta_i\del^{\mu}\vartheta_i}\textrm{ }d^2x
+\sum_a^N\cQ_{ia}\vartheta_iF_a\right)
+\int\limits_{\partial\Sigma}
(a_i-\vartheta_i)\partial_0\widetilde{u}_i\dd x^0
~\Biggr].
\label{Sdu2}
\end{equation}
Following Hori \cite{Hori1}, the bulk portion of the full mirror action is %found to be 
given by \eqref{mirrorsuper} (modulo boundary terms that arise from putting the scalar kinetic terms in  \eqref{mirrorsuper} in their standard form),
%ORY
% and the relationship   
%\begin{equation}\label{fieldduality2A}
%Y_i+\overline{Y}_i=2\overline{\Phi}_i e^{\sum_a^{N} \mathcal{Q}_{ia}V_a}\Phi_i,
%\end{equation}
%between the fields of the mirror theories, which was discussed in Section 3 for closed worldsheets, holds in this case. The expression \eqref{fieldduality2A} implies the following relationships between superfield components:
%\begin{equation}
%\begin{aligned}\label{bcchiA}
%y_i&=\varrho_i-i\vartheta_i,
%\left\{
%\begin{array}{l}
%\varrho_i=|\phi_i|^2,\\
%\partial_{\pm}\vartheta_i=
%\pm 2\Bigl(-|\phi_i|^2
%(\partial_{\pm}\varphi_i+\sum_a^N\mathcal{Q}_{ia}A_{\pm a})
%+\ov{\psi}_{\pm i}\psi_{\pm i}\Bigr),
%\end{array}\right.\\
%\chi_{+i}&=2\ov{\psi}_{+i}\phi_i,\textrm{  }\chi_{-i}=-2\ov{\psi}_{-i}\phi_i,
%\\
%\ov{\chi}_{+i}&=2\ov{\phi}_i\psi_{+i},\textrm{  }\ov{\chi}_{-i}=-2\ov{\phi}_i\psi_{-i},\\
%E_i&=-2\ov{\psi}_{-i}\psi_{+i}-2|\phi_i|^2\sum_a\mathcal{Q}_{ia}\ov{\si}_a,
%\end{aligned}
%\end{equation}
%where %$\varphi_i$ is the phase of $\phi_i$,
%$\del_ {\pm}=\del_0\pm \del_1$, $Y_i=y_i+\theta^+\ov{\chi}_{+i}+\ov{\theta}^-\chi_{-i}+\theta^+\ov{\theta}^-E_i$.
% and $\varphi_i$ is the $2\pi$-periodic phase of  $\phi_i=|\phi_i|\,e^{i\varphi_i}$. %It is now apparent 
%The relationship between the periodic fields $\vartheta_i$ and $\varphi_i$ is in fact evidence that mirror symmetry of the two theories arises from 
%T-duality on the phase of the charged chiral superfields $\Phi_i$, whereby the neutral twisted chiral superfields $Y_i$ are periodic, i.e., $Y_i\equiv Y_i + 2\pi i$ \cite{hori2003mirror}. 
while the mirror boundary action takes the form\footnote{As explained in \cite{Hori1}, unlike the bulk superpotential $\sum_{i=1}^N\e^{-Y_i}$ which is generated by vortices, no boundary F-terms can be generated by such effects.} 
\begin{equation}\label{gnlsmuvafullmirror}
\begin{aligned}
S_{\del\Sigma} 
=&{1\over 2\pi}\sum_{i=1}^N
\int\limits_{\partial\Sigma}\dd x^0\,{\rm Re}\!\int \dd\theta
\, (s_i-Y_i)
\til{\Upsilon}_i\\
&+(\textrm{Additional boundary terms required to cancel bulk SUSY variation})\\
&+\sum_a^N\frac{1}{2e_a^2}\int\dd\theta\dd\ov{\theta}\textrm{Re}\big[\Xi_a(D_+\Sigma_a-\ov{D}_-\Sigma_a)\big],
\end{aligned}
\end{equation}
%boundary terms required to preserve A-type supersymmetry
where the boundary term in \eqref{Sdu2} is contained in the first term.
%The full mirror action is given by the bulk action \eqref{mirrorsuper}, together with a boundary action which contains the term  
%\begin{equation}
%{1\over 2\pi}\sum_{i=1}^N
%\int\limits_{\partial\Sigma}\dd x^0\,{\rm Re}\!\int \dd\theta
%\, (s_i-Y_i)
%\til{\Upsilon}_i.
%\label{redBount}
%\end{equation}

Here, $\til{\Upsilon}_i$ is the `field strength' $\ov{D}\til{U}_i$ of the boundary superfield $\til{U}_i$, whose only difference from $U_i$
% and $U_i'$ 
 is that its lowest component is $\tilde{u}_i$. Integrating out  
$\til{\Upsilon}_i$, we find the boundary condition 
\begin{equation}
Y_i=s_i,
\label{Acon}
\end{equation}
at A-boundary, which is 
\begin{equation}
\begin{aligned}
y_i&=s_i\\
\ov{\chi}_{+i}-{\chi}_{-i}&=0
\end{aligned}
\end{equation}
in components. In fact, integrating out \textit{all} the boundary auxiliary fields in \eqref{gnlsmuvafullmirror} imposes the boundary conditions \eqref{Acon} and \eqref{cpngaugebcOri}, which result in the entire boundary action vanishing.

As %explained 
in Section 3, taking the $\hat{e}_b\to \infty$ limit allows us to integrate $\hat{\Sigma}_b$ out  of the action, and imposes the constraint 
\begin{equation}\label{glgconstraintA}
\sum_j^N\hat{Q}_{jb} Y_j - \hat{t}_b=0,
\end{equation}
giving us the gauged Landau-Ginzburg theory with holomorphic twisted superpotential 
\begin{equation}\label{twsuperpA}
\til{W} =\sum_c^{k}\Big(\sum_{j=1}^{N}\til{Q}_{jc} Y_j - \til{t}_c \Big) \til{\Sigma}_c
\ +\ \sum_{j=1}^{N} e^{-Y_j}.
\end{equation} 
We recall that the constraint \eqref{glgconstraintA} fixes the target space of the gauged Landau-Ginzburg theory to be the algebraic torus $(\C^{\times})^{k}$.  It is solved (c.f. Section 3) by  
\begin{equation}\label{mirrorparametA}
Y_j=\hat{s}_j+\sum^k_{c=1}v_{cj}\Theta_c,
\end{equation}
where $\hat{s}_j$ is any solution of $\hat{Q}^b_j\hat{s}^j=\hat{t}^b$.
%where $s_j=c_j-ia_j$ obeys \eqref{Gbubost}. 
Note that with ${\Theta}_c=\theta_c+\theta^+\ov{\chi}^{\ts}_{+c}+\ov{\theta}^-\chi^{\ts}_{-c}+\theta^+\ov{\theta}^-E^{\ts}_c$, the full mirror action, expanded in components is 
\begin{equation}\label{expandedmirroraction}
\begin{aligned}
&S=\frac{1}{2\pi}\int d^2x\Big[\sum_c^k\sum_d^k (-g_{cd}\del_{\mu}\theta_c\del^{\mu}\ov{\theta}_d+\frac{i}{2}g_{cd}\ov{\chi}^{\ts}_{-c}(\overleftrightarrow{\del_+})\chi^{\ts}_{-d}+\frac{i}{2}g_{cd}\ov{\chi}^{\ts}_{+c}(\overleftrightarrow{\del_-}){\chi}^{\ts}_{+d}+ g_{cd}E_c^{\ts} \ov{E}_d^{\ts})\\+&\sum^{k}_c{1 \over 2\til{e}_c^2}\Big((\til{F}_{01c})^2 
 - \partial_{\mu}\til{\sigma}_c \partial^{\mu} 
\overline{\til{\sigma}}_c   +(\til{D}_c)^2  + {i \over 2} \overline{\til{\lambda}}_{+c}
(\overleftrightarrow{\del_-})
\til{\lambda}_{+c} +{i \over 2} \overline{\til{\lambda}}_{-c}
(\overleftrightarrow{\del_+})
\til{\lambda}_{-c}  \Big)\\
+&\frac{1}{2}\Big(\sum^N_j\sum^k_c\sum^k_d\til{Q}_{jc}v_d^j(\til{\si}_c E_d^{\ts}-i \ov{\til{\lam}}_{+c}\chi^{\theta}_{-d}-i\til{\lam}_{-c}\ov{\chi}^{\ts}_{+d}+(\til{D}_c-i\til{F}_{01c})\theta_d)
\\+&\sum_c^k(\sum_j^N\til{Q}_{jc}\hat{s}^j-\til{t}_c)(\til{D}_c-i\til{F}_{01c})+\sum^N_j e^{-\sum_c v^j_c \theta_c-\hat{s}^j}
(-\sum_c^k v_c^j\ov{\chi}^{\theta}_{+c}\sum^k_d v^j_d{\chi}^{\theta}_{-d}-\sum_c^k v_c^j E_c^{\theta} )+c.c.\Big)\Big]\end{aligned},
\end{equation}
where ($\theta_c$,$\ov{\theta}_d$) (the lowest components of $(\Theta_c,\ov{\Theta}_d$)) parametrize the mirror target space $(\C^{\times})^{k}$,
%\footnote{Note the presence of the terms proportional to $\sum_j^N \til{Q}_{jc}s^j-\til{t}_c$, which actually vanish due to the constraint \eqref{Gbubost}.} 
on which %$g_{cd}$ is 
the flat K\"ahler metric is
\begin{equation}
ds^2=\sum_c^k\sum_d^k\frac{1}{4}\frac{\sum_j^N v_c^jv_d^j}{\textrm{log}(\Lambda_{UV}/\mu)}d\theta_c d\ov{\theta}_d=\sum_c^k\sum_d^kg_{cd}d\theta_c d\ov{\theta}_d.
\end{equation}

Now, the boundary condition \eqref{Acon} on $Y_i$  implies the boundary condition 
\begin{equation}\label{D0branepositionsuper}
%\Theta_c=0,
\sum^k_{c=1}v_{cj}\Theta_c=s_j-\hat{s}_j,
\end{equation}
and this means that the $U(1)^k$-equivariant A-brane in $X=\C^N//U(1)^{N-k}$ is mapped to a B-brane which is a D0-brane in the mirror Landau-Ginzburg model located at $\theta_c$, where $\theta_c$ is a solution of $\sum^k_{c=1}v_{cj}\theta_c=s_j-\hat{s}_j$.\footnote{\label{58}In the case where the two boundaries of the strip are mapped to different equivariant A-branes, labelled by $s^{\pi}_j$ and $s^{0}_j$, the positions of the mirror D0-branes are determined by $\sum^k_{c=1}v_{cj}\theta_c=s^{\pi}_j-\hat{s}_j$ and $\sum^k_{c=1}v_{cj}\theta_c=s^{0}_j-\hat{s}_j$ respectively.}
% where $\theta_c$ is the lowest component of $\Theta_c$ which parametrizes the mirror target space $(\C^{\times})^{k}$. 
  Let us investigate this D0-brane further, by studying how it is described in the  $\til{W}$-plane. In particular, we would like to find the mirror of the $U(1)^{k}$-equivariant structure on the A-brane. 
  
  Firstly, we note that the twisted superpotential \eqref{twsuperpA}
%  , which is a function of $\theta_c$ and $\til{\sigma}_c$,  
  can be rewritten as 
\begin{equation}\label{separateA}
\til{W}=\til{W}_{equiv}+\til{W}_{{X}}.
\end{equation}
where the first and second term of \eqref{twsuperpA} correspond respectively to the first and second term of  \eqref{separateA}.
The image of the D0-brane in the $\til{W}_X$-plane is 
\begin{equation}
\widetilde{W}_X=\sum_{i=1}^N \e^{-s_i},
\end{equation}
which is the mirror condition found %in \cite{Hori1} 
when studying the mirrors of A-branes without equivariant structure. However, turning to $\til{W}_{equiv}$, we find that the boundary condition \eqref{Acon} implies that the image of the D0-brane in the $\til{W}_{equiv}$-plane is $\til{W}_{equiv}=0$, and thus we require further analysis to identify the mirror of the $U(1)^k$-equivariant structure on the A-brane.
%IM W.

Now, for the D0-brane mirrors of ordinary A-branes, there is an additional requirement  which is necessary to prevent spontaneous supersymmetry breaking, %preservation of supersymmetry, 
 that is the D0-brane %should correspond to a critical point of the superpotential  $\sum_{j=1}^{N} e^{-Y_j}$ \cite{Hori1}.
 should %correspond to 
 %A W-plane.
 be at a \textit{critical point} of the twisted superpotential  $\widetilde{W}_X(\theta)=\sum_{j=1}^{N} e^{-\hat{s}_j-\sum^k_{c=1}v_{cj}\theta_c}$  \cite{Hori1,hori2003mirror,Hori2,Kapustinli}. This condition is necessary for the potential energy of the mirror Landau-Ginzburg model (with twisted superpotential $\til{W}_X$) to have a vanishing vacuum expectation value.
%TE:  V=0.
  We shall generalize this analysis to the \textit{gauged} Landau-Ginzburg model with neutral matter \eqref{expandedmirroraction} which we are presently concerned with. Here, the twisted superpotential terms can be expanded as
\begin{equation}
\begin{aligned}
&\frac{1}{2\pi}\int d^2x \frac{1}{2} \Big(\int d^2\widetilde{\theta} \widetilde{W} (\Theta,\widetilde{\Sigma})+c.c.\Big)\\=&\frac{1}{2\pi}\int d^2x \frac{1}{2}\Bigg(\sum_c^k\Big({E}^{{\theta}}_c\frac{\del \til{W}}{\del {\theta}_c} +(\til{D}_c-i\til{F}_{01c})\frac{\del \til{W}}{\del \til{\si}_c}\Big)+c.c.\\&+\sum_c^k\sum_d^k \Big(\chi^{\ts}_{-c}\ov{\chi}_{+d}^{\ts}\frac{\del^2\til{W}}{\del\theta_c \del\theta_d}+\til{\lam}_{-c}\ov{\til{\lam}}_{+d}\frac{\del^2\til{W}}{\del\til{\si}_c \del\til{\si}_d}+i\ov{\chi}^{\theta}_{+c}\til{\lam}_{-d}\frac{\del^2\til{W}}{\del\theta_c \del\til{\si}_d}+i{\chi}^{\theta}_{-c}\ov{\til{\lam}}_{+d}\frac{\del^2\til{W}}{\del\theta_c \del\til{\si}_d}\Big)+c.c.\Bigg),
\end{aligned}
\end{equation}
where $\widetilde{W}$ is given by \eqref{holtwistsup}. Taking into account the presence of the auxiliary field terms
\begin{equation}
\frac{1}{2\pi}\int d^2x (\sum_c^k\sum_d^k g_{cd}E_c^{\ts} \ov{E}_d^{\ts}+\sum_c^k\frac{1}{2\til{e}_c^2}\til{D}_c\til{D}_c)
\end{equation}
in the action,
% (where $g_{cd}$ is the flat K\"ahler metric of $(\C^{\times})^k$),
 upon integrating out the auxiliary fields $\til{D}_c$ and ${E}^{\theta}_c$, the potential energy becomes 
 \begin{equation}\label{potentialenergy}
 V=\frac{1}{2\pi}\int dx^1\Bigg(\frac{1}{4} g^{cd}\frac{\del{\til{W}}}{\del \theta_c}\frac{\del\ov{{\til{W}}}}{\del \ov{\theta}_{{d}}} +\frac{1}{2}\sum_c^k \til{e}_c^2\textrm{Re}\Big(\frac{\del{\til{W}}}{\del \til{\si}_c}\Big)\textrm{Re}\Big(\frac{\del{\ov{\til{W}}}}{\del \ov{\til{\si}}_c}\Big)\Bigg).
\end{equation}   
Now, in the non-gauged case, $\frac{\del{\til{W}}_X}{\del \theta_c}$ is a constant at 
%each
the boundaries, and therefore supersymmetry would be broken for any classical configuration unless the %corresponding
 D0-brane is located at the critical point $\frac{\del{\til{W}}_X}{\del \theta_c}=0$. However, in \eqref{potentialenergy}, 
 \begin{equation}
 \frac{\del{\til{W}}}{\del \theta_c}=\sum_j^N\langle \til{\si} ,\til{Q}_j\rangle v^j_c- \sum_j^N v^j_c e^{-\langle v^j,\theta\rangle-\hat{s}^j},
 \end{equation} 
 is not a constant at the boundaries (since $\til{\si}_c$ obeys a Neumann boundary condition ($\del_1\til{\si}_c=0$), unlike $\theta_c$), and hence 
 %zero-energy
  classical configurations  where $\frac{\del{\til{W}}}{\del \theta_c}=0$ at the boundaries can be achieved  without any additional constraint on the position of the D0-brane. Next, the second term in \eqref{potentialenergy} implies that $\textrm{Re}\Bigg(\frac{\del{\til{W}}}{\del \til{\si}_c}\Bigg)$
  %configurations whereby
   %\begin{equation}
%\textrm{Re}\Bigg(\frac{\del{\til{W}}}{\del %\til{\si}_c}\Bigg)=  \textrm{Re}\Bigg(\ \sum_j^N \til{Q}_{jc}  \Big(\langle v^j , \theta \rangle  +  \hat{s}^j \Big) 
%\ -\; \til{t}_c \Bigg)
 %+\til{Q}_{jc}\ov{s}^j-\ov{\til{t}}_c
 %\end{equation}
%vanishes ought to exist
ought to vanish at each boundary in order to prevent spontaneous breaking of supersymmetry. Indeed,
 \begin{equation}\label{rhsof}
\textrm{Re}\Bigg(\frac{\del{\til{W}}}{\del \til{\si}_c}\Bigg)=  \textrm{Re}(\sum_j^N \til{Q}_{jc}s^j-\til{t}_c)
 %+\til{Q}_{jc}\ov{s}^j-\ov{\til{t}}_c
 \end{equation}
at the boundaries, which is identically zero because it is the real part of the condition 
$\sum_j^N \til{Q}_{jc}s^j-\til{t}_c=0$, which is implied by $\sum_j^N \mathcal{Q}_{ja}s^j-{t_a}=0$.  The latter holds since it was necessary for the  A-type supersymmetry and %$U(1)^k$ 
gauge symmetry of the $U(1)^{N-k}\times U(1)^k$ GLSM (see \eqref{Gbubost}). Therefore, spontaneous supersymmetry breaking does not occur in the mirror theory, since zero-energy classical configurations can always be achieved at the boundaries. The condition $\sum_j^N \til{Q}_{jc}s^j-\til{t}_c=0$ is a new condition which did not appear in the non-gauged case, and %In %particular
%fact, the new condition $\til{Q}_{jc}s^j-\til{t}_c=0$ %was not present in the non-equivariant case, and in fact 
in fact constrains the position of the D0-brane (defined by $s^j$ via \eqref{D0branepositionsuper}). 
%and therefore can be interpreted as the mirror of the  $U(1)^k$-equivariant structure of the equivariant A-brane on $\C^N//U(1)^{N-k}$. 
In conclusion, unlike the mirrors of ordinary A-branes, we have found that 
\vspace{1em}

 \noindent\fbox{ \parbox{\textwidth}{
The mirrors of $U(1)^k$-equivariant A-branes on $\C^N//U(1)^{N-k}$ do not need to satisfy the critical point condition $\frac{\del{\til{W}_X}}{\del \theta_c}=0$, but instead their position must be further constrained by $\sum_j^N \til{Q}_{jc}s^j-\til{t}_c=0$. }}
 
 %\noindent
 \vspace{1em} 
\noindent

In this section, we have restricted ourselves to equivariant A-branes whose mirrors are D0-branes. However, there are A-branes whose mirrors are higher-dimensional branes holomorphically embedded in the mirror target space. In Hori's construction \cite{Hori1}, these can be studied by promoting the parameter $s_i$ to a superfield $S_i$. It would be interesting to study equivariant structure on these branes.

\subsection{Nonabelian Equivariant A-branes}

We may use the insights obtained from analyzing the equivariant A-branes for abelian groups 
which we have found thus far %to %infer
to find the description of equivariant A-branes for general nonabelian groups. We shall use the GNLSM notation of Section 2 in this subsection. 

Firstly, the 
%remaining 
terms in \eqref{dSgauthe} (except the terms proportional to ($\phi^*\mu_a+r_a$) and $\theta_a$) vanish
%IMPORTANT NOTE:5TH APRIL 2017, THIS HAS BEEN CHECKED
 using the 
 %gauge invariant 
 boundary conditions
\begin{equation}\label{nonabelianvectormultipletbc}
\begin{aligned}
\lampua-\lambmua&=0,\\
%\lammua-\lambpua&=0,\\
\del_1\si_a&=0,\\
%\nabla^A_1\ov{\si}_a&=0,\\
A_{1a}&=0,\\
\del_1 A_{0a}&=0,\\
D_a&=0,\\
\del_1(\lammua+\lambpua)&=0.\\
%\nabla^A_1(\lampua+\lambmua)&=0.\\
%\mu_a&=-r_a.
\end{aligned}
\end{equation}
%IMPORTANT NOTE: ON 22 MARCH 2017, ADDED D_a=0 because it is necessary to make \delta S_{gauge} vanish. Also, this makes the analysis analogous to that for nonabelian equivariant B-branes. The only place where we integrate auxiliary fields out is when we derive the matter boundary conditions below. But actually we haach other in \deltaS_matter. The christoffel connection terms also vanish using the fermion boundary conditions obtained in section 6.3, of course we must remove bars which are over indices to show this. For B-branes we should be able to do something similar. In that case I think, that the boundary condition should be $\psi^i_+=\psi^i_-$ based on the arguments below equations 4.123 (i.e., where \delta\phi is considered for B-type susy). This boundary condition can also be shown to be correct by modifying the argument in section 6.3 suitably. From the auxiliary field equations of motion, this implies that the boundary conditions should be F=0 and \ov{F}=0 for B-type susy. These boundary conditions can also be found from susy variation of the fermionic boundary condition $\psi^i_+=\psi^i_-$. And they make the terms proportional to F vanish in \deltaS_matter.
%IMPORTANT NOTE: THE CHRISTOFFEL SYMBOL TERMS IN \deltaS_matter vanish using the fermion boundary conditions described in the IMPORTANT NOTE above. of course in the A-type case, we must remove barred indices on all quantities, including the metric and christoffel symbols
Note that these conditions are a direct generalization of the conditions given for the example of $\CP^{N-1}$, except that $F_{01a}=0$ is replaced by the stricter conditions $A_{1a}=0$ and $\del_1A_{0a}=0$. This is necessary since 
%the field strength $F_{01a}$ now contains a nonabelian term
the supersymmetry transformations now contain nonabelian terms, and this causes A-type supersymmetry invariance of the set of boundary conditions to not hold unless we use the stricter conditions. The boundary conditions 
%on the gauge fields 
%IMPORTANT NOTE: ALL BOUNDARY CONDITIONS REQUIRE THAT \del_1\alpha_a=0
in fact imply that gauge transformations have to be restricted such that the transformation parameter $\alpha^a$ has vanishing derivative with respect to $x^1$ at the boundaries, in order for these boundary conditions to be gauge invariant.
%Just like in the example of $\CPN$, these boundary conditions (together with the constraint \eqref{requirement}) also preserve the locality of the equations of motion for vector multiplet components.
 %Before proceeding, we note that the constraint \eqref{requirement} implies that $T^J\del_JC_a=0$ 
 %%on $\gamma$ 
 %(where $T$ is a tangent vector of $\gamma$, $T\in T\gamma$), %%together with the identity \eqref{Cmapequation} 
%which in turn implies that
%\begin{equation}\label{Binteriorderivative}
%B_{IJ}\til{e}^I_aT^J=0
%\end{equation}
%%on $\gamma$, 
%via the identity \eqref{Cmapequation}.
%IMPORTANT NOTE: THE ABOVE SHOULD BE CORRECT, BUT JUST TAKEN IT OUT TO BE SAFE.
%Now, %%because of \eqref{requirement} 
%the only terms left to consider in \eqref{dSB} %the supersymmetry variation of the action 
%are the $B$-field terms. These vanish by requiring that the restriction of the two-form $B$ %%on $X$ 
%to the submanifold $\gamma$ vanishes.%%These vanish by requiring that the $B$-field on $X$ vanishes when restricted to the submanifold $\gamma$.
%IMPORTANT NOTE, FOR NLSMS ONE CAN HAVE MORE GENERAL BOUNDARY CONDITIONS THAN LAGRANGIAN SUBMANIFOLD IN THE PRESENCE OF B-FIELDS, SEE PAGE 804 OF MIRROR SYMMETRY BOOK

Next, we turn to the boundary conditions for the matter fields. We first recall that for $\CP^{N-1}$, the equivariant A-brane corresponded to a Lagrangian torus $T^{N-1}$, which was invariant under the $U(1)^{N-1}$ isometry of $\CP^{N-1}$. Let us 
%next
 consider the $\mathcal{N}=1$ subalgebra of A-type supersymmetry, which corresponds to $\eps_+=i\til{\eps}$, $\ov{\eps}_+=-i\til{\eps}$, $\eps_-=-i\til{\eps}$ and $\ov{\eps}_-=i\til{\eps}$, where $\til{\eps}$ is a real parameter. In this case, after integrating out the auxiliary fields $F^i$  and $\ov{F}^{\ov{\imath}}$, we find that \eqref{dSmatw} and the $B$-field terms in \eqref{dSB} are 
\begin{equation}\label{n1matw}
\begin{aligned}
%\delta S_{matter} =
-\frac{1}{2\pi}\frac{i\til{\eps}}{2}\int_{\del\Sigma}dx^0 \Big\{& \Big(g_{IJ} (\psi^J_- - \psi^J_+)-B_{IJ}(\psi^J_-+\psi^J_+)\Big)\del_0^A \phi^I
\\&+ \Big(g_{IJ}\del_1^A\phi^I+B_{JI}\del_0^A\phi^I\Big)(\psi^J_-+\psi^J_+)
\\&+g_{IJ}(\psi^I_+ - \psi^I_-)\textrm{Re}(\si^a)\he^J_a+
\omega_{IJ}(\psi^I_+ + \psi^I_-)\textrm{Im}(\si^a)\he^J_a\Big\},
\end{aligned}
\end{equation}
where $g_{IJ}X^I Y^J=g_{i\ov{\jmath}}(X^i Y^{\ov{\jmath}}+ X^{\ov{\jmath}}Y^i)$ and $\omega_{IJ}X^I Y^J=ig_{i\ov{\jmath}}(X^i Y^{\ov{\jmath}}- X^{\ov{\jmath}}Y^i)$, and where $(I,J,K,\ldots)$ are indices corresponding to real coordinates on $X$.
% are respectively the metric and K\"ahler form on $X$.
 In addition, if we insist on locality of the matter equations of motion (like in the $\CPN$ case), we require that 
\begin{equation}\label{localeomsss}
\begin{aligned}
\delta \phi^I(g_{IJ} \del_1^A\phi^J+B_{IJ}\del_0^A\phi^J)=0\\
g_{IJ}(\psi^I_-\delta_{\nabla}\psi^J_--\psi^I_+\delta_{\nabla}\psi^J_+)=0
\end{aligned}
\end{equation}
%IMPORTANT NOTE: WE GET TWO SEPARATE EQUATIONS BECAUSE THERE IS A FACTOR of i/2 which multiplies the second term when we do the variation. First term is multiplied by a factor of -1.
at the boundaries, where $\delta_{\nabla}\psi^J=\delta\psi^J+\Gamma^{J}_{KL}\delta\phi^K\psi^L$.
%where we have used the fact that the $B$-field vanishes at the boundaries.
%\footnote{Here, we generalize the approach of \cite{HIV} for non-equivariant A-branes. However, as noted in (\cite{Albertsson}, page 17), this is not the most general approach.}
%IMPORTANT NOTE: REMOVED FOOTNOTE BECAUSE I HAVE NOW INCLUDED THE CHRISTOFFEL SYMBOL TERMS. BUT IT IS IMPORTANT TO REMEMBER THAT WE STILL DO NOT CLAIM TO HAVE THE MOST GENERAL SOLUTION, AND WHAT IS IN THE FOLLOWING NOTE SHOULD STILL HOLD.
%IMPORTANT NOTE: THE REASON THIS FOOTNOTE IS INCLUDED IS BECAUSE in THE EQUATIONs ABOVE WE ARE IGNORING BOUNDARY TERMS WHICH CONTAIN THE CHRISTOFFEL SYMBOL IN  AND FERMION BILINEARS. this is not a problem, because the boundary condition on the fermion which makes the second equation above vanish also makes this 'christoffel connection term' vaniS TO THE PURELY SUPERYSYMMETRIC CASE WITH NO CONFORMAL SYMMETRY.
%The first constraint of \eqref{localeomsss} then implies that $\del_1^A\phi^J$ is normal to $\gamma$. 
An equivariant A-brane shall wrap a submanifold (denoted as $\gamma$)  of $X$, to which a boundary of the worldsheet is mapped via $(\phi^i,\ov{\phi}^{\ov{\imath}})$. Now, any allowed variation of $\phi$ (denoted $\delta \phi^I$ for the real coordinate $\phi^I$) along the boundary, and the %covariant 
derivative along the boundary, $\del_0 \phi^I$, ought to be tangent to $\gamma$.
%IMPORTANT NOTE: THE PHRASE THE BOUNDARY HERE IS CORRECT, DO NOT CHANGE!
Hence, taking into account the fact that $A_{1a}=0$ at the boundaries% as well as \eqref{Binteriorderivative}
, we find that \eqref{n1matw} vanishes while satisfying the first constraint of \eqref{localeomsss} if
%if and only if 
 %IMPORTANT NOTE: TOOK OUT IF AND ONLY IF BECAUSE THERE MIGHT BE MORE GENERAL SOLUTIONS, AS THE ALBERTSSON PAPER SAYS, SEE  FOOTNOTE WHERE I CITE IT.
$\del_1\phi^J$ is normal to $\gamma$, $\psi^I_- - \psi^I_+$ and $\psi^I_- + \psi^I_+$ are respectively normal and tangent to $\gamma$, $\til{e}^I_a$ is tangent to $\gamma$, %and 
the K\"ahler form vanishes against tangent vectors of $\gamma$, and the $B$-field vanishes against tangent vectors of $\gamma$. These last %two 
three conditions respectively imply that $\gamma$ is $G$-invariant, that it is an isotropic submanifold of $X$, and that the restriction of the two-form $B$ to $\gamma$ vanishes. 
%The conditions on the $B$-field and $\til{e}_a^I$ are also consistent with \eqref{Binteriorderivative}. 
In addition, we note that 
$\psi^I_- - \psi^I_+$ being normal to $\gamma$ and $\psi^I_- +\psi^I_+$ being tangent to $\gamma$ implies that 
%$\psi^I_--\psi^I_+=0$ 
\begin{equation}
\begin{aligned}
\psi^I_--\psi^I_+&=0,\textrm{ }I\textrm{ : tangent to $\gamma$},\\
\psi^I_-+\psi^I_+&=0,\textrm{ }I\textrm{ : normal to $\gamma$},
\end{aligned}
\end{equation}
%along $\gamma$ and along the worldsheet boundary, 
(for a choice of coordinates which separates the normal and tangent directions) which satisfies the second constraint of  \eqref{localeomsss}.

%Next, under A-type supersymmetry, $\phi^i$ transforms as
%\begin{equation}
%\begin{aligned}
%\delta\phi^i=&\epsilon_+\psi_-^i-\ov{\epsilon}_+\psi_+^i
%\\
%=&\epsilon_1(\psi_-^i-\psi_+^i)
%+i\epsilon_2(\psi_-^i+\psi_+^i),
%\end{aligned}
%\end{equation}
%where $\epsilon_+=\epsilon_1+i\epsilon_2$. Hence, $\til{\eps}(\psi_-^i-\psi_+^i)$
%and $i\til{\eps}(\psi_-^i+\psi_+^i)$ are holomorphic components of
%vectors which are tangent to $\gamma$, where $\til{\eps}$ is a real parameter. However, from the previous paragraph, we know that $i\til{\eps}(\psi_-^i-\psi_+^i)$ is the holomorphic component of a vector which is normal to $\gamma$. This implies that multiplication by $i$ converts tangent vectors of $\gamma$ into normal vectors of $\gamma$, and vice versa. Thus, $\gamma$ is a middle-dimensional Lagrangian submanifold of $X$. This Lagrangian submanifold also happens to be $G$-invariant, which we know from the previous paragraph.

Next, the $\mathcal{N}=(2,2)$ supersymmetry transformation of $\phi^I$ is
\begin{equation}
\delta\phi^I=i(\eps_{+2}\psi^I_--\eps_{+1}J^I_{\textrm{ }K}\psi^K_--\eps_{-2}\psi^I_++\eps_{-1}J^I_{\textrm{ }K}\psi^K_+),
\end{equation}
%IMPORTANT NOTE: OBTAINED FROM MELNIKOV, PLESSER AND RINKE, with J rescaled by a minus sign 
where $\eps_+=\eps_{+1}+i\eps_{+2}$ and $\eps_-=\eps_{-1}+i\eps_{-2}$, and where $J$ is the almost complex structure of $X$ locally 
%IMPORTANT NOTE: I HAVE SAID LOCALLY HERE BECAUSE MELNIKOV, PLESSER AND RINKE SAY THAT "Since the target space is a complex manifold, and we are working locally, we can choose a set of coordinates where the complex structure J^A_B is constant, in addition to being covariantly constant." Also see Aspinwall, equation 40, where he introduces almost complex structure instead. Hori just mentions complex structure on page 801 of the book, but does not claim that it is constant. Also see Witten-topological sigma models for a nice explanation of complex structure. Witten does not show a constant form of the complex structure but seems to use the constant form to obtain equation 2.18 from equation 2.16. Nakahara shows that the form of J given in equation 8.18 is independent of the chart, see right above equation 8.18. But i think this does not hold for the form of J given in equation  8.22 since this depends on how we complexify the coordinates. But then at the beginning of page 320 nakahara says that J is defined independently of charts. But then again at this point he is still talking about almost complex structure, not complex structure.
given by $J^i_{\textrm{ }k}=i\delta^i_{\textrm{ }k}$ and $J^{\ov{\imath}}_{\textrm{ }\ov{k}}=-i\delta^{\ov{\imath}}_{\textrm{ }\ov{k}}$. A-type supersymmetry corresponds to $\eps_{+1}=\eps_{-1}$ and $\eps_{+2}=-\eps_{-2}$, whereby
\begin{equation}
\delta\phi^I=i\big(\eps_{+2}(\psi^I_-+\psi^I_+)-\eps_{+1}J^I_{\textrm{ }K}(\psi^K_--\psi^K_+)\big).
\end{equation}
Hence, $\psi^I_-+\psi^I_+$ and $J^I_{\textrm{ }K}(\psi^K_--\psi^K_+)$ are tangent to $\gamma$. However, from the previous paragraph, we know that $\psi^I_--\psi^I_+$ is normal to $\gamma$. In addition, ${J^M}_IJ^I_{\textrm{ }K}=-\delta^M_{\textrm{  }K}$. Hence, the application of the almost complex structure, $J$,
%IMPORTANT NOTE: BORROWED TERMINOLOGY OF APPLICATION FROM ASPINWALL. WITTEN IN TOPOLOGICAL SIGMA MODELS ALSO NOTES THAT J is a section of the vector bundle End(TX), which is contains all linear transformations of TX
 converts normal vectors of $\gamma$ into tangent vectors of $\gamma$, and vice versa. Thus, $\gamma$ is a middle-dimensional Lagrangian submanifold of $X$. This Lagrangian submanifold also happens to be $G$-invariant, which we know from the previous paragraph.

Indeed, \eqref{dSmatw} and the $B$-field terms in \eqref{dSB} vanish under this boundary condition; integrating out the auxiliary fields $F^i$  and $\ov{F}^{\ov{\imath}}$, \eqref{dSmatw} and the $B$-field terms in \eqref{dSB} can be rewritten (for $\eps_+=\ov{\eps}_-=\eps$) as 
\begin{equation}
\begin{aligned}
%\delta S_{matter}=
\frac{1}{2\pi}\frac{1}{4}\int_{\partial\Sigma}dx^0\Big\{&\eps\Big(-g(\del_0^A\phi,\psi_--\psi_+)-i\omega(\del_0^A\phi,\psi_-+\psi_+)\\&-g(\del_1^A\phi,\psi_-+\psi_+)-i\omega (\del_1^A\phi,\psi_--\psi_+)\\&-\textrm{Re}(\si^a)g(\til{e}_a,\psi_+-\psi_-)-i\textrm{Re}(\si^a)\omega(\til{e}_a,\psi_++\psi_-)\\&-i\textrm{Im}(\si^a)g(\til{e}_a,\psi_+-\psi_-)+\textrm{Im}(\si^a)\omega(\til{e}_a,\psi_++\psi_-)
\\&+2B(\del_0^A\phi,\psi_-+\psi_+)-2i\omega^{-1}(g(\psi_--\psi_+),B\del_0^A\phi)\Big) + c.c. \Big\}
\end{aligned}
\end{equation}
(where $g(X,Y)=g_{IJ}X^IY^J$, $\omega(X,Y)=\omega_{IJ}X^IY^J$, $B(X,Y)=B_{IJ}X^IY^J$ and  $\omega^{-1}(X,Y)=\omega^{IJ}X_IY_J$), which vanishes using $A_{1a}=0$
% and \eqref{Binteriorderivative}
 as well as the conditions that $\del_0\phi^I$, $\psi_-^I+\psi_+^I$ and $\til{e}_a^I$ are tangent to $\gamma$ while $\del_1\phi^I$ and $\psi_-^I-\psi_+^I$ are normal to $\gamma$, together with the condition that $B|_{\gamma}=0$.\footnote{Recall that for a tangent vector, $T$, and normal vector, $N$, of a Lagrangian submanifold, $\gamma$, of the K\"ahler manifold $X$, we have $\omega (T,T)=\omega(N,N)=0$.
 % on $\gamma$.
  Also, $\omega^{-1}(gN,BT)=0$ means that the restriction of $B$ to $(T\gamma)^{\circ}\times T\gamma$ vanishes, where $(T\gamma)^{\circ}$ is the subspace of $TX$ orthogonal to $T\gamma$ with respect to $\omega$. When $\gamma$ is a Lagrangian submanifold, then $(T\gamma)^{\circ}=T\gamma$, and $B$ vanishes when restricted to $\gamma$. } 

Next, we consider the terms proportional to ($\phi^*\mu_a+r_a$) and $\theta_a$ in \eqref{dSgauthe}, as well as the term proportional to $\phi^*C_a$ in \eqref{dSB}. Now, on 
%the 
a $G$-invariant Lagrangian submanifold, 
%$\gamma$
 we have $\omega_{IJ}\til{e}_a^I T^J=0$ for any tangent vector $T$. Using \eqref{momentmapequation}, this implies $\del_J\mu_a T^J=0$, 
%on $\gamma$
 i.e., $\mu$ ought to be a constant along $\gamma$ \cite{Cieliebak}. Moreover, gauge invariance of the pull-back of this condition to $\del \Sigma$ requires that the constant be an element of $[\g,\g]^0$, via the identity $\alpha^b\mathcal{L}_{\til{e}_b}\mu_a=[\alpha,\mu]_a$ \cite{Cieliebak}. Choosing the constant to be 
\begin{equation}
\mu_a=-r_a,
\end{equation}
we find that the terms proportional to ($\phi^*\mu_a+r_a$) in \eqref{dSgauthe} vanish.
%NOTE THAT WE CAN ALSO SHOW THAT R AND C ARE constants IN $[\g,\g]^0$ along \gamma, and that \eqref{requirement} is gauge invariant.
Analogously, the fact that $B_{IJ}\til{e}_a^I T^J=0$ along $\gamma$ implies that $C$ ought to be a constant element of $[\g,\g]^0$ along $\gamma$. Choosing the constant to be 
\begin{equation}\label{requirementnoboundaryaction}
C_a=\theta_a,
\end{equation}
we find that the remaining term in \eqref{dSgauthe} and the remaining term in \eqref{dSB} cancel. Note that the boundary conditions \eqref{nonabelianvectormultipletbc} together with the constraint \eqref{requirementnoboundaryaction} preserve the locality of the equations of motion for vector multiplet components.

%Firstly,
Finally, we consider a boundary action. We note that the boundary action \eqref{gaugedWilsonline} for $\CP^{N-1}$ is an example of the GNLSM generalization of the NLSM boundary Wilson line \eqref{Abrane}. Hence, for general GNLSMs the boundary action ought to be 
\begin{equation}\label{gaugedWilsonlinenonabelian}
\begin{aligned}
S_{\del\Sigma}&=\int\limits_{\partial\Sigma}dx^0 A^X_M\del^A_0\phi^M
%+A^X_{\ov{\imath}}\del^A_0\ov{\phi}^{\ov{\imath}}
\\
&=\int\limits_{\partial\Sigma}dx^0\Bigg(A^X_M\del_0\phi^M
%+A^X_{\ov{\imath}}\del_0\ov{\phi}^{\ov{\imath}}
-\til{R}_a A_0^a\Bigg)
\end{aligned}
\end{equation}
%IMPORTANT NOTE:This agrees with the boundary action of the CP^N case because when this boundary action is A_I^X\del_0^A\phi^I using real notation. Then using coordinates that separate the lagrangian submanifold and directions normal to it, and using  lagrangian boundary conditions, it reduces to the CP^N case because \del_0 acting on |Z^i| is zero and the killing vector field along the |Z^i| direction is also zero.
where $A^X$ 
%is the
corresponds to a $G$-invariant ($\mathcal{L}_{\til{e}}A^X$=0) connection 
% is the moment of a flat
 %IMPORTANT NOTE: MUST SAY F_IJ=0, otherwise matter boundary conditions will get modified, see HIV PAGE 21, below equation 3.24 and surrounding equation 3.25. I think the difference from B-field is that B-field is defined on the entire target space, so the restriction of the component to the boundary need not vanish, but F_IJ is defined on the Lagrangian submanifold, so the vanishing of the restriction of the two-form F to the boundary implies that F_IJ=0 (I THINK). sEE THE LAST PARAGRAPH OF PAGE 804 OF THE MIRROR SYMMETRY BOOK
%, $G$-equivariant
of a flat ($F^X_{MN}=0$) $U(1)$ bundle on 
%$\gamma$
each A-brane, and where $\til{R}_a=-\iota_{\til{e}_a} A^X$
%defined on $\gamma$ IMPORTANT NOTE: I REMOVED DEFINED ON \gamma to avoid confusion with the next sentence, actually it is defined on \gamma, since it is the interior derivative of of a connection defined on \gamma
(we shall use ($M,N,\ldots$) as coordinate indices on the A-branes).\footnote{Note that the inclusion of this boundary action does not modify the constraints \eqref{localeomsss}, since it vanishes under arbitrary variations of $\phi^M$ because $F_{MN}=0$.}
Gauge invariance of this boundary action follows from the equivariant Bianchi identity 
\begin{equation}\label{NonabelianequivariantBianchiA}
d\til{R}=\iota_{\til{e}}F^X,
\end{equation}
and this implies that %$\gamma$
%an equivariant
each A-brane supports a flat, \textit{$G$-equivariant} $U(1)$ bundle, for which $\til{R}_a$ is the moment.\footnote{Note that gauge invariance of the boundary action requires the use of the identity $\alpha^b\mathcal{L}_{\til{e}_b}\til{R}_a=[\alpha,\til{R}]_a$.
%IMPORTANT NOTE: THIS IDENTITY FOLLOW SINCE THE MOMENT IS ALSO G-EQUIVARIANT LIKE THE MOMENT MAP, SEE PAGE 29 OF SZABO
} Its supersymmetry variation is 
\begin{equation}
\delta S_{\del{\Sigma}}=-\int\limits_{\partial\Sigma}dx^0 \til{R}_a(\frac{i}{2}(\eps(\ov{\lam}_+^a+{\lam}_-^a)+\ov{\eps}({\lam}_+^a+\ov{\lam}_-^a))),
\end{equation}
where we have used \eqref{NonabelianequivariantBianchiA}. Just like in the $\CP^{N-1}$ case, we require that this cancels the $C$-term in \eqref{dSB} and the ${\theta}_a$-term in \eqref{dSgauthe}, i.e., we require that 
\begin{equation}\label{requirement}
%2\pi\til{R}_a=-\theta_a+\phi^*C_a,
2\pi\til{R}_a=-\theta_a+C_a
\end{equation}
on $\gamma$, the pull-back of which is a gauge invariant condition on $\del\Sigma$.
%IMPORTANT NOTE: HERE WE MEAN ALL THE GAMMAS. THERE ARE MULTIPLE GAMMAS, ONE FOR EACH BOUNDARY
 This modification of \eqref{requirementnoboundaryaction} (together with the boundary conditions \eqref{nonabelianvectormultipletbc}) also preserves the locality of the equations of motion for vector multiplet components, just like in the example of $\CPN$.

 In conclusion, we find that in general,  
\vspace{1em}

 \noindent\fbox{ \parbox{\textwidth}{\linespread{1.1}\selectfont{Equivariant A-branes are $G$-invariant Lagrangian submanifolds of $X$, which support $G$-equivariant flat $U(1)$ bundles, and on which the restriction of the $B$-field vanishes.}}}
 
 %\noindent
 \vspace{1em} \noindent This implies that they are %the simplest 
 objects in the $G$-equivariant Fukaya category of $X$, by generalizing the definition of the equivariant Fukaya category for finite groups (\cite{ChoHong}, page 68) to any compact Lie group $G$.
As mentioned in Section 5.2, Kapustin et al. (\cite{Kapustin}, page 58) have conjectured that the category of $G$-equivariant A-branes is some sort of $G$-equivariant version of the Fukaya category. Hence, we have further verified their conjecture for nonabelian $G$. Fully proving their conjecture would require constructing the other objects in the category, which correspond to Lagrangian submanifolds that support equivariant flat unitary vector bundles, and these should correspond to the insertion of certain $G$-invariant Wilson lines in the path integral.

\section{Open Hamiltonian Gromov-Witten Invariants}
%Emphasize computation of invariants using mirror theory.
In this section, we shall use equivariant A-branes to define open Hamiltonian Gromov-Witten invariants. We shall first study the nonabelian open Hamiltonian Gromov-Witten invariants via the open topological gauged A-model, using the boundary conditions and boundary term we have found in Section 5.5. In the final two subsections, we shall focus on investigating the abelian open Hamiltonian Gromov-Witten invariants via mirror symmetry. %Although we shall still focus on the invariants which correspond to abelian gauge groups, we shall attempt to keep the analysis general in order to understand the nonabelian invariants as well.

\subsection{Open Topological Gauged A-model}
The closed topological gauged A-model was introduced by Baptista \cite{Baptista1}, and in the following we shall generalize it to the case with boundaries, i.e., the \textit{open} topological gauged A-model. This involves analytically continuing the Minkowski strip to the Euclidean one, subsequently twisting the fields in the action \eqref{action} + \eqref{FIterm} as well as its supercharges using their vector R-charges, and imposing the appropriate boundary conditions (found in Section 5.5) which are supersymmetric with respect to the scalar supercharge $Q_A=Q_- + \ov{Q}_+$. We also include the gauge invariant boundary term \eqref{gaugedWilsonlinenonabelian}.
% on each boundary. %following our result \eqref{gaugedWilsonline}.
%IMPORTANT NOTE: only can see that it is a Wilson lline from the path integral

%Proceeding impartially by alphabetical order, we start with the A-model. 
%Define formally a new set of fields by the formulae:
The twisted fields are redefined as follows:
\begin{align}
\chi^k &= \sqrt{2}\psim^k    &      \psi_z^a &= (-i \lamm^a)/\sqrt{2}     \label{3.1}  \\
\ov{\chi}^{\ov{k}} &=   \, \sqrt{2}\ov{\psi}_+^{\ov{k}}     &      \psi^a_{\zb}  &=  (i  \ov{\lam}^a_+)/\sqrt{2}   \nonumber \\
\varphi^a  &=  - \, i  2\si^a     &      \rho_{\zb}^k  &= \, \sqrt{2}\psip^k      \nonumber \\
\la^a  &=   \sib^a/4  &      \ov{\rho}^{\ov{k}}_z  &=  \, \sqrt{2}\ov{\psi}_-^{\ov{k}}    \nonumber \\
\eta^a &=  -i(\ov{\lam}^a_- + \lamp^a)/(2\sqrt{2})   &    \kpp^a &= i (\ov{\lam}_-^a - \lamp^a)/\sqrt{2}    \nonumber \\
\mathcal{H}^k_{\zb}  &=   4i\dd^A_{\zb} \phi^k + 2(F^k - \Gamma^k_{ij} \psip^i \psim^j  )  &     \mathcal{C}^a &= 2 (F_A)^a_{12} + 2 D^a   \  \nonumber, 
\end{align}
%The interpretation of the new fields as scalars or 1-forms comes, as explained before, from table (\ref{2.11}). 
%These local components can be combined to define the global fields
where the fields are now sections of the following bundles 
\begin{align}
\chi  &\in  \Omega_-^0 (\Sigma ; \phi^\ast \ver )  &      \varphi, \la , \mathcal{C}  &\in \Omega_+^0 (\Sigma ; \gp )     \\
\rho  &\in  \Omega_-^{0,1} (\Sigma ; \phi^\ast \ver )  &   \eta , \kpp  &\in \Omega_-^0 (\Sigma ; \gp )         \\
\mathcal{H}  &\in  \Omega_+^{0,1} (\Sigma ; \phi^\ast \ver )  &      \psi   &\in \Omega_-^1 (\Sigma ; \gp )  \ ,
\label{}
\end{align} 
with the remaining `barred' fields being interpreted as the local complex conjugates of the ones above. The action of the open gauged A-model is then\footnote{Here, we follow the notation of \cite{Baptista1}. 
%We also redefine the symplectic form as $\omega=\frac{i}{2}\met d\phi^i\wedge d\ov{\phi}^{\ov{\jmath}}$, which amounts to rescaling ($\mu_a + r_a$) by a factor of 2 in the action \eqref{action}+\eqref{FIterm}.
}
\begin{equation}
\begin{split}
S_A  =  \frac{1}{2\pi}  \int_\Sigma & \Bigl\{  \frac{1}{2e^2} |F_A|^2  + |\dd^A \phi|^2   + \frac{1}{2}e^2 |\mu \circ \phi+r|^2  
+ \frac{i}{e^2} ( \nabla^A \varphi , \nabla^A \la  )   +  \frac{1}{2e^2} |[\varphi , \la]|^2   \\ 
&+ \frac{1}{2e^2} [\varphi, \eta]_a \eta^a   -  \frac{1}{8e^2} [\varphi , \kpp]_a \kpp^a  
- \frac{1}{2e^2}  |\frac{1}{2} \mathcal{C}  - \ast F_A - e^2 \: (\mu \circ \phi+r) |^2     \\
&-  \frac{1}{4} |\mathcal{H} - 4i \db^A \phi |^2  
+ i g_{j\ov{k}} (\varphi^a \la^b + \varphi^b \la^a ) \he_a^j \, \ov{\he}_b^{\ov{k}}  
+ 2 i g_{j\ov{k}} (\nabla_l \he_a^j ) \la^a \chi^l \ov{\chi}^{\ov{k}}   \\  
&+ i g_{j\ov{k}} (\eta^a + \frac{1}{2} \kpp^a )\, \ov{\he}_a^{\ov{k}}  \, \chi^j  
+ i g_{j\ov{k}} (\eta^a - \frac{1}{2}\kpp^a )\, \he^j_a \, \ov{\chi}^{\ov{k}}  \,  \Bigr\} \,\, \rm{vol}_\Sigma   \\
+\frac{1}{2\pi}& \int_\Sigma   \Bigl\{\frac{i}{e^2}  \eta_a  {\nabla}^A \ast \psi^a - \frac{1}{2e^2} \kpp_a  {\nabla}^A \psi^a 
- \frac{i}{8}  R_{i\ov{\jmath}k \ov{m}}  (\rho^i \wedge \ov{\rho}^{\ov{\jmath}} ) \chi^k  \ov{\chi}^{\ov{m}}  \\
&+ \frac{i}{e^2} \la_a [\psi, \ast \psi]^a  
+ \frac{1}{2} g_{j\ov{k}}\,  \rho^j \wedge (\phi^\ast{\nabla}^A ) \ov{\chi}^{\ov{k}}
+ \frac{1}{2} g_{j\ov{k}} \, \ov{\rho}^{\ov{k}} \wedge (\phi^\ast {\nabla}^A ) \chi^j  \\
&+ \frac{i}{8} g_{j\ov{k}}  \varphi^a (\nabla_l \he^j) \rho^l \wedge \ov{\rho}^{\ov{k}}
+ \frac{1}{2} g_{j\ov{k}}\,  \he_a^j \, \psi^a  \wedge \ov{\rho}^{\ov{k}}
+ \frac{1}{2} g_{j\ov{k}} \, \ov{\he}_a^{\ov{k}}\, \psi^a  \wedge \rho^j \Bigr\} \ \\
+\frac{1}{2\pi}&i\Big(\int_{\Sigma} \phi^\ast B \; - \int_{\del \Sigma}   \phi^*C_a A^a \Big)+\frac{1}{2\pi}i\int_\Sigma ( \theta ,F_A)-i\int_{\del\Sigma} A^X_M d^A\phi^M,
\end{split}
\label{3.4}
\end{equation}
where $F_A$ is the curvature two-form of the connection $A=A_{\mu}dx^{\mu}$, and the measure on the worldsheet is $\textrm{vol}_\Sigma=dx\wedge d\tau =\frac{i}{2}dz\wedge d\bar{z}$.\footnote{We recall at this point that when $G=G_1\times G_2\times G_3 \ldots$, each factor $G_i$ has its own coupling constant, $e_i$. In the $G=U(1)^k$ case that we have focused on in the previous section, each $U(1)$ factor had its own coupling constant, previously denoted $\til{e}_c$.}\footnote{Note that we have performed integration by parts to undo the symmetrized form of the fermionic kinetic terms present in \eqref{action}, which is no longer necessary since a Euclidean action is not %hermitian
real. The resulting boundary terms vanish using boundary conditions found in Section 5.5.
%IMPORTANT NOTE: FOR THE MATTER FERMIONS, THE BOUNDARY TERM IS EQUAL TO \frac{1}{8}(\omega(T,T)+\omega(N,N)), which is equal to zero.
 For the case of $\C P^{N-1}$ which we studied extensively in the previous section, the relevant boundary conditions (given in \eqref{GbcAcpn} and \eqref{cpngaugebc}) %(appropriately twisted), %, are sufficient. 
 %we can use 
%the boundary conditions $\del_1\varphi_a =0$, $A_{1a}=0$,
 are $\kpp_a=0$, $\psi_z^a+\psi_{\bar{z}}^a=0$, $\ov{Z}^i\chi^i +{Z}^i \ov{\chi}^i=0$ and $\ov{Z}^i \rho_{\bar{z}}^i +{Z}^i \ov{\rho}_z^i=0$. %to
%are used to remove the boundary terms that appear when integrating by parts. 
} The fields $\mathcal{C}$ and $\mathcal{H}$ are auxiliary fields, which can be integrated out of the action using their equations of motion
%IMPORTANT NOTE: THE CURVATURE TWO-FORM IN THE NONABELIAN CASE CONTAINS A COMMUTATOR!
\begin{align}
\mathcal{C}^a &=   2 \ast F_A^a + 2e^2 \: (\mu^a \circ \phi +r^a)  \\
\mathcal{H}^k_{\zb} &=  4i\dd^A_{\zb} \phi^k \ .
\end{align}

The supersymmetry transformations generated by the scalar supercharge $Q_A = Q_- + \ov{Q}_+$ on the new fields follow from the supersymmetry 
transformations \eqref{susytrans} and \eqref{susytrans2}, with $\epsilonp = \epsilonbm ={ \sqrt{2}} $ and 
$\epsilonm = \epsilonbp = 0$, which gives %and finally write the result in an invariant form that 
%makes sense on any Riemann surface $\Sigma$. This procedure yields:
\begin{align}
Q_A\, \phi^k   &=  \chi^k     &     Q_A \, A  &=  \psi     \label{3.3}   \\
Q_A\, \chi^k  &=  \varphi^a \he_a^k     &   Q_A \,\psi &=  -  \nabla^A \varphi    \nonumber  \\
Q_A\, \la &= \eta     &   Q_A\, \kpp &= \mathcal{C}     \nonumber  \\
Q_A\, \eta &= [\varphi , \la]     &     Q_A\, \mathcal{C}   &=  [\varphi,\kpp]     \nonumber \\
Q_A\, \rho^k  &=  \mathcal{H}^k  -  \Gamma^k_{ij} \chi^i  \rho^j   &   Q_A\, \varphi  &=  0   \nonumber   \\
Q_A\, \mathcal{H}^k  &=  -R_{i\ov{\jmath}l\ov{m}} g^{k\ov{\jmath}}  \chi^l  \ov{\chi}^{\ov{m}} \rho^i  -   \Gamma^k_{jl} \mathcal{H}^j \chi^l
 +  \varphi^a (\nabla_j \he_a^k)  \rho^j    \ . \nonumber 
\end{align}
%The apparently random numerical factors in (\ref{3.1}) were chosen such as to render these last transformations as simple as possible.
%The result also agrees with \cite{Bap}, modulo the notations.

The action \eqref{3.4} is in fact $Q_A$-exact up to topological terms, 
\begin{equation}
\boxed{
S_A \: =\:  Q_A  \Psi \ + \ \frac{1}{2\pi}\int_\Sigma \phi^\ast ([\eta_{\omega}]+i[\eta_B])+\ \frac{1}{2\pi}i\int_\Sigma (\theta, F_A)-i\ \int_{\del\Sigma} A^X_M d^A\phi^M
\label{3.5}}
\end{equation}
with gauge fermion
\begin{equation*}
\begin{split}
\Psi \ = \ &\frac{1}{2\pi}\int_\Sigma  \Bigl\{  \frac{1}{2e^2}\, \kpp_a (\ast F_A + e^2 (\mu \circ \phi+r))^a  - \frac{1}{8e^2}\, \kpp_a \mathcal{C}^a 
+ \frac{1}{2e^2}\, \eta_a [\varphi , \la]^a  + i g_{j\ov{k}}\, \la^a (\he_a^j \,\ov{\chi}^{\ov{k}} + \ov{\he}_a^{\ov{k}}\, \chi^j )  
\Bigr\} \rm{vol}_\Sigma    \\
&+ \frac{1}{2\pi}\int_\Sigma  \Bigl\{\frac{i}{e^2}\, \la_a (\nabla^A \ast \psi^a)  - \frac{i}{16}\, g_{j\ov{k}} \,\ov{\rho}^{\ov{k}} \wedge (\mathcal{H} - 8i \db^A \phi)^j 
+ \frac{i}{16}\, g_{j\ov{k}}\, \rho^j \wedge \ov{(\mathcal{H} - 8i \db^A \phi)^k}\Bigr\} \ ,
\end{split}
\end{equation*}
where we have performed integration by parts %in the kinetic term for $\varphi$ and $\xi$ 
in \eqref{3.4} such that $\nabla_{\mu}^A\varphi_a\nabla^{A\mu} \xi^a$ becomes $-\xi_a\nabla_{\mu}^A\nabla^{A\mu} \varphi^a$. %to undo the symmetrization of the fermion kinetic terms in \eqref{3.4}.
%, and assumed that t
%IMPORTANT NOTE: THE HODGE STAR IN THE GAUGE FERMION IS NECESSARY TO GET A SUM OVER WORLDSHEET INDICES, SEE CALCULATIONS.
The resulting boundary term vanishes using the %appropriate 
boundary conditions $\del_1\varphi_a =0$ and $A_{1a}=0$ found in Section 5.5. %IMPORTANT NOTE \psi_+-\psi_-=0 along the boundary since it is normal to the Lagrangian submanifold. Integration by parts of the matter  fermion terms gives (\ov{\psi}_+\psi_+ - \ov{\psi}_-\psi_-), which vanishes using \psi_+-\psi_-=0. The gauginos vanish as in the CP^{N-1} case below. UPDATE ON 18/3/17, see latest explicit calculation
%For the case of $\C P^{N-1}$ which we studied extensively in the previous section, the boundary conditions are \eqref{GbcAcpn} and \eqref{cpngaugebc}, appropriately twisted. %, are sufficient. 
%In particular, %we can use 
%the boundary conditions $\del_1\varphi_a =0$, $A_{1a}=0$, $\kpp_a=0$, $\psi_z^a+\psi_{\bar{z}}^a=0$, $\ov{Z}^i\chi^i +{Z}^i \ov{\chi}^i=0$ and $\ov{Z}^i \rho_{\bar{z}}^i +{Z}^i \ov{\rho}_z^i=0$ %to
%are used to remove the boundary terms that appear when integrating by parts.

Let us %recall 
%the description of
elucidate the first topological term of
%The topological term on the right-hand-side of 
(\ref{3.5}). %from \cite{Baptista1}.
Here, $[\eta_{\omega}]$ and $[\eta_{B}]$ %represent  equivariant cohomology classes in $H^2_G (X)$. %%They are
%Their pullbacks 
%IMPORTANT NOTE: I TOOK OUT MENTION OF EQUIVARIANT COHOMOLOGY CLASSES HERE BECAUSE \PHI IS A SECTION, IE A MAP FROM \Sigma to E, so a pullback by phi should be with respect to an object in E.
are the cohomology classes in $H^2(E)$ represented by the two-forms
\begin{equation}
\begin{aligned}
\eta_{\omega} (A) &= \omega  - \dd ((\mu_a+r_a) A^a)  \qquad \in \Omega^2 (P \times X) \ ,\\
\eta_B (A) &= B  - \dd (C_a A^a)  \qquad \in \Omega^2 (P \times X) \ ,
\end{aligned}
\end{equation}
both of which descend to $E = P\times_G X$.
%IMPORTANT NOTE: I THINK THE DESCENDING PART SHOULD BE UNDERSTANDABLE FROM EQUATION 3.10 OF FIGUEORA AND STANCIU AS WELL AS REMARK 2.2 OF THE J-HOLOMORPHIC CURVES PAPER. 
%IMPORTANT NOTE: THE REFERENCE FOR THE ABOVE IS PAGE 92 OF SETTER, AND HIS APPENDIX
%IMPORTANT NOTE: WRITTEN ON 25TH MARCH 2017, SETTER ACTUALLY GETS A DIFFERENT TOPOLOGICAL TERM. SO WE CAN'T REALLY USE SETTER AS A REFERENCE HERE, HOWEVER FROM The symplectic vortex equations and invariants of Hamiltonian group actions PAGE 9, WE CAN SEE THAT THE TOPOLOGICAL TERM WITH OMEGA AND MU IS AN INNER PRODUCT BETWEEN THE EQUIVARIANT COHOMOLOGY CLASS REPRESENTED BY THE  EQUIVARIANT SYMPLECTIC FORM  AND AN EQUIVARIANT HOMOLOGY CLASS
%This form is manifestly closed, for the K\"ahler form $\omega_X$ on $X$ is closed,
%In particular and its cohomology class does not to depend on $A$. 
%In fact, $[\eta_E]$, as an
%element of $H^2 (E)$, is just the image by the Chern-Weil homomorphism of the class in $H^2_G (X)$ represented by
%the equivariantly closed form $\omega_X - \zeta^a \mu_a$. Besides not depending on $A$, 
%IMPORTANT NOTE: THE RELATION BETWEEN R \THETA AND C, IMPLIES THAT C IS A CONSTANT ALONG THE LAGRANGIAN SUBMANIFOLD, BUT NOT IN THE BULK, SO IN THE TOPOLOGICAL ACTION WHERE IT APPEARS IN A BULK TERM, IT IS NOT CONSTANT
In particular, this term is topological since
$\int_\Sigma \phi^\ast [\eta_{\omega}]$ and  $\int_\Sigma \phi^\ast [\eta_{B}]$ do not change under deformations of the map $\phi$, since the pull-back map is always
homotopy invariant. %\cite{Baptista1}.
%IMPORTANT NOTE: THIS IS SHOWN IN NAKAHARA PAGE 241
 In addition, the cohomology classes $[\eta_{\omega}]$ and $[\eta_{B}]$ are the pull-backs of the equivariant cohomology classes in $H^2_G(X)$ represented by $\omega - (\mu + r)$ and $B- C$.\footnote{This follows because $H^2_G(X)=H^2(EG\times_G X)$, and since $P\times_G X\rightarrow \Sigma$ is the pull-back bundle of $EG\times_G X\rightarrow BG$ via a map $\ov{U}:\Sigma \rightarrow BG$, where $EG\rightarrow BG$ is the universal bundle \cite{figueroa}.}
 %IMPORTANT NOTE: ALSO SEE SECTION 6.4.4 OF NAKAHARA ON PULLBACK OF COHOMOLOGY CLASSES
 %IMPORTANT NOTE: WHAT I WROTE BELOW IS CORRECT FOR THES AN EQUIVARIANT HOMOLOGY CLASS, AND WHETHER RELATIVE EQUIVARIANT HOMOLOGY CLASSES COME INTO PLAY.
%In addition, $\int_\Sigma \phi^\ast [\eta_{\omega}]$ and  $\int_\Sigma \phi^\ast [\eta_{B}]$ are also the inner products between equivariant cohomology classes in $H^2_G (X)$ represented by $\omega - (\mu + r)$ and $B- C$ and an equivariant homology class in $H_2^G (X)$ determined by the worldsheet $\Sigma$ (\cite{Cieliebak2}, page 9).
%IMPORTANT NOTE: Also see J-holomorphic curves paper section 2.3 for definition of equivariant homology class. The theory here is just that we can compose \phi with another map \phi_2 which takes us from E to EG\times_G X. Then \int \phi^*[\eta_\omega]=\int (\phi_2\circ\phi)^*[\omega-(\mu+r)]. The equivariant homology class is the pushforward of \Sigma by \phi_2\circ\phi. aLSO SEE EQUATION 16.60 OF MIRROR SYMMETRY BOOK
%IMPORTANT NOTE: FIGUEORA O FARRILL AND STANCIU ACTUALLY DISCUSS HOW THE TOPOLOGICAL TERM IS RELATED TO AN EQUIVARIANT COHOMOLOGY CLASS  AT LENGTH!

The open gauged A-model is topological as a \textit{quantum} theory,\footnote{In particular, the correlation functions of the theory are invariant under diffeomorphisms of the worldsheet, e.g., transforming it from a strip to a disk.}
%IMPORTANT NOTE: TQFT's are invariant under conformal transformations, such as that which transforms an infinite strip to a disk. see http://physics.stackexchange.com/questions/141408/relation-between-conformal-and-topological-field-theories/141431#141431 
 and in order to consistently quantize such a gauge theory, one ought to perform BRST gauge-fixing, which involves the inclusion of Faddeev-Popov ghost fields in the action. This can be done straightforwardly, and we shall not write down the gauge-fixing action, $S_{BRST}$, explicitly. However, the open gauged A-model is anomalous, and in Section 6.4, we shall compute this anomaly by canonically quantizing the gauged Landau-Ginzburg mirror of the \textit{abelian} open gauged A-model. Notably, in the process we shall describe $S_{BRST}$ for abelian gauge groups in detail. 
 %We shall also find the mirror of the open Hamiltonian Gromov
 %Finally, if desired, the auxiliary fields $C$ and $H$ can be eliminated from the action and the $Q_A$-transformations 
%through their equations of motion
%\begin{align*}
%C^a &=   2 \ast F_A^a + 2e^2 \: \mu^a \circ \phi  \\
%H^k_{\zb} &=  4i\dd^A_{\zb} \phi^k \ .
%\end{align*}

%One should also observe that the topological action $I_A$ is gauge invariant. The standard methods of local quantum 
%field theory therefore recommend that it be gauge-fixed through the introduction of Fadeev-Popov ghost fields. This
%can presumably be done as explained in \cite{Baul-Sin}, and would simply amount to adding to $I_A$ a further $Q_A$-exact 
%term.

%\subsection{Observables of the Open Gauged A-model}
\subsection{Observables and Open Hamiltonian Gromov-Witten Invariants}
A canonical set of bulk observables of the closed gauged A-model were  described in \cite{Baptista1}, with the path integrals over these observables eventually argued to be equal to the Hamiltonian Gromov-Witten invariants. In this section, we shall recall the description of these bulk observables, as well as introduce \textit{boundary} observables which are defined with respect to the topology of the equivariant A-branes.

%Having described the field content, the lagrangian and the $Q_A$-transformations of the theory, the next step is to 
%look for an interesting set of observables whose correlation functions we would like to compute. In the non-gauged 
In the ordinary open A-model, one can construct bulk observables from the de Rham cohomology classes of the target
$X$, as well as construct boundary observables from the de Rham cohomology classes of the A-branes, which wrap the subspaces of $X$ (i.e., Lagrangian submanifolds) to which boundaries of the worldsheet are mapped. For the open gauged A-model, one uses the $G$-equivariant cohomology classes of $X$ to define bulk operators, as well as the $G$-equivariant cohomology classes of the equivariant A-branes to define boundary operators.

%In the gauged model, of course, the analog procedure uses instead the $G$-equivariant cohomology classes of 
%$X$. This construction was first described in \cite{W2}, and then with a little more detail in \cite{Bap}.

The $G$-equivariant cohomology classes of a manifold, $M$, are defined using the $G$-equivariant complex $\Omega^\bullet_G (M)$, which is the set of $G$-invariant elements in the tensor product 
$S^\bullet (\g^\ast) \otimes \Omega^\bullet (M)$, with $S^\bullet (\g^\ast)$ being the symmetric algebra of the dual of $\mathfrak{g}$.  %A typical equivariant form $\alpha $ may thus be locally written as    

For $M=X$, an equivariant form, $\alpha$, can be written on a local patch of $X$ as
\begin{equation}
\alpha \ = \ \alpha_{a_1 \cdots a_r k_1 \cdots k_p \ov{l_1} \cdots \ov{l_q}}
(w) \ \xi^{a_1} \cdots \xi^{a_r} \; \dd w^{k_1} \wedge \cdots \wedge 
\dd w^{k_p} \wedge \dd \ov{w}^{\ov{l_1}} \wedge \cdots \wedge  \dd \ov{w}^{\ov{l_q}} \ ,
\label{localformalpha}
\end{equation}
where $(w^{k_i},\overline{w}^{\overline{l_i}})$ are the coordinates on the patch. The coefficients $\alpha_{a_1 \cdots a_r k_1 \cdots k_p \ov{l_1} \cdots \ov{l_q}}$
are symmetric with respect to the indices $a_i$, and antisymmetric with respect to the indices $k_i$ and $\ov{l_i}$. Such a local form can be associated with a bulk operator $\OO_\alpha$ in the open gauged A-model, %defined by the local formula
\begin{equation}
\OO_\alpha  \: = \: (\alpha_{a_1 \cdots a_r k_1 \cdots k_p \ov{l_1} \cdots \ov{l_q}}
\circ \phi )  \left[\prod_{j=1}^r    (\varphi + \psi - F_A
)^{a_j} \right]  \left[ \prod_{i=1}^p  (\chi^{k_i} - \dd^A \phi^{k_i} ) \right]  
\left[ \prod_{i=1}^q  (\ov{\chi}^{\ov{l_i}} - \dd^A \ov{\phi}^{\ov{l_i}} ) \right]  \, .
\label{3.6}
\end{equation}
%IMPORTANT NOTE: THE CURVATURE TWO-FORM IN THE NONABELIAN CASE CONTAINS A COMMUTATOR!
%IMPORTANT NOTE: THE ABOVE CHANGES IN SIGNS FROM BAPTISTA HAVE BEEN CHECKED
This correspondence holds globally on $X$. Moreover, we have
\begin{equation}\label{nenjeezhu}
(\dd_{\Sigma} + Q_A ) \; \OO_\alpha  =   \OO_{\dd_G \alpha}  \ ,
\end{equation}
where $\dd_\Sigma$ is the exterior derivative on the open worldsheet, $\Sigma$, while $\dd_G= 1 \otimes \dd + e^a \otimes \iota_{\he_a}$ is the Cartan operator defined on $\Omega^\bullet_G (X)$. $\OO_\alpha$ can be decomposed with respect to the form degree on the worldsheet,
\begin{equation*}
\OO_\alpha \ = \ \OO_\alpha^{(0)} \ +\ \OO_\alpha^{(1)} \ +\ \OO_\alpha^{(2)}\ ,  
\end{equation*}
where, in particular,
\begin{equation}
\OO_\alpha^{(0)}  \ = \ (\alpha_{a_1 \cdots a_r k_1 \cdots k_p \ov{l_1} \cdots \ov{l_q}}
\circ \phi )  \left(\prod_{j=1}^r   \varphi^{a_j} \right)  \left( \prod_{i=1}^p  \chi^{k_i}  \right)   
\left( \prod_{i=1}^q  \ov{\chi}^{\ov{l_i}} \right)  \ ,
\label{3.8}
\end{equation}
is a local operator.

If we assume that $\dd_G \alpha=0$, then \eqref{nenjeezhu} splits into the descent equations
% Now assume that $\alpha$ is 
%$\dd_G$-closed and decompose $\OO_\alpha$ according to the form degree over $\Sigma$,
%i.e. write 
%where for example
%\begin{equation}
%\OO_\alpha^{(0)}  \ = \ (\alpha_{a_1 \cdots a_r k_1 \cdots k_p \bar{l_1} \cdots \bar{l_q}}
%\circ \phi )  \left(\prod_{j=1}^r   \varphi^{a_j} \right)  \left( \prod_{i=1}^p  \chi^{k_i}  \right)   
%\left( \prod_{i=1}^q  \ov{\chi^{l_i}} \right)  \ .
%\label{3.8}
%\end{equation}
%Then in terms of this decomposition identity (\ref{3.7}) breaks into
\begin{align}
\dd_\Sigma \ \OO_\alpha^{(2)} \ &= \ 0 \  \\
\dd_\Sigma \ \OO_\alpha^{(1)} \ &= \ - \; Q_A \; \OO_\alpha^{(2)} \   \\
\dd_\Sigma \ \OO_\alpha^{(0)} \ &= \ - \; Q_A \; \OO_\alpha^{(1)} \   \\
Q_A \; \OO_\alpha^{(0)} \ &= \ 0. \
\label{3.9}
\end{align}
%which are the descent equations of the model.
 For a closed worldsheet, $\Sigma$, if $\beta$ is a $j$-dimensional homology cycle in 
$\Sigma$, one could define the $Q_A$-invariant operators 
\begin{equation*}
W (\alpha , \beta ) \ := \ \int_\beta \ \OO_\alpha^{(j)}  \ .  
\label{3.10}
\end{equation*}
%These are then the natural observables associated with the gauged A-model.
However, there are no 2-cycles on an open worldsheet. Hence,  $\OO_\alpha^{(2)}$ ought to be integrated over the entire open worldsheet, and it is necessary for $Q_A$-invariance of $\OO_\alpha^{(2)}$ that $\OO_\alpha^{(1)}=0$ at the boundaries.
%Actually there are no 2-cycles on an open worldsheet. Hence $O^2$ ought to be integrated over the open worldsheet, and it is necessary for $Q_A$-invariance of $O^2$ that $O^1=0$ at the boundaries.
%IMPORTANT NOTE: SEE TOPOLOGICAL OPEN P-BRANES BY JAE-SUK PARK, PAGE 24
%HOWEVER, note that unlike the gauged A-model on a closed worldsheet,  
 %In fact it follows as usual from  
%the descent equations and Stokes' theorem that $W (\alpha, \gamma )$ is $Q_A$-closed, so is indeed an observable. 
%IMPORTANT NOTE: THE FOLLOWING HOLDS IN THE CLOSED CASEMoreover, the $Q_A$-cohomology class of $W (\alpha , \gamma )$ only depends on the classes of $\alpha$ and 
%$\gamma$ in $H^\bullet_G (X)$ and $H_j (M)$, %respectively.

%We have thus far only discussed bulk operators on the worldsheet. Since the worldsheet contains, boundares, one may also include \textit{boundary} operators which are defined solely on the boundary. 

For $M=L$, where $L$ is an equivariant A-brane (to which a boundary component $\del \Sigma_L$ is mapped), an equivariant form, $\zeta$, can be written on a local patch of $L$ as
\begin{equation}
\zeta \ = \ \zeta_{a_1 \cdots a_r m_1 \cdots m_s }
(u) \ \xi^{a_1} \cdots \xi^{a_r} \; \dd u^{m_1} \wedge \cdots \wedge 
\dd u^{m_s} \ ,
\label{localformalphabrane}
\end{equation}
where $u^{m_i}$ are the coordinates on the patch. The coefficients $\zeta_{a_1 \cdots a_r m_1 \cdots m_s}$
are symmetric with respect to the indices $a_i$, and antisymmetric with respect to the indices $m_i$. Such a local form can be associated with a boundary operator $\OO_{\zeta}|_{\del \Sigma_L}$ in the open gauged A-model,
\begin{equation}\label{generalboundaryoperator}
\OO_\zeta|_{\del \Sigma_L}  \: = \: (\zeta_{a_1 \cdots a_r m_1 \cdots m_s}
\circ \gamma )  \left[\prod_{j=1}^r    (\varphi + \psi|_{\del \Sigma_L} )^{a_j} \right]  \left[ \prod_{i=1}^s  ((Q_A\gamma)^{m_i} - \dd^A \gamma^{m_i} ) \right]  \, ,
\end{equation}
where $\gamma$ %$\gamma^k$ are the local coordinates on the equivariant brane
%maps
is a section $\gamma:\del\Sigma_L\rightarrow E_L$ of the associated bundle $E_L=P_{\del \Sigma_L}\times_G L$ (which looks like a map $\gamma:\del\Sigma_L\rightarrow L$ locally on $\del\Sigma_L$), where $P_{\del \Sigma_L}$ is the principal $G$-bundle over $\del \Sigma_L$, and where $\psi|_{\del \Sigma_L} $ is the restriction of $\psi$ to the boundary in question. 
In particular, we have
\begin{equation}\label{NJ2}
(\dd_{\del\Sigma_L} + Q_A ) \; \OO_\zeta  =   \OO_{\dd_G \zeta}  \ ,
\end{equation}
where $\dd_{\partial\Sigma_L}$ is the exterior derivative on the worldsheet boundary $\del\Sigma_L$, while $\dd_G$ is the Cartan operator defined on $\Omega^\bullet_G (L)$. Just like bulk operators, $\OO_\zeta$ can be decomposed with respect to the form degree on the worldsheet boundary,
\begin{equation*}
\OO_\zeta \ = \ \OO_\zeta^{(0)} \ +\ \OO_\zeta^{(1)} ,  
\end{equation*}
%where, in particular,
%\begin{equation}
%\OO_\alpha^{(0)}  \ = \ (\alpha_{a_1 \cdots a_r k_1 %\cdots k_p \bar{l_1} \cdots \bar{l_q}}
%\circ \phi )  \left(\prod_{j=1}^r   \varphi^{a_j} \right)  \left( \prod_{i=1}^p  \chi^{k_i}  \right)   
%\left( \prod_{i=1}^q  \ov{\chi^{l_i}} \right)  \ ,
%\label{3.8}
%\end{equation}
%is a local operator.
If it is assumed that $\dd_G \zeta=0$, then \eqref{NJ2} splits into the descent equations
%Then in terms of this decomposition identity (\ref{3.7}) breaks into
\begin{align}
\dd_{\del\Sigma_L} \ \OO_\zeta^{(1)} \ &=0 \   \\
\dd_{\del\Sigma_L} \ \OO_\zeta^{(0)} \ &= \ - \; Q_A \; \OO_\zeta^{(1)} \   \\
Q_A \; \OO_\zeta^{(0)} \ &= \ 0, \
\end{align}
% Now assume that $\alpha$ is 
%$\dd_G$-closed and decompose $\OO_\alpha$ according to the form degree over $\Sigma$,
%i.e. write 
where, for example,
\begin{equation}\label{localboundaryoperator}
\OO_\zeta^{(0)}  \ = \ (\zeta_{a_1 \cdots a_r m_1 \cdots m_s }
\circ \gamma )  \left(\prod_{j=1}^r   \varphi^{a_j} \right) \left( \prod_{i=1}^s  (Q_A\gamma)^{m_i}\right)   \ .
\end{equation}
Thus, if $\nu$ is a $j$-dimensional homology cycle in $\del \Sigma_L$,\footnote{For $j=1$, $\nu$ is taken to be $\del \Sigma_L$, which is also the appropriate choice for noncompact boundaries.} 
%IMPORTANT NOTE: HERE IT IS ASSUMED, AS IN THROUGHOUT THE PAPER, THAT ALL FIELDS GO TO ZERO AT INFINITY ON THE WORLDSHEET
one can then define the $Q_A$-invariant operators
\begin{equation}
W_{\del \Sigma_L} (\zeta , \nu ) \ := \ \int_\nu \ \OO_\zeta^{(j)}  \ .  
\end{equation}

The most general correlation function based on the above bulk and boundary operators can then be written down (for $\Sigma=I\times \R$) as the following path integral
\begin{equation}
\boxed{
\int {\mathcal D} (A, \phi, \varphi,  \xi,\rho , \eta , \kpp,  \psi ,  \chi,b,c) \ \ 
e^{-(S_{A}+S_{BRST})} \ \prod_i \: W(\alpha_i , \gamma_i ) \prod_j \: W_{\del \Sigma_0}(\zeta_j , \nu_j )\prod_k \: W_{\del \Sigma_{\pi}}(\zeta'_k , \nu'_k ) \ ,
\label{3.11}}
\end{equation}
where $b$ and $c$ are ghost fields which appear in $S_{BRST}$.
% Here, the path integration is taken such that $\phi$ is restricted to a fixed topological sector,
%i.e., $\phi_{\ast} [\Sigma] \in H_2 (E)$ and $\phi_{\ast} [\del\Sigma] \in H_1 (E_L)$.
%IMPORTANT NOTE: THIS SHOULD BE CORRECT IN SOME SENSE, SEE PAGE 16 OF LERCHE'S SPECIAL GEOMETRY REVIEW, BUT TO BE ON THE SAFE SIDE, WE REMOVE IT. ALSO I DON'T THINK BAPTISTA IS ACCURATE WHEN HE SAYS PATH INTEGRATION IS TAKEN OVER A FIXED SECTOR, THERE SHOULD ACTUALLY BE A SUM OVER TOPOLOGICAL SECTORS LIKE IN THE MIRROR SYMMETRY BOOK.
%i.e., $\phi_{\ast} [\Sigma] \in H_2^G (X)$ and $\phi_{\ast} [\del\Sigma] \in H_1^G (L)$. 
%IMPORTANT NOTE:BAPTISTA IS WRONG IN THIS PART, THE FIELD PHI IS ALWAYS A SECTION OF THE ASSOCIATED BUNDLE, I.E., A MAP FROM THE RIEMANN SURFACE TO E. ANOTHER CONSESCUTIVE PUSHFORWARD IS REQUIRED TO GET AN EQUIVARIANT COHOMOLOGY CLASS OF X, SEE FIGUEORA O' FARRILL AND STANCIU EQUATION 2.10. also see see page 16 of Lerche's special geometry review.

Before %ending this subsection
proceeding, we shall return to %the
our example of $X= \CP^{N-1}$, where any equivariant A-brane wraps a Lagrangian submanifold $T^{N-1}$. Here, we note that 
\begin{equation}
Q_A \gamma^i=Q_A \frac{(\textrm{log } Z^i-\textrm{log } \ov{Z}^i)}{2i}=\frac{1}{2i}\Big(\frac{\chi^i}{Z^i}-\frac{\ov{\chi}^i}{\ov{Z}^i}\Big)=\frac{1}{2i|Z^i|^2}\Big(\chi^i\ov{Z}^i - \ov{\chi}^i{Z}^i\Big)
\end{equation}
is well-defined (since $|Z^i|^2=c_i/c_N$ at the boundaries), and is nonzero since the only fermionic Dirichlet boundary condition involving $\chi^i$ and $\ov{\chi}^i$ is $(\ov{Z}^i\chi^i +{Z}^i\ov{\chi}^i)=0$ (c.f. \eqref{GbcAcpn}). In addition, the boundary conditions \eqref{GbcAcpn} imply that $d^A\gamma^i$ is nonzero at the boundaries. Moreover, the boundary condition $(\ov{\til{\lam}}_{+c}-\til{\lam}_{-c})=0$ translates to $(\psi_{zc}+\psi_{\bar{z}c})=\psi_{1 c}=0$, and $\psi_c |_{\ds}=i(\psi_{zc}-\psi_{\bar{z}c})dx^2=\psi_{2c}dx^2$ is nonzero and well-defined. Finally, the boundary condition for $\varphi_c$ is $\del_{1}\varphi_c=0$ (which comes from $\del_1 \til{\si}_c =0$), and thus, $\varphi_c$ is nonzero at the boundaries. Hence,  boundary operators of the form \eqref{generalboundaryoperator} are nonzero and well-defined.
% and %%when $d_{G}\zeta =0$, 
%those which belong to the $Q_A$-cohomology can be identified with elements of $H^{\bullet}_{U(1)^{N-1}}(T^{N-1})$.  
%IMPORTANT NOTE: WE DID NOT ACTUALLY EXPLAIN HOW THE OBSERVABLES ARE ELEMENTS OF THE COHOMOLOGY EARLIER, WE ONLY SHOWED THAT THE d_G invariant forms lead to Q_A invariant observables. But it can be seen that the action of Q_A is essentially the action of d_G, at least for local observables. BUT to be safe I shall take it out.

%\textit{$Q_A^2\neq 0$ anomaly} 

%GLSM does not happen so won't happen in GNLSM. The two different types of vortices correspond when we take the limit, but one can argue that since the vortex equations have changed their effects might be different. Maybe can use LG condition for $Q^2=0$ somehow, see chapter 39.

%\subsection{Localization and Moduli Space of Open Symplectic Vortices}

Now, any supersymmetric path integral localizes to the bosonic field configurations that are fixed points of the supersymmetry \cite{hori2003mirror}. For the open gauged A-model, these field configurations can be read from the $Q_A$ variations of the fermionic fields in \eqref{3.3}, after integrating out the auxiliary fields. They correspond to the solutions of 
\begin{align}
&\db^A \phi = 0          \label{3.12}   \\
&\ast F_A  +  e^2 (\mu \circ \phi +r )= 0    \nonumber  \\
&\nabla^A \varphi = \varphi^a  (\he_a \circ \phi) = 0  \ . \nonumber 
\end{align} 
%IMPORTANT NOTE: There is one more equation from  Q_A\eta= [\phi^a T_a, \xi^b T_b], I think this is not considered since \phi^a T_a is zero if \phi^a \til{e}_a=0, which is the second equation above. This is probably true since there is an antihomomorphism from the Lie algebra to the Killing vector fields.   
%The usual credo says that a path-integral with a fermionic symmetry localizes to the bosonic field 
%configurations that are fixed points of the symmetry. Since $Q_A$ can be regarded as a generator of
%one such  symmetry, we will be interested in the bosonic field configurations annihilated by $Q_A$. These
%field configurations can be read from (\ref{3.3}) and, after eliminating the auxiliary fields, are precisely 
%the solutions of

The first two equations are known as 
the symplectic  vortex equations on an infinite strip, and were introduced by Cieliebak et al. in \cite{Cieliebak}, and are a generalization of the typical Nielsen-Olsen vortex equations on a strip. In what follows, we shall refer to them as the open symplectic vortex equations. The last two equations are non-trivial, but in most interesting cases that we will consider have the trivial solution $\varphi = 0$ \cite{Baptista1}, and therefore we can ignore them in these cases. %henceforth.
%IMpaper.
 For the first two equations of \eqref{3.12}, the boundary condition used by Cieliebak et al. on the strip was that each boundary component of the strip was mapped to a $G$-invariant Lagrangian submanifold of $X$, and this is precisely the boundary condition we found in Section 5.5. In addition, for the second equation, we have found the boundary conditions  ${A}_{1a}=0$, $\del_1A_{0a}=0$ and ${\mu}_a=-{r}_a$. %and Lagrangian boundary conditions for $Z^i$.
%IMPORR. 
 %In particular, the condition that ${\mu}_a=-{r}_a$ on a $G$-invariant Lagrangian submanifold was predicted by Lemma 4.1 of \cite{Cieliebak}. %IMPORTANT NOTE: Setter showed a simpler proof in his thesis %we have been using thus far. % For abelian $G$ and toric $X$, this is precisely the boundary condition we have been using thus far, whereby the Lagrangian submanifolds are torus fibers of $X$. 
 For the example of $X=\CP^{N-1}$, the open symplectic vortex equations read 
\begin{equation}
\begin{aligned}
&\del^A_{\bar{z}} Z^i = 0           \\
&\ast \til{F}_{Ac}  +  \til{e}_c^2 (\til{\mu}_c\circ Z  +\til{r}_c)= 0 . \end{aligned}  
\end{equation}
Recall that the boundary conditions in this case are Lagrangian boundary conditions for $Z^i$ which map each boundary to a $U(1)^{N-1}$-invariant Lagrangian torus $T^{N-1}$ as well as $\til{\mu}_c=-\til{r}_c$ and  $\ast \til{F}_{Ac}=0$. %In particular, the condition $\til{\mu}_c=-\til{r}_c$ agrees with Lemma 4.1 of \cite{Cieliebak}.

% They were first
%written down in \cite{C-G-S;MiR} and generalize the usual Nielsen-Olsen vortex equations. The two equations
%involving $\varphi$, although in general non-trivial, in many cases of interest only have the 
%$\varphi = 0$ solution, and so in these cases can be discarded. It can be shown, for example, that if
%$0$ is a regular value of the moment map $\mu$, then given any fixed homotopy class of sections of $E$, 
%for a sufficiently big value of the constant $e^2\, (\text{Vol }\Sigma)$ any solution of (\ref{3.12}) with $\phi$
%in that class has zero $\varphi$ \cite[lem. 4.2]{C-G-M-S}. Another instance, in the abelian case: if $G$ is
%a torus, $X$ is compact connected and $(\int_\Sigma F_A)/ (e^2 \text{Vol }\Sigma)$ is a regular value 
%of $\mu$, then any solution of (\ref{3.12}) has zero $\varphi$ \cite{Bap}.
%Nonetheless, even after discarding the last line of (\ref{3.12}), the two remaining (vortex) equations are very non-trivial.
%For example, unlike monopoles or instantons, no explicit non-trivial solution of these equations is known, and this for any $\Sigma$, 
%$X$ or $G$, including the non-compact  $\Sigma =\C$. 

%For the topological field theory, however,
%the main objects of interest are not the solutions themselves, but rather the spaces of all solutions,
%or more precisely the moduli spaces of solutions up to gauge equivalence.

The localization of supersymmetric path integrals of the form \eqref{3.11} thus reduce them to ordinary integrals of differential forms over the moduli spaces of open symplectic vortices, which are the spaces of solutions to the open symplectic vortex equations up to gauge equivalence. These moduli spaces are finite-dimensional, though they may be noncompact and contain singularities. The (infinite-dimensional) path integrals thus reduce to finite-dimensional integrals, which are well-defined mathematically (modulo issues related to the aforementioned noncompactness and singularities of the moduli spaces). These finite-dimensional integrals give us numbers which can be identified with the open version of the Hamiltonian Gromov-Witten invariants of $X$ in \cite{Cieliebak, Mundet, Cieliebak2}.  

 %These vortex moduli spaces are
%in general finite-dimensional, have a natural K\"ahler structure, but may contain singularities and be
%non-compact. Their virtual complex dimension, as given by elliptic theory, is
%\begin{equation}
%({\rm dim}_\C X - {\rm dim}G )(1-g) +   \langle  c_1^G (TX)\: ,\: \phi (\Sigma) \rangle \ ,
%\label{3.13}
%\end{equation}
%and is basically just the difference of the indices of the operators in (\ref{2.7}) \cite{C-G-M-S}.

%The standard heuristic arguments of TFT \cite{W1, W2} then say that, in favourable cases, the path-integrals 
%(\ref{3.11}) reduce to finite-dimensional integrals of differential forms over the vortex moduli spaces. These
%finite-dimensional integrals are completely classical objects and, modulo (in fact very difficult) problems
%related to the singularities and non-compactness of the moduli spaces, make sense in the realm of traditional 
%mathematics, as opposed to the path-integrals. The numbers provided be these finite-dimensional integrals 
%can in fact be identified with the so-called Hamiltonian Gromov-Witten invariants of $X$, which have been
%defined using a very different, rigourous, universal construction. All this story is analogous to the well 
%known case of the non-gauged sigma-model, which leads to the Gromov-Witten invariants; it is spelled out
%in detail in \cite{Bap}.

%Before proceeding to compute the dimension of the moduli spaces, 
We note that in the limit where $e^2 \rightarrow +\infty$, a dynamically gauged sigma model with target $X$
flows to an ordinary sigma model with target $X/\!/G$ \cite{Baptista1}. Hence, in analogy with the closed case \cite{Gaio}, it is predicted that there is a relationship between the open Hamiltonian Gromov-Witten invariants of $X$ and the open Gromov-Witten invariants of $X/\!/G$ \cite{Wang}.
%IMPORTANT NOTE: PHYSICALLY THIS IS JUST DUE TO THE CORRELATION FUNCTIONS BEING INVARIANT UNDER THE RG FLOW SINCE THE MODEL IS TOPOLOGICAL 
%consequence one expects some relation to exist between the HGW-invariants of $X$ and the GW-invariants of 
%$X/\!/G$ \cite{Ga-Sa}.

% We now end this section with a few references. Regarding the vortex moduli spaces, there has been a longstanding 
%interest in them. Starting with the simplest case of the abelian Higgs models --- where $X = \C$ and 
%$G = U(1)$ --- about thirty years ago, the structure of these spaces has been investigated in several 
%particular examples, mainly with $X$ a vector space. A hectic set of references is for example
%\cite{Ba} within the more mathematical literature and \cite{E-I-N-O-S, W4, M-P} within theoretical physics. 
%The Hamiltonian Gromov-Witten invariants, in comparison, have only recently been defined \cite{C-G-M-S, C-G-S;MiR}. They have been 
%furthermore studied in \cite{C-S, Ga-Sa}. 
\vspace{0.2cm}
\mbox{}\par\nobreak
\noindent
%\textit{R-anomaly and Dimension of Moduli Space of Open Symplectic Vortices on a Riemann Surface with Boundaries} 
\subsection{Dimension of Moduli Space of Open Symplectic Vortices and R-anomaly}
The boundary axial R-anomaly has been previously used to compute the %(real) 
dimension of moduli spaces of holomorphic maps from an open Riemann surface to a K\"ahler manifold whereby the boundaries are mapped to Lagrangian submanifolds \cite{hori2003mirror}. One can %also use the axial R-anomaly 
%use similar techniques to 
also compute the dimension of moduli spaces of symplectic vortices on a closed Riemann surface \cite{Baptista1}. 
%IMPORTANT NOTE: I said similar techniques, because in Baptista the axial R-anomaly isn't directly used to compute the dimension of the moduli space. Baptista says that the virtual dimension is the difference between the indices from the matter and gauge parts.
Using insights from these results, we may attempt to compute the boundary axial R-anomaly for the open gauged A-model and find the dimension of a moduli space of open symplectic vortices on an open Riemann surface. In what follows, we shall assume that we have a compact open Riemann surface, $\Sigma$, with arbitrary genus and an arbitrary number of boundary circles.

The axial R-anomaly can be deduced by investigating the zero-modes of the fermionic fields via their kinetic terms 
\begin{equation}\label{zeromodekineticterms}
\frac{1}{2\pi}\int_{\Sigma}dz\wedge d \zb\Bigg(
%\frac{i}{e^2}  \eta_a  
%{\nabla}^A \ast \psi^a - \frac{1}{2e^2} \kpp_a  {\nabla}^A \psi^a 
\frac{i}{e^2}\frac{1}{\sqrt{2}}\lam_a\nabla^A_z\psi^a_{\zb}+\frac{i}{e^2}\frac{1}{\sqrt{2}}\ov{\lam}_a\nabla^A_{\zb}\psi^a_{\z}
-\frac{1}{2}g_{j\ov{k}} \rho^j_{\zb}{(\phi^\ast {\nabla}^A )_{\z}}\ov{\chi}^{\ov{k}}
+\frac{1}{2} g_{j\ov{k}}\,  \ov{\rho}^{\ov{k}}_{\z} (\phi^\ast {\nabla}^A )_{\zb} {\chi}^{j}
%+ \frac{1}{2} g_{j\ov{k}}\,  \rho^j \wedge (\phi^\ast {\nabla}^A ) \ov{\chi}^{\ov{k}}
%+ \frac{1}{2} g_{j\ov{k}} \, \ov{\rho}^{\ov{k}} \wedge (\phi^\ast {\nabla}^A ) \chi^j 
\Bigg),
\end{equation}
where we have defined the fields 
\begin{equation}\label{kppetalam}
\begin{aligned}
\lam_a&=\frac{i2\sqrt{2}\eta_a+i\sqrt{2}\kpp_a}{2}\\
\ov{\lam}_a&=\frac{i2\sqrt{2}\eta_a-i\sqrt{2}\kpp_a}{2}.
\end{aligned}
\end{equation}

In order to evaluate the anomaly, we ought to double the open worldsheet as well as the bundles on it, as in \cite{hori2003mirror}, in order to form a closed worldsheet, on which the indices of the relevant operators can be evaluated. This is done by taking the metric on the worldsheet close to each component of $\ds$ to be that of a flat cylinder, and gluing $\Sigma$ with its orientation reversal, $\Sigma^*$. The resulting closed Riemann surface is denoted $\Sigma\#\Sigma^*$. 

The corresponding bundles over $\Sigma$ and $\Sigma^*$ shall be glued using the relevant boundary conditions. 
To demonstrate this, let us first consider the index of the twisted Dirac operator $\til{\mathcal{D}}$ which acts on the fermionic fields in the first two terms in the parantheses of \eqref{zeromodekineticterms},
\begin{equation}
\textrm{Index }\til{\mathcal{D}}=\#[(\psi_z,\overline{\psi}_{\zb})\textrm{ zero modes}]-\#[(\overline{\lam},{\lam})\textrm{ zero modes}].
\end{equation}
%for $\nabla^A_{{\zb}}$. 
The boundary conditions (c.f. Section 5.5) for these fermionic fields are
\begin{equation}\label{gluingconditions}
\begin{aligned}
\psi_z^a&=-\psi_{\zb}^a\\
\ov{\lam}_a&=\lam_a.
\end{aligned}
\end{equation}
%Next, note that the index of $\nabla^A_{{\zb}}$ is equal to the index 
Now, let us consider $\ov{\lam}_a$ and $\psi^a_z$ as fields on $\Sigma$, and $\lam_a$ and $-\psi_{\zb}^a$ as fields on $\Sigma^*$. By the boundary conditions above, $\ov{\lam}_a$ on $\Sigma$ and ${\lam}_a$ on $\Sigma^*$ continuously glue along $\del \Sigma$, and define a continuous section of $\mathfrak{g}_P\#\mathfrak{g}_P$ which we denote $\ov{\lam}_a\#{\lam}_a$. Likewise,  $\psi^a_z$ and $-\psi_{\zb}^a$ define a continuous section $\psi^a_z\#(-\psi_{\zb}^a)$ of $(\mathfrak{g}_P\#\mathfrak{g}_P) \otimes K_{\Sigma\#\Sigma^*}$. Then, if  $\nabla^A_{\zb}\ov{\lam}_a=0$, 
$\ov{\lam}_a\#{\lam}_a$ is holomorphic on $\Sigma\subset\Sigma\#\Sigma^*$. If $\nabla^A_{{z}}{\lam}_a=0$, ${\lam}_a$ is holomorphic on $\Sigma^*$ due to orientation reversal, and hence
$\ov{\lam}_a\#{\lam}_a$ is holomorphic on $\Sigma^*\subset\Sigma\#\Sigma^*$. Thus, if $\nabla^A_{\zb}\ov{\lam}_a=0$ on $\Sigma$ and $\nabla^A_{{z}}{\lam}_a=0$ on $\Sigma^*$, then $\ov{\lam}_a\#{\lam}_a$ is entirely holomorphic on $\Sigma\#\Sigma^*$.

 Similarly, if $\nabla^A_{\zb}{\psi}_z^a=0$ on $\Sigma$ and $\nabla^A_{z}{\psi}_{\zb}^a=0$ on $\Sigma^*$, then $\psi^a_z\#(-\psi_{\zb}^a)$  
 is entirely holomorphic on $\Sigma\#\Sigma^*$. %According to \cite{hori2003mirror}, t
 This implies that the index of 
 %$\nabla^{A}_{\bar{z}}$
 $\til{\mathcal{D}}$ is the 
index of the Dolbeault operator of $(\mathfrak{g}_P\#\mathfrak{g}_P)$, i.e.,
%which is calculated to be 
%and therefore
\begin{equation}\label{gauginoanomaly}
\begin{aligned}
{\rm Index }\  \til{\mathcal{D}}&=\textrm{dim }H^0(\Sigma\#\Sigma^*,\mathfrak{g}_P\#\mathfrak{g}_P)-\textrm{dim }H^0(\Sigma\#\Sigma^*,(\mathfrak{g}_P\#\mathfrak{g}_P)\otimes K_{\Sigma\#\Sigma^*})\\&=c_1(\mathfrak{g}_P\#\mathfrak{g}_P)+\textrm{dim}(G)(2-2g-h)
\end{aligned}
\end{equation} 
where $g$ is the genus and $h$ is the number of boundary circles of the worldsheet $\Sigma$. For compact $G$, $c_1(\mathfrak{g}_P\#\mathfrak{g}_P)=0$.

The index for 
%$\phi^\ast \nabla^A_{\zb}$ 
the twisted Dirac operator $\til{\mathcal{D}}'$ which acts on the fermionic fields in the last two terms in the parantheses of \eqref{zeromodekineticterms}, 
\begin{equation}
\textrm{Index }\til{\mathcal{D}}'=\#[(\chi,\overline{\chi})\textrm{ zero modes}]-\#[(\overline{\rho}_z,{\rho}_{\zb})\textrm{ zero modes}],
\end{equation}
can analogously be determined.\footnote{Setter \cite{Setter} studied a specialized version of the non-dynamical open $U(1)$-gauged A-model, where the gauge field is the spin connection on the open Riemann surface, and performed a similar anomaly computation. In comparison, we are considering a dynamically gauged theory with arbitrary compact nonabelian gauge group.} 
%The proof is a generalization of that found in Section 39.3.3 of \cite{hori2003mirror} for the non-gauged case. % and thus we shall only outline it. 
 Before A-twisting, we found boundary conditions (c.f. Section 5.5) which map each boundary to a $G$-invariant Lagrangian submanifold, $L$. In what follows, we shall assume each boundary component is mapped to the same Lagrangian submanifold. In other words, at $\del \Sigma$, the scalar fields $\phi^I$ constitute a section $\phi :\del\Sigma \rightarrow E_L$ of the associated bundle $E_L=P_{\del \Sigma}\times_G L$. %and we have 
%Hence,
We shall \textit{fix} a bosonic background of the open Riemann surface $\Sigma$, i.e., a map
\begin{equation}
\phi : (\Sigma, \del \Sigma)\rightarrow (P\times_G X,P_{\del \Sigma}\times_G L).
\end{equation}
%labelled by $\phi_{\ast} [\Sigma] \in H_2 (E)$ and $\phi_{\ast} [\del\Sigma] \in H_1 (E_L)$.
%IMPORTANT NOTE: TOOK OUT THE ABOVE TO BE ON THE SAFE SIDE, BECAUSE \Sigma has a boundary, so it's likely that relative cohomology enters here. but the statement which was there previously should be correct, see page 16 of Lerche's special geometry review.
%IMPORTANT NOTE: I ADDED THE ABOVE BECAUSE JUST LIKE THE FIRST CHERN CLASS, THE MASLOV INDEX SHOULD ALSO DEPEND ON HOMOLOGY CLASSES. WE ARE NOT TAKING THE INNER PRODUCT OF THE MASLOV INDEX WITH SOME HOMOLOGY CLASS IN THE FINAL ANSWER (THIS IS DONE IN BAPTISTA, ALSO IN TAN AND YAGI), SO IT IS UNDERSTOOD THAT IN THE FINAL ANSWER THE MASLOV INDEX IS A NUMBER. THEREFORE FOR IT TO BE A NUMBER WE MUST FIX SOME HOMOLOGY CLASS FOR THE MAP PHI. HORI DOES NOT EXPLICITLY TALK ABOUT THE HOMOLOGY CLASS FOR THE BOUNDARY WRAPPING A 1-CYCLE IN THE LAGRANGIAN SUBMANIFOLD, BUT IT SHOULD BE UNDERSTOOD FROM PAGE 813 ABOVE EQN 39.220 AND SECTION 39.3.4. ALSO SEE KEVIN SETTER'S THESIS
Next, recall from Section 2 that before A-twisting, the matter fermionic fields are the sections  $\Psi_{\pm}  \in  \Gamma (\Sigma ;  S_{\pm} \otimes \phi^\ast \ver )$.\footnote{For this subsection, we shall follow the real notation of \cite{hori2003mirror} whereby $\psi^I_{\pm}$ is written as $\Psi^I_{\pm}$.} The boundary conditions for the fermion fields  were stated in Section 5.5 for a Minkowski worldsheet  as ($\Psi^I_-+\Psi^I_+$) is tangent to $L$ and ($\Psi^I_--\Psi^I_+$) is normal to $L$.
%IMPORTANT NOTE: IN SECTION 5.5 THE BOUNDARY CONDITIONS WERE FOUND FOR AN INFINITE STRIP, BUT HERE WE HAVE AN ARBITRARY COMPACT OPEN RIEMANN SURFACE, HOWEVER HORI MAKES AN ANALOGOUS ASSUMPTION; HE SHOWS THE BOUNDARY CONDITIONS FOR THE HALF PLANE, THEN CARRIES THEM OVER TO THE CASE OF OF AN ARBITRARY COMPACT OPEN RIEMANN SURFACE. ADDED THE FOLLOWING ON 12TH APRIL, IT ALSO SEEMS THAT HORI (AND I) MAP ALL THE BOUNDARY COMPONENTS TO A SINGLE LAGRANGIAN SUBMANIFOLD (IN MY CASE LOCALLY ON THE WORLDSHEET) 
 This can be restated for a Euclidean worldsheet. First, as in \cite{hori2003mirror}, the spin bundles of opposite chirality ($S_+$ and $S_-$) can be identified at each boundary.  Then, the boundary condition can be written succinctly as
\begin{equation}\label{tauboundarycondition}
\tau(\Psi_-)=\Psi_+,
\end{equation} 
where the map $\tau: \ver|_L\rightarrow  \ver|_L$ is the identity on  $\textrm{ker } d\pi_{E_L} \rightarrow E_L$ (which  
%IMPORTANT NOTE: MADE THE FOLLOWING CHANGE ON 4TH APRIL 2017, SEE SETTER PAGE 98 AND 144.
%locally looks like $\Sigma \times TL \rightarrow \Sigma \times L$
looks like $TL$ on a local patch of $\del \Sigma$), and is $(-1)\times$ the identity on $\textrm{ker } d\pi_{E_{L}}|_{\perp} \rightarrow E_{L}$ (which  
%IMPORTANT NOTE: MADE THE FOLLOWING CHANGE ON 4TH APRIL 2017, SEE SETTER PAGE 98 AND 144.
%locally looks like $\Sigma \times NL \rightarrow \Sigma \times L$
looks like $NL$ on a local patch of $\del \Sigma$).\footnote{
%The vector bundles $\textrm{ker } d\pi_{E_L} \rightarrow E_L$ and $\textrm{ker } d\pi_{E_{L\perp}} \rightarrow E_L$ are respectively the kernels of the derivatives $d\pi_{E_L}:TE_L\rightarrow T\Sigma$ and $d\pi_{E_{L\perp}}:NE_L\rightarrow T\Sigma$. 
The vector bundle $\textrm{ker } d\pi_{E_L} \rightarrow E_L$ is the kernel of the derivative $d\pi_{E_L}:TE_L\rightarrow T\del\Sigma$, while the vector bundle $\textrm{ker } d\pi_{E_{L}}|_{\perp} \rightarrow E_L$ 
%IMPORTANT NOTE: THE COMPLEMENT IS ALSO A BUNDLE OVER E_L, because NL is a bundle over L, see ncatlab. As an example of the latter, imagine a vertical line in R^2, and compare the restriction of the tangent bundle of R^2 to the line, the tangent bundle of the line, and the restriction of the tangent bundle to the line mod the tangent bundle of the line (which is  the normal bundle).
is the orthogonal complement of $\textrm{ker } d\pi_{E_L}$ in $\textrm{ker } d\pi_{E}|_L$, i.e., $\textrm{ker } d\pi_E|_L=\textrm{ker } d\pi_{E_L}\oplus \textrm{ker }d\pi_{E_{L}}|_{\perp}$, which looks like $TX|_L=TL\oplus NL$ locally on $\del\Sigma$. %IMPORTANT NOTE: NCATLAB DESCRIBES THIS DECOMPOSITION
 The orthogonal complement is defined with respect to the metric on $\textrm{ker } d\pi_{E}$, which is inherited from the metric on $X$ (\cite{Setter}, Appendix C).}

Next, $\Psi^I_{\pm}$ is decomposed into $\psi^i_{\pm}$ and $\ov{\psi}^{\ov{\imath}}_{\pm}$, which are valued in $\phi^*\textrm{ker } d\pi_E^{(1,0)}$ (which looks like $\phi^* T^{(1,0)}X$ locally on $\Sigma$) and $\phi^*\textrm{ker } d\pi_E^{(0,1)}$ (which looks like $\phi^*T^{(0,1)}X$ locally on $\Sigma$), respectively.\footnote{
%IMPORTANT NOTE: THE FOLLOWING SHOULD STILL BE CORRECT, DESPITE THE FACT THAT I HAVE HIDDEN IT. THIS IS BECAUSE THE HOLOMORPHIC AND ANTIHOLOMORPHIC SUBBUNDLES EACH EXIST INDEPENDENTLY AS BUNDLES OVER THE WORLDSHEET, AND THEREFORE THERE EXIST PROJECTIONS FOR EACH OF THEM, AS WELL AS DERIVATIVES OF THESE PROJECTIONS The vector bundles $\textrm{ker } d\pi_{E}^{(1,0)} \rightarrow E$ and $\textrm{ker } d\pi_{E}^{(0,1)} \rightarrow E$ are respectively the kernels of the derivatives $d\pi_{E}^{(1,0)}:T^{(1,0)}E\rightarrow T\Sigma$ and $d\pi_{E}^{(0,1)}:T^{(0,1)}E\rightarrow T\Sigma$. 
The vector bundle $\textrm{ker } d\pi_{E}$ inherits a complex structure from that of $X$ (\cite{Setter}, Appendix C), and therefore its  complexification can be decomposed into holomorphic and antiholomorphic subbundles as $\textrm{ker } d\pi_E\otimes \C=\textrm{ker } d\pi_E^{(1,0)}\oplus \textrm{ker } d\pi_E^{(0,1)}$.} The map $\tau$ acts linearly on $\textrm{ker } d\pi_{E}^{(1,0)}|_L \oplus \textrm{ker } d\pi_{E}^{(0,1)}|_L$, whereby the $(1,0)$ and $(0,1)$ components are exchanged. The reason for this is that if $t\in \textrm{ker } d\pi_{E_L}$, then $Jt\in \textrm{ker } d\pi_{E_{L}}|_{\perp}$, and therefore $\tau:(t-iJt)\in \textrm{ker } d\pi_{E}^{(1,0)}|_L \rightarrow (t+iJt)\in \textrm{ker } d\pi_{E}^{(0,1)}|_L$ (this follows from the definition of $\tau$ below \eqref{tauboundarycondition}).
%IMPORTANT NOTE: SEE PAGE 125 OF INTRODUCTION TO SYMPLECTIC TOPOLOGY TO UNDERSTAND WHY WE HAVE THE ABOVE SPECIFIC FORMS FOR THE HOLOMORPHIC AND ANTIHOLOMORPHIC VECTOR FIELDS
 Hence,
\begin{equation}
\tau: \phi^*\textrm{ker } d\pi_{E}^{(1,0)}|_{\del \Sigma}\rightarrow \phi^*\textrm{ker } d\pi_{E}^{(0,1)}|_{\del \Sigma}.
\end{equation}
The boundary condition \eqref{tauboundarycondition} can then be written as 
\begin{equation}
\tau(\psi_-)=\ov{\psi}_+\textrm{,  }\tau(\ov{\psi}_-)={\psi}_+.
\end{equation}

After A-twisting, these become
\begin{equation}
\tau(\chi)=\ov{\chi}\textrm{,  }\tau(\ov{\rho}_{\z})=\rho_{\zb}.
\end{equation}
These boundary conditions are analogous to those given 
in \eqref{gluingconditions}, and can be used to continuously glue $\phi^*\textrm{ker } d\pi_{E}^{(1,0)}$ over $\Sigma$ with $\phi^*\textrm{ker } d\pi_{E}^{(0,1)}$ over the orientation reversal $\Sigma^*$, by considering $\chi$ and $\ov{\rho}_{\z}$ as fields on $\Sigma$ and $\ov{\chi}$ and ${\rho_{\zb}}$ as fields on $\Sigma^*$. In this way, we obtain a continuous section of $\phi^*\textrm{ker } d\pi_{E}^{(1,0)}\#\phi^*\textrm{ker } d\pi_{E}^{(0,1)}$ (denoted $\chi \#\ov{\chi}$) and a continuous section of $(\phi^*\textrm{ker } d\pi_{E}^{(1,0)}\#\phi^*\textrm{ker } d\pi_{E}^{(0,1)})^*\otimes K_{\Sigma \# \Sigma^*}$ (denoted $\ov{\rho}_{\z}\#\rho_{\zb}$). 

%Now, note that the index of $\phi^\ast \nabla^A_{\zb}$ is 
Now, if 
$\phi^\ast \nabla^A_{\zb}\chi=0$, $\chi \#\ov{\chi}$ is holomorphic on $\Sigma\subset \Sigma\#\Sigma^*$, and if 
$\phi^\ast \nabla^A_{\z}\ov{\chi}=0$, $\chi \#\ov{\chi}$ is holomorphic on $\Sigma^*\subset \Sigma\#\Sigma^*$ due to orientation reversal. Hence, if $\phi^\ast \nabla^A_{\zb}\chi=0$ and $\phi^\ast \nabla^A_{\z}\ov{\chi}=0$, then $\chi\#\ov{\chi}$ is entirely holomorphic on $\Sigma\#\Sigma^*$. Analogously, if $\phi^\ast \nabla^A_{\zb}\ov{\rho}_{\z}=0$ and $\phi^\ast \nabla^A_{\z}{\rho_{\zb}}=0$, $\ov{\rho}_{\z}\#\rho_{\zb}$ is entirely holomorphic on $\Sigma\#\Sigma^*$. This implies that the index of 
%$\phi^\ast \nabla^A_{\zb}$
$\til{\mathcal{D}}'$ is the index of the Dolbeault operator of $\phi^*\textrm{ker } d\pi_{E}^{(1,0)}\#\phi^*\textrm{ker } d\pi_{E}^{(0,1)}$, i.e.,
%and therefore
\begin{equation}\label{matteranomaly}
\begin{aligned}
{\rm Index }\  \til{\mathcal{D}}'&=\textrm{dim }H^0(\Sigma\#\Sigma^*,\mathcal{E}\#\mathcal{E}^*)-\textrm{dim }H^0(\Sigma\#\Sigma^*,(\mathcal{E}\#\mathcal{E}^*)^*\otimes K_{\Sigma\#\Sigma^*})\\&=c_1(\phi^*\textrm{ker } d\pi_{E}^{(1,0)}\#\phi^*\textrm{ker } d\pi_{E}^{(0,1)})+\textrm{rank}(\phi^*\textrm{ker } d\pi_{E}^{(1,0)})(2-2g-h),
\end{aligned}
\end{equation}
where $\mathcal{E}=\phi^*\textrm{ker } d\pi_{E}^{(1,0)}$.

In addition, we note that the map $\tau$ is associated 
with the orthogonal decomposition 
\begin{equation}
(\phi^*\textrm{ker } d\pi_E^{(1,0)})|_L=\phi^*[\textrm{ker } d\pi_{E_L}]^{(1,0)}\oplus \ i\phi^*[\textrm{ker } d\pi_{E_L}]^{(1,0)},
\end{equation}
which is obtained from $(\phi^*\textrm{ker } d\pi_E)|_L=\phi^*\textrm{ker } d\pi_{E_L}\oplus \phi^*\textrm{ker }d\pi_{E_{L}}|_{\perp}$ via the projection $(1-iJ):\phi^*\textrm{ker } d\pi_E\rightarrow \phi^*\textrm{ker } d\pi^{(1,0)}_E$. 
%IMPORTANT NOTE: MULTIPLICATION BY i on $\phi^*\textrm{ker }d\pi_{E_L}$ takes it to $\phi^*\textrm{ker }d\pi_{E_{L\perp}}$
Following the general argument in \cite{hori2003mirror}, this allows us to identify $c_1(\phi^*\textrm{ker } d\pi_{E}^{(1,0)}\#\phi^*\textrm{ker } d\pi_{E}^{(0,1)})$ with $\mu (\phi^*\textrm{ker } d\pi_E^{(1,0)},\phi^*[\textrm{ker } d\pi_{E_L}]^{(1,0)})$ which is known as the Maslov index of the pair $(\phi^*\textrm{ker } d\pi_E^{(1,0)},\phi^*[\textrm{ker } d\pi_{E_L}]^{(1,0)})$. Thus, we obtain 
\begin{equation}\label{matteranomaly2}
{\rm Index }\  \til{\mathcal{D}}'= \mu (\phi^*\textrm{ker } d\pi_E^{(1,0)},\phi^*[\textrm{ker } d\pi_{E_L}]^{(1,0)})+\textrm{dim}_{\C}(X)(2-2g-h),
\end{equation} 
where we have used $\textrm{rank}(\phi^*\textrm{ker } d\pi_{E}^{(1,0)})=\textrm{dim}_{\C}(X)$.

%After A-twisting, the quantum number of the fermions change so that $\chi$ is valued in $\phi^*\textrm{ker } d\pi_E^{(1,0)}$, where the vector bundle $\textrm{ker } d\pi_E^{(1,0)}$ is the kernel of the derivative $d\pi_E^{(1,0)}:T^{(1,0)}E\rightarrow T\Sigma$. The homomorphism 
%\begin{equation}
%\tau : (\phi^* \textrm{ker } d\pi^{(1,0)})\rightarrow (\phi^* \textrm{ker } d\pi^{(1,0)})^*
%\end{equation}
%defined by the Lagrangian boundary conditions which map the boundary to a $G$-invariant Lagrangian submanifold, $L$, is associated with the orthogonal decomposition 
%\begin{equation}
%(\phi^*\textrm{ker } d\pi_E^{(1,0)})|_L=\phi^*[\textrm{ker } d\pi_{E_L}]^{(1,0)}\oplus \ i\phi^*[\textrm{ker } d\pi_{E_L}]^{(1,0)},
%\end{equation}
%where the vector bundle $\textrm{ker } d\pi_{E_L} \rightarrow E_L$ is the kernel of derivative $d\pi_{E_L}:TE_L\rightarrow T\Sigma$, where $E_L=P\times_G L$. %These boundary conditions 

%The above information implies that
%\begin{equation}
%\textrm{index } \phi^\ast \nabla^A_{\zb}= \mu (\phi^*\textrm{ker } d\pi_E^{(1,0)},\phi^*[\textrm{ker } d\pi_{E_L}]^{(1,0)})+\textrm{dim}_{\C}(X)(2-2g-h),
%\end{equation} 
%where $\mu (\phi^*\textrm{ker} d\pi_E^{(1,0)},\phi^*[\textrm{ker } d\pi_{E_L}]^{(1,0)})$ is known as the Maslov index. 

We find from \eqref{gauginoanomaly} and \eqref{matteranomaly2} that for compact $G$, the axial R-anomaly is 
\begin{equation}\label{theboundaryRanomaly}
\boxed{
\begin{aligned}
%{\mathcal A} \: &=   {\rm index }\ \phi^\ast \nabla^A_{\zb} + 
%{\rm index }\  \nabla^A_{\zb} \ \\
\mu (\phi^*\textrm{ker } d\pi_E^{(1,0)},\phi^*[\textrm{ker } d\pi_{E_L}]^{(1,0)})+(\textrm{dim}_{\C}(X)+\textrm{dim}(G))(2-2g-h).
\end{aligned}}
\end{equation} 
Hence, in order for correlation functions to be nonzero, an appropriate number of boundary operators with 
suitable 
%whose 
axial R-charges 
%sum to $\mathcal{A}$  
should be inserted into the path integral, such that axial R-symmetry is preserved at the boundaries.
%IMPORTANT NOTE: AS HORI MENTIONS ON PAGE 810 OF THE MIRROR SYMMETRY BOOK, THE BULK ANOMALY PERSISTS, SO WHAT WE ARE ACTUALLY COMPUTING IS THE BOUNDARY R-ANOMALY. ALSO SEE PAGE 815 WHERE THE COMPOSITION WITH THE BULK ANOMALY IS DISCUSSED. 
 %IMPORTANT NOTE: This gives us the dimension of the moduli spaces of symplectic  vortices, since the path integral is nonzero and axial R-symmetry is preserved only if we include the correct number of fermionic operators to absorb the zero modes.
%IMPORTANT NOTE: ACTUALLY THIS IS NOT ACCURATE, IT IS POSSIBLE TO HAVE R-ANOMALY ZERO, BUT FERMION ZERO MODES PRESENT, IF THEIR NUMBERS MATCH IN SUCH A WAY TO MAKE THE INDICES ZERO.
%IMPORTANT NOTE: THE VECTOR R-SYMMETRY IS BROKEN AT THE BOUNDARY, SEE HORI, LINEAR MODELS. HOWEVER, IN THE BULK THE VECTOR R-SYMMETRY IS PRESERVED, SO IT GIVES ANOTHER SELECTION RULE FOR THE BULK OPERATORS, SEE PAGE 411 OF MIRROR SYMMETRY BOOK

%As in the closed case \cite{Baptista1},Cieliebak2}, the %virtual complex 
%IMPORTANT NOTE: TOOK OUT "VIRTUAL COMPLEX" BECAUSE HORI REFERS TO THIS IN THE NONGAUGED CASE AS THE REAL DIMENSION
The virtual real 
%IMPORTANT NOTE: THE EXPLANATION OF VIRTUAL IS IN WITTEN- MIRROR MANIFOLDS AND TOPOLOGICAL FIELD THEORY. REAL IS BECAUSE IN CIELIEBAK ET AL. THEY ARE EVALUATING THE REAL DIMENSION, AND IT CAN BE SEEN THAT IT IS TWICE THE EXPRESSION GIVEN IN BAPTISTA'S PAPER, WHICH HE CALLS THE VIRTUAL COMPLEX DIMENSIONS. ALSO, IT REDUCES FOR TRIVIAL G TO WHAT HORI CALLS THE REAL DIMENSION OF THE MODULI SPACE OF HOLOMORPHIC MAPS.
dimension of the moduli space of open symplectic vortices is given by the \textit{difference} of 
%the indices 
\eqref{matteranomaly2} and \eqref{gauginoanomaly} for compact $G$, which is
%, which, for compact $G$, is
%IMPORTANT NOTE: THE IDEA BEHIND THIS IS THAT THE ARGUMENTS OF CIELIEBAK ET AL ON PAGE 28 OF THEIR PAPER, AT THE TOP, HOLDS EVEN IN OUR CASE. THERE THEY COMPUTE THE INDEX OF THE LINEARIZED OPERATOR. IN OUR CASE THE ONLY NEW THING IS THE BOUNDARY CONDITIONS.
%IMPORTANT NOTE: 5TH APRIL 2017, THE LOGIC HERE IS THAT WE HAVE GLUED TWO OPEN RIEMANN SURFACES TO FORM A CLOSED RIEMANN SURFACE, AND EVALUATED INDICES ON THEM. FOR A CLOSED RIEMANN SURFACE, IT HAS BEEN ANALYZED IN \cite{Baptista1,Cieliebak2}THAT THE INDEX OF THE LINEAR OPERATOR WHICH GIVES THE DIMENSION OF THE MODULI SPACE IS EQUAL TO THE DIFFERENCE OF THE INDICES OF THE OPERATORS MENTIONED ABOVE. SO THIS HOLDS IN OUR CASE AS WELL. NOTE THAT ON PAGE 812 OF MIRROR SYMMETRY HORI EXPLAINS HOW (1-g_{\Sigma # \Sigma^*})=(2-2g-h). So what we have for the dimension agrees exactly with Baptista's expression in equation (24) of his paper.
\begin{equation}\label{virtualcomplexdim}
\boxed{
\mu (\phi^*\textrm{ker } d\pi_E^{(1,0)},\phi^*[\textrm{ker } d\pi_{E_L}]^{(1,0)})+(\textrm{dim}_{\C}(X)-\textrm{dim}(G))(2-2g-h).}
\end{equation} 
%IMPORTANT NOTE: AS in the mirror symmetry book, the maslov index should be related to the two winding numbers for bulk and boundary. therefore we actually have many moduli SPACES of symplectic vortices, but for consistency i will just use moduli space everywhere (unlike baptista)
The reason for this is that the 
%elliptic 
%IMPORTANT NOTE: TOOK OUT ELLIPTIC JUST TO BE SAFE, BUT IT SHOULD BE CORRECT
%index of the 
\textit{linearized} operator (whose index is the %virtual complex 
%which is equal to the 
dimension of the moduli space \cite{mcduff1998introduction})
%IMPORTANT NOTE:THERE SEEMS TO BE A RELATED DISCUSSION AT http://mathoverflow.net/questions/78788/dimension-of-moduli-space-in-lagrangian-floer-homology
 one derives from the 
 %open
%IMPORTANT NOTE: TOOK OUT OPEN BECAUSE WE ARE TALKING ABOUT TWO OPEN RIEMANN SURFACES GLUED TOGETHER 
  symplectic vortex equations is a compact perturbation of the direct sum of
  %essentially the difference 
%sum 
%of the indices for the Dolbeault operators of 
%of the indices for 
the operator $\phi^\ast \nabla^A_{\zb}$ (which has the same index as $\til{\mathcal{D}}'$) and an operator whose index can be evaluated to be $-\chi_{\Sigma}\textrm{dim }G $ \cite{Cieliebak2}, where $\chi_{\Sigma}=2-2g-h$ is the Euler characteristic of $\Sigma$.
%$\nabla^A_{\zb}$
%, where the index of the latter is (${\rm index }\  \nabla^A_{\z}=-{\rm index }\  \nabla^A_{\zb}$) 
%on $\Sigma \# \Sigma^*$ \cite{Cieliebak2}.
 The Maslov index in \eqref{theboundaryRanomaly} and \eqref{virtualcomplexdim} can be regarded as the equivariant Maslov index for the pair $(X,L)$, since for trivial $G$ it reduces to $\mu(\phi^*T^{(1,0)}X,\phi^*[TL]^{(1,0)})$.
\subsection{$\hat{Q}_A^2\neq 0$ Anomaly}

We have previously defined open Hamiltonian Gromov-Witten invariants as integrals over the moduli spaces of open symplectic vortices. However, as mentioned, we have in fact ignored problems related to singularities
%and noncompactness of this moduli space. 
in such a moduli space.
%However, we have in fact
In particular, we have ignored %problems related to
 the singular boundary strata which have codimension one in the moduli space, which occur due to disk bubbling \cite{Xu}. 
 %AA
%MK
This phenomenon obstructs integration over the moduli space. 
%In particular,
%noncompactness of the moduli space of symplectic vortices occurs due to disk bubbling \cite{Xu}. 

This is also a problem for ordinary open Gromov-Witten invariants, since disk bubbling also causes %the noncompactness of 
singular codimension one boundary strata in the moduli spaces of open worldsheet instantons of the non-gauged open A-model \cite{Georgieva}. 
Disk bubbling manifests itself in the open A-model as a nonpertubative instanton effect which causes the violation of the nilpotency of the scalar supercharge, i.e., $Q^2_A\neq 0$ (\cite{hori2003mirror}, page 833). Moreover, this anomaly of the supersymmetry algebra also spoils the cohomological structure 
%PORok
of the space of supersymmetric ground states of the open A-model, %\footnote{In fact, the anomaly implies that there can be no supersymmetric ground states, since if $Q_A|0\rangle =0$, then $Q^2_A|0\rangle =0$. %MP12.} 
which are %usually 
identified with elements of the Floer cohomology group %groups
for a pair of intersecting Lagrangian submanifolds. In fact, the anomaly implies that there are no supersymmetric ground states, and therefore supersymmetry is \textit{broken}.

Now, the fact that open symplectic vortices are open worldsheet instantons when $G$ is trivial means that open symplectic vortices cause $Q^2_A\neq 0$ and therefore supersymmetry breaking, for trivial $G$. Thus, for nontrivial $G$, we expect that open symplectic vortices will cause an analogous effect in the open $G$-gauged A-model, i.e., $\hat{Q}_A^2\neq 0$ (where $\hat{Q}_A=Q_A+Q_{BRST}$, %is the generator of gauge transformations
with $Q_{BRST}$ being the BRST charge), 
%in which case %
indicating singular codimension one boundary strata in the moduli spaces of open symplectic vortices, %will %be noncompact
%have 
 and
implying that the supersymmetric ground states of the open gauged A-model (which %we have argued 
we expect to be elements of the vortex Floer cohomology group \cite{Xu2,Cieliebak} for a pair of $G$-invariant Lagrangian submanifolds)
%Mk
 would not only lose their cohomological structure, but would cease to be supersymmetric, implying \textit{supersymmetry breaking}.%MAL

For the non-gauged open A-model, it is difficult to directly compute the violation of $Q_A^2 =0$ in general; one can only do so for specific examples, e.g., $X=S^2$ \cite{Hori2}. %due to instanton effects. 
Fortunately, at least for toric manifolds with $c_1(X)\geq 0$, one is able to use the mirror theory
%, which does not contain solitonic objects, 
%IMPORTANT NOTE: IT SEEMS LIKE OUR MIRROR THEORY DOES CONTAIN SOLITONIC OBJECTS, SEE SUSY VARIATIONS OF GAUGINOS. 
to compute this violation in general (and identify the condition whereby it vanishes) via canonical quantization, as shown by Hori \cite{hori2003mirror,Hori2}. The condition found was that for a pair of Lagrangian submanifolds supporting flat $U(1)$ bundles, the $Q_A^2 \neq 0$ anomaly vanishes if and only if the value of the superpotential on the mirror B-branes match each other, and in such a case supersymmetry is manifest.

It is thus natural to investigate the $\hat{Q}_A^2 \neq 0$ anomaly due to open symplectic vortices in the open gauged A-model via canonical quantization of its mirror theory. We shall do this for toric target spaces $X=\C^N//U(1)^{N-k}$, i.e., by topologically A-twisting the 
 $U(1)^k$-GNLSM on an infinite strip whose boundaries are mapped to different equivariant A-branes in $X$,\footnote{These equivariant A-branes are labelled by the GLSM parameters $s_j^{\pi}$ and $s_j^{0}$, which determine the position of their respective mirror D0-branes (see footnote \ref{58}).} whose mirror (c.f. Section 5.4) has the (Euclidean) action
\begin{equation}\label{expandedmirroractionEuc2}
\begin{aligned}
&S_E=\frac{1}{2\pi}\int  d^2x\Big[\sum_c^k\sum_d^k (g_{cd}\del_{\mu}\theta_c\del^{\mu}\ov{\theta}_d-\frac{i}{2}g_{cd}\ov{\chi}^{\ts}_{-c}(\overleftrightarrow{\del_+})\chi^{\ts}_{-d}-\frac{i}{2}g_{cd}\ov{\chi}^{\ts}_{+c}(\overleftrightarrow{\del_-}){\chi}^{\ts}_{+d}- g_{cd}E_c^{\ts} \ov{E}_d^{\ts})\\+&\sum^{k}_c{1 \over 2\til{e}_c^2}\Big((\til{F}_{12c})^2 
 +\partial_{\mu}\til{\sigma}_c \partial^{\mu} 
\overline{\til{\sigma}}_c   -(\til{D}_c)^2  - {i \over 2} \overline{\til{\lambda}}_{+c}
(\overleftrightarrow{\del_-})
\til{\lambda}_{+c} -{i \over 2} \overline{\til{\lambda}}_{-c}
(\overleftrightarrow{\del_+})
\til{\lambda}_{-c}  \Big)\\
-&\frac{1}{2}\Big(\sum^N_j\sum^k_c\sum^k_d\til{Q}_{jc}v_d^j(\til{\si}_c E_d^{\ts}-i \ov{\til{\lam}}_{+c}\chi^{\theta}_{-d}-i\til{\lam}_{-c}\ov{\chi}^{\ts}_{+d}+(\til{D}_c-\til{F}_{12c})\theta_d)+\sum^k_c(\sum^N_j\til{Q}_{jc}\hat{s}^j-\til{t}_c)(\til{D}_c-\til{F}_{12c})\\+&\sum^N_j e^{-\sum_c v^j_c \theta_c-\hat{s}^j}
(-\sum_c^k v_c^j\ov{\chi}^{\theta}_{+c}\sum^k_d v^j_d{\chi}^{\theta}_{-d}-\sum_c^k v_c^j E_c^{\theta} )\\+&\sum^N_j\sum^k_c\sum^k_d\til{Q}_{jc}v_d^j(\ov{\til{\si}}_c \ov{E}_d^{\ts}-i {\til{\lam}}_{+c}\ov{\chi}^{\theta}_{-d}-i\ov{\til{\lam}}_{-c}{\chi}^{\ts}_{+d}+(\til{D}_c+\til{F}_{12c})\ov{\theta}_d)+\sum^k_c(\sum^N_j\til{Q}_{jc}\ov{\hat{s}}^j-\ov{\til{t}}_c)(\til{D}_c+\til{F}_{12c})\\+&\sum^N_j e^{-\sum_c v^j_c \ov{\theta}_c-\ov{\hat{s}}^j}
(-\sum_c^k v_c^j\ov{\chi}^{\theta}_{-c}\sum^k_d v^j_d{\chi}^{\theta}_{+d}-\sum_c^k v_c^j \ov{E}_c^{\theta} )\Big)\Big],\end{aligned}
\end{equation}
where $d^2x=dx^1 dx^2$ and $\del_{\pm}=i\del_2\pm \del_1$.
 Performing the topological A-twist for the mirror theory amounts to the following field redefinitions:
\begin{align}
\til{\varphi}_c&=-i2\til{\si}_c   &  \til{\xi}_c&=\ov{\til{\si}}_c/4  \nonumber \\
{\til{\lam}}_c&={\til{\lam}}_{+c} &   \ov{\til{\lam}}_c&=\ov{\til{\lam}}_{-c}  \nonumber  \\
\til{\psi}_{+c}&=\frac{2i}{\sqrt{2}}\ov{\til{\lam}}_{+c}    &   \til{\psi}_{-c}&=\frac{2i}{\sqrt{2}}{\til{\lam}}_{-c} \nonumber \\  
\chi^{\theta c}&={\chi}^{\theta c}_+   &  \ov{\chi}^{\theta c}&=\ov{\chi}^{\theta c}_-  \nonumber \\ \mx^{\theta c}_+&=2\ov{\chi}^{\theta c}_+   &  \mx^{\theta c}_-&=2{\chi}^{\theta c}_-,  
\end{align}
where $\til{\lam}_c$, $\ov{\til{\lam}}_c$, $\chi^{\theta c}$ and $\ov{\chi}^{\theta c}$ are scalars, while  $\til{\psi}_{\pm c}=i\til{\psi}_2\pm \til{\psi}_1$ and  $\mx_{\pm}^{\theta c}=i\mx^{\theta c}_2\pm \mx^{\theta c}_1$ are one-forms.
Hence, the mirror action of the open $U(1)^k$-gauged A-model with toric target $X=\C^N//U(1)^{N-k}$ is 
\begin{equation}\label{expandedmirroractionEuc2topological}
\begin{aligned}
&S_E=\frac{1}{2\pi}\int  d^2x\Big[\sum_c^k\sum_d^k (g_{cd}\del_{\mu}\theta_c\del^{\mu}\ov{\theta}_d-\frac{i}{4}g_{cd}\ov{\chi}^{\ts}_{c}(\overleftrightarrow{\del_+})\mx^{\ts}_{-d}-\frac{i}{4}g_{cd}{\mx }^{\ts }_{+c}(\overleftrightarrow{\del_-}){\chi}^{\ts}_{d}- g_{cd}E_c^{\ts} \ov{E}_d^{\ts})\\+&\sum^{k}_c{1 \over 2\til{e}_c^2}\Big((\til{F}_{12c})^2 
 +i2\partial_{\mu}\til{\varphi}_c \partial^{\mu} 
{\til{\xi}}_c   -(\til{D}_c)^2  - {\sqrt{2} \over 4} {\til{\psi}}_{+c}
(\overleftrightarrow{\del_-})
\til{\lambda}_{c} -{\sqrt{2} \over 4} \overline{\til{\lambda}}_{c}
(\overleftrightarrow{\del_+})
\til{\psi}_{-c}  \Big)\\
-&\frac{1}{2}\Big(\sum^N_j\sum^k_c\sum^k_d\til{Q}_{jc}v_d^j(\frac{i}{2}\til{\varphi}_c E_d^{\ts}- \frac{\sqrt{2}}{4}\til{\psi}_{+c}\mx^{\theta }_{-d}-\frac{\sqrt{2}}{4}\til{\psi}_{-c}{\mx}^{\ts }_{+d}+(\til{D}_c-\til{F}_{12c})\theta_d)\\+&\sum^k_c(\sum^N_j\til{Q}_{jc}\hat{s}^j-\til{t}_c)(\til{D}_c-\til{F}_{12c})+\sum^N_j e^{-\sum_c v^j_c \theta_c-\hat{s}^j}
(-\frac{1}{4}\sum_c^k v_c^j{\mx}^{\theta}_{+c}\sum^k_d v^j_d{\mx}^{\theta }_{-d}-\sum_c^k v_c^j E_c^{\theta} )\\+&\sum^N_j\sum^k_c\sum^k_d\til{Q}_{jc}v_d^j(4{\til{\xi}}_c \ov{E}_d^{\ts}-i {\til{\lam}}_{c}\ov{\chi}^{\theta}_{d}-i\ov{\til{\lam}}_{c}{\chi}^{\ts}_{d}+(\til{D}_c+\til{F}_{12c})\ov{\theta}_d)\\+&\sum^k_c(\sum^N_j\til{Q}_{jc}\ov{\hat{s}}^j-\ov{\til{t}}_c)(\til{D}_c+\til{F}_{12c})+\sum^N_j e^{-\sum_c v^j_c \ov{\theta}_c-\hat{s}^j}
(-\sum_c^k v_c^j\ov{\chi}^{\theta}_{c}\sum^k_d v^j_d{\chi}^{\theta}_{d}-\sum_c^k v_c^j \ov{E}_c^{\theta} )\Big)\Big],\end{aligned}
\end{equation}
which is invariant under the supersymmetry transformations
\begin{align} 
\delta_{Q_A}& \til{A}_{1c} \ =\frac{\sqrt{2}}{2}\eps \til{\psi}_{1c}  &   \delta_{Q_A}& \til{\psi}_{-c} \ =\ \, -\frac{1}{\sqrt{2}}\eps\partial_- \til{\vphi}_c   \nonumber  \\
\delta_{Q_A}& \til{A}_{2 c} \ =\frac{\sqrt{2}}{2}\eps \til{\psi}_{2c}  &   \delta_{Q_A}& {\til{\psi}}_{+c} \ =\  \, -\frac{1}{\sqrt{2}}\eps\partial_{+} \til{\vphi}_c \nonumber  \\
\delta_{Q_A}& \til{\varphi}_c \ = \ 0  &  \delta_{Q_A}&  \ov{\til{\lam}}_{c}  \ = \ -i\eps(\til{F}_{12c}+\til{D}_c) \nonumber \\
\delta_{Q_A}& {\til{\xi}}_c \ = \ - \frac{i\eps (\ov{\til{\lam}}_{c} + \til{\lam}_{c})}{4}   &  \delta_{Q_A}&   \til{\lam}_{c} \ = \ i\eps(\til{F}_{12c}+\til{D}_c)  \nonumber \\
\delta_{Q_A}& \til{D}_c=-\frac{\sqrt{2}}{2}\eps(\del_1{\til{\psi}}_{2c}-\del_2{\til{\psi}}_{1c})  &   \delta_{Q_A}& \mx_-^{\theta c} \ =\ - 2 i \eps\, \partial_- \theta^c  \nonumber  \\
\delta_{Q_A}& \theta^c \ = \ 0     &   \delta_{Q_A}& \mx_+^{\theta c} \ =\  2i\eps \, \partial_{+} \theta^c \nonumber \\  
\delta_{Q_A}& \ov{\theta}^c \ = \ \eps (\chi^{\theta c} - \ov{\chi}^{\theta c} )  &  \delta_{Q_A}& \ov{\chi}^{\theta c} \ =\eps \ov{E}^{\theta c} \nonumber \\ \delta_{Q_A}& E^{\theta c}=i\eps\del^{\mu}\mx_{\mu}^{\theta c}   &  \delta_{Q_A}& {\chi}^{\theta c} \ =\eps \ov{E}^{\theta c}  \nonumber  \\
\delta_{Q_A}& \ov{E}^{\theta c}=0 \label{mirrorsusy}
\end{align}
generated by the supercharge $Q_A$. However, this mirror theory %of concern 
 is in fact a gauge theory, and any consistent quantization procedure should include gauge fixing, in order to remove unphysical degrees of freedom. 

To this end, we shall choose the Lorentz gauge 
\begin{equation}\label{lorentzgauge}
\langle \psi' |\del_{\mu}\til{A}_c^{\mu}| \psi \rangle=0
\end{equation}
(where $|\psi \rangle$ and $|\psi' \rangle$) are physical states), 
%which is effected by including 
and include the following BRST gauge fixing action 
\begin{equation}\label{gfaction}
S_{BRST}=\frac{1}{2\pi}\sum_c^k\frac{1}{2\til{e}_c^2}\int d^2x(-iB_c\del_{\mu}\til{A}^{\mu}_c-(B_c)^2+\del_{\mu}\til{b}_{c}\del^{\mu}\til{c}_{c}-i\frac{\sqrt{2}}{2}\del_{\mu}\til{b}_c\til{\psi}^{\mu}_c),
\end{equation}
where $\tib_c$ and $\tic_c$ are fermionic ghost fields, while $B_c$ is a bosonic auxiliary field. As expected, the first term explicitly breaks the $U(1)^k$ gauge symmetry of the gauged LG model.  

Now, note that the gauge fixing action \eqref{gfaction} can be rewritten as 
\begin{equation}\label{gfactionexact}
S_{BRST}=\eps^{-1}(\delta_{Q_A}+\delta_{BRST})\Big\{\sum_c^k\frac{1}{2\til{e}_c^2}\int d^2x(-i\til{b}_c\del_{\mu}\til{A}_c^{\mu}+\til{b}_cB_c)\Big\},
\end{equation}
where we have performed integration by parts, and used the boundary condition $({\til{\lam}}_{-c}-\ov{\til{\lam}}_{+c})=0$ (which is equivalent to $\til{\psi}_{1c}=0$) that we have previously imposed, as well as the boundary condition 
\begin{equation}\label{cboundc}
\del_1\til{c}_c=0,
\end{equation}
which we impose at present. Here, $\delta_{BRST}$ is the standard BRST symmetry variation given by 
\begin{equation}
\begin{aligned}
\delta_{BRST} \til{A}_{\mu c}&=i\eps\del_{\mu}\til{c}_c\\
\delta_{BRST} \tib_c&=\eps B_c\\
\delta_{BRST} \tic_c&=0\\
\delta_{BRST} B_c&=0,
\end{aligned}
\end{equation}
with the BRST variations of all other fields being equal to zero. %while $\delta_{Q_A}$ is the supersymmetry variation of the A-model previously defined for the physical fields. 
For the unphysical fields used for gauge fixing, the supersymmetry transformations are $\delta_{Q_A}\tib_c=0$ and $\delta_{Q_A}B_c=0$ while\footnote{Note that with respect to \eqref{csigma}, the boundary condition \eqref{cboundc} obeys A-type supersymmetry, since the boundary condition on $\til{\vphi}_c$ is $\del_1\til{\vphi}_c=0$.} 
 \begin{equation}\label{csigma}
 \delta_{Q_A}\tic_c=-\frac{i}{2}\eps\til{\vphi}_c.
 \end{equation}
Now, $\delta^2_{Q_A}\propto\delta_G(\til{\vphi})$ and $\delta^2_{BRST}=0$ on all fields.\footnote{The transformation $\delta_G(\til{\vphi})$ is a $U(1)^k$ gauge transformation whose local parameter is $\til{\vphi}_c$.} %wherefrom 
In addition, we can show that 
\begin{equation}
(\delta_{Q_A}+\delta_{BRST})^2=0
\end{equation}
on all fields. This implies that the BRST gauge fixing action \eqref{gfactionexact} is in fact invariant under 
$\hat{\delta}=\delta_{Q_A}+\delta_{BRST}$.
Since the physical action  \eqref{expandedmirroractionEuc2topological} is also invariant under $\delta_{BRST}$, this further implies that the entire action $S_E+S_{BRST}$ is invariant under $\hat{\delta}=\delta_{Q_A}+\delta_{BRST}$. This suggests that the relevant symmetry of the action after gauge fixing is that which is generated by $\hat{Q}_A= {Q_A}+ Q_{BRST}$. 

The conjugate momentum for any field, denoted $X$, is defined as 
\begin{equation}
\pi_{X}=\frac{\del L_E}{\del (\del_2 X)},
\end{equation}  
with the convention that derivatives are taken from the right for fermionic fields.
The canonical conjugate momenta are\footnote{Note that in order to derive a consistent set of anticommutation relations, we convert the fermionic kinetic terms with $\del_2$ derivatives to a form identical to that found in closed theories via integration by parts, e.g. $-\frac{i}{4}g_{cd}\ov{\chi}^{\ts}_{c}i(\overleftrightarrow{\del_2})\mx^{\ts}_{-d}\rightarrow-\frac{i}{2}g_{cd}\ov{\chi}^{\ts}_{c}({i\del_2})\mx^{\ts}_{-d}$. The symmetric form of the kinetic terms are in fact not necessary  because a Lagrangian on a Euclidean worldsheet is not real.} %of the dynamical fields are 
%IMPORTANT NOTE: THE CONJUGATE MOMENTA DEPEND ON THE POSITIONS OF THE DERIVATIVE \DEL_2, THEREFORE WE REFER TO THE FOLLOWING AS THE CANONICAL ONES
\begin{equation}
\begin{aligned}
\pi_{\theta^c}&=\frac{1}{2\pi} g_{cd}\del_2\ov{\theta}^d\\
\pi_{\ov{\theta}^c}&=\frac{1}{2\pi} g_{cd}\del_2\theta^d\\
\pi_{\chi^{\theta c}}&=\frac{1}{4\pi} g_{cd}{\mx}_{+}^{\theta d}\\
\pi_{\mx_{-}^{\theta c}}&=\frac{1}{4\pi} g_{cd}\ov{\chi}^{\theta d}\\
\pi_{\til{\vphi}_c}&=\frac{1}{2\pi}\frac{1}{2\til{e}_c^2}i2\del_2{\til{\xi}}_c\\
\pi_{\til{\xi}_c}&=\frac{1}{2\pi}\frac{1}{2\til{e}_c^2}i2\del_2{\til{\vphi}}_c\\
\pi_{\til{\lam}_{ c}}&=\frac{1}{2\pi}\frac{1}{2\til{e}_c^2}\frac{\sqrt{2}}{2i}{\til{\psi}}_{+ c}\\
\pi_{\til{\psi}_{- c}}&=\frac{1}{2\pi}\frac{1}{2\til{e}_c^2}\frac{\sqrt{2}}{2i}\ov{\til{\lam}}_{ c}\\
%\pi_{\til{A}_{1c}}&=-\frac{1}{2\pi}\big(\frac{1}{2 %\til{e}_c^2}2\til{F}_{12c}-i\sum_j^N \sum_d^k \til{Q}%_{jc} v^j_d \textrm{Im}(\theta_d)\big)\\
\pi_{\til{A}_{1c}}&=-\frac{1}{2\pi}\Big(\frac{1}{\til{e}_c^2}\til{F}_{12c} +i \textrm{Im}\Big(\ \sum_j^N \til{Q}_{jc}  \Big(\langle v^j , \theta \rangle  +  \hat{s}^j \Big) 
\ -\; \til{t}_c  \Big)\Big)\\
\pi_{\til{A}_{2c}}&=\frac{1}{2\pi}\frac{1}{2 \til{e}_c^2}(-iB_c)\\
\pi_{\til{b}_c}&=\frac{1}{2\pi}\frac{1}{2 \til{e}_c^2}\big(\del_2\til{c}_c-\frac{\sqrt{2}}{4}(\til{\psi}_{+c}+\til{\psi}_{-c})\big)\\
\pi_{\til{c}_c}&=\frac{1}{2\pi}\frac{1}{2 \til{e}_c^2}\big(\del_2\til{b}_c\big).\\
\end{aligned}
\end{equation}
The equal-time canonical commutation relations are 
\begin{equation}
\begin{aligned}
{}[X_i(x^1),\pi_{X_j}(y^1)]&=\delta_{ij}\delta(x^1-y^1)\\
[X_i(x^1),X_j(y^1)]&=0\\
[\pi_{X_i}(x^1),\pi_{X_j}(y^1)]&=0
\end{aligned}
\end{equation}
%when $X$ is a bosonic field and 
for $X_i=\{\theta^c,\ov{\theta}^c, \til{\vphi}_c, {\til{\xi}}_c,\til{A}_{1c},\til{A}_{2c}\}$, and the equal-time canonical anticommutation relations are
\begin{equation}\label{anticommutators}
\begin{aligned}
\{X_i(x^1),\pi_{X_j}(y^1)\}&=\delta_{ij}\delta(x^1-y^1)\\
\{X_i(x^1),X_j(y^1)\}&=0\\
\{\pi_{X_i}(x^1),\pi_{X_j}(y^1)\}&=0
\end{aligned}
\end{equation}
for $X_i=\{\chi^{\theta c},\mx_-^{\theta c},\til{\lam}_{c},\til{\psi}_{-c},\til{b}_c,\til{c}_c\}$. 
%IMPORTANT NOTE:KEEP THE FOLLOWING IN MIND %(here, we have considered ${\chi}^{\theta c}$ and ${\lam}_{\pm c}$ to be fields and $\ov{\chi}^{\theta c}$ and $\ov{\lam}_{\pm c}$ to be their conjugate momenta, the opposite choice also leads to %\eqref{anticommutators})
%the same anticommutation relations via \eqref{anticommutators}). 
In addition, all commutators between a bosonic operator and a fermionic operator vanish, and the commutation relations obeyed by 
%the auxiliary fields 
$E^{\theta}_c$ and $\til{D}_c$ are 
%governed
specified by their equations of motion, which hold as operator equations due to Ehrenfest's theorem. Then, the non-vanishing canonical commutation and anticommutation relations are
%\begin{equation}
\begin{align}\label{aligncommutators}
\lbrack \theta^c(x^1),g_{de}\del_2\ov{\theta}^e(y^1)\rbrack&=2\pi \delta^c_{d} \delta(x^1-y^1)\nonumber\\
\lbrack \ov{\theta}^c(x^1),g_{de}\del_2{\theta}^e(y^1)\rbrack&=2\pi \delta^c_{d} \delta(x^1-y^1)\nonumber\\
\{\chi^{\theta c}(x^1),{\mx}^{\theta d}_{+}(y^1)\}&=4\pi g^{cd}\delta(x^1-y^1)\nonumber\\
\{\mx^{\theta c}_{-}(x^1),\ov{\chi}^{\theta d}(y^1)\}&=4\pi g^{cd}\delta(x^1-y^1)\nonumber\\
\lbrack \til{\vphi}_c(x^1), \frac{1}{2\til{e}_d^2}i2 \del_2{\til{\xi}}_d (y^1)\rbrack&= 2\pi\delta_{cd} \delta(x^1-y^1)\nonumber\\
\lbrack \til{\xi}_c(x^1), \frac{1}{2\til{e}_d^2}i2 \del_2{\til{\vphi}}_d (y^1)\rbrack&= 2\pi\delta_{cd} \delta(x^1-y^1)\nonumber\\
\{\til{\lam}_{c}(x^1),\frac{1}{2\til{e}_d^2} \frac{\sqrt{2}}{2i}{\til{\psi}}_{+ d}(y^1)\}&=2\pi \delta_{cd}\delta(x^1-y^1)\nonumber\\
\{\til{\psi}_{- c}(x^1),\frac{1}{2\til{e}_d^2}\frac{\sqrt{2}}{2i} \ov{\til{\lam}}_{ d}(y^1)\}&=2\pi \delta_{cd}\delta(x^1-y^1)\nonumber\\
\lbrack \til{A}_{1c}(x^1),-\Big(\frac{1}{\til{e}_c^2}\til{F}_{12c}(y^1) \Big)\rbrack&=2\pi \delta_{cd}\delta(x^1-y^1)\nonumber\\
%\lbrack \til{A}_{1c}(x^1),-\frac{1}{2 %\til{e}_d^2}\big(2\til{F}_{12d}(y^1)-i\sum_j^N \sum_e^k %\til{Q}_{jd} v^j_e \textrm{Im}(\theta^e(y^1))\big)%\rbrack&=2\pi \delta_{cd}\delta(x^1-y^1)\nonumber\\
\lbrack \til{A}_{2c}(x^1),\frac{1}{2 \til{e}_d^2}(-iB_d(y^1))\rbrack&=2\pi\delta_{cd}\delta(x^1-y^1)\nonumber\\
\{ \til{b}_{c}(x^1),\frac{1}{2 \til{e}_d^2}\big(\del_2\til{c}_d(y^1)\big)\}&=2\pi\delta_{cd}\delta(x^1-y^1)\nonumber\\
\{ \til{c}_{c}(x^1),\frac{1}{2 \til{e}_d^2}\big(\del_2\til{b}_d(y^1)\big)\}&=2\pi\delta_{cd}\delta(x^1-y^1).
\end{align}
%The integral form of the supercharge $\hat{Q}_A$ is %given by
%\begin{equation}
%\hat{Q}_A=\int dx^1(\hat{J}_A^2),
%\end{equation}
%where $\hat{J}^{\mu}_A$ is the supercurrent defined via
%\begin{equation}
%\delta S_E= \int d^2x \del_{\mu} \epsilon \hat{J}_A^{\mu}.
%\end{equation}
%Its %is given 
The integral form of the supercharge $\hat{Q}_A$ is %given by
\begin{equation}
\hat{Q}_A=\int dx^1(\hat{J}_A^2),
\end{equation}
where $\hat{J}^{\mu}_A$ is the supercurrent defined via
\begin{equation}
\delta S_E= \int d^2x \del_{\mu} \epsilon \hat{J}_A^{\mu}.
\end{equation}
Its %is given 
explicit form is\footnote{Charge conservation follows from $\del_2 \hat{Q}_A=\int dx^1 \del_2 \hat{J}^2_A=-\int dx^1 \del_1 \hat{J}^1_A$, which can be shown to be zero using the boundary conditions we have previously imposed, as well as the boundary condition $\del_1\til{b}_c=0$, which we impose at present.}
\begin{equation}\label{conservedsupercharge}
\begin{aligned}
\hat{Q}_A=&\frac{1}{2\pi}\int dx^1\Big(g_{cd}\del_2\theta^c(\chi^{\theta d}-\ov{\chi}^{\theta d})-ig_{cd}\del_1\theta^c(\chi^{\theta d}+\ov{\chi}^{\theta d})-\sum_c^k \frac{\sqrt{2}}{4\til{e}_c^2 }\til{F}_{12c}({\til{\psi}}_{+c}-{\til{\psi}}_{-c})\\&
-\sum_c^k\frac{i}{\til{e}_c^2}\til{F}_{12c}\del_1\til{c}_c +\sum_c^k \frac{1}{4\til{e}_c^2}({\til{\lam}}_{c}+\ov{\til{\lam}}_{c})\del_2\til{\vphi}_c +\sum_c^k \frac{i}{4\til{e}_c^2}(\ov{\til{\lam}}_{c}-{\til{\lam}}_{c})\del_1\til{\vphi}_c \\&
+\frac{1}{4}\sum_c^k({\mx}^{\theta c}_++{\mx}^{\theta c}_-)\frac{\del \til{W}}{\del \theta^c}+\sum_c^k\frac{\sqrt{2}}{4}({\til{\psi}}_{+c}-{\til{\psi}}_{-c})\frac{1}{2}\Big(\frac{2}{i}\frac{\del\til{W}}{\del \til{\vphi}_c}+\frac{1}{4}\frac{\del\ov{\til{W}}}{\del \til{\xi}_c}\Big)\\&+\sum_c^k\del_1\til{c}_c\frac{1}{2i}\Big(\frac{2}{i}\frac{\del\til{W}}{\del \til{\vphi}_c}-\frac{1}{4}\frac{\del\ov{\til{W}}}{\del \til{\xi}_c}\Big)
%-\frac{i}{2}\sum_j^N\sum_c^k\sum_d^k \til{Q}_{jc}v^j_d%\del_1\tic_c(\theta^d-\ov{\theta}^d)
\\&
-\sum_c^k\frac{1}{2\til{e}_c^2}\frac{\sqrt{2}}{4}B_c({\til{\psi}}_{+c}+{{\til{\psi}}}_{-c}) +\sum_c^k\frac{1}{2\til{e}_c^2}B_c\del_2\til{c}_c+\sum_c^k\frac{i}{4\til{e}_c^2}\del_2\tib_c \til{\vphi}_c
\Big),
\end{aligned}
\end{equation}
where
\begin{equation}
\begin{aligned}
\frac{\del\til{W}}{\del \theta^c}\ &= \ \frac{i}{2}\sum_j^N\langle \til{\vphi} , \til{Q}_j  \rangle v^j_c 
   \ -\  \sum_{j=1}^{N}  v^j_c e^{- \langle v^j , \theta \rangle - \hat{s}^j} \ ,\\
\frac{2}{i}\frac{\del\til{W}}{\del \til{\vphi}_c} &= \ \sum_j^N \til{Q}_{jc}  \Big(\langle v^j , \theta \rangle  +  \hat{s}^j \Big) 
\ -\; \til{t}_c  ,\\
\frac{1}{4}\frac{\del\ov{\til{W}}}{\del \til{\xi}_c} &= \ \sum_j^N \til{Q}_{jc}  \Big(\langle v^j , \ov{\theta} \rangle  +  \ov{\hat{s}}^j \Big) 
\ -\; \ov{\til{t}}_c. 
\end{aligned}
\end{equation}
Then,
\begin{equation}\label{anomaly}
\begin{aligned}
\hat{Q}_A^2=\frac{1}{2}\{\hat{Q}_A,\hat{Q}_A\}
%=&\frac{1}{2}\int dx^1\int dy^1\{\hat{J}^2(x^1),\hat{J}%^2(y^1)\}\\
=&\frac{1}{2\pi}\int dx^1 \Big( (-i)\sum_c^k\del_1\theta^c \frac{\del \til{W}}{\del \theta^c} 
+(-i)\sum_c^k \del_1\til{\vphi}_c  \frac{\del\til{W}}{\del \til{\vphi}_c} -\sum_c^k \big(\frac{i}{4\til{e}_c^2}\big)\del_2 \til{\vphi}_c B_c   \\&-\sum_c^k\sum_d^k\frac{\del^2\ov{\til{W}}}{\del \ov{\theta}^d \del {\til{\xi}}_c}(\chi^{\theta d}-\ov{\chi}^{\theta d})(\frac{i}{8})(\frac{\sqrt{2}}{4i}({\til{\psi}}_{+c}-{\til{\psi}}_{-c})+\del_1\til{c}_c )\Big)
 \end{aligned}
\end{equation}
where we have used the boundary conditions $\til{F}_{12c}=0$ and $\sum^k_{c=1}v_{cj}\Theta_c=s_j-\hat{s}_j$
(where $s_j=s_j^{\pi}$ at $x^1=\pi$ and $s_j=s_j^{0}$ at $x^1=0$), as well as the constraint $\sum_j^N\til{Q}_{jc}s_j-\til{t}_c=0$. 
%PE

%Now, the second and third terms cancel, and this is related to the effect of the gauge-fixing procedure which removes the term proportional to $\delta_G(\til{\vphi})$ %$G(\til{\vphi})$ 
%in %$Q_A^2$.
%$\delta^2_{Q_A}$. Next, the terms proportional to $\frac{\del\ov{\til{W}}}{\del \til{\xi}_c}$ cancel. Then, the remaining
The terms with first-order derivatives of the superpotential can be written as
\begin{equation}\label{derivethedifference}
\begin{aligned}
&\frac{(-i)}{2\pi}\int dx^1 \Big( \sum_c^k\del_1\theta^c \frac{\del \til{W}}{\del \theta^c}+\sum_c^k \del_1\til{\vphi}_c \frac{\del \til{W}}{\del \til{\vphi}_c}\Big)\\=&\frac{(-i)}{2\pi}\int dx^1 \del_1 \til{W}(\theta,\til{\vphi})\\
=&\frac{(-i)}{2\pi}(\til{W}(\theta,\til{\vphi})_{\pi}-\til{W}(\theta,\til{\vphi})_{0}).
\end{aligned}
\end{equation} 
From the analysis below \eqref{separateA}, we know that this is equal to $\frac{(-i)}{2\pi}(\sum_{i=1}^N \e^{-s^\pi_i}-\sum_{i=1}^N \e^{-s^0_i})$.
%POROR

The remaining terms to consider are then 
\begin{equation}\label{operatorexpression}
\begin{aligned}
&\frac{1}{2\pi}\int dx^1 \Big(-\sum_c^k \big(\frac{i}{4\til{e}_c^2}\big)\del_2 \til{\vphi}_c B_c  -\sum_c^k\sum_d^k\frac{\del^2\ov{\til{W}}}{\del \ov{\theta}^d \del {\til{\xi}}_c}(\chi^{\theta d}-\ov{\chi}^{\theta d})(\frac{i}{8})(\frac{\sqrt{2}}{4i}({\til{\psi}}_{+c}-{\til{\psi}}_{-c})+\del_1\til{c}_c )\Big).
\end{aligned}
\end{equation}
%OENT
%We note that the rightmost factors of these terms are proportional to the following $\hat{Q}$-exact expressions, 
%\begin{equation}\label{threecommutators}
%\begin{aligned}
%\frac{i( \ov{\til{\lam}}_{+c}-{\til{\lam}}_{-c})}{2}+i\del_1\til{c}_c&=\hat{\delta}\til{A}_{1c}=i\eps[\hat{Q},\til{A}_{1c}]\\ \frac{( \ov{\til{\lam}}_{+c}+{\til{\lam}}_{-c})}{2}+i\del_2\til{c}_c&=\hat{\delta}\til{A}_{2c}=i\eps[\hat{Q},\til{A}_{2c}] \\
%B_c&=\delta \til{b}_c= i\eps\{\hat{Q},\til{b}_c\}.
%\end{aligned}
%\end{equation}
%Now, the ghost operator $\til{b}_c$ %%cannot 
%only gives rise 
%to unphysical excitations. %%and hence it ought to annihilate physical states. 
Unlike \eqref{derivethedifference}, these cannot be written in terms of boundary data, and hence are bulk terms which occur even for closed worldsheets. However, as in the non-anomalous closed case, %has been shown rigorously 
these bulk terms ought to be equal to zero.
The vanishing of these terms can also be understood as follows. The auxiliary field $B_c$ obeys its equation of motion $B_c=-\frac{i}{2}\del_{\mu}\til{A}^{\mu}_c$ as an operator equation due to Ehrenfest's theorem, and hence the matrix elements  of the first term in the integrand with respect to the physical Hilbert space vanish  due to the Lorentz gauge condition \eqref{lorentzgauge}.\footnote{This statement follows from the fact that the Lorentz gauge condition can equivalently be written as $\del_{\mu}(\til{A}_c^{\mu})^+|\psi \rangle=0$ or $\langle\psi |\del_{\mu}(\til{A}_c^{\mu})^-=0$ (where $\til{A}_c^{\mu}=(\til{A}_c^{\mu})^++(\til{A}_c^{\mu})^-$ is the decomposition with respect to positive and negative momenta)%in the interaction picture of quantum mechanics
, as well as the fact that  $\del_{\mu}\til{A}_c^{\mu}$ commutes with $\del_2\til{\varphi}_c$.} Next, note that we are dealing with an A-twisted theory, whose topological correlation functions are invariant under $\hat{Q}_A$-exact deformations of the action. Therefore, we should be able to deform the action such that the second term in \eqref{operatorexpression} vanishes.
%OR)$
%the conjugate twisted F-terms such that the second term in \eqref{operatorexpression} vanishes.
 Indeed, this can be achieved by adding the following term which is %both $Q_A$-exact and 
 $\hat{Q}_A$-exact to the action, i.e.,
\begin{equation}
\begin{aligned}
&\eps^{-1}\delta_{\hat{Q}_A}\Bigg[\frac{1}{2\pi}\int d^2x\frac{1}{2}\Bigg(\sum_c^k i\til{\lam}_c\Big(\sum_j^N\til{Q}_{jc}(\sum_d v^j_d\ov{\theta}^d +\ov{\hat{s}}^j)-\ov{\til{t}}_c\Big)+\sum_c^k 4 \til{\xi}_c\Big(\sum_j^N\sum_d^k\til{Q}_{jc}v^j_d\ov{\chi}_d\Big)\Bigg)\Bigg]\\
=&\frac{1}{2\pi}\int d^2x\frac{1}{2}\Bigg(\sum^N_j\sum^k_c\sum^k_d\til{Q}_{jc}v_d^j(4{\til{\xi}}_c \ov{E}_d^{\ts}-i {\til{\lam}}_{c}\ov{\chi}^{\theta}_{d}-i\ov{\til{\lam}}_{c}{\chi}^{\ts}_{d}+(\til{D}_c+\til{F}_{12c})\ov{\theta}_d)\\+&\sum^k_c(\sum^N_j\til{Q}_{jc}\ov{\hat{s}}^j-\ov{\til{t}}_c)(\til{D}_c+\til{F}_{12c})\Bigg),
\end{aligned}
\end{equation}
%+&\sum^N_j\sum^k_c\sum^k_d\til{Q}_{jc}v_d^j(4{\til{\xi}}_c \ov{E}_d^{\ts}-i {\til{\lam}}_{c}\ov{\chi}^{\theta}_{d}-i\ov{\til{\lam}}_{c}{\chi}^{\ts}_{d}+(\til{D}_c+\til{F}_{12c})\ov{\theta}_d)\\+&\sum^k_c(\sum^N_j\til{Q}_{jc}\ov{\hat{s}}^j-\ov{\til{t}}_c)(\til{D}_c+\til{F}_{12c})
 which upon doing so, the terms proportional to $\del{\ov{\til{W}}(\ov{\theta},\til{\xi})}/{\del \til{\xi}_c}$ in \eqref{conservedsupercharge} vanish, whence the second term in \eqref{operatorexpression} also vanishes. 
%This does not affect our previous calculation, since the terms proportional to $\del{\ov{\til{W}}(\ov{\theta},\xi)}{\del \xi_c}$ in \eqref{anomaly} 

%Ms)

% Next, we note that since it is impossible for physical photons, i.e., those with transverse polarizations, to exist on a manifold with only one spatial dimension, the gauge fields $\til{A}_{1c}$ and  $\til{A}_{2c}$ have %Ps
%This can be shown to be true as a result of the Lorentz gauge condition. In fact, the form of the gauge fixing action also implies that there is a gauge fixing condition on the fermionic fields....
%Then, the $\hat{\delta}$-transformations of the gauge fields, i.e., 
%\begin{equation}
%\begin{aligned}
%\hat{\delta}\til{A}_{1c}&=\frac{i( \ov{\til{\lam}}_{+c}-{\til{\lam}}_{-c})}{2}+i\del_1\til{c}_c\\ \hat{\delta}\til{A}_{2c}&=\frac{( \ov{\til{\lam}}_{+c}+{\til{\lam}}_{-c})}{2}+i\del_2\til{c}_c,
%\end{aligned}
%\end{equation}
%should also have unphysical excitations, and precisely these expressions appear in the second and third terms in the integrand. Since they act on states to produce unphysical excitations, the resulting states cannot give rise to nonzero physical amplitudes. Thus, the operator expression \eqref{operatorexpression} vanishes with respect to the physical spectrum of the quantum field theory at hand, and does not contribute to the quantum anomaly. 

Hence, we find that 
\begin{equation}
\boxed{
\hat{Q}_A^2=\frac{(-i)}{2\pi}(\til{W}(\theta,\til{\vphi})_{\pi}-\til{W}(\theta,\til{\vphi})_{0}),}
\end{equation}
i.e., the $\hat{Q}_A^2\neq 0$ anomaly (which occurs due to the nonpertubative quantum effects of open symplectic vortices in the open gauged A-model) vanishes when the value of the superpotential $\til{W}(\theta,\til{\vphi})$ is equal on both boundaries, i.e., $\sum_{i=1}^N \e^{-s^\pi_i}=\sum_{i=1}^N \e^{-s^0_i}$. In other words, there is no anomaly when each boundary %is mapped to 
ends on a D0-brane %, 
such that both D0-branes %must be
are mapped to the same value of $\til{W}(\theta,\til{\vphi})$. %alternatively,
One way this can occur is when %both boundaries %can %be mapped to 
the boundaries end on %the \textit{same}
\textit{coincident} D0-branes. %This is a generalization of 
Although the condition $\sum_{i=1}^N \e^{-s^\pi_i}=\sum_{i=1}^N \e^{-s^0_i}$ seems identical to the condition (found by Hori \cite{hori2003mirror,Hori2}) for the vanishing of the ${Q}_A^2\neq 0$ anomaly %(which occurs due to the effects of open worldsheet instantons) 
of the open A-model, this is in fact not true, as %and that 
%the critical point condition on the D0-branes in our case is generalized to \eqref{condition1expli}
 the D0-branes do not have to be located at a critical point where $\del_{\theta_c}\til{W}_X=0$ in our case (where $\til{W}_X$ is the superpotential in the non-gauged case, which only depends on $\theta$ in the bulk), and the position of each D0-brane (defined by $s_i$ via \eqref{D0branepositionsuper}) is in our case constrained by $\sum_i^N \mathcal{Q}^i_{a}s_i-{t_a}=0$ instead of just $\sum_i^N \hat{Q}^i_{b}s_i-\hat{t}_b=0$.  In conclusion, for abelian $G$, we have found that for a pair of $G$-invariant Lagrangian %submanifolds 
tori of a toric manifold supporting flat $G$-equivariant $U(1)$ bundles, the quantum anomaly of $\hat{Q}_A^2\neq 0$ (which indicates an obstruction to integration over the moduli spaces of open symplectic vortices) vanishes if and only if the values of the superpotential $\til{W}(\theta,\til{\vphi})$ on the mirror B-branes are the same, and in this case, supersymmetry is manifest. 

\subsection{Mirror Computation of Abelian Invariants}

%We may also use the mirror Landau-Ginzburg theory to compute open Hamiltonian Gromov-Witten invariants for abelian gauge groups. 
%It is in principle much 
In principle, it is simpler to use the mirror gauged Landau-Ginzburg description of the open gauged A-model to compute open Hamiltonian Gromov-Witten invariants for abelian gauge groups and toric target spaces with $c_1(X)\geq 0$, since there are no open symplectic vortices in this gauged LG model. 

We shall focus on the mirror computation of invariants that come from path integrals over the $Q_A$-invariant \textit{local}  observables associated with equivariant cohomology classes, i.e., those given by \eqref{3.8} (where $d_G\alpha=0$) and \eqref{localboundaryoperator} (where $d_G\zeta=0$).\footnote{Note that the $G$-invariance of these physical observables implies that they are invariant under $Q_{BRST}$, and therefore also invariant under $\hat{Q}_A$.} %\footnote{Note that in order for the path integrals over these operators to be nonzero, there cannot be any gaugino zero modes.} 
%IMPORTANT NOTE: THIS IS NOT TRUE, SEE PAGE 414 OF MIRROR SYMMETRY BOOK, FERMION ZERO MODES CAN COME FROM INTERACTION TERMS. 
After integrating out the auxiliary fields, the supersymmetry transformations %\eqref{mirrorsusy} 
(generated by $\hat{Q}_A$) of the physical fields of the mirror theory %\eqref{expandedmirroraction} 
on a Euclidean worldsheet parametrized by complex coordinates ($z,\ov{z}$) are 
 (with $\eps=\sqrt{2}$)\footnote{The fermionic fields $\til{\eta}_c$ and $\til{\kpp}_c$ are related to the fields $\til{\lam}_c$ and $\ov{\til{\lam}}_c$ defined in the previous section via a field redefinition of the form given in \eqref{kppetalam}.}
 % become 
\begin{align} \label{lastsusy}
\delta_{\hat{Q}_A}& \til{A}_{zc} \ =\ \til{\psi}_{zc}  +i\sqrt{2}\del_{z}\til{c}_c &   \delta_{\hat{Q}_A}& \til{\psi}_{zc} \ =\ -\, \partial_z \til{\varphi}_c   \nonumber   \\
\delta_{\hat{Q}_A}& \til{A}_{\zb c} \ =\  \til{\psi}_{\zb c}+i\sqrt{2}\del_{\zb}\til{c}_c &   \delta_{\hat{Q}_A}& \til{\psi}_{\zb c}\ =\  -\, \partial_{\zb} \til{\varphi}_c \nonumber  \\
\delta_{\hat{Q}_A}& \til{\varphi}_c \ = \ 0  &  \delta_{\hat{Q}_A}&  \til{\eta}_c \ = \ 0  \nonumber \\
\delta_{\hat{Q}_A}& \til{\xi}_c \ = \ \til{\eta}_c  &  \delta_{\hat{Q}_A}&   \til{\kpp}_c \ = \ 2 [ \ast \til{F}_{Ac} - 
 \til{e}_c^2  (-i\del_{\til{\varphi_c}}\til{W}+(1/8)\del_{\til{\xi_c}}\ov{\til{W}})]  \nonumber \\
\delta_{\hat{Q}_A}& \theta^c \ = \ 0    &   \delta_{\hat{Q}_A}& \mx_z^{\theta c} \ =\ - 2\sqrt{2} i \, \partial_z \theta^c  \nonumber  \\
\delta_{\hat{Q}_A}& \ov{\theta}^c \ = \  \sqrt{2}(\chi^{\theta c}- \ov{\chi}^{\theta c} )    &   \delta_{\hat{Q}_A}& {\mx}_{\zb} ^{\theta c} \ =\  2\sqrt{2}i \, \partial_{\zb} \theta^c \nonumber \\  
\delta_{\hat{Q}_A}& (\chi^{\theta c} - \ov{\chi}^{\theta c} )  \ =\ 0  &  \delta_{\hat{Q}_A}& [ \, g_{cd}(\chi^{\theta d} + \ov{\chi}^{\theta d} )] \ =\ -\sqrt{2}\, \partial_{\theta^c} \til{W}. \ 
\end{align}
 The bulk physical operators of this theory were studied by Baptista \cite{Baptista2}, where he showed that the bulk chiral ring is given by 
\begin{equation}
\C [ \til{\varphi}^1, \ldots , \til{\varphi}^k , (x^1)^{\pm 1} , \ldots , (x^k)^{\pm 1} ] \ / \ D(\til{W}) \ ,
\label{chiralring}
\end{equation}  
i.e., holomorphic functions of $\til{\varphi}_c$ and $(x^c)^{\pm 1} := \exp{(\mp \theta^c)}$, modulo the ideal $D(\til{W})$, where $D(\til{W})$ is %the ideal 
generated by the derivatives
\begin{equation}
\partial_{\theta^c}\til{W} \ = \ -x^c\, \partial_{x^c} \til{W} \ = \ \sum_{j=1}^{n}\; \til{Q}^c_j\: \Big[\frac{i}{2}\langle \til{\varphi} , v^j \rangle \  -\ e^{-\hat{s}^j} 
\prod_{d=1}^{k}  (x^d)^{v^j_d} \Big]\ . 
\label{5.222}
\end{equation}
In addition, one ought to restrict the bulk physical operators to finite-degree polynomials, since in the equivariant de Rham complex one only considers finite-degree forms and polynomials in the Lie algebra. 
%We shall denote an arbitrary element of the chiral ring \eqref{chiralring} as $W^{mirror}$.

Let us now find the elements of the boundary chiral ring, concentrating first on boundary physical operators which come from the matter multiplets.
Now, at each boundary, we know that $\theta^c$ and $\ov{\theta}^c$ are constants which determine the position of the mirror D0-brane (see \eqref{D0branepositionsuper} and footnote \ref{58}). Then, via \eqref{lastsusy}, we find that $(\chi^{\theta c}- \ov{\chi}^{\theta c} )=0$ on each boundary, and thus $\chi^{\theta c}- \ov{\chi}^{\theta c}$  cannot be an operator in the boundary chiral ring. Even $\theta^c$ and $\ov{\theta}^c$ cannot be elements of the ring since they are not fields along each boundary; rather, they are constants. On the other hand, $\chi^{\theta c}+\ov{\chi}^{\theta c}$ is a nonzero field at each boundary. For the mirror of the non-gauged open A-model, due to the critical point condition $\del_{\theta_c}\til{W}_X=0$ at the boundaries, the $k$ fermionic fields $\chi^{\theta c}+\ov{\chi}^{\theta c}$ are $Q_A$-invariant at each boundary and in fact form the boundary chiral ring  \cite{Hori1}. However, recall from Section 5.4 that we do not have such a  critical point condition, implying that 
$\chi^{\theta c}+\ov{\chi}^{\theta c}$ is \textit{not} $\hat{Q}_A$-invariant at the boundaries, and therefore is not an element of the boundary chiral ring for the mirror of the open gauged A-model.

%Moreover, % which implies, 
%because of the critical point condition ($\partial_{\theta^c}\til{W}=0$), %that 
%$\chi^{\theta c}+\ov{\chi}^{\theta c}$ is %a valid 
%$\hat{Q}_A$-invariant. Hence, $\chi^{\theta c}+\ov{\chi}^{\theta c}$ is a valid $\hat{Q}_A$-invariant operator defined on each boundary. 
%IMPORTANT NOTE: a boundary operator must be a valid FIELD at the boundary, \theta^c does not qualify because it is a constant at the boundary.

%For the non-gauged open A-model \cite{Hori1}, the $k$ fermionic fields $\chi^{\theta c}+\ov{\chi}^{\theta c}$ form the boundary chiral ring, and also provide a basis of the de Rham cohomology of the torus A-branes with topology $T^k$. Hence, for the open gauged A-model, the boundary chiral ring ought to be formed by the $k$ fermionic fields $\chi^{\theta c}+\ov{\chi}^{\theta c}$, as well as the  $k$ bosonic fields $\til{\varphi}_c$, since these provide a basis of the \textit{equivariant} de Rham cohomology of a torus A-brane. 

%Indeed, there are no other valid scalar $\hat{Q}_A$-invariant fields in the vector multiplets which can be elements of the boundary chiral ring. 
Since there are no $\hat{Q}_A$-invariant boundary operators which can be obtained from the matter fields, let us now turn to the vector multiplet fields.
From the boundary condition $ (\ov{\til{\lam}}_{-c} -\til{\lam}_{+c})=0$, we know that $\til{\kpp}_c=0$ at each boundary, so it cannot be such an operator. The operator $\til{\eta}_c$ is nonzero at each boundary,
% but
and is  $\hat{Q}_A$-invariant%for abelian $G$
, but it was not included in \eqref{localboundaryoperator} for abelian $G$ since its anticommuting %character
behaviour implies that it cannot be associated with equivariant cohomology classes, and hence it should not be included as a mirror boundary observable.
%IMPORTANT NOTE I TOOK OUT THE WORDS FOR ABELIAN G BECAUSE EVEN \varphi is not $\hat{Q}_A$-invariant for nonabelian G due to the Q_{BRST} transformation
%and thus, we do not consider it since we are concerned with the mirrors of the $\hat{Q}_A$-invariant observables \eqref{localboundaryoperator} which are valid even in the nonabelian case, and which are associated with equivariant cohomology classes on an equivariant A-brane.
%IMPORTANT NOTE: ADDING ETA WILL SPOIL THE ASSOCIATION WITH EQUIVARIANT FORMS, SINCE ETA ANTICOMMUTES, AND DOESN'T COMMUTE LIKE \varphi. There must be association of the observables with equivariant forms, because that is the definition of Hamiltonian Gromov-Witten invariants, according to Baptista's twisting gauged nonlinear sigma models at least. 
%IMPORTANT NOTE: IT IS POSSIBLE THAT THERE ARE GAUGINO ZERO MODES, BUT FOR THAT WE WILL ASSUME THERE IS A GENERALIZATION LIKE PAGE 414 OF THE MIRROR SYMMons
 On the other hand, $\til{\varphi}_c$ is $\hat{Q}_A$-invariant, and obeys a Neumann boundary condition, and as such is a valid bosonic boundary operator. 
 %However, note that the possible values that $\til{\varphi}_c$ can take along the boundaries are constrained by the critical point condition ($\partial_{\theta^c}\til{W}=0$); see \eqref{5.222}.
Thus, %an arbitrary element of 
the boundary chiral ring at a particular boundary component $\del\Sigma_L$ is given by
\begin{equation}
\boxed{
\C [ \til{\varphi}^1, \ldots , \til{\varphi}^k ] \ / ( D(\til{W})|_{\del \Sigma_L}) \ .
\label{boundarychiralring}}
\end{equation}  
Here, we have taken into account the fact that $\del_{\theta^c}\til{W}$ is a $\hat{Q}_A$-exact function of $\til{\varphi}_c$ at the boundaries. Moreover, we ought to restrict the boundary physical operators to finite-degree polynomials, as we did for the bulk physical operators. 
%Thus, an arbitrary element of 
%the boundary chiral ring is denoted as
%\begin{equation}\label{mirrorboundaryobs}
%W^{mirror}_{\del\Sigma}=
%%\left(\prod_{j=1}^r   \til{\varphi}^{c_j} %\right) 
%\mathcal{F} ( \til{\varphi}^1, \ldots , %\til{\varphi}^k )
%%\left( \prod_{i=1}^s  (\chi^{\theta }+\ov{\chi}^{\theta %})^{c_i}\right)
%\end{equation} 
%%where %$r$ and 
%%$s$ can take any value from 1 to $k$, and 
%where $\mathcal{F}$ is a polynomial of finite degree.
%%while $r$ can take any finite value 
%%(since we must restrict the boundary physical operators to finite-degree polynomials as we did for the bulk physical operators).  

Denoting an arbitrary element of the bulk chiral ring \eqref{chiralring} as $W^{mirror}$, and an arbitrary element of a boundary chiral ring \eqref{boundarychiralring} as $W^{mirror}_{\del\Sigma_L}$,
the most general correlation function of local bulk and boundary observables in the gauged Landau-Ginzburg model is therefore written (for $\Sigma=I\times \R$) as 
\begin{equation}
\boxed{
\int {\mathcal D} (\til{A}, \theta, \til{\varphi},  \til{\xi} , \til{\eta} , \til{\kpp},  \til{\psi} ,  \chi^{\theta},\til{b},\til{c}) \ \ 
e^{-(S_{A}+S_{BRST})} \ \prod_i \: W^{mirror}_i \prod_j \: W^{mirror}_{\del \Sigma_0|j} \prod_k W^{mirror}_{\del \Sigma_{\pi}|k} \ .}
\end{equation}
This is the mirror correlation function which computes \eqref{3.11} for local observables. 

\mbox{}\par\nobreak
\noindent

\acknowledgments
We would like to thank Sushmita Venugopalan for explaining to us many relevant mathematical points regarding the moduli space of open symplectic vortices. We would also like to thank Jo\~ao Manuel Baptista, Sebastian Goette, Suresh Govindarajan, Yuan Luo, %Ted Shifrin, 
 %Constantin Teleman, 
Daniel S. Park, Petr Va\v{s}ko, Junya Yagi, Masaya Yata and Qin Zhao for helpful comments and discussions. This work is supported by NUS Tier 1 FRC Grant R-144-000-316-112.
%\printbibliography

%\bibliography{eqbranesbib}{}

\begin{thebibliography}{10}

%\bibitem{hitchin1987}
%N.J.~Hitchin, A. ~Karlhede, U. ~Lindstr{\"o}m, and M. ~Ro\u{c}ek, {\it Hyper-K{\"a}hler Metrics and Supersymmetry},
%{\em Comm. Math. Phys.} {\bf 108} (4) (1987) 535-589

%Polchinski, J., 1995. Dirichlet branes and Ramond-Ramond charges. Physical Review Letters, 75(26), p.4724
\bibitem{Polchinski}
J.~Polchinski, {\it Dirichlet Branes and Ramond-Ramond Charges}, {\em Physical Review Letters} {\bf 75} (26) (1995) 4724 [\href{http://arxiv.org/abs/hep-th/9510017}{{\tt arXiv:hep-th/9510017}}]

\bibitem{Kontsevich}
M. ~Kontsevich, {\it Homological Algebra of Mirror Symmetry}, in {\em Proceedings of the
International Congress of Mathematicians} (Z¨urich, 1994), Birkh¨auser, Boston,
(1995) 120-139 [\href{http://arxiv.org/abs/alg-geom/9411018}{{\tt arXiv:alg-geom/9411018}}]

\bibitem{Baptista1}
J.M.~Baptista, {\it Twisting Gauged Non-linear Sigma-models}, {\em Journal of High Energy Physics} {\bf 2} (2008) 096 [\href{http://arxiv.org/abs/0707.2786}{{\tt arXiv:0707.2786}}]

\bibitem{mcduff1998introduction}
 D.~McDuff, D.A.~Salamon, {\it Introduction to Symplectic Topology}, Oxford University Press, Oxford U.K. (1998)

\bibitem{libermann1987symplectic}
 P.~Libermann,  and C.-M.~Marle, {\it Symplectic Geometry and Analytical Mechanics}, Springer Science \& Business Media (1987)

\bibitem{yano1957theory}
 K.L.~Yano, {\it The Theory of Lie Derivatives and its Applications}, North-Holland (1957)

\bibitem{hori2003mirror} K.~Hori et al., {\it Mirror Symmetry}, Clay Mathematics Monographs \textbf{1}, AMS (2003)
%@book{hori2003mirror,
 % title={{Mirror Symmetry}},
 % author={Hori, Kentaro and others},
  %volume={1},
  %year={2003},
  %publisher={American Mathematical Soc.}
%}
\bibitem{WittenPhases}
E.~Witten, {\it Phases of N= 2 Theories in Two Dimensions}, {\em Nuclear Physics B} {\bf 403} (1) (1993) 159-222 [\href{http://arxiv.org/abs/hep-th/9301042}{{\tt arXiv:hep-th/9301042}}]

%Witten, Edward. "Phases of N= 2 theories in two dimensions." Nuclear Physics B 403.1 (1993): 159-222.
\bibitem{HoriVafa} K.~Hori, C.~Vafa, {\it Mirror Symmetry, ArXiV High-Energy Physics-Theory e-prints} (February, 2000) [\href{http://arxiv.org/abs/hep-th/0002222}{{\tt arXiv:hep-th/0002222}}]


%Hori, Kentaro, and Cumrun Vafa. "Mirror symmetry." arXiv preprint hep-th/0002222 (2000)

\bibitem{HIV} K.~Hori, A.~Iqbal, C.~Vafa, {\it D-branes and Mirror Symmetry, ArXiV High-Energy Physics-Theory e-prints} (May, 2000) [\href{http://arxiv.org/abs/hep-th/0005247}{{\tt arXiv:hep-th/0005247}}]

\bibitem{Hori1} K.~Hori, {\it Linear Models of Supersymmetric D-branes}, in {\it Symplectic Geometry and Mirror Symmetry} (Seoul, 2000) (K. Fukaya et al., eds.), World Scientific, River Edge, NJ (2001) 111-186 [\href{http://arxiv.org/abs/hep-th/0012179}{{\tt arXiv:hep-th/0012179}}]
%@inproceedings{Hori1,
%      author         = "Hori, Kentaro",
 %         title = "{{Linear Models of Supersymmetric D-branes}}",
  %    booktitle      = "{Symplectic Geometry and Mirror Symmetry (Seoul, 2000) (K. Fukaya et al., eds.)}",
   %   year           = "2001",
%      pages        ={111-186}
 %     publisher=''{{World Scientific, River Edge, NJ}}'',	
  %    eprint         ={hep-th/0012179},
   %   archivePrefix  = "arXiv",
%}
\bibitem{Baptista2}
J.M.~Baptista, {\it The Quantum Equivariant Cohomology of Toric Manifolds through Mirror Symmetry}, {\em Journal of High Energy Physics} {\bf 4} (2009) 017 [\href{http://arxiv.org/abs/0806.2091}{{\tt arXiv:0806.2091}}]


\bibitem{Govinda}
S.~Govindarajan, T.~Jayaraman, T.~Sarkar, {\it On D-branes from Gauged Linear Sigma Models}, {\em Nuclear Physics B} {\bf 593} (2001) 155-182 [\href{http://arxiv.org/abs/hep-th/0007075}{{\tt arXiv:hep-th/0007075}}]

\bibitem{Kapustin}
A.~Kapustin, K.~Setter, K.~Vyas, {\it Surface Operators in Four-dimensional Topological Gauge Theory and Langlands Duality, ArXiV High-Energy Physics-Theory e-prints} (February, 2010) [\href{http://arxiv.org/abs/1002.0385}{{\tt arXiv:1002.0385}}] 

\bibitem{Vyas}
K.~Vyas, {\it  Topics in Topological and Holomorphic Quantum Field Theory} (Doctoral dissertation, California Institute of Technology, 2010)

\bibitem{Setter}
K.~Setter, {\it  Topological Quantum Field Theory and the Geometric Langlands Correspondence} (Doctoral dissertation, California Institute of Technology, 2013)

\bibitem{GSW2}
M.B.~Green, J.H.~Schwarz, E.~Witten, {\it Superstring Theory: Volume 2, Loop Amplitudes, Anomalies and Phenomenology}, Cambridge University Press (2012)

\bibitem{nappi}
A.~Abouelsaood, C. G.~Callan, C. R.~Nappi, S. A.~Yost, {\it Open Strings In Background
Gauge Fields, Nuclear Physics B} \textbf{280} (1987) 599

\bibitem{aspinwall}
P.S.~Aspinwall, {\it D-branes on Calabi-Yau Manifolds,} in {\it Progress in String Theory: TASI 2003} (J. Maldacena, ed.), World Scientific (2005) 1-152 [\href{http://arxiv.org/abs/hep-th/0403166}{{\tt arXiv:hep-th/0403166}}] 

\bibitem{BGV}
N.~Berline, E.~Getzler, M.~Vergne, {\it Heat Kernels and Dirac Operators}, Springer Science and Business Media (1992)
%Berline, N., Getzler, E., & Vergne, M. (1992). Heat kernels and Dirac operators. Springer Science & Business Media.
\bibitem{szabo}
R.J.~Szabo, {\it Equivariant Cohomology and Localization of Path Integrals}, Springer Science and Business Media (2003)

%\bibitem{HoriPhases} 
%M.~Herbst, K.~Hori, D.~Page,  {\it Phases of N= 2 Theories in 1+ 1 Dimensions with Boundary, ArXiV High-Energy Physics-Theory e-prints} (March, 2008) [\href{http://arxiv.org/abs/0803.2045}{{\tt arXiv:0803.2045}}]

\bibitem{cautiskamnitzer} 
S.~Cautis, J~Kamnitzer, {\it Knot Homology via Derived Categories of Coherent Sheaves, I: The $\mathfrak{sl}(2)$-case,} {\em Duke Mathematical Journal} {\bf 142} (3) (2008) 511-588 [\href{http://arxiv.org/abs/math/0701194}{{\tt arXiv:math/0701194}}]

%\bibitem{Albertsson}
%C.~Albertsson, U.~Lindstr\"om, M.~Zabzine, {\it $\mathcal{N}= 1$ Supersymmetric Sigma Model with Boundaries, I.,} {\em Communications in Mathematical Physics} \textbf{233} (3) (2003) 403-421 [\href{http://arxiv.org/abs/hep-th/0111161}{{\tt arXiv:hep-th/0111161}}]

\bibitem{Zaslow}
B.~Fang, C.-C.M.~Liu, D.~Treumann, E.~Zaslow, {\it T-duality and homological mirror symmetry for toric varieties}, {\it Advances in Mathematics} {\bf 229} (3) (2012) 1873 - 1911
[\href{http://arxiv.org/abs/0811.1228}{{\tt arXiv:0811.1228}}]

\bibitem{Futaki}
M.~Futaki, K.~Ueda, {\it Tropical Coamoeba and Torus-Equivariant Homological Mirror Symmetry for the Projective Space}, {\it Communications in Mathematical Physics} {\bf 332} (1) (2014) 53-87
[\href{http://arxiv.org/abs/1001.4858}{{\tt arXiv:1001.4858}}]


\bibitem{aspinwall2009Dmirror} P.~Aspinwall et al., {\it Dirichlet Branes and Mirror Symmetry}, Clay Mathematics Monographs \textbf{4}, AMS (2009)
%Aspinwall, Paul. Dirichlet branes and mirror symmetry. Vol. 4. American Mathematical Soc., 2009.
%Szabo RJ. Equivariant cohomology and localization of path integrals. Springer Science & Business Media; 2003 Jul 1.
\bibitem{ChoHong} 
C.H.~Cho, H.~Hong,  {\it Finite Group Actions on Lagrangian Floer Theory, ArXiV Mathematics e-prints} (July, 2013) [\href{http://arxiv.org/abs/1307.4573}{{\tt arXiv:1307.4573}}]
%Cho, C. H., & Hong, H. (2013). Finite group actions on Lagrangian Floer theory. arXiv preprint arXiv:1307.4573.

\bibitem{Hori2}
K.~Hori, {\it Mirror Symmetry and Quantum Geometry}, in {\it Proceedings of ICM 2002, Vol. III} (T. Li, ed.), Higher Education Press, Beijing (2002) 431-443 [\href{http://arxiv.org/abs/hep-th/0207068}{{\tt arXiv:hep-th/0207068}}]

\bibitem{Kapustinli}
A.~Kapustin, Y.~Li, {\it D-branes in Landau-Ginzburg Models and Algebraic Geometry}, {\em Journal of High Energy Physics} {\bf 12} (2003) 005 [\href{http://arxiv.org/abs/hep-th/0210296}{{\tt arXiv:hep-th/0210296}}]


\bibitem{Cieliebak}
K.~Cieliebak, A.R.~Gaio, D.A.~Salamon, {\it J-holomorphic Curves, Moment Maps, and Invariants of Hamiltonian Group Actions}, {\em International Mathematics Research Notices} {\bf 16} (2000) 831-832 [\href{https://arxiv.org/abs/math/9909122}{{\tt arXiv:math/9909122}}]
%IMPORTANT NOTE: IT SEEMS THAT PRIOR TO CIELIEBAK2, Hamiltonian gromov-witten invariants were only defined for flat target spaces, this includes CIELIEBAK where the case with boundaries was introduced. But note that sushmita said that the general Hamiltonian Gromov-Witten invariants were discussed by Mundet, though his abstract only talks about $G=S^1$. Baptista's introduction only refers to Cieliebak2 when introducing Hamiltonian Gromov-Witten invariants, not Mundet. But later he cites [10] and [11] in his paper.

\bibitem{figueroa}
J.M.~Figueroa-O'Farrill and S. ~Stanciu, {\it Equivariant Cohomology and Gauged Bosonic
Sigma-models, ArXiV High-Energy Physics-Theory e-prints} (July, 1994) [\href{http://arxiv.org/abs/hep-th/9407149}{{\tt arXiv:hep-th/9407149}}]


\bibitem{Mundet} 
I.~Mundet i Riera, {\it Hamiltonian Gromov-Witten Invariants}, {\em Topology} {\bf 42} (2003)
525-553 [\href{https://arxiv.org/abs/math/0002121}{{\tt arXiv:math/0002121}}]

\bibitem{Cieliebak2}
K.~Cieliebak, A.R.~Gaio, I.~Mundet i Riera, D.A.~Salamon,{\em The Symplectic Vortex Equations and Invariants of Hamiltonian Group Actions}, {\em Journal of Symplectic Geometry} {\bf 1} (3) (2002) 543-646
[\href{https://arxiv.org/abs/math/0111176}{{\tt arXiv:math/0111176}}]

\bibitem{Gaio}
A.R.~Gaio, D.A.~Salamon, {\it Gromov-Witten Invariants of Symplectic Quotients
and Adiabatic Limits}, {\em Journal of Symplectic Geometry} {\bf 3} (2005) 55-159 [\href{https://arxiv.org/abs/math/0106157}{{\tt arXiv:math/0106157}}]

\bibitem{Wang}
D.~Wang, G.~Xu, {\it Compactness in the adiabatic limit of disk vortices}, {\it Mathematische Zeitschrift} (2016) [\href{https://arxiv.org/abs/1505.05945}{{\tt arXiv:1505.05945}}]


%\bibitem{LuoTan}
%Y.~Luo, M.C.~Tan, {\it A Topological Chern-Simons Sigma Model and New Invariants of Three-Manifolds}, {\em Journal of High Energy Physics} {\bf 2} (2014) 067 [\href{http://arxiv.org/abs/1302.3227}{{\tt arXiv:1302.3227}}]
\bibitem{Xu}
G.~Xu, {\it The Moduli Space of Twisted Holomorphic Maps with Lagrangian Boundary Condition: Compactness}, {\em Advances in Mathematics} {\bf 242} (2013) 1-49 [\href{https://arxiv.org/abs/1202.4096}{{\tt arXiv:1202.4096}}]

\bibitem{Georgieva}
P.V.~Georgieva, {\it  Orientability of Moduli Spaces and Open Gromov-Witten Invariants} (Doctoral dissertation, Stanford University,  2011)
%Kapustin, Anton, and Yi Li. "D-branes in Landau-Ginzburg models and algebraic geometry." Journal of High Energy Physics 2003.12 (2004): 005.
%Georgieva, Penka Vasileva, et al. Orientability of moduli spaces and open Gromov-Witten invariants. Stanford University, 2011.


\bibitem{Xu2}
G.~Xu, {\it Gauged Hamiltonian Floer Homology I: Definition of the Floer Homology Groups}, {\em Transactions of the American Mathematical Society} {\bf 368} (4) (2016) 2967-3015 [\href{https://arxiv.org/abs/1312.6923}{{\tt arXiv:1312.6923}}]
\bibitem{mcduff2}
D.~McDuff, D.A.~Salamon, {\it J-holomorphic Curves and Symplectic Topology}, American Mathematical Society (2012)
%\bibitem{Witten:1}
%E.~Witten, {\it Five-brane effective action in M-theory},
 % {\em Journal of Geometry and Physics} {\bf 22 (2)} (1997) 103-133
%  [\href{http://arxiv.org/abs/1006.0977}{{\tt arXiv:1006.0977}}].
  
%\bibitem{Dimofte:33}
%T.~Dimofte, D.~Gaiotto and S.~Gukov, {\it 3-manifolds and 3d indices},
 % [\href{http://xxx.lanl.gov/abs/1112.5179}{{\tt arXiv:1112.5179}}].
%\bibitem{Dimofte:GT}T.~Dimofte, D.~Gaiotto and S.~Gukov, {\it Gauge Theories Labelled by  Three-manifolds}, {\em Commun. Math. Phys.}  {\bf 325} (2014) 367-419   [\href{http://xxx.lanl.gov/abs/1108.4389}{{\tt arXiv:1108.4389}}].    \bibitem{Alday:LC}L.~F. Alday, D.~Gaiotto and Y.~Tachikawa, {\it {Liouville Correlation Functions from Four-dimensional   Gauge Theories}}, {\em Lett. Math. Phys.}  {\bf 91} (2010) 167 [\href{http://xxx.lanl.gov/abs/0906.3219}{{\tt  arXiv:0906.3219}}].    \bibitem{Yagi:3T}J.~Yagi, {\it 3d TQFT from 6d SCFT}, {\em JHEP} {\bf 08} (2013) 017  [\href{http://arxiv.org/abs/1305.0291}{{\tt arXiv:1305.0291}}].
\end{thebibliography}
%\bibliographystyle{JHEP}

\end{document}